\documentclass[prl,twocolumn,superscriptaddress,amsfonts,amssymb]{revtex4-2}
\usepackage{graphicx}
\usepackage{dcolumn}
\usepackage{xcolor}
\usepackage{color} 
\usepackage{bm}
\usepackage{times}
\usepackage{color,soul}

\begin{document}

\newcommand{\de }{$^{\circ}$}
\newcommand{\IPZ}[1]{{\textcolor{blue} {\bf{IPZ: #1}}}}
\newcommand{\JL}[1]{\textcolor{red}{{\bf{JL: #1 }}}} 
\newcommand{\jvb}[1]{\textcolor{purple}{{\bf{jvb: #1 }}}}

\title{Engineering interfacial quantum states and electronic landscapes by molecular nanoarchitectures}

\author{Ignacio Piquero-Zulaica}
\email {ipiquerozulaica@gmail.com}
\affiliation{Centro de F\'{\i}sica de Materiales CSIC/UPV-EHU-Materials Physics Center, Manuel Lardizabal 5, E-20018 San Sebasti\'an, Spain}
\affiliation{Donostia International Physics Center, Paseo Manuel Lardizabal 4, E-20018 Donostia-San Sebasti\'an, Spain}
\affiliation{Physics Department E20, Technical University of Munich, 85748 Garching, Germany}

\author{Jorge Lobo-Checa}
\email{jorge.lobo@csic.es}
\affiliation{Instituto de Nanociencia y Materiales de Arag\'on (INMA), CSIC-Universidad de Zaragoza, Zaragoza 50009, Spain}
\affiliation {Departamento de F\'{\i}sica de la Materia Condensada, Universidad de Zaragoza, E-50009 Zaragoza, Spain}

\author{Zakaria M. Abd El-Fattah}
\affiliation{Physics Department, Faculty of Science, Al-Azhar University, Nasr City E-11884 Cairo, Egypt}

\author{J. Enrique Ortega}
\affiliation{Centro de F\'{\i}sica de Materiales CSIC/UPV-EHU-Materials Physics Center, Manuel Lardizabal 5, E-20018 San Sebasti\'an, Spain}
\affiliation{Donostia International Physics Center, Paseo Manuel Lardizabal 4, E-20018 Donostia-San Sebasti\'an, Spain}
\affiliation{Departamento de F\'{\i}sica Aplicada I, Universidad del Pa\'{\i}s Vasco UPV/EHU, E-20018 Donostia-San Sebasti\'an, Spain} 

\author{Florian Klappenberger}
\affiliation{Physics Department E20, Technical University of Munich, 85748 Garching, Germany}

\author{Willi Auw{\"a}rter}
\affiliation{Physics Department E20, Technical University of Munich, 85748 Garching, Germany}

\author{Johannes V. Barth}
\email{jvb@tum.de}
\affiliation{Physics Department E20, Technical University of Munich, 85748 Garching, Germany}

\date{\today}

\begin{abstract}
Surfaces are at the frontier of every known solid. They provide versatile supports for functional nanostructures and mediate essential physicochemical processes. Being intimately related with two-dimensional materials, interfaces and atomically thin films often feature distinct electronic states with respect to the bulk, which are key for many relevant properties, such as catalytic activity, interfacial charge-transfer, or crystal growth mechanisms. Of particular interest is reducing the surface electrons’ dimensionality and spread with atomic precision, to induce novel quantum properties via lateral scattering and confinement. Both atomic manipulation and supramolecular principles provide access to custom-designed molecular superlattices, which tailor the surface electronic landscape and influence fundamental chemical and physical properties at the nanoscale. Herein, we review the confinement of surface state electrons focusing on their interaction with molecule-based scaffolds created by molecular manipulation and self-assembly protocols under ultrahigh vacuum conditions. Starting from the quasi-free two-dimensional electron gas present at the (111)-terminated surface planes of noble metals, we illustrate the enhanced molecule-based structural complexity and versatility compared to simple atoms. We survey low-dimensional confining structures in the form of artificial lattices, molecular nanogratings or quantum dot arrays, which are constructed upon appropriate choice of their building constituents. Whenever the realized (metal-)organic networks exhibit long-range order, modified surface band structures with characteristic features emerge, revealing intriguing physical properties, such as discretization, quantum coupling or energy and effective mass renormalization. Such collective electronic states can be additionally modified by positioning guest species at the voids of open nanoarchitectures. The necessary insight into these scattering potential landscapes can be obtained through semiempirical models, which brings closer the prospect of total control over surface electron confinement and quantum engineering.\\
\end{abstract}


\maketitle
\tableofcontents

\section{0. Introduction}

Reaching atomistic control and understanding of matter has been a long-nourished desire of mankind, receiving widespread attention once the foundations of atomic and molecular sciences were established. Enormous efforts were dedicated to develop fabrication, characterization, and manipulation procedures providing access to novel expressions of materials, useful electronic, photonic and magnetic properties, or collective atomic states with ever increasing precision and complexity~\cite{Klitzing1980, Shechtman1984, Amano1986, Heeger1988, Binasch1989, Davis1995, Kane2005, Bloch2008, Muhlbauer2009}. 
Multiple instrumental developments were crucial to grant direct insights into the nature and behaviour of atomic and molecular species down to the \AA{}ngstr{\"o}m regime, whereby the imaging capabilities by electron, field-ion, scanning tunneling microscopy (EM, FIM, STM) techniques proved to be extremely valuable. 
Striking direct visualization of crystal surface atomic lattices and adsorbed species  could be firstly achieved by FIM and later on with the more versatile STM, generating not only revolutionary scientific insights but also inspiration and opportunities for generations of researchers~\cite{Muller1965,Binnig1987}.
Using cryogenic STM the paradigmatic advance of addressing and positioning individual atoms became a reality~\cite{Stroscio1991} and was immediately recognized as an emblematic achievement in nanoscale science.\\
\indent
Likewise, insights into the essential electronic structure of materials became accessible by means of local scanning tunneling spectroscopy (STS) and space-averaging high-resolution spectroscopies, notably including angle-resolved photoemission (ARPES) {~\cite{Damascelli2003}. The combination of these powerful tools, ideally complemented with theoretical modeling is an asset to fully characterize electronic properties and their implications.~\cite{Lobo2009, Gambardella2003, Klappenberger2014, Galeotti2020, Yin2021}.\\
\indent
Bottom-up construction procedures were developed to design a wide variety of nanosystems amenable to scanning probe and space-averaging scrutiny. In particular, since the turn of this century, supramolecular chemistry principles have been increasingly employed to create low-dimensional functional nanostructures at well-defined interfaces~\cite{Barth2000, Beton2003, Yokoyama2001, Stepanow2004}. They are readily achieved by selecting appropriate molecular species with defined endgroups favoring self-assembly into purely organic or metal-organic nanoarchitectures~\cite{Barth2005, Kuhnle2009, Kudernac2009, NianLin2016, Goronzy2018, Xing2019}. In this context, supramolecular nanoporous networks were utilized as host lattices for the preferential trapping of guest species or as confining arrays for molecular motion~\cite{Beton2003, Stohr2007, Barth2010, Pivetta2013, Nowakowska2016, NianLin2015, Nowakowska2015, DeFeyter2016}. Moreover, they provide significant potential for incorporation of molecular switches in nanoelectronic circuits~\cite{Repp2019} or to host ligands suitable for gas sensing applications~\cite{Kern2015, Stepanow2016, Ecija2018}. In addition, metal-organic networks are appealing for the exploration of novel magnetic properties~\cite{Umbach2012, Stepanow2013, Gao2020}, catalytic effects~\cite{Kern2015, Stepanow2016, Hotger2019}, oxidation states~\cite{Gottfried2012}, and exotic tessellation patterns (\textit{e.g.}, quasicrystals or Archimedean tilings)~\cite{Urgel2016, Zhang2018, NianLin2017}, with the added prospect of exhibiting tunable quantum phases and topological properties~\cite{Zhang2016, Kumar2018, Gao2019, Sun2018, Dong2016, Wang2013, Hernandez2021, Jiang2021}.\\
\indent
The extensive activities devoted to the exploration of quantum confinement and quasiparticle scattering at nanostructures on metal surfaces promoted the advance of condensed matter research. This progress is revisited here taking the surface states of coinage metals as a canonical playground, whereby the design of appropriate nanoarchitectures tailors the electronic structure and topology of the outermost layers. We also introduce the generally relevant aspects underpinning the scattering and confinement phenomena of two-dimensional electron gases (2DEGs) existing at appropriate surfaces listing prominent cases where exquisite control over such 2DEGs is exerted. 
We cover the ground-breaking quantum corral and resonator structures built by atomic manipulation and move into more complex molecule-based arrangements and supramolecular self-assemblies. The latter enable an upscaling of the quantum properties due to the mesoscopic templating, which is a requirement for (opto-)electronic devices manufacturing. The electronic structure of these systems is explored mainly by two complementary experimental techniques: STM/STS at the atomic level, and ARPES whenever large and homogenous domains exist. Semiempirical methods are recurrently used to simulate the molecular scattering potential landscape responsible for such 2DEG modification. \\
\indent
Moreover, we discuss electronic structure modifications by artificial lattices yielding exotic properties such as Dirac cones or flat bands~\cite{Gomes2012, Slot2017}, topological edge states~\cite{Kempkes2019a, Swart2020}, fractal behaviour~\cite{Kempkes2019} or Penrose tiling quasicrystals~\cite{Collins2017}, which are at the foundation of emerging fields such as twistronics ~\cite{Cao2018a, Cao2018b} and mirror concepts established for Bose-Einstein condensates ~\cite{Ketterle2002, Zapf2016, Louie2009, Polini2013, Leykam2018, Greiner2002}. Such emerging properties often exist for natural materials, such as graphene, but are here induced, and importantly, tuned by the proper nanostructuring of the featureless dispersion of the 2DEGs at metal surfaces.\\
\begin{figure} [b!]
\begin{center}
  \includegraphics[width=0.5\textwidth,clip]{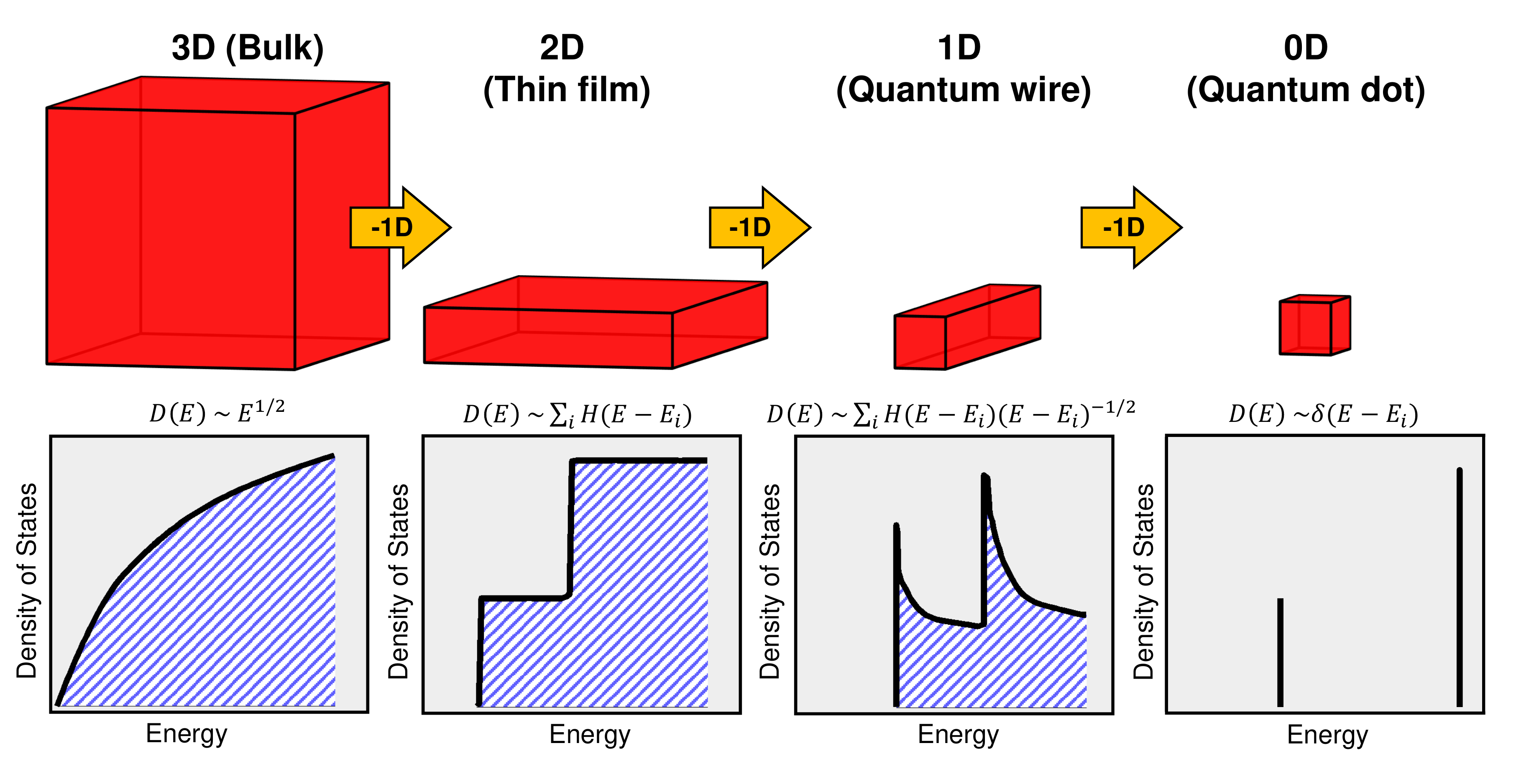}
 \end{center}
\vspace*{-5mm}
\caption[] {Modification of the DOS distribution in a free-electron gas upon dimensionality reduction. The 3D smooth relation becomes progressively quantized as system dimensions are decreased.
}
\label{figure1}
\end{figure}
%
\section{I. Harnessing surface states to explore quantum mechanical phenomena}
\indent
Well-defined surfaces offer a plethora of intriguing properties and have been studied extensively for more than a century. With the development of ultrahigh-vacuum technology, systematic investigations of the surface atomic arrangement and electronic structure became possible~\cite{Duke2003}. In later years, atomistically clean metal surfaces served as versatile platforms for the construction of precisely defined nanoscale architectures, the dimension of which falls below the wavelength of Fermi electrons whence the quantum regime is entered. Electron scattering with confining attributes on surfaces emerge at natural defects or reconstructions, as well as nanostructures realized by STM tip manipulation ~\cite{Gross2004, Pennec2007, Barth2009, Lobo2009, Bartels2010, NianLin2013b, Stohr2016, Martinez2019}.\\ 
\indent
As illustrated in Figure~\ref{figure1}, lowering the spatial dimensions has a conspicuous effect on the density of states (DOS) of a free-electron gas system~\cite{Ashcroft1976}. The smooth three dimensional (3D) DOS distribution (E$^{1/2}$) evolves into energy-independent step-like staircases  in two-dimensions (2D) and reaches fully discretized ``delta function'' [$\delta$($E$-$E_i$)] lineshapes in zero-dimensional (0D) systems, while in one-dimensional (1D) structures a convolution of the previous distributions is found.\\
\indent
The relevance of DOS modification upon dimensional reduction in electronic systems can be gauged from its extended application in the semiconductor industry that relies on ultrathin, quasi-2D electron systems for building electronic and computational devices~\cite{Brennan1999, Yoffe2001, Floris2013, Vanwees1988, Kanisawa2001}. Even future quantum computation realization may come in reach using tailored low-dimensional superconducting structures ~\cite{Sato2017, Nayak2008, Choi2019}. 
Because high-technology developments were stimulated by the fabrication of quantum wire (QW) and quantum dot (QD) architectures, understanding the plethora of fundamental properties of low-dimensional systems is important to explore new applications fields~\cite{Barth2005, Harrison2005, Vanderwiel2002, Hanson2007, Pekola2013, Wharam1988, Ohnishi1998, Tarucha1996}. Particularly, material engineering, molecular electronics, and quantum computation require the construction and exploration of artificial coupled or hybrid quantum materials~\cite{Kagan2015, Kagan2016, Broome2018, Leon2020, Walkup2020, Tokura2017, Keimer2017}. Possible candidates may emerge from 2D materials that exhibit  long-range magnetic order ~\cite{Huang2017, Gong2017}, flat bands or low-temperature superconductivity~\cite{Cao2018a, Cao2018b, Andrei2020, Liljeroth2020, Crommie2021}, providing an outstanding platform to obtain distinct quantum states of matter.\\
\begin{figure}
\begin{center}
  \includegraphics[width=0.5\textwidth,clip]{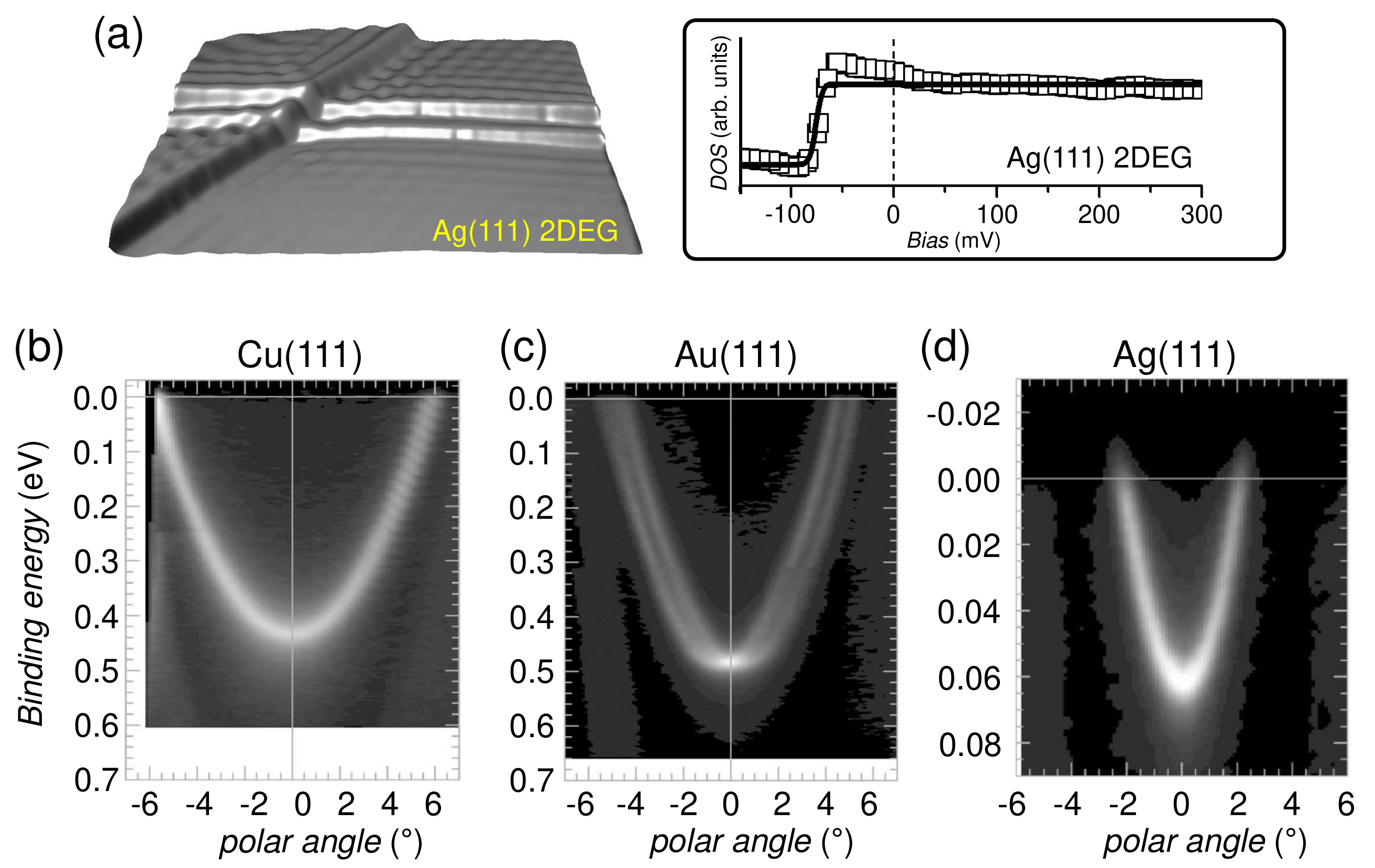}
 \end{center}
\vspace*{-5 mm}
\caption[]{DOS and band dispersions from 2DEGs on (111)-terminated noble metal surfaces. ({\bf a}) (\textit{Left}) Differential conductance (d$I$/d$V$) map overlaid with a topography image of a local region of Ag(111) hosting few monoatomic steps. The electron scattering  at the step edges produces quasiparticle interference standing wave patterns. (\textit{Right}) The conductance spectrum of the Ag surface state reveals the typical step-like DOS distribution of a 2DEG (adapted from~\cite{Pennec2007}). ({\bf b}-{\bf d}) Parabolic dispersion relations of copper, silver and gold fcc(111) Shockley states that are characteristic of 2D free-electron gases (adapted from~\cite{Reinert2001}). The data for Au(111) shows the spin-orbit splitting. 
}
 \label{figureNobleMetal}
\end{figure}
\subsection{A. Surface states providing canonical 2DEGs}
The model systems chosen in this review to address the scattering and confinement properties of diverse artificial or self-assembled nanostructures are the well-known Shockley states present at the (111)-terminated coinage metal surfaces (Cu, Ag and Au)}~\cite{Shockley1939}. These 2D surface states exist in the  $\Gamma$-L projected bulk band gap~\cite{Kevan1987,Paniago1995b}. The pertaining electrons, principally residing at the outermost surface layers, behave as a quasi-free 2DEG, propagating unrestricted (with plane wave characteristics) in a regular and pristine surface plane~\cite{Reinert2001, Reinert2005, Malterre2007, Hirofumi2014, Tamai2013} [Fig.~\ref{figureNobleMetal}(a)]. Their quasi-free character is mirrored in the parabolic dispersion [see Fig.~\ref{figureNobleMetal}(b-d)]~\cite{Reinert2001,Paniago1995} that follows the relation $E(k_{\parallel})$=$E_0$+$\frac{\hbar^{2}k_{\parallel}^{2}}{2m^*}$, where $E_0$ is the fundamental binding energy of the surface state ($i.e.$, the band minimum), $m^*$ the electron's effective mass, and $k_{\parallel}$ the wave-vector parallel to the surface (cf. Table \ref{TableSurfaceStates}). Energy and wave-vector can be directly probed using ARPES that can nowadays access the spin-orbit splitting, encountered for instance for Au(111)~\cite{LaShell1996,Tamai2013} and even their proposed topological properties in combination with two-photon photoemission (2PPE)~\cite{Yan2015}.\\
\indent
The isotropic dispersion of these Shockley states results in step-like DOS distributions characteristic for 2D systems (Figure~\ref{figure1}). The relating electronic characteristics can be directly accessed at the atomic (local) scale by scanning tunneling spectroscopy (STS) using differential conductance spectra ($dI/dV$s)~\cite{Tersoff1985} [see Fig.~\ref{figureNobleMetal}(a)]. The observed local density of states (LDOS) correlates with the spatially integrated ARPES signals since the STS onset of the Shockley states matches the band bottom (fundamental) energies of these low-dimensional systems [cf. Fig.~\ref{figureNobleMetal}(a) and (d)].\\ 
\begin{table}
	\begin{center}
		\begin{tabular}{|c|c|c|c|c|}
                        \hline 
			 & $E_0$~[meV] & $m^*$/$m_e$ & $k_F$ [\AA$^{-1}$] & ${\lambda}_F$ [\AA{}]  \\
			\hline 
			Ag(111) & 63$\pm$1 & 0.397& 0.080 & 78.5 \\
			\hline
			Au(111) & 487$\pm$1 & 0.255& 0.167/0.192 & 32.7/37.6 \\
			\hline
			Cu(111) & 435$\pm$1 & 0.412& 0.215 &29.2\\
			\hline
		\end{tabular}
	\end{center}
	\caption[]{Parameters from the parabolic fits of the fcc(111) Shockley state dispersions shown in Fig.~\ref{figureNobleMetal}. $E_0$ corresponds to the fundamental binding energy (energy minimum), $m^*$ to the effective mass, $k_F$ to the Fermi momentum and ${\lambda}_F$ to the Fermi wavelength. The electron wave vector $k_{\parallel}$ is directly linked to the kinetic energy of the photoelectron ($E_{kin}$) and the polar angle ($\theta$)  through $k_{\parallel}\approx0.512\sqrt{E_{kin}} \sin(\theta)$. Values taken from ~\cite{Reinert2001}.
}
\label{TableSurfaceStates}
\end{table}
\begin{figure}
\begin{center}
  \includegraphics[width=0.5\textwidth,clip]{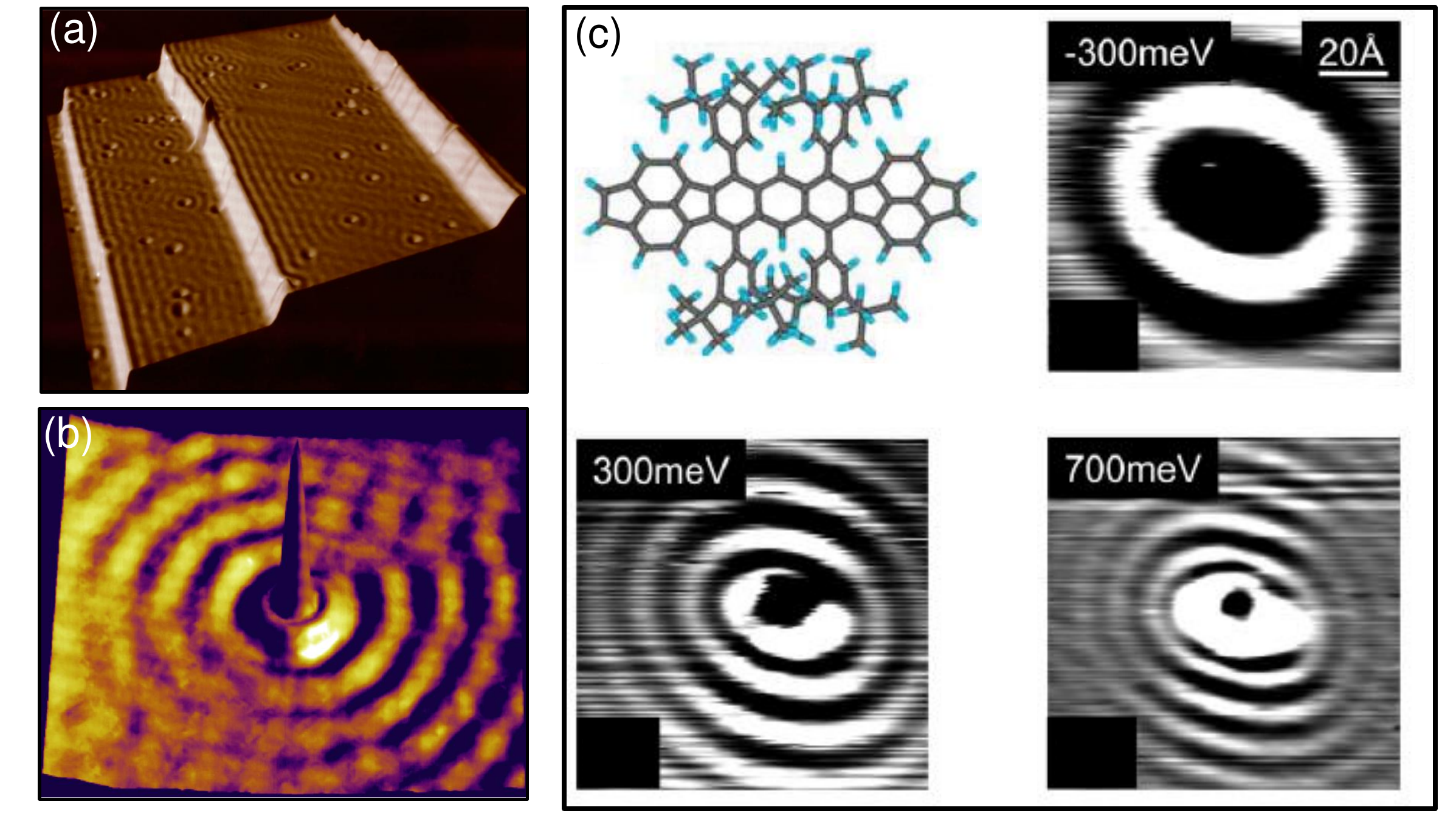}
 \end{center}
\vspace*{-5 mm}
\caption[]{2DEG scattering with atomic and molecular adsorbates. ({\bf a}) Three monoatomic steps and about 50 point defects are visible on the Cu(111) surface. Spatial quasiparticle interference (QPI) patterns from the 2DEG scattering are clearly visible (STM topographic image 50 $\times$ 50 nm$^2$, adapted from~\cite{Crommie1993}). ({\bf b}) Single Fe adatom on Cu(111). The concentric rings surrounding the Fe site correspond to a standing wave pattern due to the scattering of surface state electrons (STM topographic image with 13 $\times$ 13 nm$^2$, adapted from~\cite{Crommie1993a}). ({\bf c}) Molecules can efficiently scatter the surface electrons, as evidenced from the interference patterns of a so-called lander molecule recorded at different energies (adapted from~\cite{Gross2004}). 
}
 \label{figureScattering}
\end{figure}
\indent
Metal surface 2DEGs reside at the outermost atomic planes and are therefore very sensitive to impurities in the form of adsorbed atoms or molecules~\cite{Memmel1996,Kulawik2005},  structural defects such as atomic steps or vacancy islands~\cite{Crommie1993, Hasegawa1993, Berndt1999, Rodary2007, Avouris1994, Burgi1998} and thin overlayers comprising \textit{e.g.}, rare gases or alkali metals~\cite{Forster2004}. In STM topographic data these effects become manifest and can be directly resolved when applying small bias voltages at low temperatures, whereby Fermi level electrons dominate the tunneling current  [Fig.~\ref{figureScattering}(a, b)]. The pertaining electron density oscillations exhibit standing wave character and are related to Friedel oscillations generated by point-like charge impurities in bulk systems~\cite{Friedel1958}, that can be rather pronounced, depending on material characteristics~\cite{Sprunger1997}.
Such standing wave oscillations are recognized as quasiparticle interference (QPI) patterns, originating from the constructive interference of electron waves after scattering at surface obstacles. Importantly, when Fourier transforming the QPI patterns taken at different bias voltages, the full energy and momentum space  (band dispersion) of these electrons can be accessed. Similar to single atoms, molecular adsorbates equally scatter interfacial 2DEGs and produce QPI patterns. In large molecular species typically a repulsive potential  is generated by the charge distribution through the molecular backbone. Remarkably, as shown for isolated species, not all molecular moieties necessarily scatter in the same way~\cite{Gross2004} [cf. Fig.~\ref{figureScattering}(c)]. It also must be mentioned that surface-state-mediated interactions exist, which can influence the spatial distribution of adsorbates at surfaces~\cite{Repp2000,Knorr2002,Ternes2010}. Under suitable conditions these adatoms are placed at positions reflecting an oscillatory interaction arising due the surface state electrons, with ${\lambda_F}/2$ periodicity and 1/$r^2$ decay, whereby $r$ turns out to be the distance between neighboring adatoms~\cite{Weiss2012, Ding2007, Knorr2002, Repp2000, Silly2004}. These indirect interactions are also the driving force for the creation of superlattices of individual adsorbed atoms when the adatom concentration, the sample temperature and the adatom diffusion barrier are in subtle balance [Fig.~\ref{figureCeSuperlattice}(a)].\\\\
\begin{figure} 
\begin{center}
  \includegraphics[width=0.5\textwidth,clip]{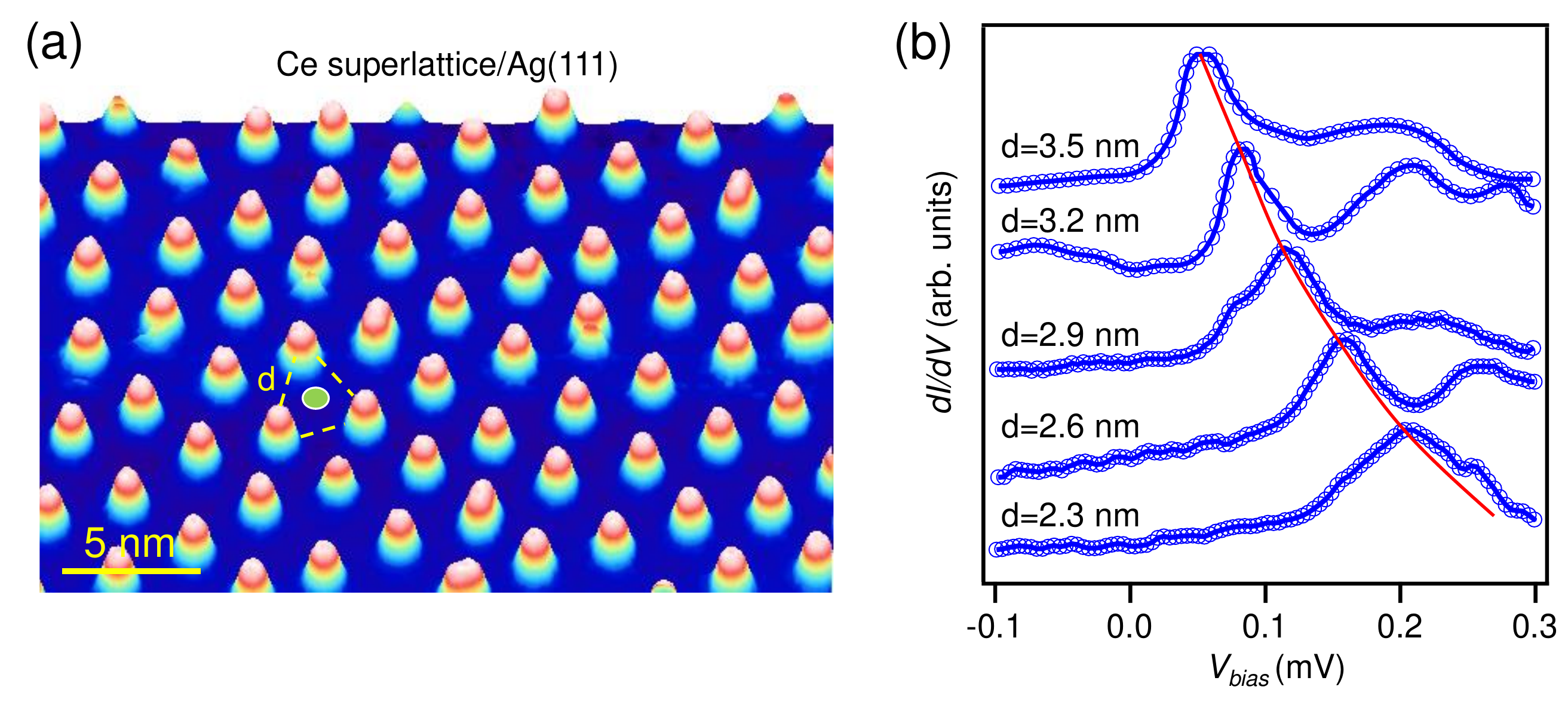}
 \end{center}
\vspace*{-7 mm}
\caption[] {Formation of surface-state-mediated superlattices: Self-organized Ce adatom array on Ag(111). ({\bf a}) Constant-current STM image of 0.01 ML Ce on Ag(111) displaying an average distance of 3.2 nm, obtained at T=3.9 K. ({\bf b}) Set of spectra taken in the center of a triangle formed by 3 Ce adatoms with next-neighbor distance  $d$. The adatom-adatom distance decreases with increasing Ce coverage, whereby a quadratic energy upshift of the first resonance level is observed with shrinking QD size (adapted from~\cite{Ternes2010}). 
}
 \label{figureCeSuperlattice}
\end{figure}
\subsection{B. Principles of 2DEG confinement by atomic steps, corrals and superlattices}
\begin{figure}
\begin{center}
  \includegraphics[width=0.5\textwidth,clip]{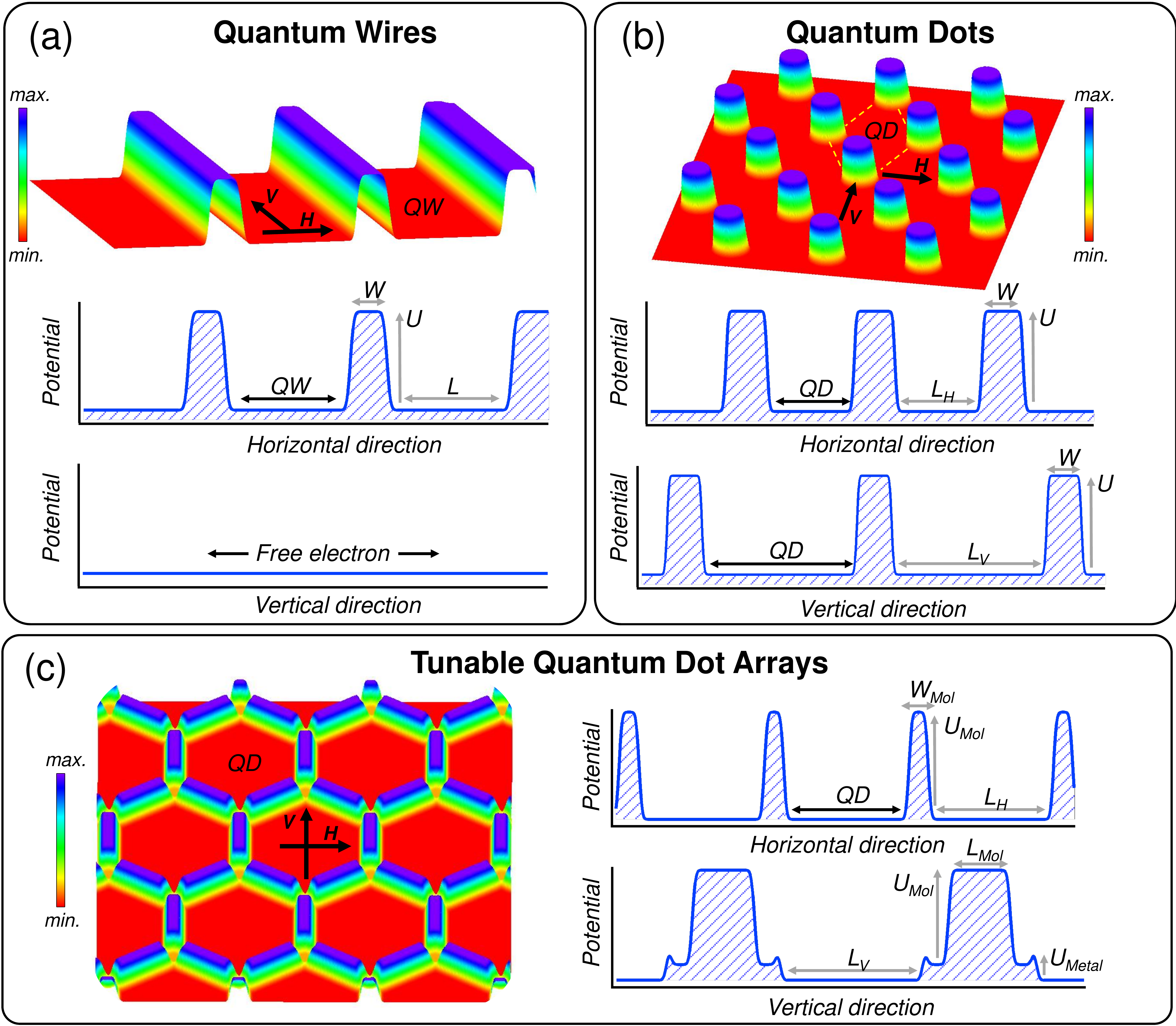}
 \end{center}
\vspace*{-5 mm}
\caption[]{Engineering freely propagating surface state electrons via scattering superlattices. ({\bf a}) One-dimensional electron grating made of scattering barriers separated by a distance $L$ producing QWs. While surface electrons are confined in the horizontal direction, they freely propagate in the vertical direction. ({\bf b}) Building an array of QDs via the precise positioning of 2DEG scattering point-like (0D) barriers. This hexagonal superlattice produces an array of QDs, where surface electrons are mainly scattered along the indicated directions, while being less restricted in other orientations. As a result, an efficient coupling to adjacent QDs is obtained. ({\bf c}) Well-encapsulated QD array. Shown is a situation with anisotropic barriers (compare both profiles) affecting both 2DEG confinement and QD intercoupling, as will be discussed for metal-organic nanoporous networks. 
}
 \label{figurePotentials}
\end{figure}

\indent 
To systematically engineer and confine freely propagating surface electrons it is neccessary to reduce the system's dimensionality to one or zero dimensions by building QW or QD structures. Since adsorbed atoms and vacancies~\cite{Crommie1993, Crommie1995, Kliewer2001, Berndt1998, Berndt1999, Jensen2005}, molecules~\cite{Gross2004, Wackerlin2014} and  step edges~\cite{Burgi1998,Hasegawa1993,Avouris1994} efficiently scatter the surface electrons, three approaches were established to produce well-defined 2DEG confining nanostructures that are schematically reproduced in  Fig. \ref{figurePotentials}:  (a) step arrays and nanogratings, (b) discrete artificial nanostructures via atomic or molecular manipulation, and (c) self-assembled atomic superlattices and molecular nanoporous networks. 1D-QWs  in the form of step arrays and nanogratings confine the surface electrons  in the 'horizontal' direction by periodically spaced scattering barriers, whereas, except for an energy shift, a free-electron behavior is kept perpendicular to it (in the 'vertical' direction) [Fig. \ref{figurePotentials}(a)]. More restrictive are QDs that confine the electrons in all directions. In particular, QD superlattices emerge whenever the confinement occurs on the surface from point-like (0D) scattering units periodically distributed in the nanoscale regime [Fig. \ref{figurePotentials}(b)]. These scattering superlattices are often realized artificially via atomic or molecular manipulation, but also by self-assembly. A very promising engineering approach to obtain coupled QDs exploits the spontaneous self-assembly of simple organic building blocks on surfaces. Thus molecular nanoporous networks can be fabricated being purely organic or comprising metal-organic coordination motifs [Fig. \ref{figurePotentials}(c)]. Note that the embedded metal centers and the molecules may scatter surface electrons differently, whence an interesting heterogenous scattering potential landscape for the 2DEG confinement and QD intercoupling emerges, the consequence of which will be explained later in section IV. Until the end of this section, we summarize relevant cases of QW and QD systems built from inorganic materials.\\
\begin{figure*}
\begin{center}
  \includegraphics[width=0.8\textwidth,clip]{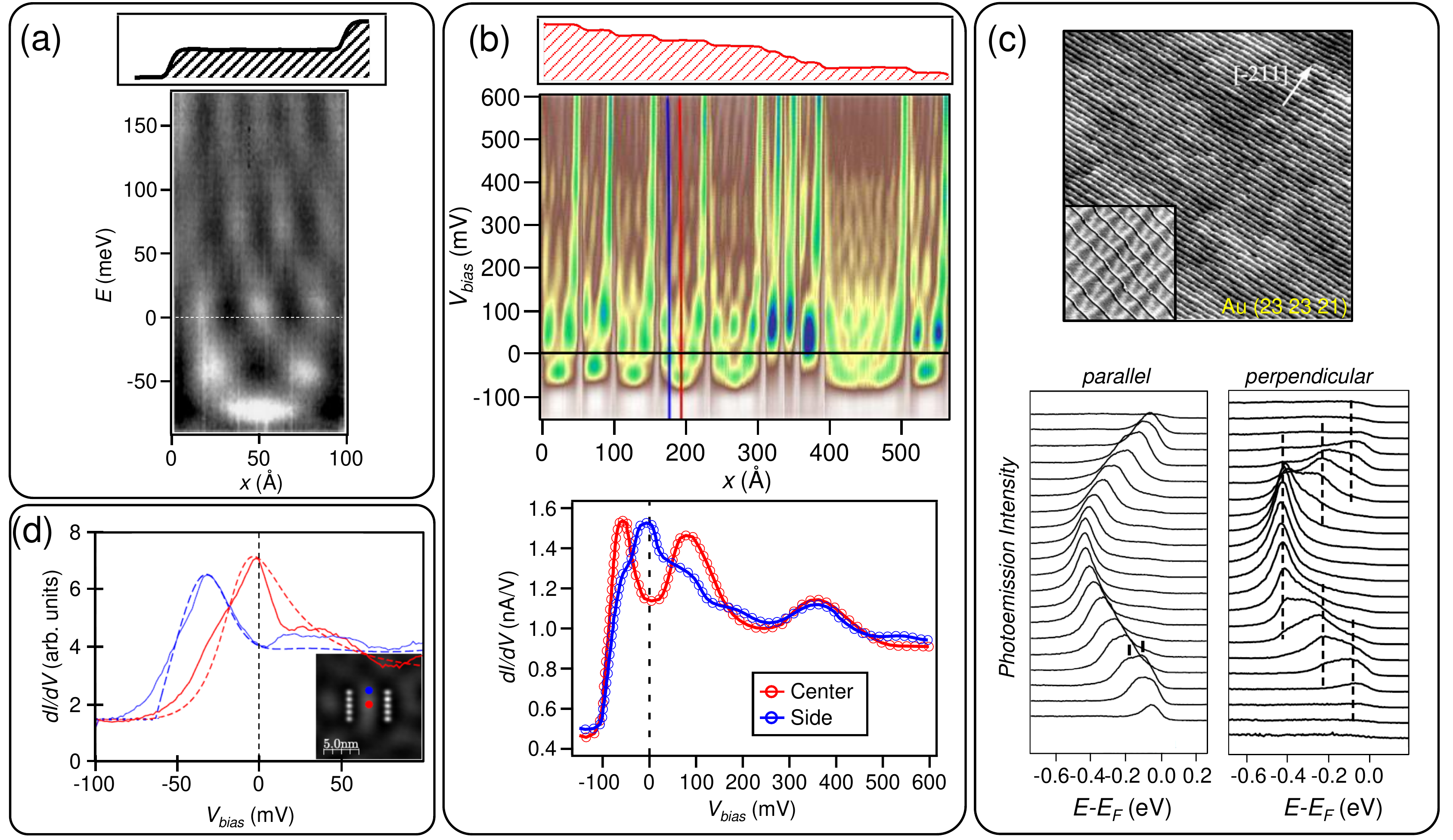}
 \end{center}
\vspace*{-5 mm}
\caption[]{1D confinement of 2DEGs via step arrays and atomic linear resonators. ({\bf a}) Constant current line scan over 104~\AA{} wide asymmetric resonator formed by two parallel steps (adapted from~\cite{Burgi1998}); the lower panel depicts the pertaining LDOS across a terrace which is characterized by a series of QWS resonances as a function of the energy resulting from the 1D electronic confinement. ({\bf b}) $dI/dV$ linespectra across an inhomogeneous step array formed by a 5 ML Ag film grown on a curved Au(111) substrate. The topographic profile is shown on the panel top. The bottom panel displays two selected tunneling spectra exhibiting different quantum well resonances at the center and the border of the same terrace, respectively (adapted from~\cite{Ortega2018}). ({\bf c}) STM topography of a regularly stepped Au(23 23 21) vicinal surface. Below, the ARPES EDCs show the dispersion parallel (\textit{left}) and perpendicular to the step direction (\textit{right}). Clearly, a free-electron-like behavior (parabolic dispersion) appears parallel to the steps, whereas quasi-1D, weakly dispersing peaks (indicated by dashed lines) emerge perpendicular to the steps (adapted from~\cite{Mugarza2003}). ({\bf d}) Two strings of Co adatoms generated by tip manipulation on Ag(111) provide a resonator element. The diagram shows the experimental tunneling spectroscopy (full line) and calculated LDOS signature at the center (red color) and end of the resonator (blue color) (adapted from~\cite{Fernandez2016}).
}
 \label{figureVicinal}
\end{figure*}
\indent
1D electronic states spontaneously evolve on noble metal surfaces whenever parallel steps define single terraces [Fig. \ref{figureVicinal}(a)]. Alternatively, they can be created artificially by arranging adspecies into parallel atomic wires [Fig. \ref{figureVicinal}(d)]. The atomic steps efficiently back-scatter the surface electrons [see Fig.\ref{figureNobleMetal} (a)], producing confined states --commonly known as lateral quantum well states (QWS)-- when the average terrace width falls below the surface state electron coherence length~\cite{Burgi1998,Hirofumi2014} [Fig. \ref{figureVicinal}(a)]. These QWS can be modeled using infinite confining potential barriers at steps~\cite{Burgi1998,Ortega2013}, so that electron motion becomes restricted within the terrace with freedom to travel parallel to the steps~\cite{Ortega2018} (analogous to Fabry-P\'{e}rot resonators known from optics). As observed in Fig. \ref{figureVicinal}(b), the energy of such QWS can be tuned by altering the terrace width or the atomic-row spacing~\cite{Negulyaev2008, Fernandez2016}.\\
\indent
This scenario becomes even more intriguing whenever a periodic (regular) step array is created on the surface, going beyond two parallel steps or a set of terraces with different widths. Such regular vicinal crystals~\cite{Mugarza2006,Baumberger2002,Baumberger2004,Kawai2004} provide step superlattices giving rise to extended quasi-1D band structures that can be probed by non-local techniques such as ARPES. The electron confinement within the terraces leads to gapped and anisotropic dispersions accomplished by upward shifts of the 2DEG fundamental energy scaling inversely with the terrace size. Importantly, some electronic coupling between adjacent terraces emerges, which translates into finite step potential barriers that have to be taken into account~\cite{Mugarza2006,Ortega2013,Mitsuoka2011}. The physical nature of the modulated electronic bands can be captured by the 1D-Kronig Penney model~\cite{Mugarza2006}. Herein, the steps are considered repulsive, in the form of square-shaped finite potential barriers of magnitude $U_0$ $\times$ $b$, where $U_0$ corresponds to the height of the barrier and $b$ to its width~\cite{Ortega2011}.
A beautiful example of the resulting quasi-1D band structure is shown in Fig. \ref{figureVicinal} (c) for the case of Au(23 23 21) featuring a miscut angle of $\alpha = 2.4$\de{} from the (111) plane~\cite{Mugarza2003}. By analysing the energy distribution curves (EDCs), weakly dispersive peaks separated by energy gaps are observed in the direction perpendicular to the steps, whereas the parallel direction exhibits the expected (unconfined) parabolic dispersion~\cite{Mugarza2003}.\\
\begin{figure*} 
\begin{center}
  \includegraphics[width=0.8\textwidth,clip]{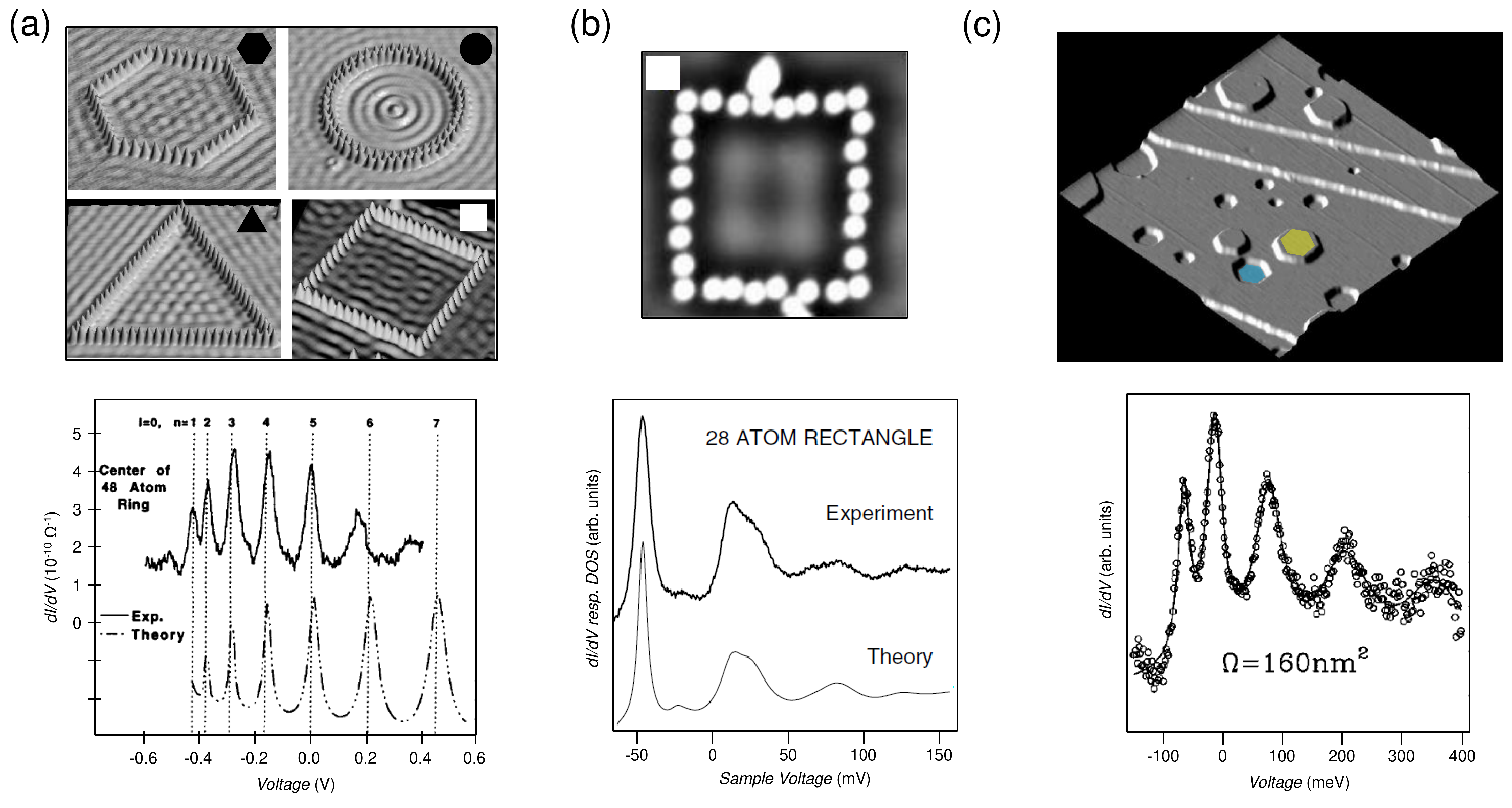}
 \end{center}
\vspace*{-5 mm}
\caption[]{2DEG confinement via artificial quantum corrals and spontaneously formed nanoislands. ({\bf a}) Hexagonal, circular, triangular and square-shaped quantum corrals constructed by tip manipulation of Fe atoms on a Cu(111) surface (adapted from~\cite{Crommie1993a}). A characteristic STS spectrum at the center of one of these structures (48 Fe atom ring) is shown below (adapted from~\cite{Crommie1995a}). 
({\bf b}) Rectangular structure made up of 28 Mn atoms and its corresponding STS spectrum and calculated LDOS at the corral center (adapted from~\cite{Kliewer2001}). ({\bf c}) STM constant current topographic image of Ag adatoms (yellow) and vacancy (blue) islands on Ag(111) (adapted from~\cite{Berndt1999}) accompanied with a typical STS acquired at a hexagonal  island center (adapted from~\cite{Berndt1998}). In all cases, STS data indicate a broadened 0D LDOS distribution (\textit{cf}. Fig.\ref{figure1}).
}
 \label{figureCorrals}
\end{figure*}
\indent
Nanostructures enclosing specific areas provide striking electron confining properties (Fig.~\ref{figureCorrals}). Atomic arrangements constructed by tip manipulation methods were introduced as ``quantum corrals'' ~\cite{Crommie1993a, Heller1994, Crommie1995, Crommie1995a, Kliewer2001} that exhibit distinct standing wave patterns, \textit{i.e.}, QPI phenomena clearly revealing the 2DEG response to the confining geometries (\textit{e.g.}, hexagons, circles, squares or triangles)~\cite{Braun2002,Lagoute2005,Kumagai2008,Kliewer2001}. For some cases, the confining properties are almost element-independent, whence different adatom species (\textit{e.g.} Fe and Mn) result in similar STS signatures [Fig.~\ref{figureCorrals}(a), (b)]~\cite{Crampin1996}. Moreover, at the corral center the DOS may feature a multipeak lineshape, stemming from confined resonances drifting apart as the enclosed area is reduced. 
The adatoms defining the quantum corrals can be modelled as an impenetrable boundary, \textit{i.e.}, a hard-wall barrier and thus, the eigenstates and the eigenenergies can be determined using a particle in a 2D box model of the corresponding geometry~\cite{Heller1994, Kumagai2008,Kumagai2009,Tatsumi2018}. However, experimentally finite peakwidths (more evident as the energy separates from the 2DEG onset) are detected instead of delta functions expected for perfectly isolated quantum boxes. Thus, the nanostructure walls require the consideration of leaky barriers that reflect only a fraction of the incident amplitude~\cite{Heller1994, Crommie1995a, Kliewer2001, Harbury1996, Rahachou2004, Fiete2003}. The existence of inelastic absorptive channels has been ascribed to the coupling of surface electrons with bulk states~\cite{Crommie1995, Fiete2003, Kliewer2001}. To avoid specific assumptions or empirical input data, \textit{ab initio} calculations based on density functional theory and a multiple scattering approach employing the Korringa-Kohn-Rostocker Green's function method were used to describe the electronic structure of quantum corrals~\cite{Niebergall2006}. Alternatively, the elastic scattering theory approach (finite-height potential barriers) turned out to be a simpler model capable of reliably reproducing the experimental findings~\cite{Harbury1996, Rahachou2004}. Furthermore, similar LDOS distributions to those encountered for quantum corrals were  found in nanovacancies or adatom islands created by gentle sputtering on Ag(111), \textit{i.e.}, without involving direct atom manipulation~\cite{Berndt1998, Berndt1999, Jensen2005, Crampin2005} [see Fig.~\ref{figureCorrals}(c)].\\
\indent
The surface-state-mediated superlattices  also exhibit 0D partial confinement of the 2DEG. An example is the case of Ce adatoms on Ag(111), where the atoms arrange in perfectly ordered triangular QD units, the size of which can be altered with careful coverage control~\cite{Silly2004,Ternes2004,Ternes2010} [Fig.~\ref{figureCeSuperlattice}(a)]. With increasing Ce concentration, the position of the first confined state shifts quadratically to higher energies with the average distance $d$, as shown in Fig.~\ref{figureCeSuperlattice}(b). Note that the Ag surface state becomes depopulated in these cases and appears in the unoccupied regime ($E$ $>$ $E_F$), where resonances in the form of LDOS peaks appear. Moreover, not only ordered superlattices can confine the 2DEG, but also disordered ones. In this regard, researchers succeeded in generating and probing wavefunctions of a disordered 2DEG multifractal scaling characteristics, using the mixed surface alloy BiPb/Ag(111)~\cite{Jack2020}.
\\
\indent
The quantum corrals can also be regarded as artificial atoms, in view of their discrete electronic levels. In recent work their bonding properties were explored using atomic force microscopy. The measured interactions are very weak, but covalent attraction to metallic and repulsions for CO terminated tips could be discriminated.~\cite{Stilp2021}.\\
\begin{figure} 
\begin{center}
  \includegraphics[width=0.5\textwidth,clip]{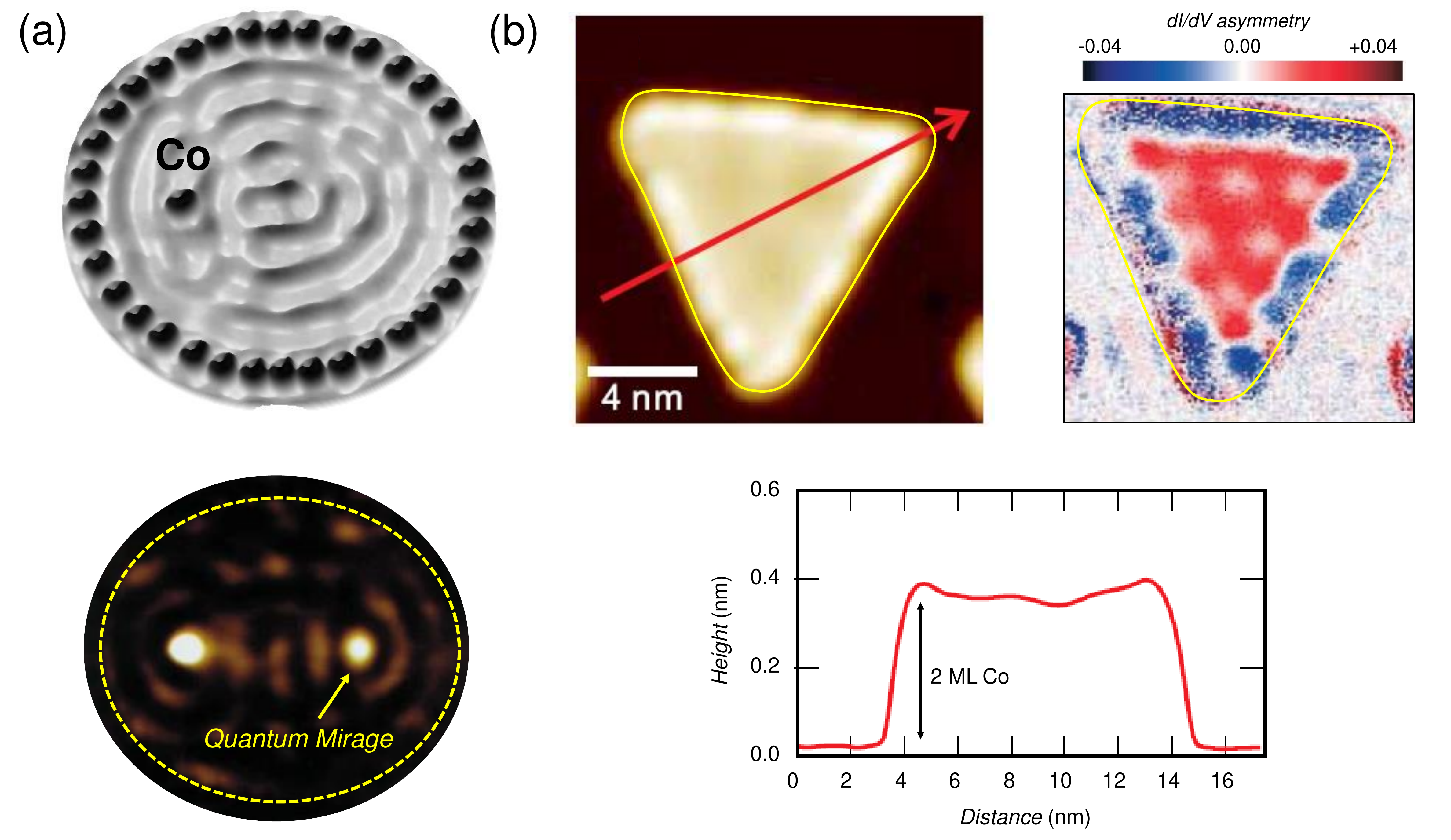}
 \end{center}
\vspace*{-5 mm}
\caption[]{ Magnetism effects induced in quantum corrals and thin films. ({\bf a}) Elliptical electron resonator with a Co atom at the left focus. Associated $dI/dV$ maps showing the Kondo effect projected to the empty right focus (adapted from~\cite{Manoharan2000}). ({\bf b}) Constant-current STM image of a 2 ML Co triangular island on Cu(111). Its line profile along the red arrow evidences the two atomic layer height. Experimental $dI/dV$-asymmetry map measured near the Fermi level at V=0.03 V and B=-1.1 T showing a clear magnetic contrast between the center and edge of the island (adapted from~\cite{Oka2010,Hirofumi2014}).
}
 \label{figureMagnetism}
\end{figure}
\indent 
It is possible to go beyond the ``standard'' electronic confinement control by constructing specially configured quantum corrals enclosing magnetic impurities. For instance, the so-called quantum mirage gives rise to remarkable nanomagnetic effects by positioning a single Co adatom at one focus of an ellipsoidal quantum corral~\cite{Manoharan2000,Stepanyuk2005} [Fig.~\ref{figureMagnetism}(a)]. The Co atom Kondo signature becomes replicated (with reduced intensity) at the second, pristine (Co-free) focal point. Therefore, the corral acts as a quantum mechanical resonator and the 2D confined Cu surface state electrons form the medium through which the  magnetic moment of the Co adatom is projected to the opposite focus~\cite{Stepanyuk2005,Rossi2006,Morr2019}. Recently the Kondo-free mirage effect was also demonstrated using Fe and Ag single atoms~\cite{Li2020}. Overall, the quantum size effects in corrals or nanoislands promote important effects, such as the observation of quantum-guided diffusion and adatom self-assembly, the control of statistical fluctuations, the tunability of the Kondo temperatures or the build up of atomic logic gates~\cite{Li2020b}.\\ 
\indent
Likewise, adsorbed magnetic clusters provide additional magnetic effects triggered by confinement. For example, triangular 2 ML thick Co islands on Cu(111) not only display efficient confinement of the Cu 2DEG, but also spin-dependent quantum interference effects [Fig.~\ref{figureMagnetism}(b)]. Differential tunneling conductance ($dI/dV$) asymmetry maps show strong position-dependent signals within the Co islands. A rim state localized at the edges (in blue) originates from minority $d$ states, while the modulation pattern at the inner part (in red) shows the opposite (majority) spin character, which is ascribed to the electron quantum confinement of the free electron-like $sp$ surface state (\textit{i.e.}, the Cu 2DEG)~\cite{Oka2010, Pietzsch2006, Hirofumi2014}. A similar spin-confinement interplay effect has been demonstrated from the analysis of QPI phenomena on Bi(110), where spin-orbit interactions entail a marked surface state splitting, emphasizing that their spin character cannot be neglected~\cite{Pascual2004}.\\
\indent
Over recent years, 2DEG scattering and confinement were examined for a series of prototypical topological insulator materials exhibiting insulating bulk and  topologically protected spin-split surface states~\cite{Zhang2020b, Sobota2020}. Due to their intriguing quantum nature, remarkable effects in striking contrast to the previous scenarios described occur in these materials, as for example the transmission through steps~\cite{Yazdani2010}. Moreover, an analysis of confinement patterns for the scattering of topological states from Ag impurities and step edges on the Bi$_2$Te$_3$(111) surface indicated a complete suppression of electron backscattering, due to the manifestation of time-reversal symmetry~\cite{Zhang2009}. However, more recent studies on such substrates featuring triangularly shaped Bi quantum corrals atop displayed quasi-bound states, indicative of topological surface state confinement effects~\cite{Chen2019}. Based on these findings, selection rules were suggested, governed by the shape and spin texture of the surface state constant energy contour upon the strong hexagonal warping in the substrate~\cite{Zhang2020b, Sobota2020}. In more recent work, even the dual (weak-crystalline) topological material  Bi$_2$TeI was spectroscopically investigated, displaying distinct 2D Dirac surface states behaving differently in the vicinity of atomic steps and susceptible to mirror symmetry breaking~\cite{Avraham2020}. Related to these findings is the demonstration of spin-polarized midgap states at odd atomic step edges of stoichiometrically controlled PbSnSe alloys featuring topological crystalline insulator surfaces~\cite{Sessi2016}. Interestingly, also the Shockley states of Au(111) and other noble metals can be interpreted as topologically derived surface states of a topological insulator~\cite{Yan2015}.\\ 
\indent
A remarkable effect in the noble metal surface states is that they can be made superconducting by the proximity effect~\cite{Lee2012,Wei2019}. Using a hybrid material platform consisting of a thin Au(111) film on superconducting vanadium substrate and patterned EuS, to additionally magnetize the surface state electrons via exchange coupling, the signatures of so-called Majorana zero modes could be accessed~\cite{Manna2020}.\\
\indent
In the following we focus on the electronic confinement and scattering properties induced by molecular nanoarchitectures fabricated on (111)-terminated metals featuring Shockley-type 2DEGs. We discuss prominent examples that exhibit control over the surface state confinement using molecular arrays published in the last two decades. We start with finite confining structures obtained by tip manipulation protocols affording CO artificial lattices (\textit{e.g.}, hexagonal, Lieb and fractals) and vacancy arrangements in closed-packed porphyrin-based molecular layers. Then we move on to extended molecular structures obtained by supramolecular and metallo-supramolecular self-assembly protocols. These arrange into nanogratings, polymorphic structures (\textit{e.g.}, kagom\'{e}, rectangle and rhombic lattices), triangular and fractal structures, and tunable hexagonal nanoporous networks. These nanoporous networks are very versatile and provide independent tunability of all confining parameters: pore size, interpore separation and scattering potential barriers. Importantly, the addition of coordinating metal centers induces prominent effects on the 2DEG confinement, to the extent that  they can strongly modify (renormalize) the pristine surface state and open interaction channels thereby enhancing the QD coupling. Finally, we elaborate on the placement of guest species into open supramolecular grid structures, providing an extra route for self-assembly and altering surface electronic features.\\
\section{II. Defining quantum states by molecular manipulation}
\indent
Molecular adsorbates, apart from scattering interfacial 2DEGs similarly to single atoms, they provide increased complexity and tunability to the generated nanoarchitectures, which can be advantageous for confinement control. Notably, CO atomic positioning on Cu(111) was extensively used to manipulate electron waves in closed geometries~\cite{Moon2008, Manoharan2009}. This construction scheme is highly interesting since prescient theoretical considerations suggested that, under an appropriate external periodic potential of hexagonal symmetry, massless Dirac fermions could be bestowed on 2DEGs near the corners of the supercell Brillouin zones~\cite{Park2009}. Using a superimposed nanofabricated quantum well lattice on gallium arsenide, this was validated at the mesoscopic scale ~\cite{Singha2011}, and at the nanometer scale by creating an artificial honeycomb ``molecular graphene'' sheet of CO molecules on the Cu(111) surface~\cite{Gomes2012}. The latter findings evidenced that atomically precise DOS engineering provides access to new physics, and many further remarkable results have been obtained over the following years~\cite{Yan2019, Khajetoorians2019, Yan2019a}.\\
\indent
At  the current stage, two manipulation protocols exist for the design of scattering barriers and geometries to modify the surface 2DEG: i) the unit by unit construction of artificial lattices over pristine surface regions, and ii) the removal of adspecies from an extended and periodic molecular layer to generate vacancies that expose well-defined patches of the substrate.
\begin{figure} 
\begin{center}
  \includegraphics[width=0.5\textwidth,clip]{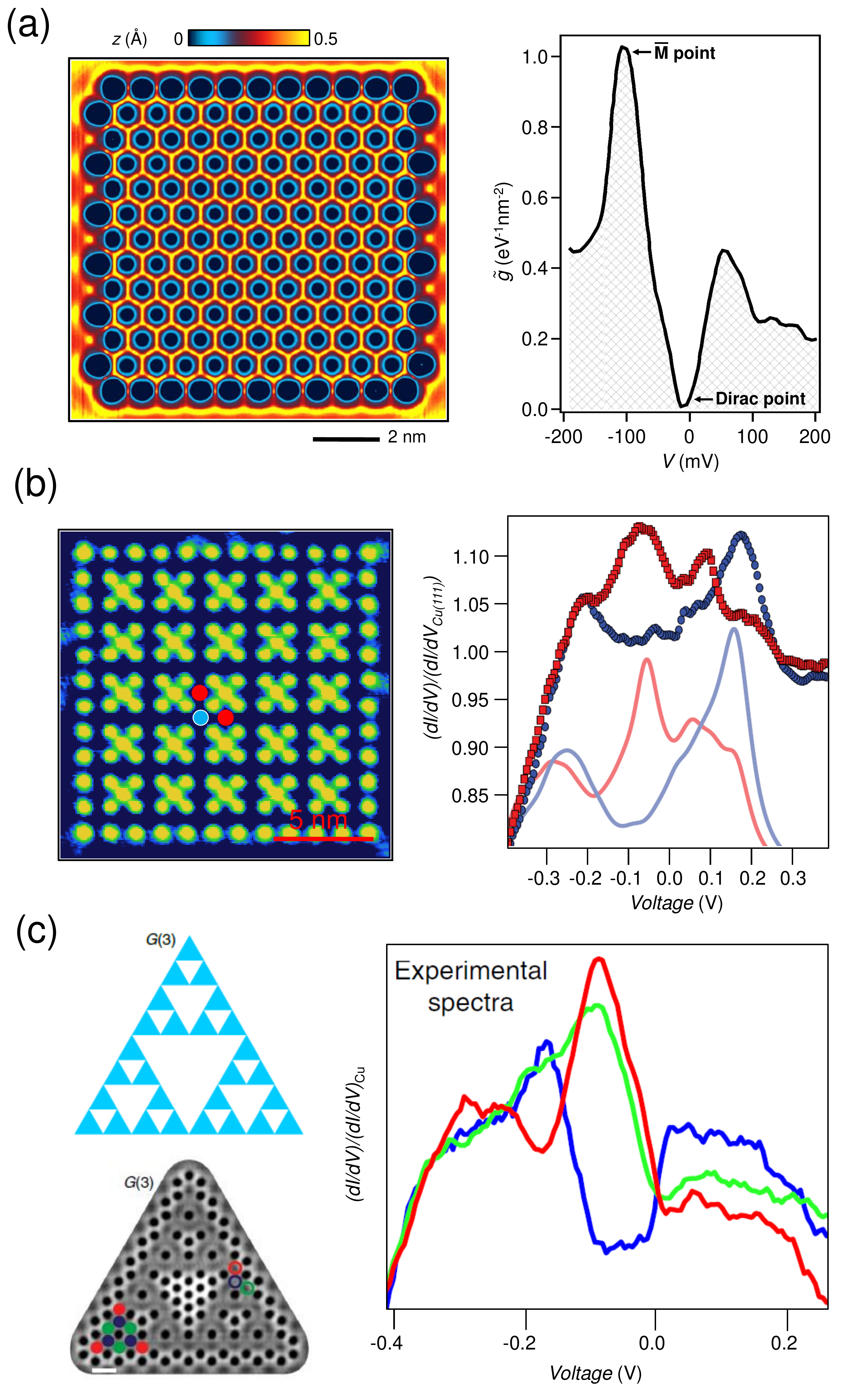}
 \end{center}
\vspace*{-5 mm}
\caption[]{2D lattices created by serial positioning of CO on Cu(111). ({\bf a}) Artificial graphene lattice generated by manipulation of 149 CO molecules and corresponding spatially averaged STS that features a V-shaped curve characteristic of a Dirac-like band structure (adapted from~\cite{Gomes2012}). ({\bf b}) Artificial Lieb lattice (dark regions) imposed on Cu(111) 2DEG (CO molecules marked as green circles). The normalized STS curves acquired above corner (light blue dot) and edge (red dot) sites are shown on the right (adapted from~\cite{Slot2017}). ({\bf c}) Model and constant-current STM image of the G(3) Sierpi\'{n}ski triangles lattice generated by CO molecules on Cu(111). The atomic sites of one G(1) building block (bottom left) are indicated. The normalized differential conductance spectra acquired above the positions of red, blue and green open circles are reproduced with the same color coding (adapted from~\cite{Kempkes2019}).
}
 \label{figureMolec}
\end{figure}
\subsection{A. Artificial 2D lattices}
\indent
A pioneering study, demonstrating quantum state control by an artificial molecular lattice was realized in a triangular lattice designated ``molecular graphene'' constructed by tip manipulation of CO molecules on a Cu(111) surface~\cite{Gomes2012}. Indeed, the resulting band structure features Dirac cones analogous to graphene, which is derived from the modified 2DEG [see Figure~\ref{figureMolec}(a)]. Each CO unit acts as a repulsive barrier to the surface electrons that get confined within the honeycomb grooves left between the molecules and create a strong depletion of states (conductance dip in STS), characteristic of  Dirac-like band structures. Interestingly, closely related  artificial graphene layers at the mesoscale were recently reported by growing regular C$_{60}$ monolayer superlattices on Cu(111)~\cite{Yue2020}. Moreover, by arranging coronene molecules on Cu(111), artificial electronic kagome lattices~\cite{Pavel2021} and graphene nanoribbons were fabricated, the latter of which display new states at the zigzag edges~\cite{Wang2014,Hla2021}.\\ 
\indent
Using the established CO manipulation strategy, a series of lattice configurations were systematically explored, \textit{e.g.}, Lieb lattices~\cite{Slot2017}, Sierpi\'{n}ski triangle fractals~\cite{Kempkes2019}, Penrose tiling quasicrystals~\cite{Collins2017} and topological state hosting nanostructures~\cite{Kempkes2019a, Swart2020}. The case of a Lieb lattice was addressed by adding extra CO molecules at the center of alternating squares of a CO square lattice [visualized as yellow X's in Figure~\ref{figureMolec}(b)]. The repulsive CO potentials deplete the surface electrons in their surrounding and yield a band structure composed of Dirac cones at the Brillouin zone edges and an extended flat band at the Dirac energy. This electronic structure is reflected in the position dependent STS: at the corner sites (blue point) the spectrum exhibits two peaks (lowest and highest energies of the dispersive bands) with a strong attenuation in between that marks the Dirac point, which converts into a maximum in the spectra at the edge sites (red points), evidencing the existence of the flat band. Note that such Lieb lattices were recently predicted to exist similarly as a molecular covalent organic framework (COF)~\cite{Jiang2020}.\\ 
\indent
In essence, the desired band structures are generated by manipulating the molecules into the inverted artificial 2D lattices. In this way, fractal arrangements as the one shown in Figure~\ref{figureMolec}(c) could be created~\cite{Kempkes2019}. The realized Sierpi\'{n}ski triangles bestow non-integer or fractional dimensions to the surface electrons that inherit the fractional dimension of the spaces left by the molecules, so that their delocalized wavefunctions decompose into self-similar parts at higher energies.\\
\begin{figure} 
\begin{center}
  \includegraphics[width=0.5\textwidth,clip]{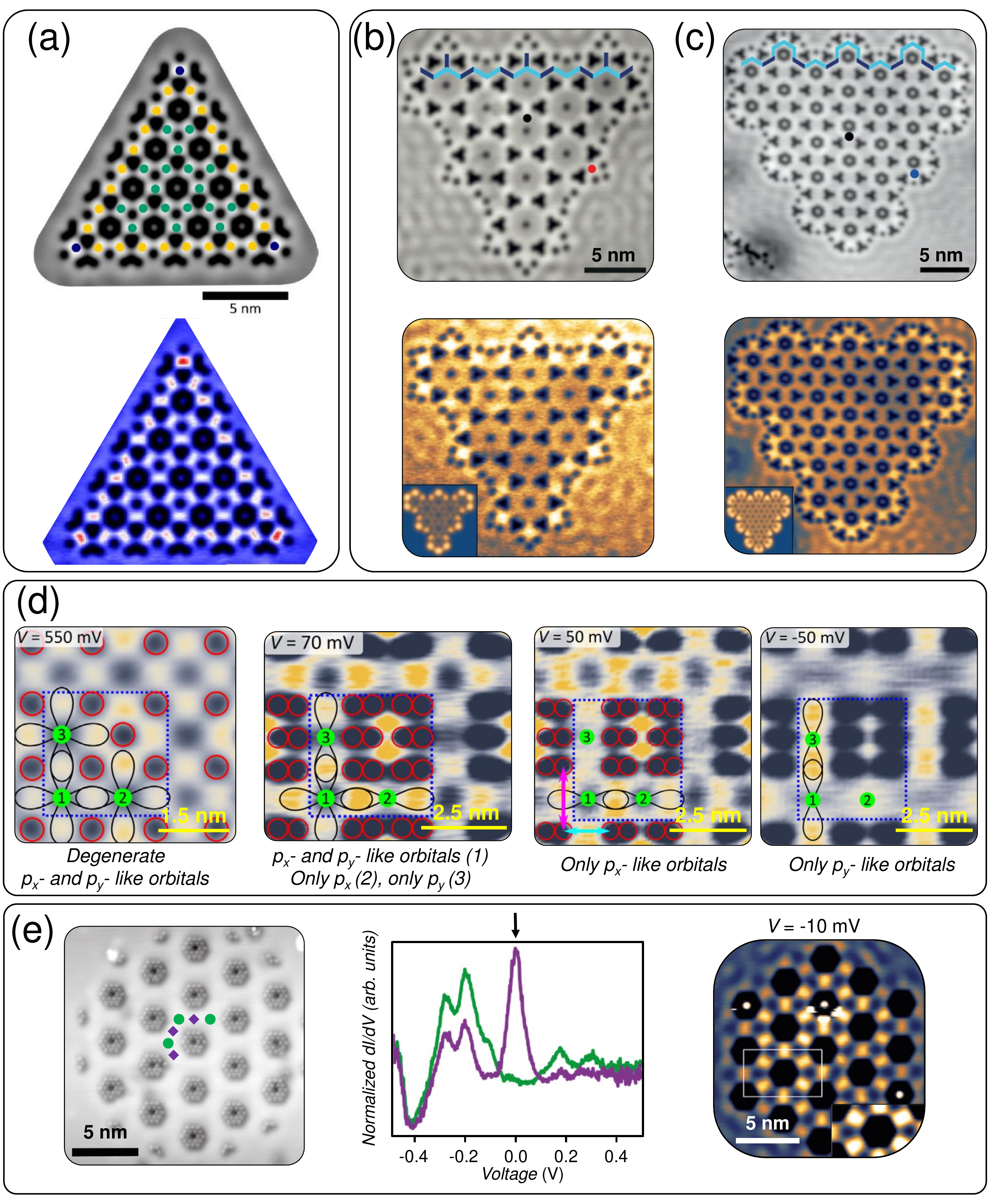}
 \end{center}
\vspace*{-5 mm}
\caption[]{Design of quantum states using artificial 2D lattices of precisely arranged CO on Cu(111). ({\bf a}) Robust corner localized zero-energy modes in an electronic higher-order topological insulator (adapted from~\cite{Kempkes2019a}). ({\bf b,c}) Topological edge modes arising at Kekulé lattices with partially bearded and molecular zigzag edge terminations (adapted from~\cite{Swart2020}). ({\bf d}) Manipulation of higher-energy $p$-orbital bands in artificial lattices with fourfold rotational symmetry. By tuning the lattice, the degeneracy of $p_x$ and $p_y$ orbitals can be lifted (adapted from~\cite{Slot2019}). ({\bf e})  Enlarged honeycomb lattice with practically unmixed orbital bands: the double peak at lower energies corresponds to the $s$ Dirac cone, the sharp peak marked by the black arrow relates to a $p$-orbital flat band (visualized in the conductance map on the right) while the highest energy double peak corresponds to a $p$-orbital Dirac cone (adapted from~\cite{Gardenier2020}).
}
 \label{figureCOtopology}
\end{figure}
\indent
The geometrical flexibility of  artificial lattices can be exploited to study fundamental aspects of intercoupling and topologically protected edge states at the local scale. For instance, triangular lattices constructed following breathing kagome geometries by alternating weak and strong bonds, have been reported to exhibit electronic structures with Dirac-cones and flat bands with topological states at their corners~\cite{Kempkes2019a} [Fig.~\ref{figureCOtopology}(a)]. Importantly, these corner modes are protected by a generalized chiral symmetry, which leads to a particular robustness against perturbations. In stark contrast to conventional topological insulators, this lattice has been defined as a high-order topological insulator (HOTI) since the  topological state has two dimensions less than the bulk.
However, it was found that not all edge geometries and intercouplings in triangular Kekule-type lattices [Fig.~\ref{figureCOtopology}(b,c)] are able to host topological edge modes~\cite{Swart2020}.\\
\indent
Generally, in these lattices the resulting artificial-atom sites have $s$-like character in their lowest energy signatures. By extending the studies to higher states, $p$-band engineering becomes accessible, \textit{i.e.}, manipulation of fourfold and threefold rotational symmetry of $p$-orbital bands~\cite{Slot2019,Gardenier2020}. As shown for the Lieb lattice in Fig.~\ref{figureCOtopology}(d)\, by introducing asymmetries, the degeneracy of $p_x$- and $p_y$-like orbitals can be lifted~\cite{Slot2019}. Moreover, these higher $p$-orbitals can  give rise to distinct electronic structures with flat bands and Dirac cones in a honeycomb lattice configuration [see Fig.~\ref{figureCOtopology}(e)]~\cite{Gardenier2020}.\\
\indent
A further question of interest is the systematic study of coupling effects between quantum corrals. Recently, rectangular and triangular structures in dimer and trimer arrangements were fabricated using the base units of CO/Cu(111) platforms. These exhibited differences in their QPI with respect to totally closed structures. Their electronic features could be understood using tight binding models, often applied to describe the coupling of atoms or molecules~\cite{Freeney2020}.\\
\indent
In addition, such closed nanoarchitectures where also explored towards overcoming the single-atom limit for information storage density. Precisely, using the coherence of the 2DEG of Cu(111), quantum holograms comprised of individually manipulated CO molecules where fabricated, which projected an electron pattern onto a portion of the surface. This innovative idea was further developed theoretically by introducing the quantum spin holography which additionally allowed to store information in two spin channels independently~\cite{Manoharan2009,Stepanyuk2012}.
\begin{figure} 
\begin{center}
  \includegraphics[width=0.5\textwidth,clip]{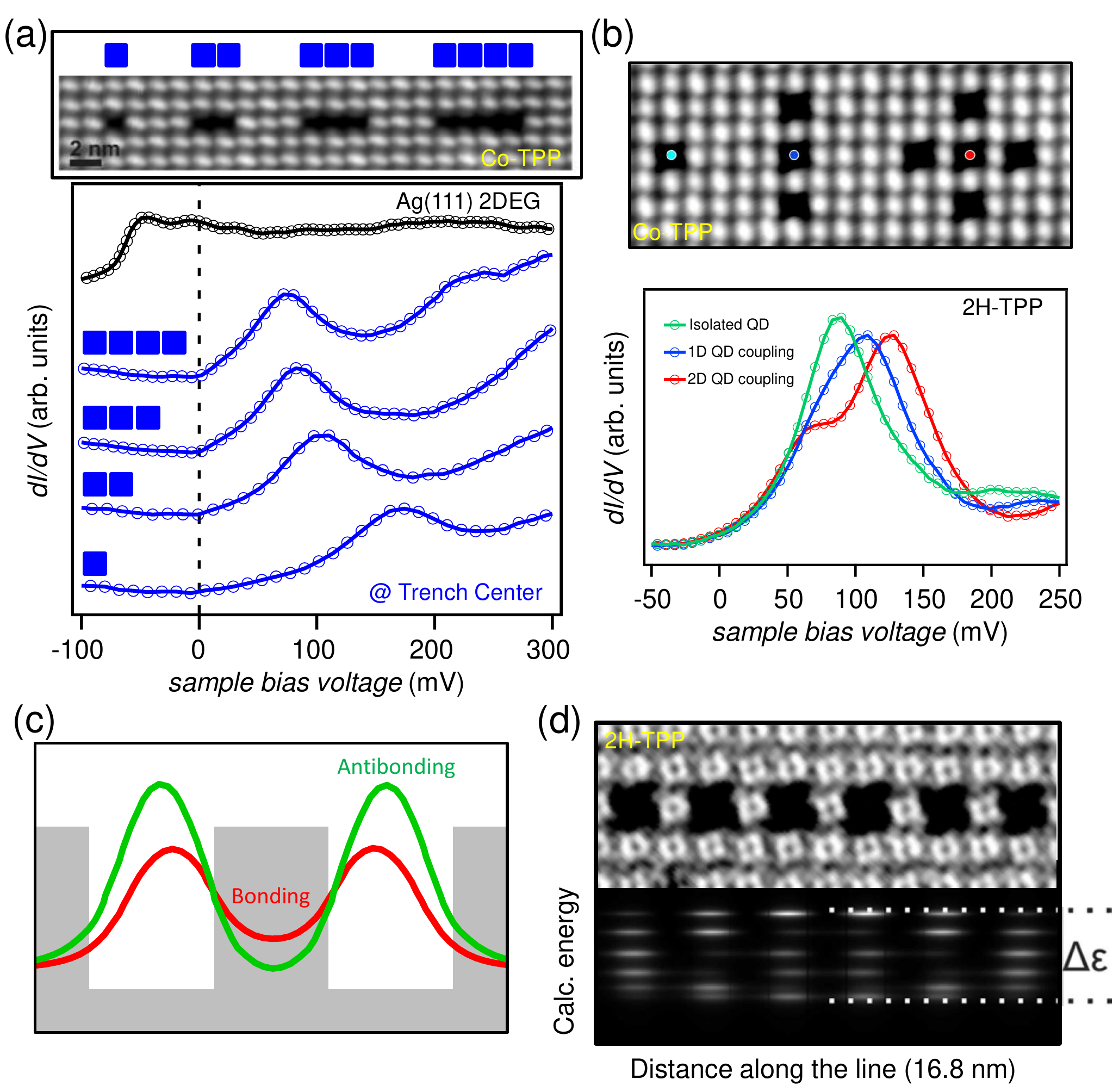}
 \end{center}
\vspace*{-5 mm}
\caption[]{Examples of molecular trenches created by tip removal of single units in tetraphenyl porphyrin films self-assembled on Ag(111). ({\bf a}) (\textit{top}) STM image showing four different length quantum wells generated by removal of 1 to 4 molecules. The STS spectra below, acquired at the trench centers, display a continuous shift of the first bound resonance, as the trench extends in length, towards the Ag(111) surface state onset at $-65$~mV (black curve). ({\bf b}) STM image presenting three monomer vacancies (single QDs) surrounded by a different number of identical neighbors: none on the left, two in line at the center and four in a cross-like geometry on the right. The STS acquired at the central site (colored dots in STM image) display spectroscopic differences that relate to QD intercoupling effects through the single molecular barriers. 
({\bf c}) Simplified model of two identical and coupled quantum wells with a bonding (red) and antibonding (green) state. The antibonding wavefunction protrudes further into the vacuum than its bonding counterpart.
({\bf d}) EBEM simulation (bottom) of six coupled QDs (top) that shows an intermediate stage to the formation of continuum states (1D band). The top and bottom energies are defined by the antibonding and bonding states displayed in (c). Figure adapted from~\cite{Seufert2013}.
}
 \label{figureSeufert}
\end{figure}
\subsection{B. Molecular vacancies and trenches}
\indent
The second manipulation protocol uses molecules that are weakly bound to the surface, which frequently aggregate into extended periodic islands stabilized by attractive intermolecular interactions~\cite{Huang2011}. These molecular films can be locally disrupted by controlled removal of single units using a STM tip, creating artificial nanostructured vacancies or molecular trenches~\cite{Seufert2013}. Within these voids, the underlying surface electrons emerge, \textit{i.e.}, 2DEG engineering can be carried out at the local scale by playing with the geometries that induce the electron confinement. \\
\indent
A nice example is provided by the tetraphenyl porphyrin (TPP) monolayer assembled on Ag(111)~\cite{Seufert2013}. This film exhibits an interface state with quasi-free electron-like character~\cite{Auwarter2010, Hofer2013,Caplins2014}. Linear structures with the desired length in multiples of 1.4 nm can be fabricated~\cite{Auwarter2010}, as shown in Figure~\ref{figureSeufert}(a). The molecules defining these trenches act as scattering barriers to the surface electrons that get confined within, as extracted from STS. Increasing the length of these linear structures  shows that the first confined resonance shifts towards the Ag(111) reference, in agreement with a particle-in-a-box case. In the longest chain (with 4 removed molecules) the second confined state peak is visible at $\sim{200}$~mV. Note that the confinement signature is present in the unoccupied region ($E$ $>$ $E_F$)~\cite{Karina2002,Horn1997}. \\
\indent
This manipulation method is ideally suited to engineer specific configurations as 1D chains or 2D artificial lattices and also to perform fundamental studies on QD intercoupling phenomena. Figure~\ref{figureSeufert}(b) compares a single molecular vacancy, analogous to an isolated QD, with a linear trimeric structure (1D) and a cross-like assembly (2D). The STS recorded at the central cavities become significantly modified by the number of nearest-neighbor QDs. This is due to the electron intercoupling through the leaky molecular wall. The overlap between neighboring electronic states results in asymmetric and broadened spectra reflecting a  wavefunction delocalization. Periodically repeating these coupled QDs structures gives rise to bonding and antibonding continuum states [Figure~\ref{figureSeufert}(c) and (d)]. These will generate a defined band structure whose fundamental energy is established by the bonding states and the overall peak-width ($\Delta \epsilon$)  (which is proportional to the QD interaction) by the antibonding ones. The wavefunction shape for the bonding state is more spread out than the antibonding one that peaks out much more into the vacuum~\cite{Seufert2013} [see Figure~\ref{figureSeufert}(c)]. Consequently, the STS technique probes more efficiently the antibonding state than the bonding one~\cite{Piquero2017}.\\
\indent
We envision that similar artificial lattices to the ones discussed previously could also be built by removal of molecules in square or hexagonal self-assembled monolayers~\cite{Seufert2013, Udhardt2017}, whereby the hopping parameters could be further tuned with the use of ``blends'' (mixture of two building blocks)~\cite{Wintjes2010, Stadtmuller2014, Goiri2015, Girovsky2017}.\\
\indent
Following a similar strategy, the vacancy engineering approach was explored using atomic platforms, whereby Cl monolayer islands on the Cu(100) surface proved very versatile. In this case, the atomic manipulation produces defect arrays with localized electronic vacancy states. Particularly, an automated digital atomic-scale memory could be realized~\cite{Kalff2016} as well as topological states within Lieb lattices~\cite{Drost2017, Huda2020}. The significant potential of this elegant method is recognized and was successfully exploited to study emergent band formations in lattices of varying structure, density and size~\cite{Otte2017}.\\
%
%
\section{III. Quantum resonators in supramolecular grids}
\indent
The molecular manipulation experiments confirm that the generated barriers and their geometry can be used for ultimately tuning of 2DEG confinement, giving rise to novel physical phenomena. Moreover, the interwell coupling can be tuned by the deliberate choice of the molecules and geometries. However, the underlying serial processes require very specific conditions, \textit{e.g.}, manipulable building units and cryogenic environments, typically requiring extensive construction times for the desired nanoarchitectures. Additionally, the overall lateral area of the designed structures is too small to meet the micro/macro scales (mesoscale) required for practical applications. Thus, it is advantageous to explore alternate fabrication routes from the formation, upscaling and stability points of view. Supramolecular building protocols are an ideal choice since they rely on self-assembly and self-correction processes that can repeatedly extend simple structural units and produce robust, highly regular, extended homoarchitectures with exquisite control while simultaneously engineering the system's electronic structure at the mesoscopic level.\\
 \begin{figure} 
\begin{center}
  \includegraphics[width=0.5\textwidth,clip]{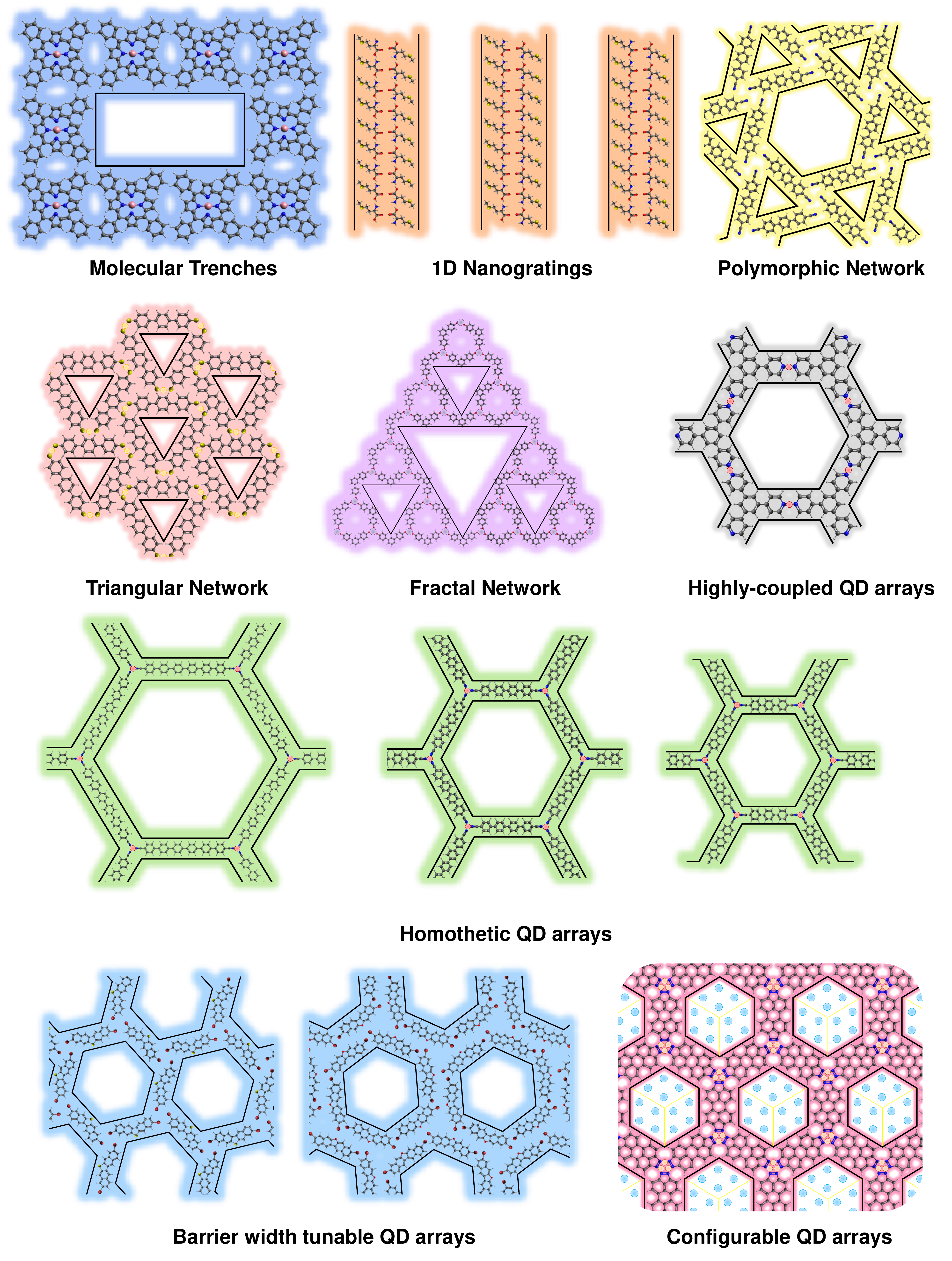}
 \end{center}
\vspace*{-5 mm}
\caption[]{Schematic representation of exemplary supramolecular motifs discussed throughout this review. This compilation provides an idea of the molecular nanostructure versatility that can be generated on suitable smooth metal surfaces to tune and engineer their 2DEGs. 
}
 \label{figureMolecularmodels}
\end{figure}
 \begin{figure} 
\begin{center}
  \includegraphics[width=0.5\textwidth,clip]{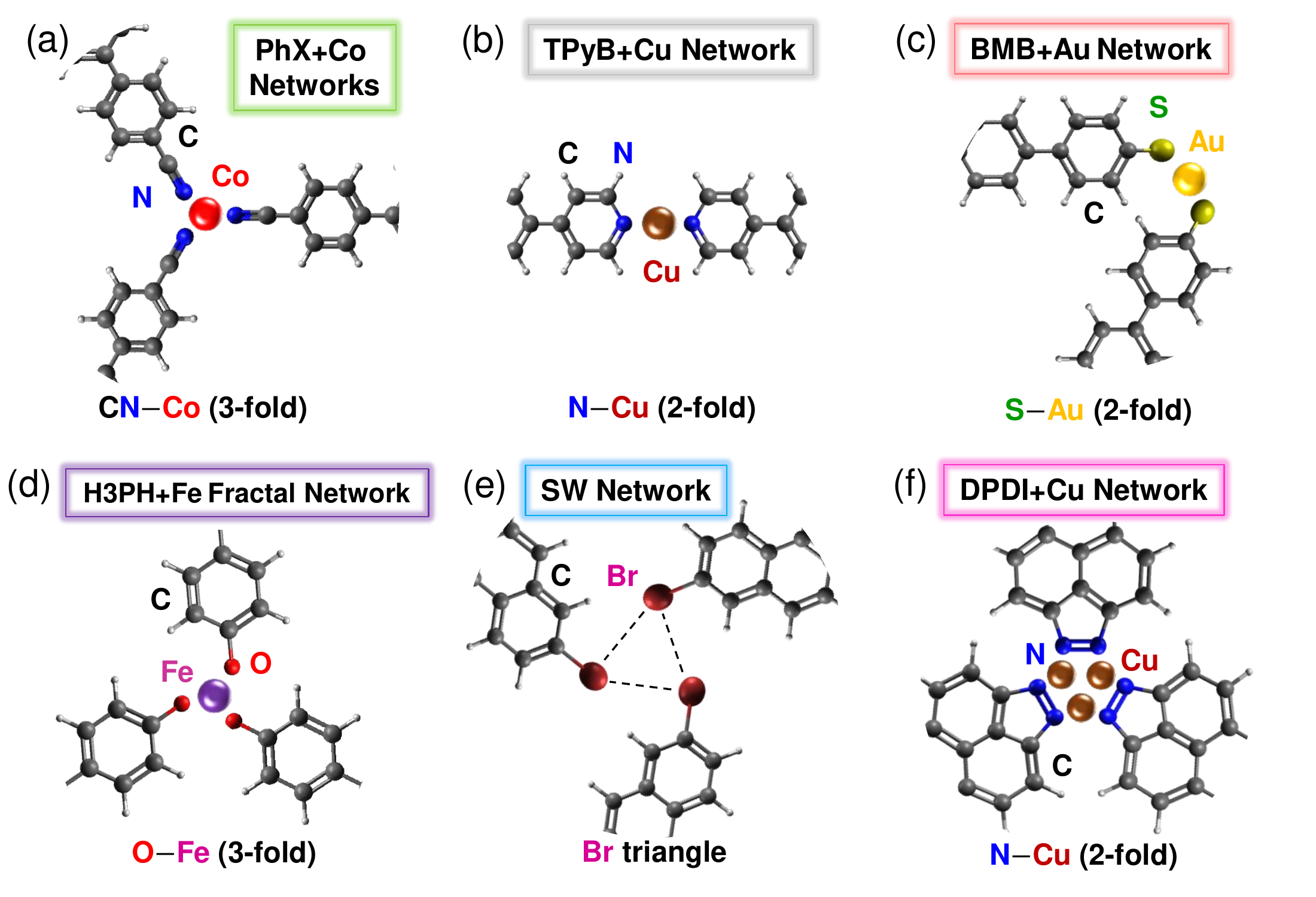}
 \end{center}
\vspace*{-5 mm}
\caption[]{Schematic representation of exemplary metal-organic and halogen-bonds, underpinning the formation of extended molecular nanostructures on smooth fcc(111) metal surfaces delineated in Fig.~\ref{figureMolecularmodels}.
}
 \label{figureMolecularbonds}
\end{figure}
\indent
Indeed, nowadays we can realize molecular structures in the form of 1D nanogratings and 2D networks that efficiently scatter and confine surface 2DEGs~\cite{Barth2007,Stohr2016,NianLin2016}. Thus, engineering the LDOS and the interpore-coupling is feasible and results in unprecedented band structures. As demonstrated in Fig.~\ref{figureMolecularmodels} and Fig.~\ref{figureMolecularbonds}, a careful design and choice of pre-synthesized molecular building blocks allows for precise control over the size and shape of the nanoarchitectures, providing command over the overall electronic properties. In the following paragraphs key examples of self-assembled geometries with demonstrated 2DEG confining capabilities will be highlighted.
\begin{figure}
\begin{center}
  \includegraphics[width=0.5\textwidth,clip]{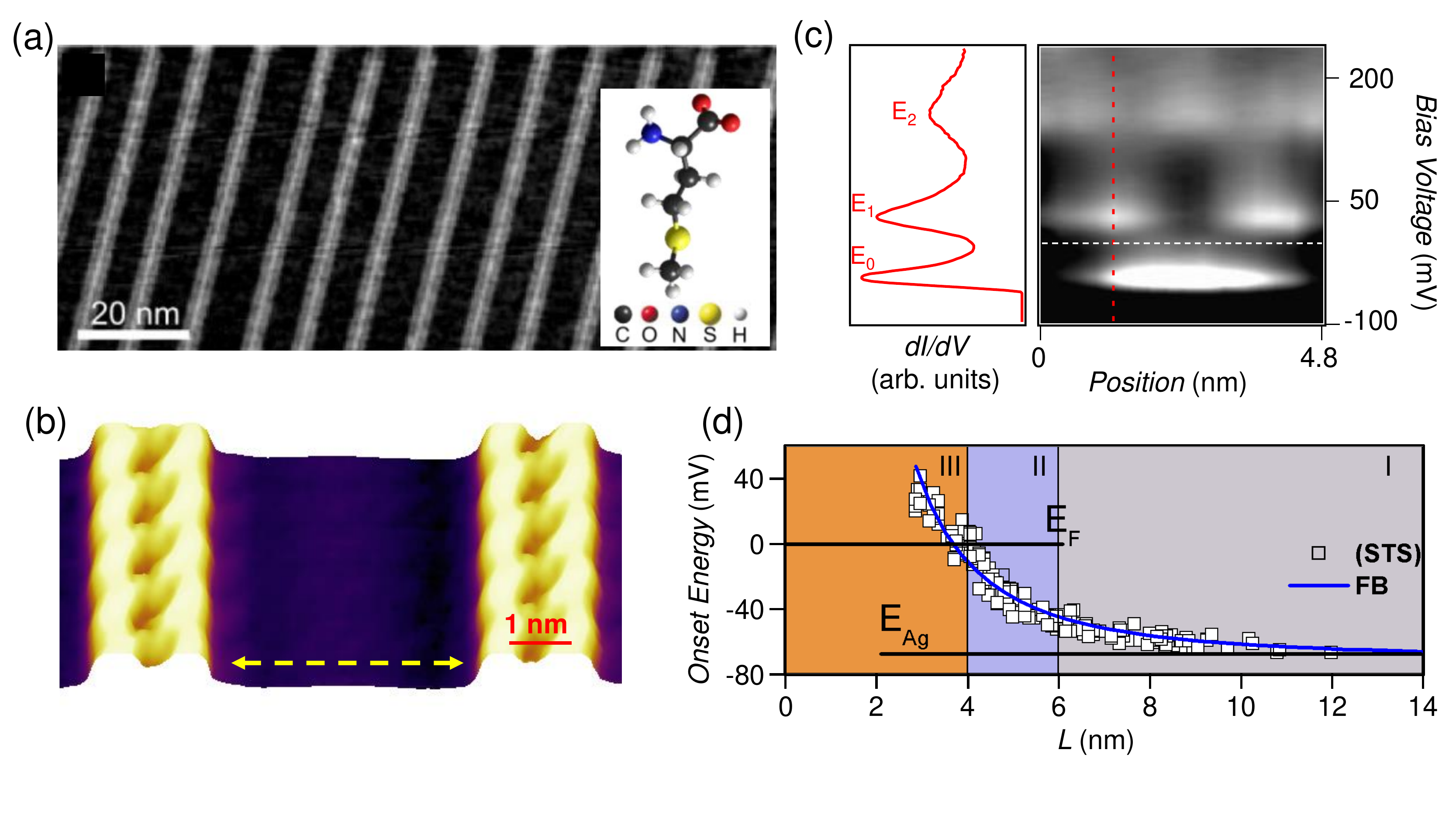}
 \end{center}
\vspace*{-8 mm}
\caption[]{1D nanograting created by zwitterionic self-assembly of $L$-methionine on  Ag(111). ({\bf a})  Overview topography of regularly spaced nanogratings. ({\bf b}) High resolution 3D perspective of a 4.8 nm wide quantum well  and ({\bf c}) characteristic STS acquired perpendicular to the cavity that beautifully exhibits the first three 1D-QWS resonances. ({\bf d}) The energy of the first QWS resonance varies in an inverse quadratic way with the chain separation, as expected for a particle in a box model. Figure adapted from~\cite{Pennec2007}.
}
 \label{figureMethionine}
\end{figure}	
\subsection{A. 1D organic nanogratings}
Straight molecular chains expressing hydrogen-bonded 1D nanogratings were obtained by self-assembly processes using the suitable precursors on mildly reactive and smooth metal surfaces~\cite{Pennec2007, Barth2000, Kern2001}. The linear structure formation is driven by the interplay between molecule-molecule and molecule-substrate interactions. For the case of $L$-methionine molecules on Ag(111) [see Fig.~\ref{figureMethionine}(a), (b)] ammonium and carboxylate groups interact and form zwitterionic dimer units~\cite{Schiffrin2007}. Similar to inorganic step arrays at vicinal surfaces, the molecular chains scatter the 2DEG and 1D confinement takes place. Accordingly, the differential conductance scan in Figure~\ref{figureMethionine}(c), acquired along a line perpendicular to the chains [Fig.~\ref{figureMethionine}(b)], reproduces the three lowest resonator states (featuring zero, one and two nodes). The energy of these QWS follow a quadratic inverse relation with the quantum well size [as shown in Figure~\ref{figureMethionine}(d) for the $n=1$ resonance],  which is well reproduced using the Fabry-P\'{e}rot model that describes quantization effects between finite (parallel) potential barriers~\cite{Burgi1998}.\\ 
\indent
The 2DEGs scattering can be analogously exerted by the rims of compact molecular islands. Particularly, regular tetracyanoquinodimethane species in periodic arrangement can discretize the electron momentum parallel to the island edge, which effect is ascribed to the Bragg scattering from the periodic and corrugated 1D edge~\cite{Martinez2019}.\\
\indent
As a drawback, supramolecular nanogratings~\cite{Barth2000,Kern2001} are difficult to control in terms of extended regularity and cannot be strictly considered periodic~\cite{Pennec2007, Urgel2016a}. This is due to the weakness of the long-range repulsive interactions mediating the grid formation of the self-assembled molecular twin chain constituents. Although surface reconstruction patterns or the like can be employed to guide such assemblies~\cite{Kern2001,Clair2005}, highly regular coupled 1D-quantum systems based on molecules require improved fabrication procedures.
%
\begin{figure} 
\begin{center}
  \includegraphics[width=0.45\textwidth,clip]{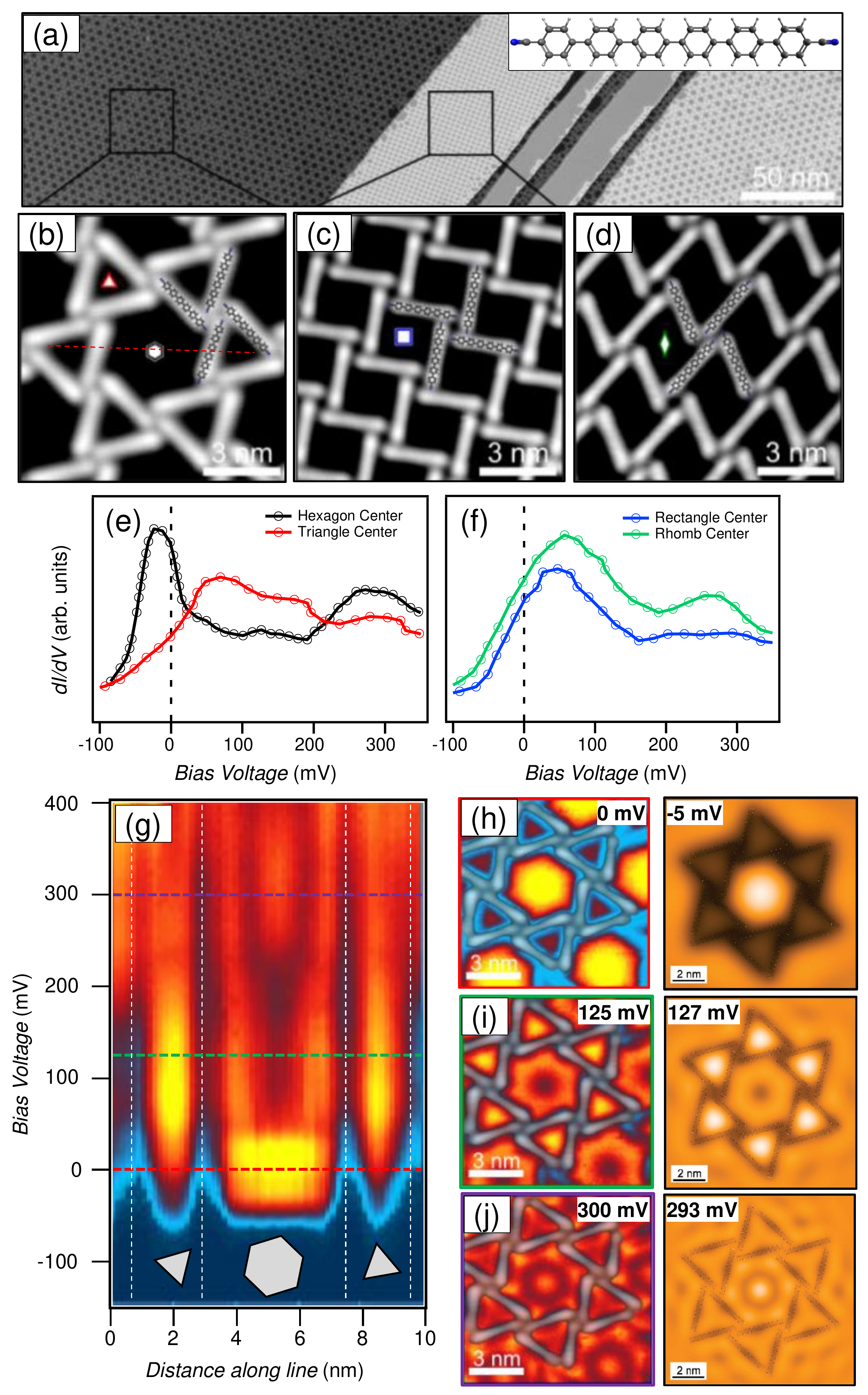}
 \end{center}
\vspace*{-5 mm}
\caption[]{Example of polymorphic organic nanoporous networks acting as self-assembled interwoven quantum corral structures. ({\bf a}) Overview STM image showing coexistence of several network geometries generated by dicyano-hexa(p-phenylene) (Ph6) molecules after deposition on Ag(111). The pure organic assemblies form ({\bf b}) Kagom\'{e}, ({\bf c}) rectangular and ({\bf d}) rhombic structures. ({\bf e}) and ({\bf f}) show differential conductance spectra recorded at the pore centers found in ({\bf b}) to ({\bf d}). ({\bf g}) $dI/dV$ line scan acquired along the red line marked in the STM image shown in ({\bf b}). Up to three (two) confined states can be distinguished at the quasi-hexagonal (triangular) pore. ({\bf h}) - ({\bf j}) Conductance maps  (left) and EBEM simulated LDOS maps (right) at the indicated energies showing the spatial distribution of the 2DEG confined states. Figure adapted from~\cite{Barth2009} and \cite{Barth2013}.
}
 \label{figPh6}
\end{figure}
\subsection{B. Organic nanoporous networks as QD arrays}
\indent
2D nanoporous networks frequently present well-defined arrays commensurate with the substrate and comprising atomically precise pores. We will show that within these nanocavities the substrate's 2DEG gets confined and follows the same scattering mechanisms occurring at the inorganic quantum corrals and nanoislands described earlier in Fig.~\ref{figureCorrals}. Particularly, we discuss the LDOS observed at the pores of nanoporous networks created using dicyano-poly(p-phenylene) molecules. The structural morphologies displayed by this molecular family depend on the cyano-aryl endgroup interaction and molecular backbone length. 
In the case of the dicyano-hexa(p-phenylene) (Ph6) molecules, self-assembly on Ag(111) affords several nanoporous structures featuring a common four-fold bonding motif (see Figure~\ref{figPh6})~\cite{Barth2009, Barth2013, Chung2011}, which allows for a straightforward comparative assessment. When classified by the pore shape, three networks are identified:  kagom\'{e} (with triangular and quasi-hexagonal pores), rectangular and rhombic [Fig.~\ref{figPh6}(b-d)]. These polymorphic nanoporous networks are commensurate with the substrate and display two chiral orientations for each structure. The conductance spectra at the center of these pores contain several peaks with maxima at different energies [see Fig.~\ref{figPh6}(e), (f)]. At first glance, these spectra seem unrelated to the highly featured quantum corral lineshapes displayed in Fig.~\ref{figureCorrals}. However, an order of magnitude difference exists between the area defined by the hexagonal nanoisland in Fig.~\ref{figureCorrals}(c) and the quasi-hexagon of Fig.~\ref{figPh6}(b). The severe pore size reduction significantly separates the resonance peaks of these networks when compared to the atom corrals. Consequently, only the lowest resonances can be accessed within the energy window probed~\cite{Berndt1998}. In the STS spetra from Fig.~\ref{figPh6}(e) and (f) the first peak's energy position is, as expected, mainly dominated by the pore size: $\sim-20$~meV for the quasi-hexagonal (largest), $\sim50$~meV for the rectangular, $\sim60$~meV for the rhombic and $\sim80$~meV for the triangular (smallest) pores.\\
\indent
To study the impact that molecular nanopore size and shape have on the quantum confinement, a series of conductance spectra and maps were acquired for the kagom\'{e} lattice. The line scan in Fig.~\ref{figPh6}(g) and the conductance maps in Fig.~\ref{figPh6}(h)-(j) show the energy and spatial variations of the LDOS associated with the different pore shapes~\cite{Barth2009,Barth2013}. For the quasi-hexagon, the lowest confined state ($n=1$) exhibits a dome shape (without nodes) [Fig.~\ref{figPh6}(h)] while the second confined state ($n=2$) features one node with roughly toroidal shape [Fig.~\ref{figPh6}(i)]. Note that at that particular energy, the triangular pores now exhibit their first confined state. As expected, the third state ($n=3$) of the quasi-hexagons display a sombrero-shape structure with two nodes [Fig.~\ref{figPh6}(j)]~\cite{Barth2009,Barth2013}. Remarkably, the geometric chirality signatures of the pores is recognized in the confined state LDOS distribution. At the highest energy we also find a three-fold symmetry in the triangular nanocavities, indicative of energy proximity to their $n=2$ confined state~\cite{Schouteden2012}.\\
\indent
For a deeper understanding of the 2DEG confinement resonances, a modeling by the so-called electron boundary element method (EBEM) is very helpful. This semiempirical method  has been extensively  employed for solving Maxwell's equation and determining the optical response for arbitrary shapes~\cite{Abajo2008}. It can accurately reproduce the electron confinement effects in molecular nanostructures by parameterizing the 2DEG~\cite{Barth2009, Florian2011} and generally considers the molecules as repulsive  barriers. The EBEM simulations nicely reproduce the experimental confinement features [Fig.~\ref{figPh6}(h-j)] using a molecular potential of $V_{mol}=500$ meV [see Table~\ref{TableEPWE}]. The fact that these peaks bear an intrinsic broadening (they are not delta functions) allows a certain energy range for the visualization of these confined states in the conductance maps and also leads to a mixing of eigenstates~\cite{Florian2011,  Wang2018}. Contrary to the polymorphic structure of Ph6, the self-assembly of dicyano-penta(p-phenylene) (Ph5) or other shorter species (ter- or quater-phenylene) on Ag(111) provides long-range ordered chiral kagom\'{e} lattice~\cite{Schlickum2008} or other homogenous structures~\cite{Klyatskaya2011}. In a similar fashion, the on-surface synthesis and assembly of circumcoronene on Cu(111) has been recently used to create an extended chiral electronic kagom\'{e}-honeycomb lattice where the 2DEG is confined into two emergent electronic flat bands~\cite{Pavel2021}. Moreover, other organic-based corrals, in the form of molecular nanohoops, honeycombene oligophenylene macrocycles and porous COFs show similar 2DEG confinement capabilities~\cite{Nazin2016, Gottfried2016, Hao2019, NianLin2013b}. 
Associated to this, on-surface synthesized organic nanowires and rings can show interesting intramolecular electronic confinement (different from the 2DEG), such that the nanorings  act as whispering gallery mode resonators for the oligomeric states~\cite{Reecht2013}.
\begin{figure} 
\begin{center}
  \includegraphics[width=0.35\textwidth,clip]{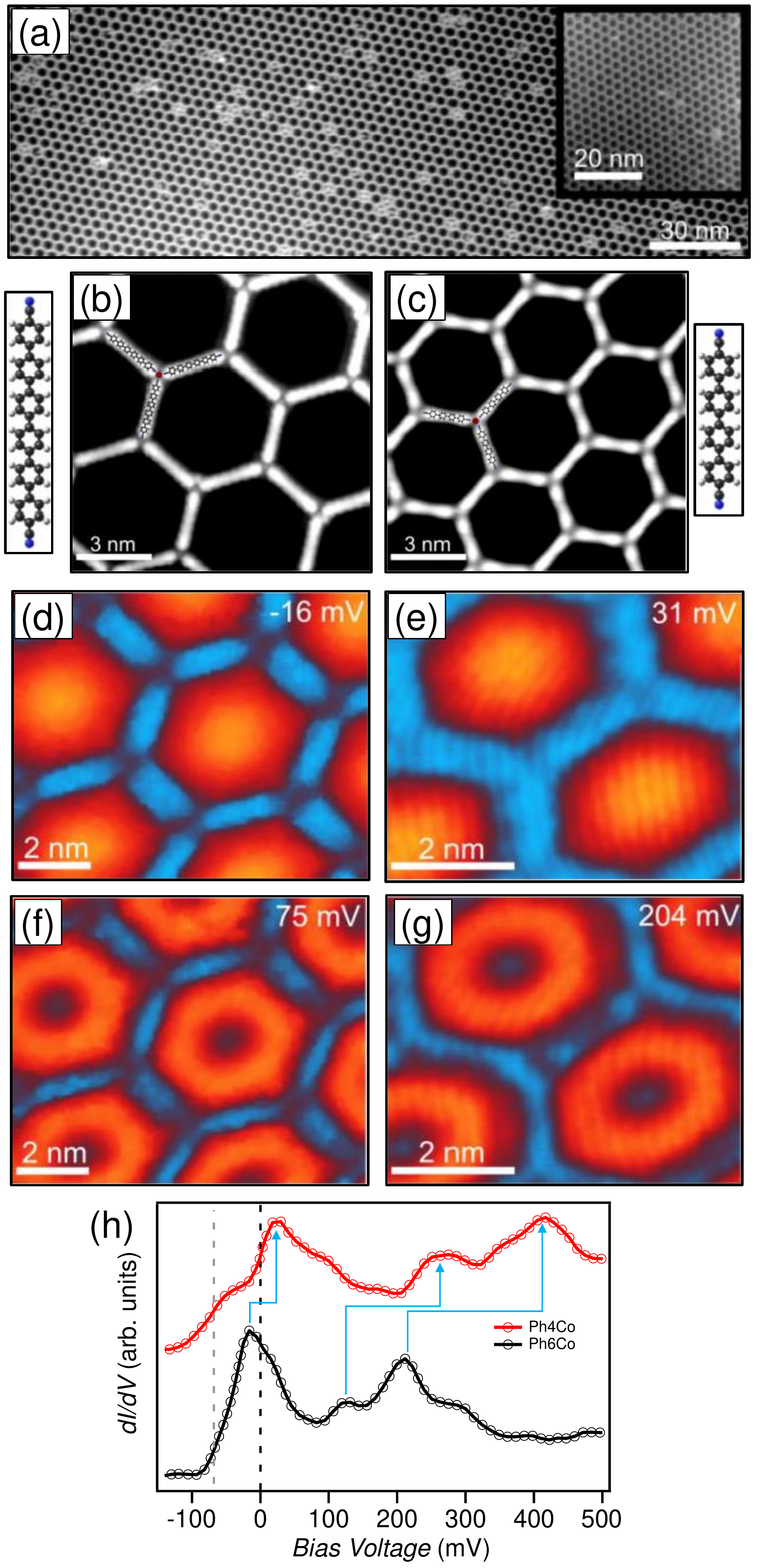}
 \end{center}
\vspace*{-5 mm}
\caption[]{2DEG confinement by metal-organic coordination of dicyano-poly(p-phenylene) molecules.  ({\bf a}) Overview image of metal-organic Ph6+Co hexagonal network obtained on Ag(111). The inset shows an homothetic (scalable) network using shorter molecules  (Ph4+Co),  also grown on Ag(111).
({\bf b}) and ({\bf c}) display high resolution topographs of these networks and include their corresponding molecular structures. ({\bf d}) and ({\bf f}) show conductance maps of the first ($n=1$) and second ($n=2$) confined states in the Ph6+Co network, whereas ({\bf e}) and ({\bf g}) show the same information for the Ph4+Co array. ({\bf h}) Conductance spectra at the center of both hexagonal pores exhibit a larger energy shift as the pore size is reduced. Figure adapted from~\cite{Florian2011}.
}
 \label{figPh46Co}
\end{figure}
\subsection{C. Metal-Organic QD arrays}
\indent
Metal-directed coordination protocols provide an additional control knob on metallo-supramolecular assemblies~\cite{Schlickum2007,Kuhne2009,Pacchioni2015}, while simultaneously enabling the formation of robust metal-organic networks. Particularly, when Co atoms are co-deposited with dicyano-poly(p-phenylene) molecules on Ag(111), the formation of crystal quality, monodomain, hexagonal nanoporous networks occurs~\cite{Schlickum2007,Kuhne2009}.  As shown in Fig.~\ref{figPh46Co}(a) to (c), these metal-organic coordination networks (MOCNs) require a 3:2 stoichiometry of dicyano-poly(p-phenylene) molecules with Co atoms~\cite{Florian2011}. Isostructural CN$\cdots$Co coordination nodes at specific substrate positions prevail for both dicyano-tetra(p-phenylene) (Ph4) and Ph6~\cite{Schlickum2007} [see Fig.~\ref{figureMolecularbonds}(a)]. Thus homothetic (\textit{i.e.}, scalable) geometries are available, serving as ideal systems for studying the nanopore size dependence on the 2DEG confinement.\\
\indent
As the pores are regular in shape and size, the overall 2DEG confinement must be repeatedly spread throughout the surface. This is confirmed by the conductance maps  for Ph6+Co and Ph4+Co, shown in Fig.~\ref{figPh46Co}(d) to (g), and their corresponding STS acquired at the hexagonal pore centers [Fig.~\ref{figPh46Co}(h)]. The $dI/dV$ spectra display the first confined states ($n=1$) at $\sim-6$~meV and $\sim15$~meV for the Ph6+Co and Ph4+Co networks, respectively. As expected, these are visualized as domes when conductance maps are acquired close to these energies. The second confined state ($n=2$), visualized as a torus in the conductance maps, cannot be clearly distinguished in STS since it coincides with a node at the pore center. We must reach the third ($n=3$) resonance to observe a conspicuous peak again at the pore center.\\
\indent
Overall, the conductance maps of Fig.~\ref{figPh46Co}(d) to (g) exhibit the same spatial LDOS distribution throughout these two networks, except for and energy shift (and a slight broadening) dictated by the different pore size. Indeed, the energy position of these resonances are identical when scaled using a reduction factor R=1.74 for the Ph4+Co, very close to the nanopore area ratio (R*=1.83)~\cite{Florian2011}.
\\
\indent
These networks scatter the Ag(111) 2DEG through the finite hexagonal barriers that have been simulated using EBEM~\cite{Florian2011}. The calculations suggested an inhomogeneous scattering potential landscape whereby molecules and adatoms scatter electrons differently (see Table~\ref{TableEPWE}). A successful Ansatz implies that the metal centers behave as slightly attractive regions ($V_{Co}=-50$~meV) whereas the molecules are strongly repulsive ($V_{mol}=500$~meV) for the substrate 2DEG. However, the assignment for the coordination nodes is debatable, since an attractive potential should host bound states~\cite{Limot2005,Crommie2001,Liu2006,Olsson2004,Silly2004}, which were not-to-date experimentally identified in the nanoporous networks~\cite{Florian2011,Piquero2019b}. By contrast, chemisorbed closed-packed arrays of Au-TCNQ and Mn-TCNQ can display such states at lower energies than the surface state onsets~\cite{Faraggi2012}, though the absence of open pores prevents any expression of confined states. The scattering potential of the coordination nodes requires further scrutiny and this issue will be discussed in detail later.\\
\begin{figure}
\begin{center}
  \includegraphics[width=0.45\textwidth,clip]{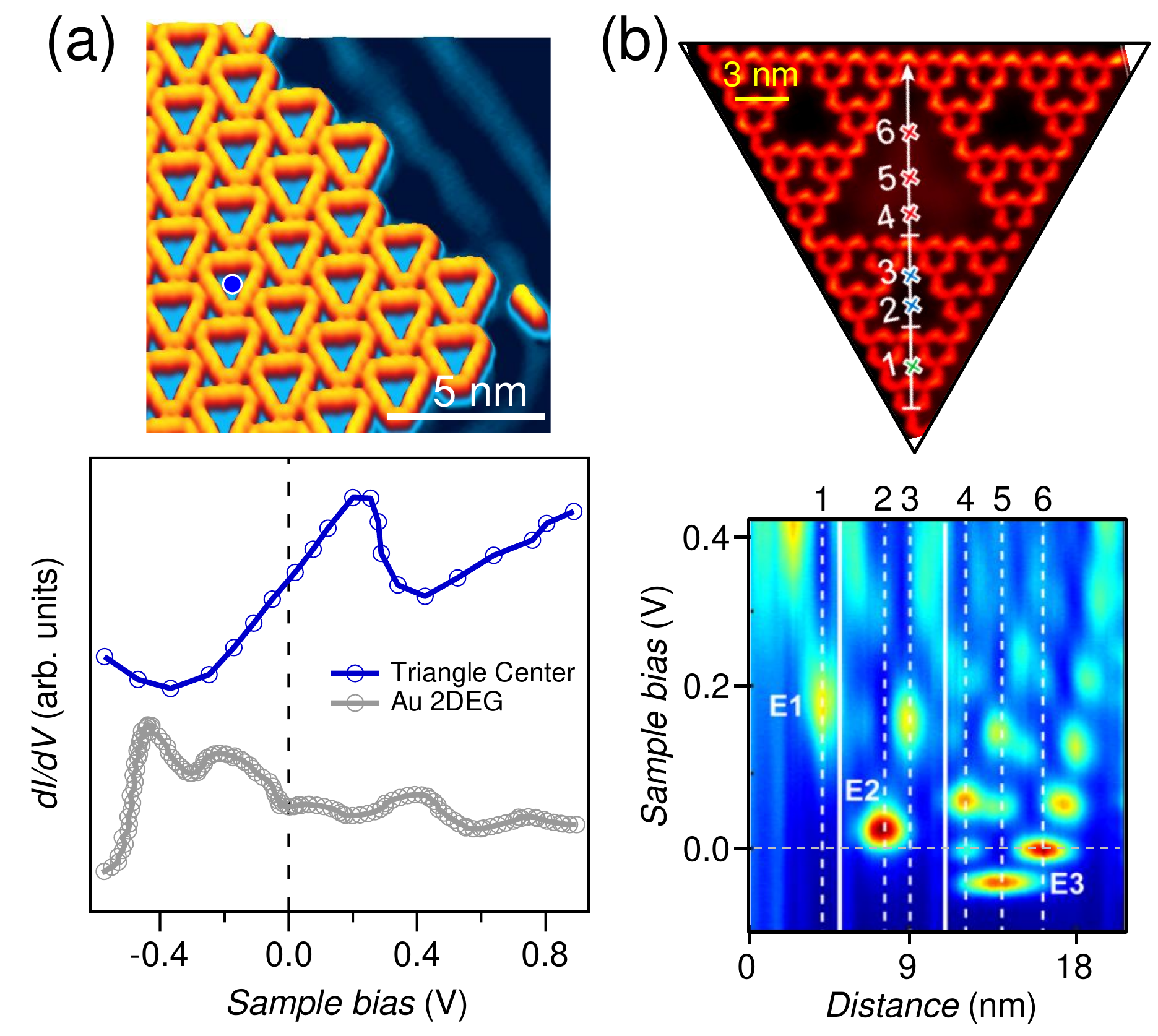}
 \end{center}
\vspace*{-5 mm}
\caption[]{Confinement properties of triangular and Sierpi\'{n}ski triangle fractals generated by metal-directed assembly. ({\bf a}) STM topograph of an array of condensed triangular Au$_3$BMB$_3$ complexes. STS data acquired at the triangular nanopore (blue dot) show a resonance strongly shifted with respect to the pristine Au 2DEG (adapted from~\cite{Colazzo2019}). ({\bf b}) STM image of self-assembled H3PH-Fe Sierpi\'{n}ski triangles and corresponding spatially resolved STS map along the white arrow (adapted from~\cite{Wang2018}).
}
 \label{figuretriangles}
\end{figure}
\indent
Other pore shapes are also feasible by metal-directed assembly protocols. A prominent example is the formation of a 2D triangular MOCN stabilized by Au-thiolate bonds [see Fig.~\ref{figureMolecularbonds}(c)]. It could be obtained simply via the deposition of 1,4-bis(4-mercaptophenyl)benzene (BMB) on the Au(111) surface, providing intrinsic adatoms that engage in the asembly of the coordination superlattice~\cite{Colazzo2019}. The Au$_3$BMB$_3$ units forming this array consist in embedded triangular nanopores that strongly confine the Au(111) surface state [Fig.~\ref{figuretriangles}(a)]. The significant peak shift of $\sim700$~mV is well reproduced by EBEM that for this particular system required one of the largest repulsive potentials reported for nanopourous networks using this semiempirical method ($V_{mol}$= 600 meV, see Table~\ref{TableEPWE}).\\
\indent 
Combined triangular units can be even more interesting when generating Sierpi\'{n}ski lattices by self-assembly methods. In particular, the codeposition of 4,4''-dihydroxy-1,1':3',1''-terphenyl (H3PH) and Fe atoms on Ag(111) followed by a mild annealing to 380 K affords the fractal structures shown in Fig.~\ref{figuretriangles}(b)~\cite{Wang2018,Wang2019} [see the bonding motif in Fig.~\ref{figureMolecularbonds}(d)]. Similar to the CO manipulated counterpart already discussed and depicted in Fig.~\ref{figureMolec}(c), the Ag surface state is confined at the scalable triangular nanopores and leads to an area dependency of the 2DEG eigenstates. Other interesting network geometries in the form of demi-regular lattices, Kepler tilings and quasicrystals have been recently achieved and are expected to present related complex confinement capabilities~\cite{NianLin2017, Piquero2019c, Ecija2013, Urgel2016}.\\
\section{IV. Emergence and engineering of band structures from coupled quantum dot arrays}
\indent
Since molecular superlattices with leaky barriers and well-defined nanopores can homogeneously carpet the surface with minute defect concentrations, a coherent electronic signal stemming from coupled QDs can be measured using space-averaging techniques such as ARPES. In the following we visit a series of studies highlighting the formation of genuine band structures that can be engineered by the choice of the employed molecular building blocks.
\begin{figure*} 
\begin{center}
  \includegraphics[width=0.9\textwidth,clip]{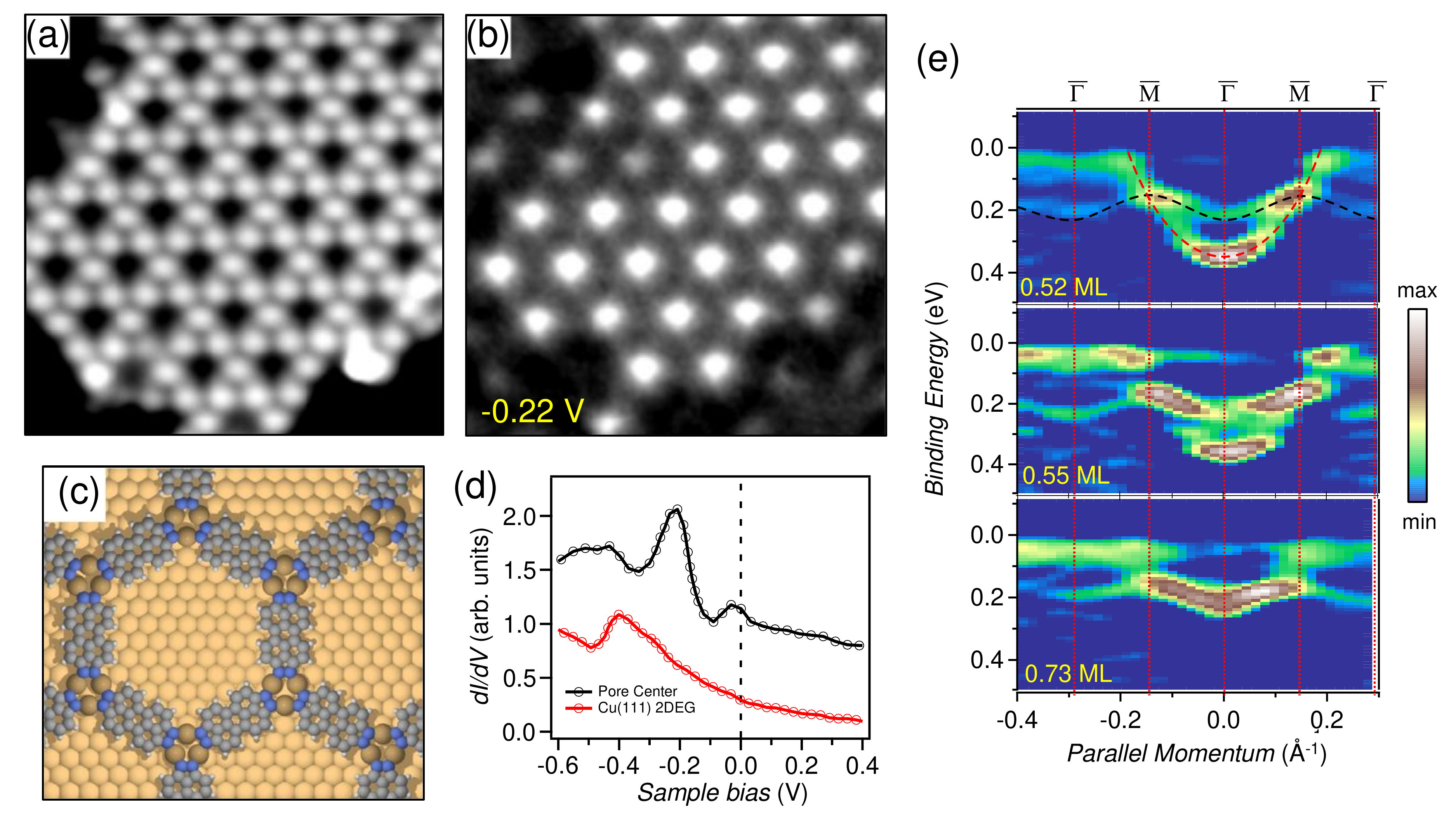}
 \end{center}
\vspace*{-5 mm}
\caption[]{0D electron confinement and emergence of a 2D band structure by the 3deh-DPDI-based metal-organic coordination network on Cu(111). ({\bf a},{\bf a}) STM image of the hexagonal network and simultaneously acquired conductance map at -0.22 V. At this energy the electrons are confined ($n=1$ resonance) within the pores. ({\bf c}) Structural model of the tri-metal-coordinated perilene-based nanoporous network. ({\bf d}) STS spectra measured at the center of a hexagonal pore (black) and pristine Cu (red). ({\bf e}) Band structure of the extended nanoporous network as the surface is progressively covered (from top to bottom) with the molecular array. The cosine-shaped black dashed line at the top panel (lowest coverage) marks the first network confined band, whereas the red dashed line marks the pristine surface state stemming from the network-free regions. As the coverage increases the emission from the pristine surface state gradually disappears while the band originating from the coupled QDs gains intensity, until only the shallow dispersive network band remains (bottom panel). Note that the energy of this band matches (after normalization of the measurement temperature) the dominant STS peak position in ({\bf d}), which is related to the $n=1$ confined state displayed in ({\bf b}). Figure adapted from~\cite{Lobo2009}.
}
 \label{figureDPDI}
\end{figure*}
\subsection{A. Emergence of dispersive band structures}
\indent
The first band structure from a QD array was measured for the Cu(111) surface state confined by the DPDI+Cu extended network~\cite{Lobo2009}. This hexagonal metal-organic nanoporous network  is formed after thermal  dehydrogenation of 4,9-diaminoperylene quinone-3,10-diimine (DPDI) molecules on the metal surface. The three-fold symmetric array is characterized by a unit cell composed of three molecules and six Cu adatoms [see Fig.~\ref{figureMolecularbonds}(f)] with a periodicity of 2.55~nm~\cite{Matena2014, Gade2014, Piquero2016, Piquero2019a} [see Fig.~\ref{figureDPDI}(a), (c)], which has been also used to host guest species ($e.g.$, octaethylporphyrins and C$_{60}$) within the voids~\cite{Stohr2007}.\\
\indent
As shown by the conductance map and STS curves in Fig.~\ref{figureDPDI}(b) and (d), this molecular network confines the Cu(111) surface state electrons within its pores. The first confined state ($n=1$) is found at $\sim-0.22$~V and the second one ($n=2$) at $\sim-0.08$~V. The conductance map at $n=1$ shows the characteristic dome-like shape centered at the network pores, so that each pore acts as a single QD. However, the width of the STS peak suggests the possibility of coupling between neighboring pores, fulfilling the extended scenario described above and depicted in Fig. \ref{figureSeufert}(d).\\
\indent
To check for the existence of a band, ARPES measurements were performed close to the Fermi energy with the network covering a significant part of the surface [Fig.~\ref{figureDPDI}(e)]. This metallo-supramolecular array is ideal because it is monodomain, commensurate with the substrate ($10\times10$), affords large domains (laterally exceeding $50$~nm) and homogeneously covers the surface with a relative small amount of defects. Consequently, the DPDI+Cu network can be conceived as a periodic superlattice of QDs, where the Cu(111) surface state (the 2DEG) gets confined by their building units (molecules and Cu adatoms) that strongly modify the initial surface potential landscape. 
The ARPES signal is displayed in  [Fig.~\ref{figureDPDI}(e)] and exhibits the progressive extinction of the 2DEG parabolic band related to the pristine surface state, which is replaced by a cosine-shaped  band centered at $\sim-0.2$~eV as the network fills the surface (note that the nanoporous network covers the surface completely at $\sim0.73$~ML). Therefore, the 2DEG is engineered by the network into new electronic bands whose dispersion relates to the QD coupling strength and pore size. The fundamental energy and bandwidth matches the STS peak observed in Fig.~\ref{figureDPDI}(d) and it has been proven to originate from the pristine substrate's Shockley state~\cite{Piquero2016}.\\
\begin{figure}
\begin{center}
  \includegraphics[width=0.5\textwidth,clip]{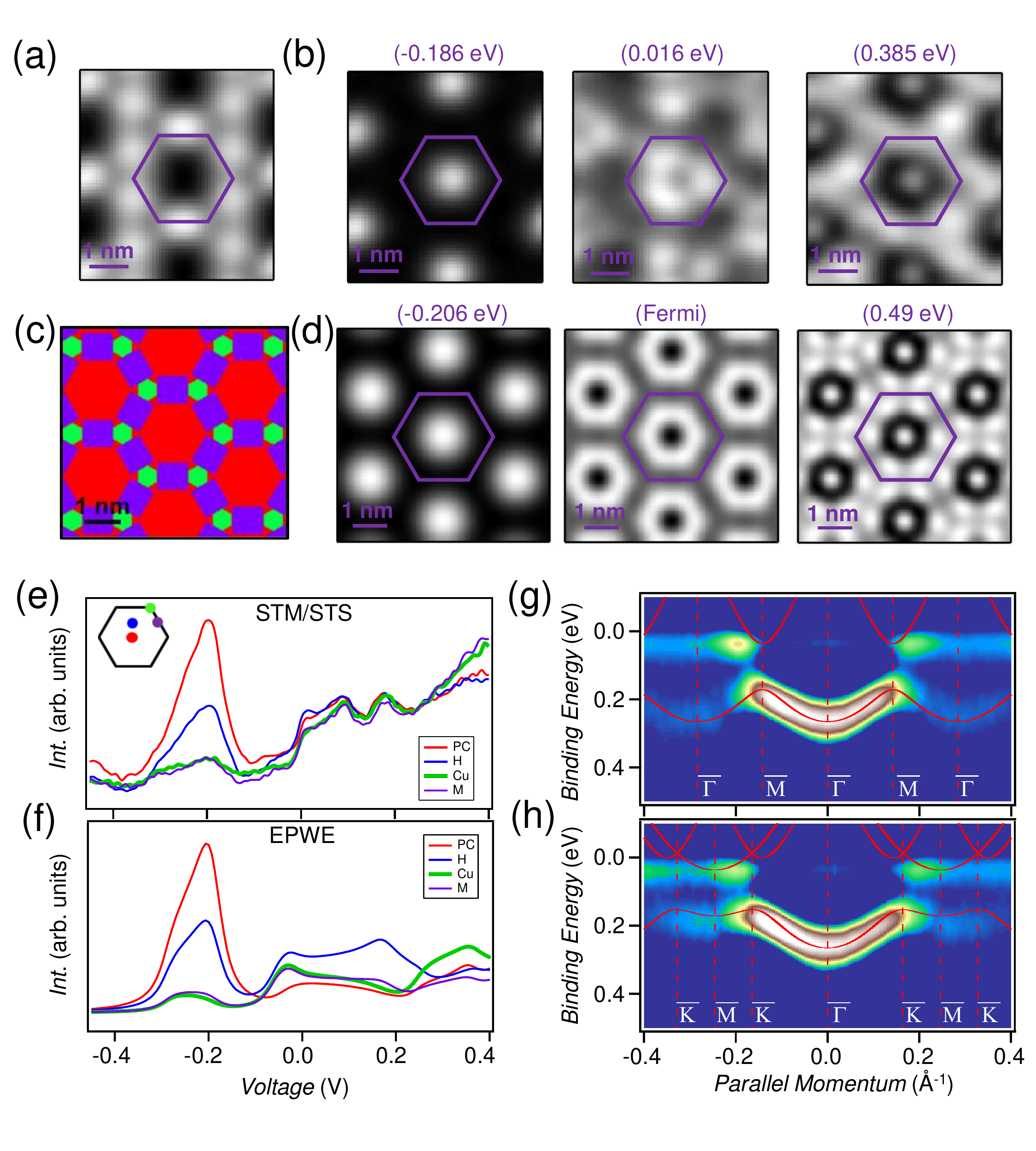}
 \end{center}
\vspace*{-8 mm}
\caption[]{Comparison between experimental datasets (LDOS and electronic band structures) with the EPWE semiempirical simulations of the DPDI+Cu network.  ({\bf a}) STM image and ({\bf b}) three constant height conductance maps at selected energies close to the first, second and fourth confinement resonances.  ({\bf c}) Potential landscape constructed for the EPWE simulations. It consists of three different regions: Cu substrate (red hexagons), molecules (purple rectangles) and metal centers (green hexagons). The potential values given to molecules and adatoms are provided in Table~\ref{TableEPWE}. ({\bf d}) Calculated LDOS at the indicated energies with the scattering geometry in ({\bf c}) showing similar spatial distributions as the upper experimental cases. ({\bf e}) $dI/dV$ spectra acquired at different unit cell positions (see inset)  and corresponding EPWE simulated LDOS ({\bf f}). ({\bf g}, {\bf h}) Colorplots of the experimental band structure (second derivative of the ARPES intensity) for the two high symmetry directions $\overline{\Gamma\rm{M}}$ ({\bf g})  and $\overline{\Gamma\rm{K}}$  ({\bf h}) and simulated EPWE bands superimposed as red lines. The high quality of the network allows to observe faint replica bands in adjoining Brillouin zones that are  matched by the calculations. Figure adapted from~\cite{Piquero2019a}.     
}
 \label{figureDPDI1}
\end{figure}
\indent
This extraordinary band structure is the natural extension of the artificial 2D lattices and quantum corrals without the requirement of molecular manipulation. The modeling of the potential landscape generated by this DPDI+Cu network~\cite{Barth2015,Piquero2019a} was realized using the  semiempirical EPWE method, which uses linear combinations of plane waves (the 2DEG) that are scattered by the potential barriers in an infinite periodic lattice (mimicking our metal-organic network). It is most accurate whenever the simulation starts from a realistic scattering geometry and different scattering potentials are assigned to molecules and adatoms to account for the 2DEG electron barriers~\cite{Piquero2019a, Piquero2019b}. The combination of experimental datasets (STS \& ARPES) as inputs for  these semiempirical simulations allow us to capture the intricacies of the scattering potential landscape generated by these networks and to establish systematic modeling procedures (see subsection E).\\
\indent
A comparison between modeling and experiments for the DPDI+Cu network is shown in Figure~\ref{figureDPDI1}. The agreement between  STM/STS and ARPES datasets with the EPWE simulations (LDOS and band structure) is remarkable. The potential landscape [Figure~\ref{figureDPDI1}(c)] was constructed using the structural model of the network~\cite{Matena2014} that emulates the experimental STM [Figure~\ref{figureDPDI1}(a)] and noncontact atomic force microscopy (nc-AFM) images~\cite{Kawai2016}. It was found that the scattering at the molecular backbones is homogeneous and similar to the coordination nodes ($V_{DPDI} = V_{Cu} = 390$~meV)~\cite{Piquero2019a}. The scattering character of the metal centers is repulsive in the DPDI+Cu network, which agrees with other atom-based 2DEG confining entities, such as step edge adatoms, quantum corral barriers or dislocation networks~\cite{Crommie1993a,Mugarza2006,Malterre2011}. Importantly, to match the experimental datasets the pristine surface state requires to be renormalized. In other words, the network's presence modifies the 2DEG parabolic dispersion, affecting both the effective mass ($m^*$) and the reference energy ($E_{0}$, also referred as the onset energy). The potential values and 2DEG renormalization used for the simulations are collected in Table~\ref{TableEPWE}. This renormalization process of 2DEGs becomes evident for most nanoporous MOCNs studied so far with photoemission~\cite{Piquero2017,Piquero2019a, Piquero2019b,Piquero2019c}, and its underlying physics are explained in more detail in the following subsections.
\begin{figure}
\begin{center}
  \includegraphics[width=0.5\textwidth,clip]{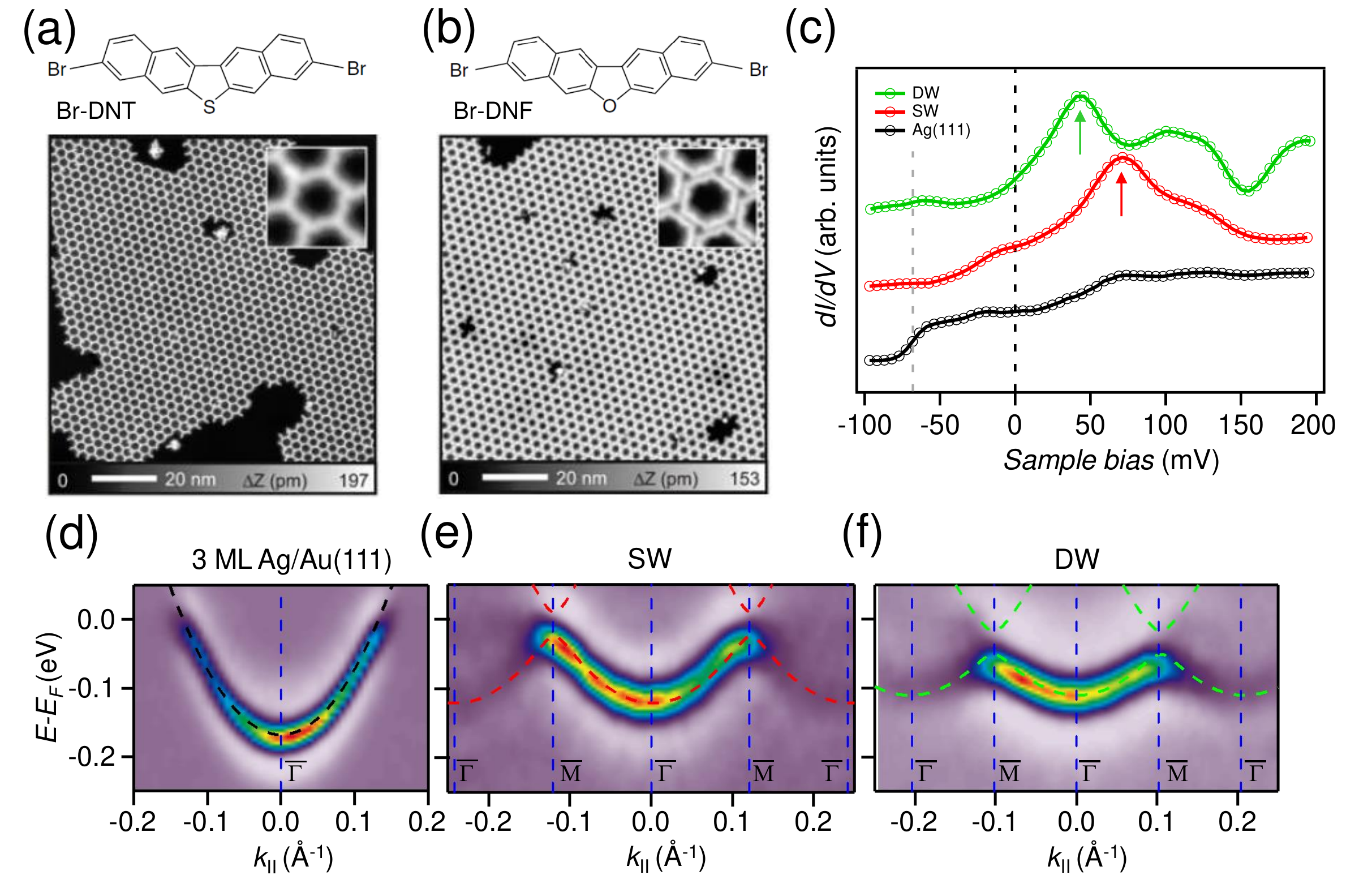}
 \end{center}
\vspace*{-5 mm}
\caption[]{QD band engineering by tunable barrier widths: Self-assembled  single-wall (SW) and double-wall (DW) QD arrays showing different interpore coupling of confined quantum states.  
Large-scale STM topographies for the SW network generated with Br-DNT ({\bf a}) and the DW network with Br-DNF ({\bf b}). Insets depict constituent arrangements for each network. ({\bf c}) STS at the pore center of the SW and DW networks compared to the Ag substrate's typical stepwise DOS increase. ({\bf d-f}) QD band structure of the $n=1$ confined state induced by the SW and DW networks on Ag (3ML)/Au(111) along $\overline{\Gamma{\rm{M}}}$. The match with the EPWE calculated electronic bands (black, red and green dashed lines) is excellent. From these experiments it was concluded that the analysis of STS and confinement trends should not be limited to inspection of peak shifts. Figure adapted from~\cite{Piquero2017}.
}
 \label{figureSWDW}
\end{figure}
\subsection{B. Regulating QD crosstalk through the barrier width}
\indent
The network barriers condition the degree with which individual QDs couple to each other and ultimately define the energy position and shape of the electronic bands. Hence, the control over the potential barriers between neighboring QDs turns out to be essential to engineer the 2DEGs as they alter the crosstalk (interaction) between their confining units.\\
\indent
Experimentally it is nowadays possible to  tune the confinement and intercoupling properties of individual QDs by engineering the network barrier widths without affecting the pore size~\cite{Piquero2017}. This was achieved in molecular networks using the halogen bond versatility~\cite{ Kawai2015, Ho2017, Feyter2019, Shang2015} without the need of metal coordination. Particularly, two haloaromatic compounds (Br-DNT and Br-DFN), that differ in just a single atom at their center (S vs O), generate the two supramolecular networks shown in Fig.~\ref{figureSWDW} on Ag(111) and on thin Ag monolayers (MLs) grown on Au(111). The extended organic arrays enclose identical pore areas with a relatively small amount of defects, but their pores  are separated by one (single wall, SW) or two  (double  wall, DW) molecules. The condensation of Br-DNT into a nanoporous network happens solely through trigonal halogen bonding [see Fig.~\ref{figureMolecularbonds}(e)], whereas in Br-DNF the furan group electronegativity  introduces  O$\cdots$Br bonds, increasing the interaction complexity and leading to the double-rim formation~\cite{Piquero2017}.\\
\indent
The LDOS at the pore centers [Fig.~\ref{figureSWDW}(c)] evidences the quantum confinement lineshape. However, when inspecting merely the peak locations, the first resonances ($n=1$) seem unexpectedly energy inverted, when simply assuming stronger confinement for the wider barrier case (DW) (see arrows). Particularly the energy shift is visibly larger for the SW case (peak maximum towards the right of the green one). To understand this behavior and resolve the QD arrays' band structure, ARPES measurements were performed on both networks. Figure~\ref{figureSWDW}(e,f) displays weakly dispersive cosine-shaped bands, typical of QD arrays. As expected for confined 2DEG electrons, their fundamental energy (onset of the band) shifts to higher energy and deviates from the initial parabolic dispersion [cf. Figure~\ref{figureSWDW}(d-f)]. In addition, evidencing a stronger confinement, the DW onset is further away from the 2DEG reference compared to the SW array.\\
\indent
EPWE model calculations of these two arrays were carried out to gain further insights. Matching simultaneously the STS and band structure was possible  using an effective potential $V_{mol}$=140 meV (see Table~\ref{TableEPWE}). This evidenced that both experimental techniques must be probing the same electronic states, but in a different way. The origin of this discrepancy can be understood with the knowledge gained in the previously described molecular trenches (Fig. \ref{figureSeufert}): STM is more sensitive to the antibonding states that transform into the top of the coupled QDs band. Indeed, the STS peak maxima coincides with the energy of the $\overline{\rm{M}}$ points in the ARPES datasets. Contrarily, the band onset (fundamental energy) found by ARPES shows up as a weak shoulder in the STS at the pore center. This renders asymmetric the LDOS peak lineshapes with maxima displaced towards the top of the band in periodic QD arrays.\\
\indent
In essence, the 2DEG confinement strength of DW exceeds that of SW networks, showing a reduced bandwidth in ARPES and peakwidth in STS compared to the SW. This decrease of the bandwidth with identical potential scattering barriers ($V_{mol}=140$~meV) relates to a lower interpore coupling imposed by a wider set of barriers~\cite{Seufert2013,Piquero2017}, which translates into a reduction of the electron wavefunction overlap with QD separation.\\
\indent
To understand the electron confinement phenomena we require STM/STS and ARPES experimental datasets. The EPWE simulations can then determine effective scattering potential barriers that the surface electrons experience. This allows us to understand the nature of the intercoupling processes and quantify the 2DEG renormalization occurring by the presence of the network on the surface, induced by overlayer-substrate interactions~\cite{Piquero2019a}.
\begin{figure*} 
\begin{center}
  \includegraphics[width=0.7\textwidth,clip]{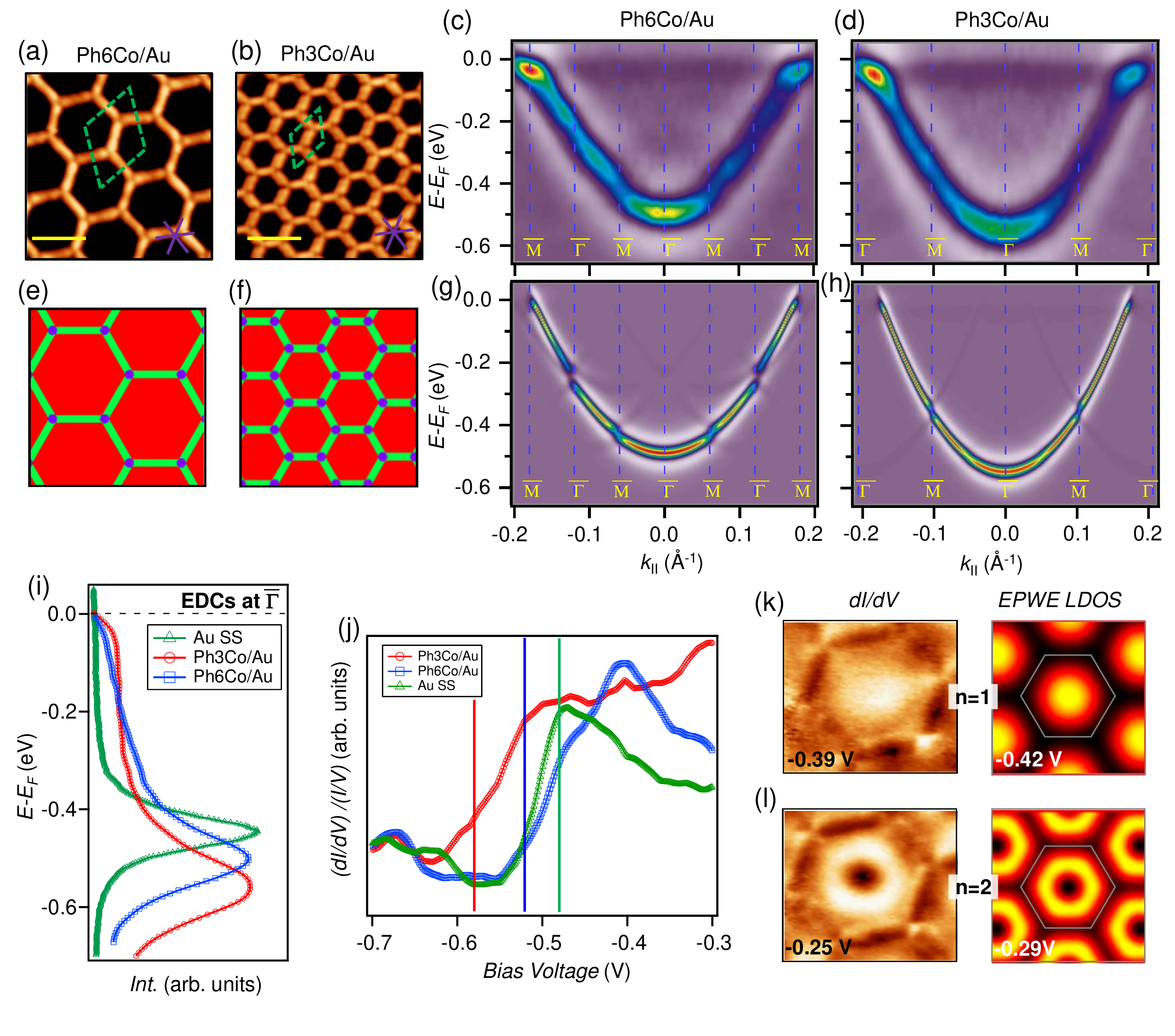}
 \end{center}
\vspace*{-5 mm}
\caption[]{Band structure engineering with tunable pore sizes: 2DEG confinement bands of dicyano-poly(p-phenylene) metal-organic coordination networks grown on Au(111). STM topographies of  single-domain Co-coordinated QD arrays using Ph6 ({\bf a}) and dicyano terphenyl (Ph3) ({\bf b}).  ARPES spectral density (second derivative) of Ph6+Co ({\bf c}) and Ph3+Co ({\bf d}) networks. 
({\bf e}) and ({\bf f}) shows the corresponding 2D potential geometry used for the EPWE modelization (molecules in green, Co atoms in purple and red for cavity regions).  ({\bf g}) and ({\bf h}) displays the simulated ARPES bands obtained by EPWE. Note that  the band structure exhibits downward shifts of the band bottom and gap openings at the superstructure symmetry points compared to the pristine Au(111). Matching the experimental data requires a significant modification of the 2DEG energy reference. ({\bf i}) Normal emission EDCs clearly display the gradual downshift of the fundamental energy as the pore size is reduced, which is also observed in the blow-up of the conductance spectra ({\bf j})  close to the pristine Au SS onset. ({\bf k,l}) Experimental (\textit{left}) and EPWE simulated (\textit{right})  conductance maps of the $n=1$ and  $n=2$ confined resonances. Adapted from~\cite{Piquero2019b}.
}
 \label{figureP3P6}
\end{figure*}
\subsection{C. 2DEG renormalization induced by the metal centers: dicyano-poly(p-phenylene) networks}
\indent
The next logical step is to engineer the coupled QD band structures by modifying the pore size while maintaining fixed potential barriers. Ideal candidates for such studies are the MOCNs generated using dicyano-poly(p-phenylene) molecules ~\cite{Florian2011, Schlickum2007,Kuhne2009}, already discussed and shown in Fig.~\ref{figPh46Co}. These homothetic Co coordinated networks cannot be properly studied with ARPES if grown on Ag(111) since the 2DEG onset is too close to the Fermi energy (see Fig.~\ref{figureNobleMetal}). For that matter these MOCNs were generated on the Au(111)  surface, where isostructural arrangements to those reported on Ag(111) evolve~\cite{Piquero2019b}. \\
\indent
Figure~\ref{figureP3P6}(a, b) shows the Ph3+Co and Ph6+Co honeycomb networks formed on Au(111) that feature pore areas of  8 nm$^2$ and 24 nm$^2$, respectively [see bonding motif in Fig.~\ref{figureMolecularbonds}(a)]. The measured band structures show very weak umklapps (replicas) of the main signal with modulated gaps [Fig.~\ref{figureP3P6}(c, d)]. The STS and the conductance maps [Fig.~\ref{figureP3P6}(j-l)] confirm 2DEG confinement at the pores, agreeing with the networks generated on Ag(111) [Fig.~\ref{figPh46Co}]. Moreover, there is clear evidence for a gradual downshift of the band bottom as the pore size is reduced [$\Delta{E_{Ph6+Co}}=-40$~meV and $\Delta{E_{Ph3+Co}}=-100$~meV with respect to the Au 2DEG, cf. Figure~\ref{figureP3P6}(i)]. This opposes the expected upward shift found for all other coupled QD bands and also inorganic scatteres shown before~\cite{Lobo2009, Piquero2017, Piquero2019a}.\\
\indent
EPWE simulations were employed once again to unravel the potential landscapes generated by the molecular networks~\cite{Piquero2019b}. The STS and ARPES datasets were matched assuming strong repulsive scattering potentials at the molecular sites ($V_{mol}=250$~meV) and weaker ones for the adatoms ($V_{Co}=50$~meV). Intriguingly, a severe 2DEG renormalization was required to account for the downshifts of the band bottom ($E_0^{Ph3+Co} = -0.52$~eV and $E_0^{Ph6+Co} = -0.56$~eV compared to the Au 2DEG $E_{0}^{Au} = -0.48$~eV, see Table~\ref{TableEPWE}).\\
\indent
To understand such strong 2DEG modification, \textit{ab-initio}  DFT calculations were carried out. In these calculations the interaction between a Co lattice (defined by the MOCN) with the Au(111) substrate was investigated and showed that the magnitude of the 2DEG downshift can be directly related to the concentration of (isolated) Co adatoms existing on the surface. However, the molecular network presence interferes since it ultimately defines the interaction strength of the adatom array with the substrate (Co-Au hybridization)~\cite{Piquero2019b, Schlickum2007}. This applies to single atom coordinated MOCNs since the adsorption height of  the adatoms increases due to the coordination with the molecules, which  effectively reduces their interaction with the substrate. As will be shown in the next section, downward energy renormalization effects occur systematically in other single-atom-coordinated MOCNs and substrates, but were previously not fully recognized since complementary photoemission experiments were required for determining the 2DEG onset~\cite{Piquero2019b}.
\begin{figure*}  
\begin{center}
  \includegraphics[width=0.9\textwidth,clip]{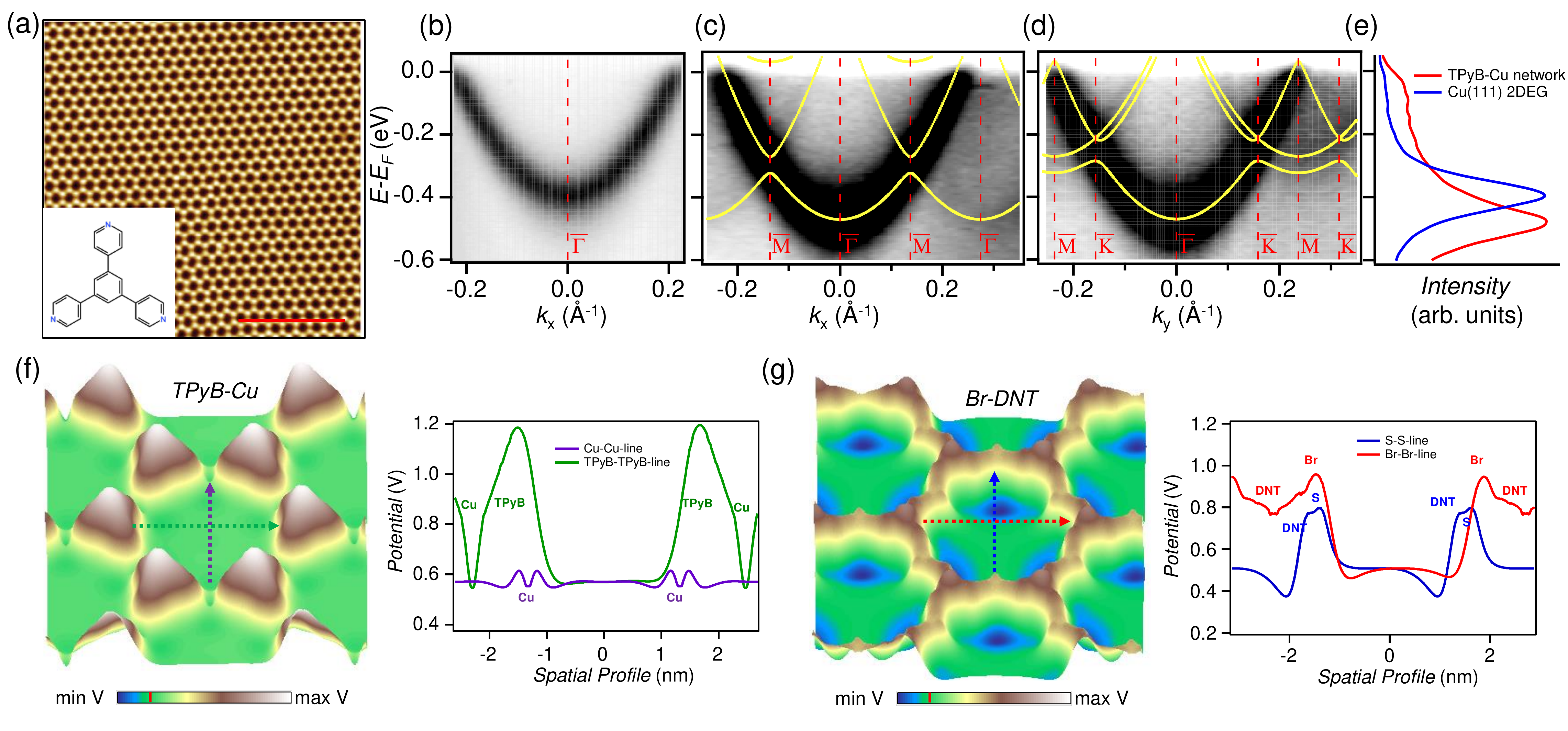}
 \end{center}
\vspace*{-5 mm}
\caption[]{Opening of electron transmission channels at confining QD arrays by the embedded metal centers. 
({\bf a}) STM image of the extended single-domain TPyB+Cu hexagonal QD array grown on Cu(111). The inset shows the precursor molecule. ARPES band structure of the pristine 2DEG  ({\bf b}) and the QD array band of the  TPyB+Cu network along the two high-symmetry directions: ({\bf c}) $\overline{\Gamma{\rm{M}}}$ and ({\bf d}) $\overline{\Gamma{\rm{K}}}$. ({\bf e}) EDCs at normal emission ($\overline{\Gamma}$ point) that exhibit -70 meV downshift of the confined state with respect to the pristine Cu 2DEG. ({\bf f})  and ({\bf g})  show on the left 3D perspectives of the electrostatic potential map calculations for TPyB+Cu and Br-DNT networks and their corresponding perpendicular potential line profiles on the right. Significant potential difference is encountered between the adatoms with respect to the molecules. Adapted from~\cite{NianLin2013b,Piquero2019c}.
}
 \label{figureTPyB}
\end{figure*}
\subsection{D. Electron transmission through coordination spheres}
\indent
The confining strength of the metal centers in dicyano-poly(p-phenylene) based MOCNs turns out to be significantly weaker than that of the molecules. This property can be further used to engineer the coupled QDs band structure aiming at increasing the crosstalk between neighboring pores. To this end, the electronic structure of the Cu-coordinated network (TPyB+Cu) was investigated~\cite{Piquero2019c}  (see Figure~\ref{figureTPyB}). This extended, monodomain honeycomb network is generated from 1,3,5-tri(4-pyridyl)-benzene (TPyB) molecules deposited on Cu(111), where the substrate provides the Cu centers bridging the TPyB pyridyl groups by a two-fold coordination~\cite{NianLin2013b} [see Fig.~\ref{figureMolecularbonds}(b)].\\
\indent
The band structure from  the TPyB+Cu MOCN displays a downshift of -70 meV at the $\overline{\Gamma}$ point with respect to the pristine Cu 2DEG [Figure~\ref{figureTPyB}(b)-(e)]. The same scenario applies here as for the just discussed Ph3+Co and Ph6+Co networks~\cite{Piquero2019b} and for T4PT+Cu MOCN~\cite{Zhou2020}, in relation to a single adatom array interacting with the 2DEG. Note however, that such downshift is not a specific property of hexagonal networks, since it is also observed in demi-regular networks~\cite{Piquero2019c}. The replica bands (umklapps) and small energy gaps observed in ARPES indicate weak 2DEG scattering from the network barriers, therefore the electron confinement (visible when probing with STS) should be relatively weak within the pores. Thus, the electronic structure corresponds to highly coupled QDs. Even though interesting robust half metallicity properties were also predicted in the free-standing TPyB+Cu~\cite{Zhao2015}, no such features appear for this and similar networks on Cu(111)~\cite{Piquero2019c,Zhou2020}.\\
\indent
EPWE simulations based on the ARPES and STS datasets of the TPyB+Cu network, show again a significant 2DEG renormalization to account for the 2DEG downshift and the existence of molecular potential barriers exceeding that of Cu atoms ($V_{TPyB}=250$~meV, $V_{Cu}=50$~meV, see Table~\ref{TableEPWE}). These heterogeneous scattering potentials were corroborated by electrostatic potential (ESP) maps obtained from DFT calculations at the largest probability density of the Cu(111) surface state region~\cite{Echenique2004} [see Figure~\ref{figureTPyB}(f)]. The potential line profiles extracted from these calculations show that at the Cu coordination nodes the outer rim is weakly repulsive, but quickly reverses its character towards its center. Since these Cu adatoms are located at network sides, they represent transmission channels between adjacent pores, yielding significant interdot coupling. This scenario does not occur in purely organic networks (without coordination atoms) where the ESP maps typically display a very homogeneous repulsive potential barrier landscape [see the Br-DNT case in Figure~\ref{figureTPyB}(g)]. Accordingly, the QD intercoupling is hindered and larger energy gaps and band flattening occurs in this SW network (cf. Fig.~\ref{figureSWDW}). Note that no downshift of the fundamental energy was found for purely supramolecular halogen bonded nanoporous networks, substantiating an origin related to metal-organic coordination nodes.
\begin{table*}
	\begin{center}
		\begin{tabular}{|c||c|c|c|c|c|c|c|}
                         \hline        
			Network Type & Substrate & V$_{molecule}$ & V$_{adatom}$ & E$_0$ & $m^*$/$m_e$ & Technique & Reference  \\
			$$[O/MO] &  & [meV] & [meV] & [eV] &  &  &\\
                        \hline \hline
			2H-TPP Trenches [O] & Ag(111)&300& ---& -0.065& 0.42 & STS &~\cite{Seufert2013} \\
			\hline 
			
			Ph6 Kagome [O] & Ag(111)&500& ---& -0.065& 0.42 & STS &~\cite{Barth2009} \\
			\hline
			Ph4+Co/Ph6+Co [MO] &Ag(111)& 500& -50 & -0.065& 0.42 & STS &~\cite{Florian2011} \\
			\hline
                                SW/DW [O] &Ag(111)& 140& --- & -0.065& 0.49/0.54 & STS/ARPES &~\cite{Piquero2017} \\
			\hline
                                 Ph3+Co/Ph6+Co [MO] &Au(111)& 250& 50 & -0.52/-0.56& 0.22/0.21 & STS/ARPES &~\cite{Piquero2019b} \\
			\hline
                                BMB+Au [MO] &Au(111)& 600& 600 & -0.48& 0.26 & STS &~\cite{Colazzo2019} \\
			\hline
                                3deh-DPDI+Cu [MO] &Cu(111)& 390& 390 & -0.44& 0.49 & STS/ARPES &~\cite{Piquero2019a} \\
			\hline
                                TPyB+Cu [MO] &Cu(111)& 250& 50 & -0.53& 0.41 & STS/ARPES &~\cite{Piquero2019c} \\
			\hline
		\end{tabular}
	\end{center}
	\caption[]{Summary of the scattering potentials and surface state renormalization used in literature for the EBEM/EPWE simulations of different organic [O] and metal-organic [MO] networks discussed here. 
}
\label{TableEPWE}
\end{table*}
\begin{figure} 
\begin{center}
  \includegraphics[width=0.5\textwidth,clip]{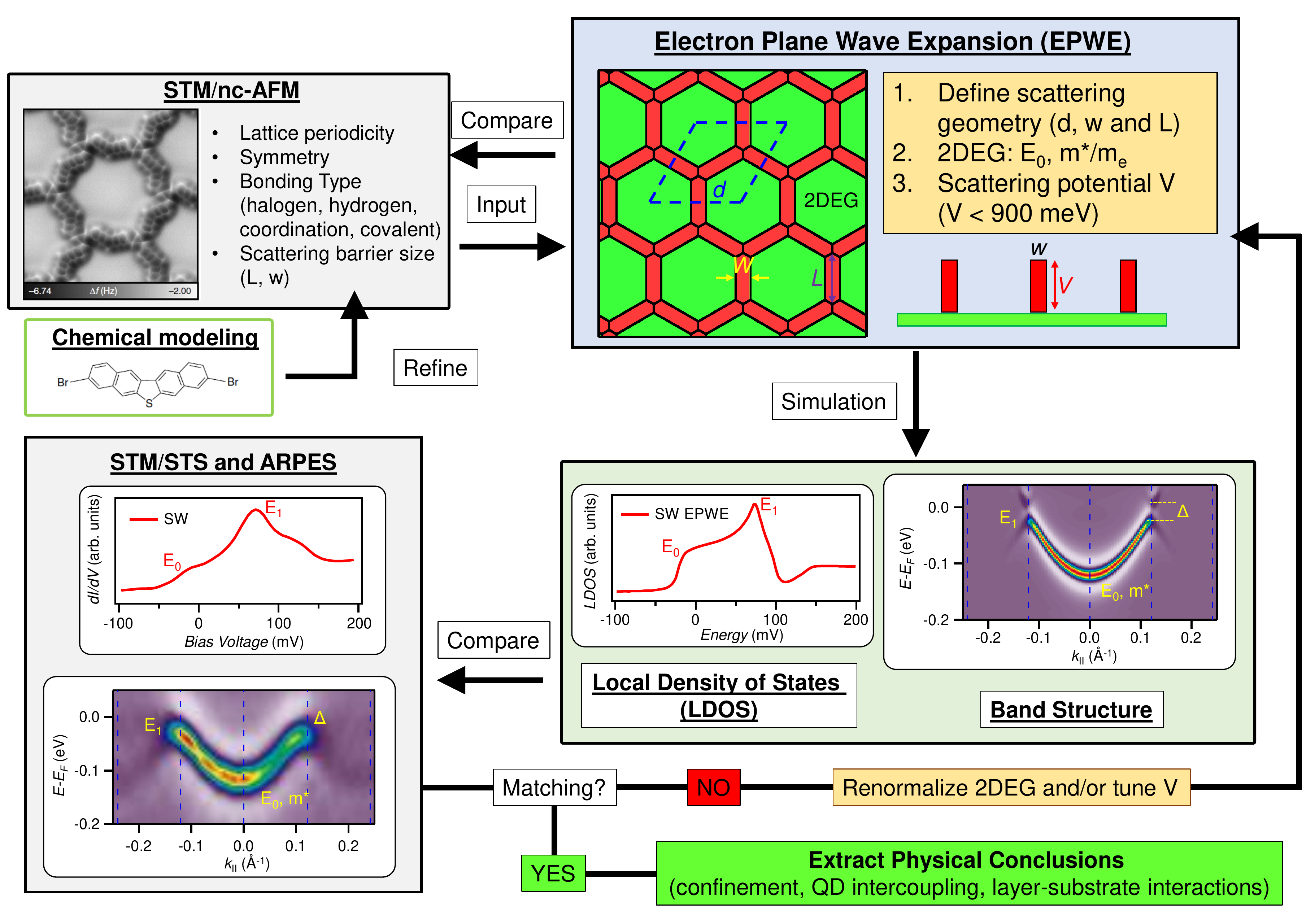}
 \end{center}
\vspace*{-5 mm}
\caption[]{Proposed iterative fitting procedure for semiempirical EPWE simulations based on surface sensitive experimental techniques (\textit{i.e.} STM/STS, nc-AFM and ARPES). This systematic simulation process should lead to physically meaningful conclusions from any molecular confining structure grown on a substrate featuring a pristine surface 2DEG.
}
 \label{figureMachinelearning}
\end{figure}
\subsection{E. Systematics of electron confinement and QD array bands using semiempirical simulations}
\indent
As we have seen repeatedly, semiempirical models are recurrently used to assess the complex scattering potential landscapes arising from these QD arrays. The analyzed organic and metal-organic systems require the use of the EBEM for finite structures  (local density of states) and/or the EPWE for periodic arrays (Bloch-wave states)~\cite{Florian2011,GarciaAbajo2010,Zakaria2019}. Note that EPWE is a very strong predictive tool  not only for single layer molecular networks on surfaces, but also for atomic 2D materials (\textit{e.g.}, graphene and \textit{h}-BN), graphene nanoribbons, polymers and single molecules~\cite{Zakaria2019,Piquero2018,Zakaria2020}.\\
\indent
EPWE and EBEM, as semiempirical methods, require several starting assumptions regarding the potential barrier strength, the repulsive/attractive condition and the geometry. At times they lead to what \textit{a priori} may look as arbitrary or un-physical conditions, such as unrealistically thin molecular backbones (resulting in excessively high potentials), attractive scattering potential regions at the metal sites or enlarged effective masses compared to the pristine 2DEG.
The assumptions considerably improve when results from different experimental techniques are at hand (\textit{i.e.}, STM/STS, nc-AFM and ARPES). In such ideal cases the simulations can accurately reproduce the LDOS (local 2DEG confinement), the QD band structure (interpore coupling) and the 2DEG energy and mass renormalization.  Several simulations based on extended experimental datasets that were presented in this review are summarized in Table~\ref{TableEPWE}.\\
\indent 
The simulation procedure follows an iterative fitting of different parameters that should systematically lead to a unique physically meaningful solution. Figure~ \ref{figureMachinelearning} displays the recursive process to obtain the scattering potential landscapes based on experimental datasets of electron confining structures:
\begin{enumerate}
\item{{\it Setting of the scattering geometry.} A model of the network (\textit{i.e.}, lattice constant and symmetry), as well as molecular barrier size (L, w) is accurately extracted from STM and nc-AFM images and refined by chemical structures.}
\item{{\it Definition of the 2DEG.} Once the geometry is fixed, parameters from the pristine case are initially used based on their fundamental energy (E$_0$) and effective mass ($m^*/m_e$) (see Table~\ref{TableSurfaceStates}).}
\item{{\it Adjusting repulsive scattering barriers.} These potential barriers are initially repulsive and limited to $V_{mol}<900$~meV for carbon-based molecules. Note that this value depends on the molecular backbone size (related to step 1). Whenever metal adatoms are present in the network, a similar potential is assigned, which will likely need reduction in following iterations. Note that $V_{adatom}$ will be positive (repulsive), unless bound states are experimentally detected, which has never been the case in porous MOCNs.}
\item{{\it Simulate the experimental LDOS and band structure.} This is the algorithm core: Vary $V_{mol}$, $V_{adatom}$ and the 2DEG renormalization ($E_0$ and $m^*/m_e$)  to match the fundamental energy ($E_0$) of the QD band, the top of the $n=1$ LDOS peak, the ARPES band ($E_1$) and the gap size between $n=1$ and $n=2$ confined states ($\Delta$). Once the 2DEG has been renormalized to a proper value, $E_0$ and gap size ($\Delta$) values depend on $V_{mol}$ and $V_{adatom}$, whereas STS peak-width and ARPES band-width (\textit{i.e.}, $E_1$-$E_0$) depend also on $m^*/m_e$. Hence, a refinement of $V_{mol}$ and 2DEG renormalization ($E_0$ and $m^*/m_e$) should be carried out iteratively until a satisfactory match with ARPES and STS is found. A more accurate fitting can be obtained whenever higher confined states are introduced by the experiments~\cite{Piquero2019a}.}
\item{{\it Extract the parameters once the iteration process does not improve the simulations further.} These parameters reflect physically meaningful information on confinement strength, QD intercoupling and energy and mass renormalization effects from the studied network.}
\end{enumerate}
\indent
Alternatively, it is tempting to envision the inversion of this iterative processes and start by proposing a set of desired electronic features to be realized. In other words, we could consider programming algorithms that would work inversely by yielding experimentally feasible geometries prone to host specific electronic structures. Particularly, theoretical methods have already been proposed that addresses such an ‘inverse approach’ for finding atomic configurations that produce a prescribed electronic structure~\cite{Franceschetti1999}. In this context, descriptors containing the demonstrated electronic scattering and coupling introduced by artificial lattices ~\cite{Gomes2012, Slot2017, Kempkes2019a, Swart2020, Kempkes2019, Collins2017} and/or quantum corrals~\cite{Li2020b, Freeney2020} could be conceived to develop simplified geometries and engineer nanoarchitectures  that target specific electronic effects. Indeed, machine learning procedures are currently at their initial stages and could steer the material research through such data-driven codes that rely on supervised training of the algorithms~\cite{Ourmazd2020,Hormann2019}. Interestingly, such computer coded methods have been recently applied to SPM-based experiments so as to decide upon the data quality and system morphology~\cite{Alldritt2020}, that are ultimately capable of autonomous operation~\cite{Krull2020}. \\
\indent
In short, we anticipate that machine learning algorithms could be implemented into the EBEM/EPWE simulations to train and develop new protocols capable of providing the desired band structures from physically meaningful scattering potential landscapes \textit{ i.e.}, feasible molecular superlattices. This could be combined with recently developed first-principles/machine-learning algorithms capable of predicting the crystal structure of certain organic molecules on metal surfaces~\cite{Hormann2019}. When such geometries are defined, synthetic chemists could fabricate the proposed molecular building blocks and conceive fabrication protocols to be tested experimentally.\\
\section{V. Mutual response of guest species and confined quantum states}
\indent
We have demonstrated that interfacial 2DEG can be confined and tailored by molecular nanostructures following a proper selection of constituent materials and realization by interfacial synthesis protocols. The energy position of the resonance states and bands, the intercoupling and the final geometry can be defined by the molecular  design and the nature of coordination nodes in metallo-supramolecular constructs. However, whenever the confining structure is already set, there is an additional route to tune these electronic states: the deliberate positioning of  adsorbed guest species within the confining structures, which can be in the form of organic molecules, single atoms or atom clusters~\cite{Negulyaev2008, Bartels2010,Pivetta2013, Beton2003,  Beton2010, DeFeyter2016, Nowakowska2015}.\\
\indent
In the following we describe exemplary cases that reveal how adsorbed species are influenced by the confined LDOS and, in turn, the confined states respond to the presence of adsorbed guest molecules and atoms, similar to single atoms placed in quantum corrals, discussed in the introduction~\cite{Stepanyuk2005,Kliewer2000,Li2020b,Stilp2021}. 
We notably address the interplay regarding organization and electronic response that transition metal atoms, simple molecules or noble gases undergo in the confined 2DEGs of 1D-nanogratings and QD arrays.
\begin{figure}
\begin{center}
  \includegraphics[width=0.45\textwidth,clip]{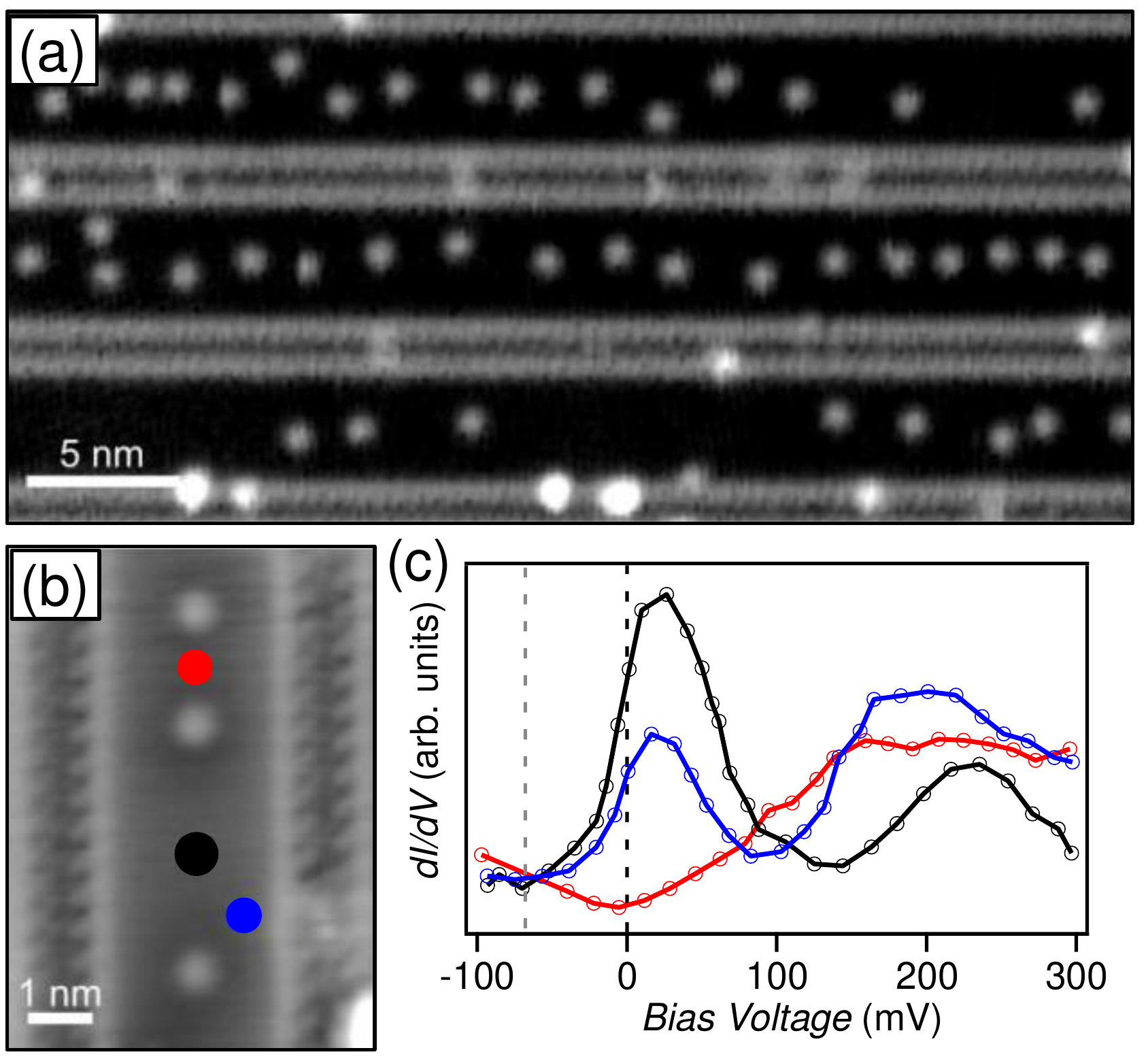}
 \end{center}
\vspace*{-5 mm}
\caption[]{Transition metal adatoms self-aligning within $L$-methionine 1D-nanogratings and modification of the electronic structure (from 1D QWS to 0D QDs). ({\bf a}) Atomic chains of Co preferentially following the central axis of $L$-methionine trenches after deposition at 8 K and subsequent annealing to 18 K. ({\bf b}) Tip manipulation of Fe atoms into selected interatomic distances within the 1D-nanogratings. ({\bf c}) The STS at the indicated position in ({\bf b}) reveal that the atoms act as scattering barriers for the 1D-confined surface state and yield 0D confinement (black and blue spectra) or quenching of the confined state (red spectrum). Adapted from ~\cite{Schiffrin2008}.
}
 \label{fGuest1D}
\end{figure}
\subsection{A. Self-alignment of adspecies in molecular nanogratings}
\begin{figure} 
\begin{center}
  \includegraphics[width=0.45\textwidth,clip]{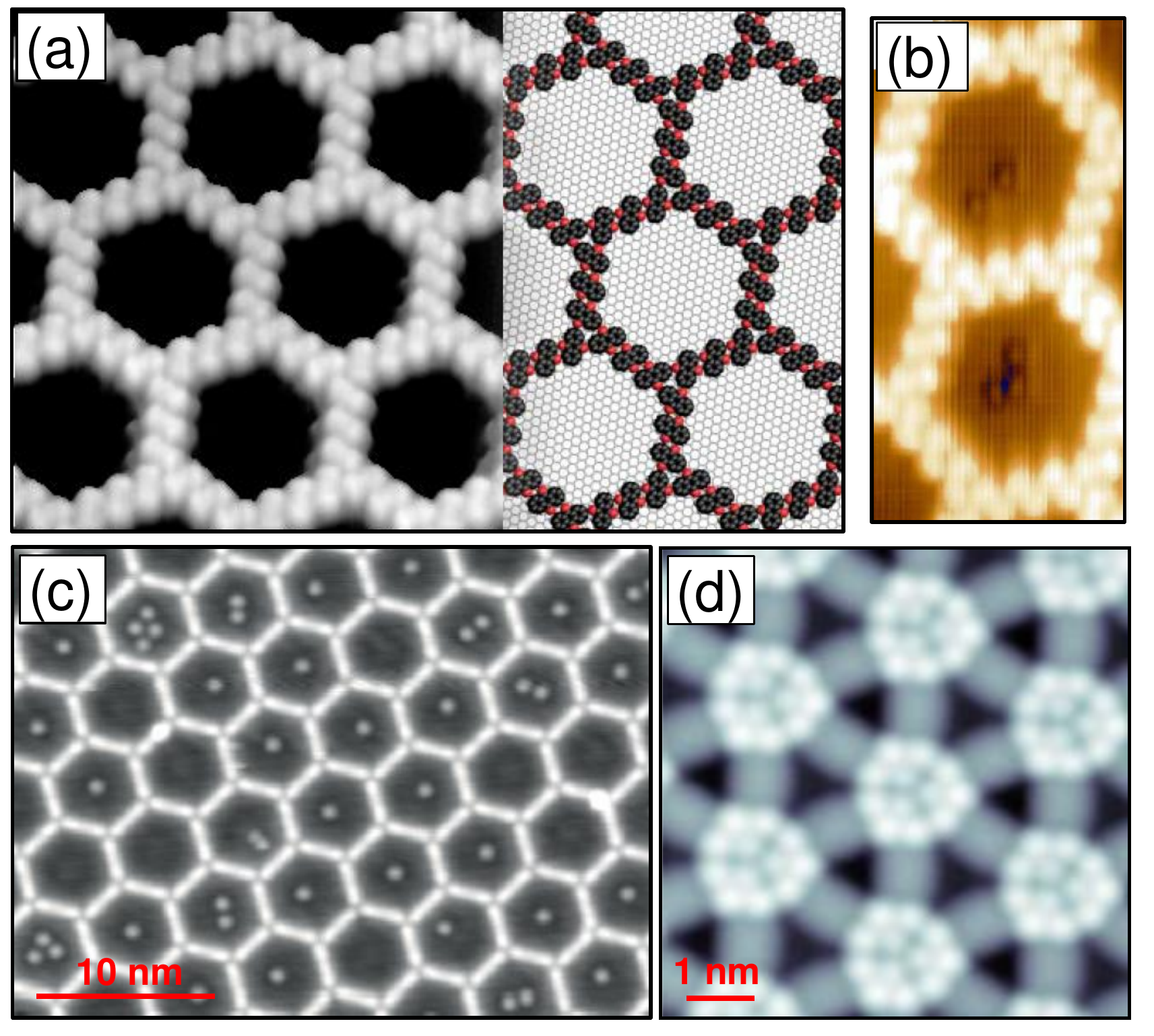}
 \end{center}
\vspace*{-5 mm}
\caption[]{Adsorbed species guided by the confined LDOS in QD arrrays. ({\bf a}) Overview topograph of the chiral anthraquinone network generated on Cu(111) (adapted from~\cite{Pawin2006}). ({\bf b}) Adsorption of CO molecules within the network pores (dark protrusions) where their diffusion becomes defined by the confined LDOS  (adapted from~\cite{Bartels2010}). ({\bf c}) Fe single atoms packing on Ph5+Cu MOCN on Cu(111)  (adapted from~\cite{Pivetta2013}). ({\bf d}) Xe adsorption on the 3deh-DPDI network, showing a maximum packing of 12-atom in three-fold bunches (adapted from~\cite{Nowakowska2015}).
}
 \label{fGuestQDs}
\end{figure}

\indent
The deposition of 3$d$-metals (Co and Fe) into the 1D-organic nanogratings previously described (cf. Fig.~\ref{figureMethionine}) was investigated at low temperatures to follow the electronic structure modifications and the dynamics of the system~\cite{Schiffrin2008}. Deposition at 8 K of Fe or Co atoms continued by a subtle annealing to 18 K, modifies the randomly distributed guest atom positions, which then self-align as atomic 1D chains at the center of the furrows in the molecular nanogratings [cf. Fig.~\ref{fGuest1D}(a)]. Accordingly, this two-stage assembly scenario merges supramolecular organization principles with spontaneous adatom positioning steered by indirect interactions. Interestingly, at sufficiently high local coverage, a preferred distance between individual atoms of 23~\AA{} and 25~\AA{} results for Fe and Co atoms, respectively. This separation reveals a long-range interaction (row-adatom and adatom-adatom) mediated via the QWS, similarly to 2D hexagonal atomic lattices existing on pristine substrates~\cite{Knorr2002, Repp2000, Silly2004,Negulyaev2009}.\\
\indent
Notably, 0D confinement is found for this system since the guest adatoms act as repulsive scattering barriers to the 1D-confined surface state electrons (the driving force of the atom self-organization). This is evidenced at the LDOS when further tip manipulation procedures are applied to systematically vary the interatomic spacings at the center of a trench, which allows for the fabrication of the tiniest quantum dots [Fig.~\ref{fGuest1D}(b)] ~\cite{Pennec2007}. The STS lineshapes in  Fig.~\ref{fGuest1D}(c) display the familiar QD signatures of Fig.~\ref{figPh6}(e) instead of the QWS of Figs.~\ref{figureSeufert} and \ref{figureMethionine}. Minima in the interaction energy between the atoms and the resonator boundaries are theoretically expected and observed for both metals when located at the center of the 1D-trenches~\cite{Schiffrin2008, Weiss2012}. In essence, employing self-aligning atomic strings, we can modify the dimensionality and onsets of the confined states. This methodology can be generally employed for nanoscale control of matter and for the positioning of single atomic or molecular species in surface-supported supramolecular architectures. In this regard, the positioning and mobility of tetracene molecules in such methionine nanogratings has been studied~\cite{Urgel2016}. In addition, it was demonstrated that the spin polarization of surface electrons caused by magnetic adatoms can be projected to a remote location by quantum states of corrals~\cite{Stepanyuk2005}. In particular, the exchange interaction between magnetic atoms is operative at appreciable distances and a similar behavior is expected for magnetic atoms organized in such 1D nanogratings or QD arrays~\cite{Pennec2007, Pivetta2013} [see Fig.~\ref{fGuest1D}(a) and Fig.~\ref{fGuestQDs}(c)].
\subsection{B. Guided adsorption of adatoms and simple molecules in QD arrays}
\indent 
The presence of a 2DEG can strongly influence the self-assembly processes of the adsorbed species~\cite{Wang09}. This is also the case whenever the surface state is confined. As a first example we show the adsorption of CO molecules at the pores of the chiral anthraquinone (AQ) network grown on Cu(111)~\cite{Bartels2010} [see Fig.~\ref{fGuestQDs}(a), (b)]. This AQ network presents one of the largest regular pore areas since the self-assembly is mediated by C-H-O interactions that involve 18 molecules~\cite{Pawin2006, Bartels2010, Wyrick2011}. 
At 40 K, the diffusion of the CO molecules is restricted to a single pore and dynamic processes confirm their repulsive character~\cite{Bartels2010}. Indeed, a hierarchy of preferred adsorption sites prevails, depending on the number of CO molecules captured within a pore. For a single CO molecule, the center of the network cavity is favored and maintained whenever a second CO molecule is added, which then sits on a midway site between the pore center and edge. Higher number of captured COs  will preferentially occupy any available halfway adsorption site. Simulations on this system conclude that the first and second confined states (in this order) drive the CO adsorption positions~\cite{Bartels2010}, so CO molecules seize adsorption sites with high LDOS. Therefore, the diffusion is not limited just by the physical barriers (molecules), but also by the modified LDOS generated by the QD array~\cite{Einstein2018}.\\
\indent
Similarly, the arrangement of Fe atoms follows a related scheme when filling the pores of dicyano-poly(p-phenylene) based MOCNs~\cite{Pivetta2013}. Particularly, the adsorption of Fe atoms onto a Ph5+Cu MOCN assembled on Cu(111) preferentially follows electron-mediated interactions between the adatoms, which are reinforced by the cavity confined surface electrons [Fig.~\ref{fGuestQDs}(c)]. Thus, a certain control over the guest species assemblies is exerted by proper selection of the network used as confining template. Similar to the atomic chains in supramolecular nanogrids described above, this process requires thermal activation, and upon overcoming the formation energy barrier discrete Fe clusters in the pores are obtained.\\
\indent
The last example considered studies the adsorption of Xe noble gas atoms  into the well characterized DPDI+Cu network~\cite{Nowakowska2015, Nowakowska2016, Jung2018, Jung2019}. The atom-by-atom condensation leads to a maximum occupation of 12 Xe guests that do not follow a single set of hierarchic filling rule, but adapt their structures with their neighbors~\cite{Nowakowska2015, Nowakowska2016} [see Fig.~\ref{fGuestQDs}(d)]. This pore saturation exhibits tetramer grouping with Xe atoms adsorbed at on-top sites of the Cu(111) atomic lattice, resulting in a ($\sqrt{3}\times\sqrt{3}$)R30\de{} overlayer structure that coincides with the direct adsorption of Xe on pristine Cu(111)~\cite{Seyller1998}. This tetramer grouping  matches the three-fold symmetry of the DPDI network with respect to the substrate~\cite{Piquero2019a}. In other words, the $n=2$ confined state --closest to the Fermi energy-- appears to guide the Xe condensation. Alternatively, these Xe atoms could be just marking the three equivalent metal coordination sites that exhibit the lowest surface potential of the network~\cite{Dil2008a, Piquero2019a}.\\
\begin{figure*} 
\begin{center}
  \includegraphics[width=0.9\textwidth,clip]{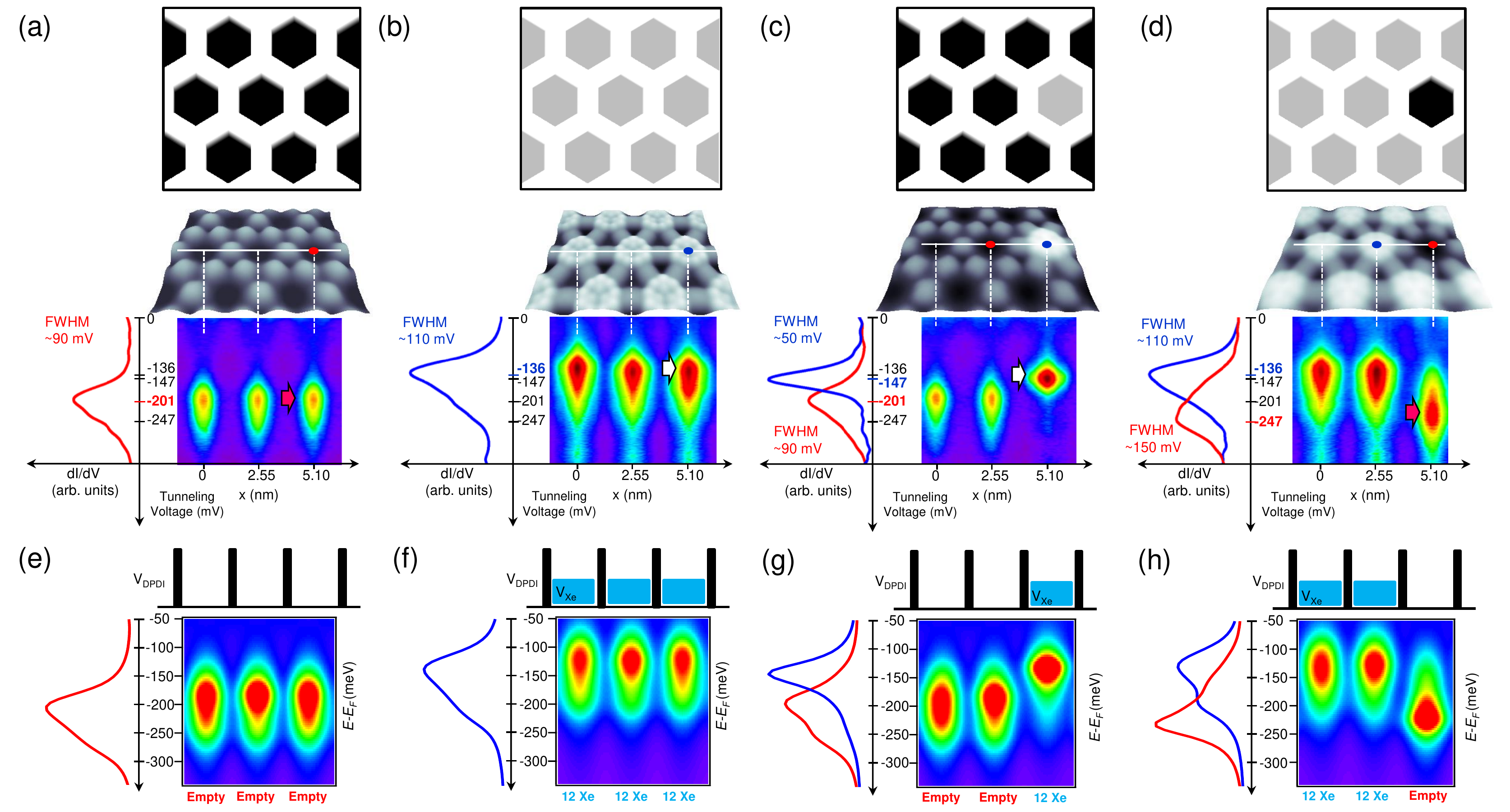}
 \end{center}
\vspace*{-5 mm}
\caption[]{QD electronic structure alteration by pore occupation in the 3deh-DPDI network. Depending on the Xe occupation of the pores, four different  configurations are shown: Empty network ({\bf a}), saturated network ({\bf b}), single filled pore  ({\bf c}) and single empty pore ({\bf d}). These panels contain the experimental information where the STM images are displayed in the top, a $dI/dV$ linescan passing through the center of three neighboring pores can be found below and position dependent STS on the left.  The bottom row (panels {\bf (e) - (h)}) display the EPWE simulated LDOS (line scans and STS) for the upper cases. A significant modification of the LDOS is evident upon Xe occupation and depends on the state of the neighboring pores, presenting an electronic breadboard. Figure adapted from~\cite{Nowakowska2016}. 
}
 \label{figureDPDIXe1}
\end{figure*}
\indent
In this context, it is also interesting to note that any mobility of hosted species entails a temporally fluctuating confinement geometry. Thus monomers or trimeric units caged in nanoporous honeycomb MOCNs are expected to present dynamic confinement patterns, which can vary rapidly and account for rather complex electronic configurations~\cite{Barth2010,Palma2014,Palma2015}.
\subsection{C. Configuring QD states by manipulation of guest adsorbates}
\indent
The Xe occupation and its effect on the QD electronic properties was investigated with respect to its adjacent neighbors on the DPDI+Cu network~\cite{Nowakowska2016}. This system was chosen because Xe is physisorbed within these pores and can be atomically manipulated using a STM tip. In this way, artificial intermediate situations can be created between empty and saturated pores, simulating an electronic breadboard [cf. Figure~\ref{figureDPDIXe1}(a)-(d)]. The empty pore case has been electronically addressed in Figs.~\ref{figureDPDI} and \ref{figureDPDI1} and exhibits a broad $n=1$ confined peak with maximum around $\sim -0.2$~V that relates to the first QD array band [cf. Fig.~\ref{figureDPDIXe1}(a)]. Upon Xe saturation, this confined state peak shifts by $\sim{60}$ mV towards the Fermi level. This upshift effect is attributed to Pauli repulsion between the rare gases and the confined states~\cite{Park2000,Hovel2001,Forster2004}. Note that the broad peak width remains since the Xe physisorption at the pores does not practically modify the interpore coupling, as evidenced in ARPES measurements~\cite{Nowakowska2016}. \\
\indent
Intermediate cases, achieved through atomic manipulation, yield a single filled pore  within an empty network [Fig.~\ref{figureDPDIXe1}(c)] and an empty pore embedded in a Xe saturated network [Fig.~\ref{figureDPDIXe1}(d)]. Different occupation dependent electronic states (QD states) are observed that display shifted energy maxima. In practice, the isolated filled pore shows a delta-like (non-coupled) confined state peaking close to the Xe filled network energy [Fig.~\ref{figureDPDIXe1}(b)], whereas the empty one displays a broad peak close to the fundamental energy of the DPDI+Cu network [Fig.~\ref{figureDPDIXe1}(a)]. Note that the neighboring pores appear electronically unaffected by these local cases, so for the overall system they can be considered as isolated defects. \\
\indent
EPWE was used to simulate the Xe-induced effect in these four cases and gain insight into confinement strength and QD coupling. The simulated LDOS depicted in the bottom row of Figure~\ref{figureDPDIXe1} uses scattering potentials of $V_{molecules} = V_{adatoms} = 390$~meV (from Table~\ref{TableEPWE}) and $V_{Xe}=80$~meV at the occupied pores to account for the Pauli repulsion effect induced by the Xe adatoms. Not only the energy shifts, but also the confined state peakwidths and shapes are accurately matched. From these simulations we conclude that  the presence of Xe at the pores generates an energy upshift of the confined states that maintain extended Bloch waves because the interpore coupling is not affected. The fact that the peak energies are markedly different between Xe filled and empty pores indicates that this system can be used as an electronic breadboard to store information with a bit areal density of $3.3\cdot 10^{14}$~bits/cm$^{2}$, by far exceeding current hard disk storage densities.\\
\section{VI. Conclusions and outlook} 
\indent
Throughout this review we have shown the wide range of possibilities that molecular nanoarchitectures provide for confining and engineering surface 2D electron gases. The chemical versatility provides a multitude of opportunities to generate molecular assemblies featuring the desired scattering potential barriers that control surface electrons and shape the interfacial electronic landscape. Ultimately, these quantum properties depend on the physicochemical nature of the selected building units. Although we have focused on the fcc(111)  surfaces of copper, silver and gold, the elaborated principles are extensible to many other system exhibiting quasi-free electrons on their surface, such as 3D topological insulators, surface alloys, 2D materials or thin film superconductors. To date, similar molecular structures onto such surfaces reemain practically unexplored.\\
\indent
The desired structures can be obtained  by tip manipulation, by molecular self-assembly protocols, or a combination of both. Such building methods have substantially expanded the available toolbox initially limited to atom corrals and vicinal surfaces. We have  shown that the molecular structures are extremely versatile, whereby all the confining parameters and the QD intercoupling degree among repeating units can be addressed. This bears the prospect of efficient fabrication of confining structures at the technological level and miniaturization for device integration.\\
\indent
The experimental and theoretical cases we reviewed reveal the deep insight into the scattering potential configurations that molecular nanoarchitectures and networks represent for surface electrons. While molecules typically exhibit strong repulsive potential barriers for the 2DEG electrons, adatoms are less predictable, and are generally weaker in their scattering strength. Nonetheless, they present the remarkable property of renormalizing significantly the pristine surface state onset and as coordination nodes can open channels to enhance the QD intercoupling. In essence, molecule-based networks with pores modulate the 2DEG through the potential barriers they produce and can generate distinct band structures that can be easily engineered through self-assembly protocols. \\
\indent
We also showed that the positioning of guest species at nanogrids or network pores provides an extra route to alter the confined electronic states. Systems with significant energy differences between occupied and empty states could be also used as breadboards to engineer artificial lattices. These electronic breadboards are envisioned as operating COFs or nanoporous graphenes (NPGs) as they are stable and manipulable at room temperature.\\
\indent
The semiempirical methods used to model these networks have significantly improved over the years and can convincingly reconstruct the scattering potential landscape responsible for the 2DEG confinement. Nowadays we can envision the use of machine learning to  propose molecular geometries that induce the electronic properties we ultimately desire. Such machine learning processes could be based on the inverse EPWE methodology discussed. This procedure could guide the design of novel artificial molecular lattices beyond the ones already explored.\\
\begin{figure} [t!]
\begin{center}
  \includegraphics[width=0.5\textwidth,clip]{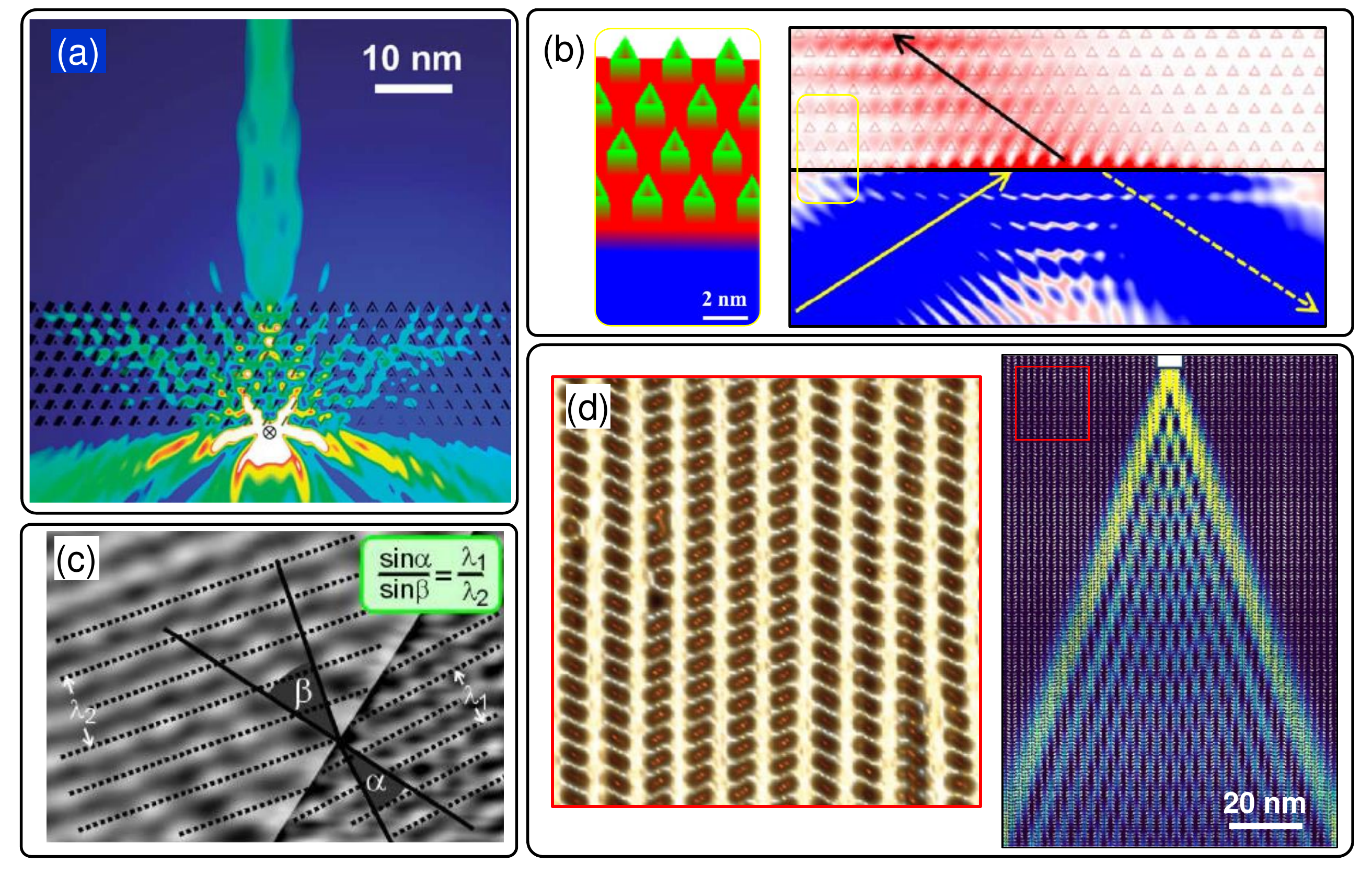}
 \end{center}
\vspace*{-5mm}
\caption[] {Mimicking analogous optical effects with electrons using organic and inorganic nanostructures. ({\bf a, b}) Simulations of electron collimation and negative refraction at Cu(111) and 1 ML Ag/Cu(111) lateral interfaces (adapted from~\cite{GarciaAbajo2010, Zaka2017}). ({\bf c}) Experimental verification of Snell's law at a Cu(111) and 2 ML NaCl/Cu(111) interface. The refraction of electrons wavefront (black dotted lines) is clearly visible in the real-space atomic resolution STM image (adapted from~\cite{Repp2004}). ({\bf d}) Diffractive Talbot effect propagation simulated on NPG arrays (adapted from~\cite{Moreno2018, Calogero2019}).
}
\label{figOptics}
\end{figure}
\indent
The survey of the many examples reveals that a rationale is about to be established in this field, which is reaching its maturity. It provides nowadays the ability, ingredients and methodology to generate well-defined molecular (and atomic) nanoarchitectures that can open up many future opportunities for interfacial electronic structure and functionality design. 
\\
A vast playground is at hand to transfer, for example, optical effects into low-dimensional quantum systems given our ability to directly image the QPI patterns generated by these nanostructures, with the benefit of a 1000-fold reduction in wavelength and structural parameters~\cite{GarciaAbajo2010}.  For instance, exotic refraction anomalies leading to few nanometer electron focusing (collimation) or negative refraction and beam splitting has been predicted for triangular superlattices [see Figure~\ref{figOptics}(a,b)]~\cite{GarciaAbajo2010, Zaka2017}. Optical-like effects, such as lateral electron refraction have already been observed in simpler 2D systems, allowing the validation of Snell's law for electronic waves~\cite{Repp2004} [see Figure~\ref{figOptics}(c)]. Likewise, diffraction and interference patterns analogous to photons in coupled waveguides (Talbot effect) have been predicted in NPGs due to the presence of Dirac cones in the band structure~\cite{Moreno2018,Calogero2019} [Figure~\ref{figOptics}(d)]. The existence of these topological signatures in artificial lattices and networks should promote similar optical-like properties in selected 2DEG scattering nanostructures. A relevant prospect is the replication of the quantum holography concept by means of artificial lattices  that overcame the single-atom limit for information storage density~\cite{Manoharan2009}. Indeed, the introduction of magnetism into the system could double the information density as compared to the volumetric quantum holographic encoding, which could be experimentally accessible using a spin-polarized STM~\cite{Stepanyuk2012}.\\
\begin{figure} [t!]
\begin{center}
  \includegraphics[width=0.5\textwidth,clip]{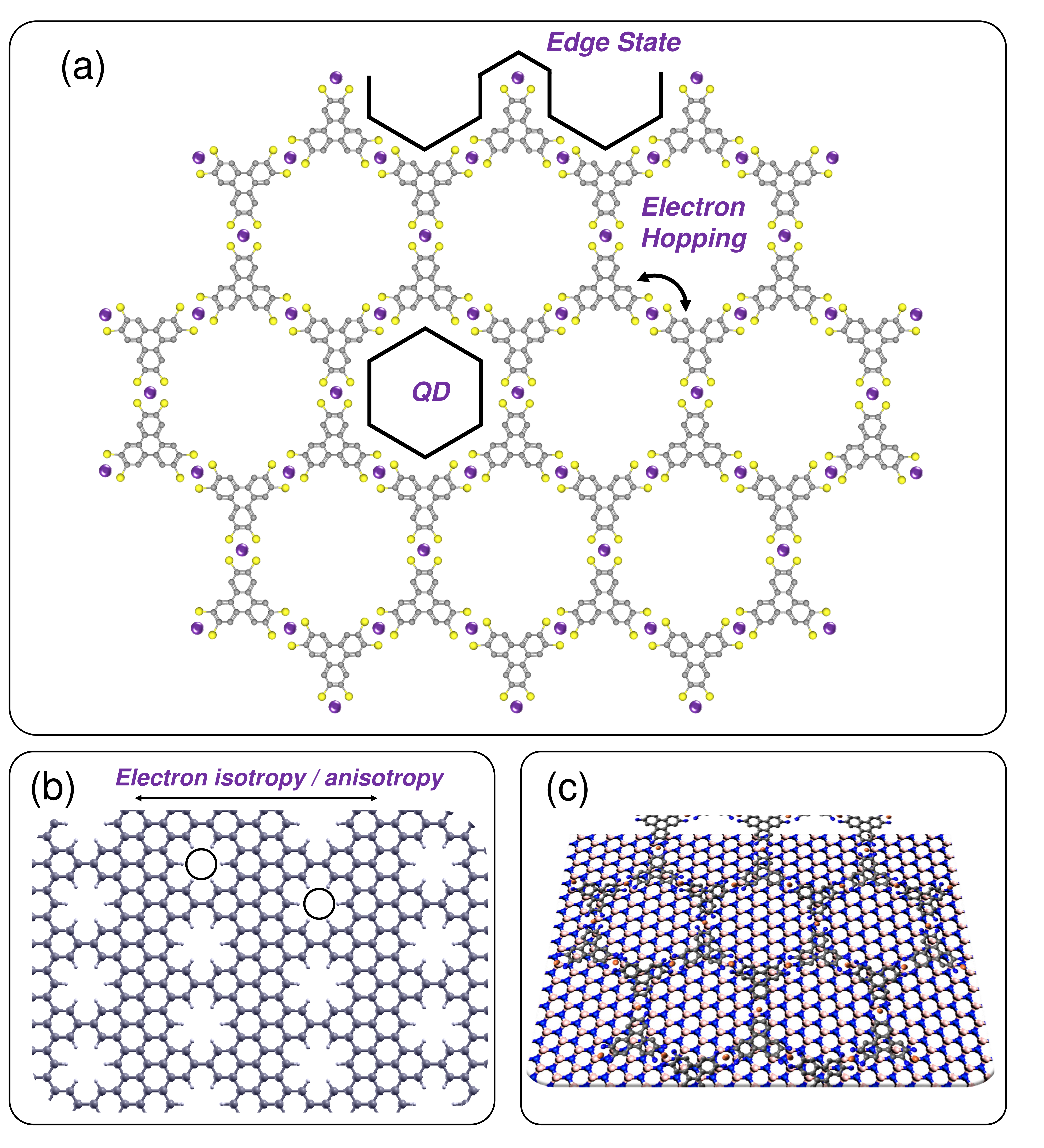}
 \end{center}
\vspace*{-5mm}
\caption[] {Perspectives of (metal-)organic nanostructures going beyond 2DEG scattering. ({\bf a}) Several MOCNs have been predicted to exhibit topological edge states and collective magnetic properties (superconductors or ferromagnets). 
({\bf b}) 2D-COFs with nanopores are expected to form from 2D-MOFs. The geometry of the pores can introduce electronical anisotropy into the system, but more importantly, the pores should bring extra functionality into devices. 
({\bf c}) Bringing these nanostructures to selected non-metallic substrates, such as topological insulators or transition metal dichalcogenides, should lead to the emergence of novel quantum states of matter.
}
 \label{figConclusion}
\end{figure}
\indent
Novel quantum properties are also envisioned by further engineering these 2DEGs. In particular, electronic scattering and confinement effects should be transferable to image potential states, that are accessible via STS or 2PPE techniques~\cite{Fauster2014, Echenique2004, Schouteden2012}. Likewise, these network modified 2DEGs have been recently proposed to play a key role in the stabilization and significant energy downshifting of super-atom molecular orbital (SAMO) states~\cite{Kawai2021}. These states of molecular origin feature a very localized and high DOS at small pores ($\sim{0.5}$ nm wide) that could be envisioned for further applications~\cite{Zhang2016a, Hieulle2018, Moreno2018,Kawai2021}.\\
\indent
From a fundamental perspective, further action is required  beyond the 2DEG scattering and confinement properties in these molecule-based networks. For example, validation of predicted MOCNs as organic topological insulators, superconductors, quantum spin liquid systems or ferromagnets still awaits~\cite{Zhang2016, Kumar2018, Zhang2020, Gao2019, Sun2018, Dong2016, Wang2013, Gao2020, Zhang2017, Hernandez2021} [Fig.~\ref{figConclusion}(a)]. Likewise, the synthesis of atomically perfect, periodic and extended nanoporous 2D-COFs is still experimentally pending~\cite{Bieri2009, Galeotti2020}. Formation of such topological array should likely originate from a particular, pre-arranged, well-defined and extended nanoporous network. The expected technological relevance of such covalent networks is huge since they could be used as 2D stamps, so that these structures could be transferred to other relevant supports. Also, as illustrated in Fig.~\ref{figConclusion}(b), the pores could introduce anisotropy into the electronic transport (as for NPG structures) or provide functionality in the form of gas selectivity or catalysing activity in industrially relevant processes~\cite{Moreno2018,Jacobse2020}.\\ 
\indent
From a practical point of view these emerging quantum states of matter should be made extensible to non-metal substrates such as 2D layered materials, topological insulators, semiconductors, superconducting substrates or hybrid structures, which are considered to be the base for future device architectures~\cite{Gobbi2018} [Fig.~\ref{figConclusion}(c)]. Steps to make this a reality are currently underway, since, for example, molecular networks have been used to alter the graphene properties by inducing an extended and atomically precise covalent coupling between layers~\cite{Yu2020}. All in all, (metal-)organic nanoporous networks as the ones here described bear the prospect of being excellent technological candidates for the quantum era ahead of us.\\ 

	
\section{Acknowledgements}
IPZ, JLC and EO acknowledge the financial support from the Spanish Ministry of Economy, Industry and Competitiveness (MINECO, Grant No. PID2019-107338RB-C6-3 and  PID2019-107338RB-C6-4) from the regional Government of Aragon (Grant No. E12-20R) and the Basque Government (IT-1255-19) and from the European Regional Development Fund (ERDF) under the program Interreg V-A España-Francia-Andorra (Contract No. EFA 194/16 TNSI).\\
WA acknowledges funding from the German Research Foundation (DFG, Heisenberg professorship) and the ERC Consolidator Grant NanoSurfs, and JVB support via DFG (BA 3395/2-1), ERC (Advanced Grant MolArt), as well as the DFG Excellence Cluster Munich Center of Quantum Science and Technology (MCQST).\\
We are grateful to all graduate students, cooperation partners and colleagues coauthoring original work.


\begin{thebibliography}{302}
\expandafter\ifx\csname natexlab\endcsname\relax\def\natexlab#1{#1}\fi
\expandafter\ifx\csname bibnamefont\endcsname\relax
  \def\bibnamefont#1{#1}\fi
\expandafter\ifx\csname bibfnamefont\endcsname\relax
  \def\bibfnamefont#1{#1}\fi
\expandafter\ifx\csname citenamefont\endcsname\relax
  \def\citenamefont#1{#1}\fi
\expandafter\ifx\csname url\endcsname\relax
  \def\url#1{\texttt{#1}}\fi
\expandafter\ifx\csname urlprefix\endcsname\relax\def\urlprefix{URL }\fi
\providecommand{\bibinfo}[2]{#2}
\providecommand{\eprint}[2][]{\url{#2}}

\bibitem[{\citenamefont{Klitzing et~al.}(1980)\citenamefont{Klitzing, Dorda,
  and Pepper}}]{Klitzing1980}
\bibinfo{author}{\bibfnamefont{K.~v.} \bibnamefont{Klitzing}},
  \bibinfo{author}{\bibfnamefont{G.}~\bibnamefont{Dorda}}, \bibnamefont{and}
  \bibinfo{author}{\bibfnamefont{M.}~\bibnamefont{Pepper}},
  \bibinfo{journal}{Phys Rev Lett} \textbf{\bibinfo{volume}{45}},
  \bibinfo{pages}{494} (\bibinfo{year}{1980}).

\bibitem[{\citenamefont{Shechtman et~al.}(1984)\citenamefont{Shechtman, Blech,
  Gratias, and Cahn}}]{Shechtman1984}
\bibinfo{author}{\bibfnamefont{D.}~\bibnamefont{Shechtman}},
  \bibinfo{author}{\bibfnamefont{I.}~\bibnamefont{Blech}},
  \bibinfo{author}{\bibfnamefont{D.}~\bibnamefont{Gratias}}, \bibnamefont{and}
  \bibinfo{author}{\bibfnamefont{J.~W.} \bibnamefont{Cahn}},
  \bibinfo{journal}{Phys Rev Lett} \textbf{\bibinfo{volume}{53}},
  \bibinfo{pages}{1951} (\bibinfo{year}{1984}).

\bibitem[{\citenamefont{Amano et~al.}(1986)\citenamefont{Amano, Sawaki,
  Akasaki, and Toyoda}}]{Amano1986}
\bibinfo{author}{\bibfnamefont{H.}~\bibnamefont{Amano}},
  \bibinfo{author}{\bibfnamefont{N.}~\bibnamefont{Sawaki}},
  \bibinfo{author}{\bibfnamefont{I.}~\bibnamefont{Akasaki}}, \bibnamefont{and}
  \bibinfo{author}{\bibfnamefont{Y.}~\bibnamefont{Toyoda}},
  \bibinfo{journal}{Appl Phys Lett} \textbf{\bibinfo{volume}{48}},
  \bibinfo{pages}{353} (\bibinfo{year}{1986}).

\bibitem[{\citenamefont{Heeger et~al.}(1988)\citenamefont{Heeger, Kivelson,
  Schrieffer, and Su}}]{Heeger1988}
\bibinfo{author}{\bibfnamefont{A.~J.} \bibnamefont{Heeger}},
  \bibinfo{author}{\bibfnamefont{S.}~\bibnamefont{Kivelson}},
  \bibinfo{author}{\bibfnamefont{J.~R.} \bibnamefont{Schrieffer}},
  \bibnamefont{and} \bibinfo{author}{\bibfnamefont{W.~P.} \bibnamefont{Su}},
  \bibinfo{journal}{Rev Mod Phys} \textbf{\bibinfo{volume}{60}},
  \bibinfo{pages}{781} (\bibinfo{year}{1988}).

\bibitem[{\citenamefont{Binasch et~al.}(1989)\citenamefont{Binasch, Gr\"unberg,
  Saurenbach, and Zinn}}]{Binasch1989}
\bibinfo{author}{\bibfnamefont{G.}~\bibnamefont{Binasch}},
  \bibinfo{author}{\bibfnamefont{P.}~\bibnamefont{Gr\"unberg}},
  \bibinfo{author}{\bibfnamefont{F.}~\bibnamefont{Saurenbach}},
  \bibnamefont{and} \bibinfo{author}{\bibfnamefont{W.}~\bibnamefont{Zinn}},
  \bibinfo{journal}{Phys Rev B} \textbf{\bibinfo{volume}{39}},
  \bibinfo{pages}{4828} (\bibinfo{year}{1989}).

\bibitem[{\citenamefont{Davis et~al.}(1995)\citenamefont{Davis, Mewes, Andrews,
  van Druten, Durfee, Kurn, and Ketterle}}]{Davis1995}
\bibinfo{author}{\bibfnamefont{K.~B.} \bibnamefont{Davis}},
  \bibinfo{author}{\bibfnamefont{M.~O.} \bibnamefont{Mewes}},
  \bibinfo{author}{\bibfnamefont{M.~R.} \bibnamefont{Andrews}},
  \bibinfo{author}{\bibfnamefont{N.~J.} \bibnamefont{van Druten}},
  \bibinfo{author}{\bibfnamefont{D.~S.} \bibnamefont{Durfee}},
  \bibinfo{author}{\bibfnamefont{D.~M.} \bibnamefont{Kurn}}, \bibnamefont{and}
  \bibinfo{author}{\bibfnamefont{W.}~\bibnamefont{Ketterle}},
  \bibinfo{journal}{Phys Rev Lett} \textbf{\bibinfo{volume}{75}},
  \bibinfo{pages}{3969} (\bibinfo{year}{1995}).

\bibitem[{\citenamefont{Kane and Mele}(2005)}]{Kane2005}
\bibinfo{author}{\bibfnamefont{C.~L.} \bibnamefont{Kane}} \bibnamefont{and}
  \bibinfo{author}{\bibfnamefont{E.~J.} \bibnamefont{Mele}},
  \bibinfo{journal}{Phys Rev Lett} \textbf{\bibinfo{volume}{95}},
  \bibinfo{pages}{146802} (\bibinfo{year}{2005}).

\bibitem[{\citenamefont{Bloch et~al.}(2008)\citenamefont{Bloch, Dalibard, and
  Zwerger}}]{Bloch2008}
\bibinfo{author}{\bibfnamefont{I.}~\bibnamefont{Bloch}},
  \bibinfo{author}{\bibfnamefont{J.}~\bibnamefont{Dalibard}}, \bibnamefont{and}
  \bibinfo{author}{\bibfnamefont{W.}~\bibnamefont{Zwerger}},
  \bibinfo{journal}{Rev Mod Phys} \textbf{\bibinfo{volume}{80}},
  \bibinfo{pages}{885} (\bibinfo{year}{2008}).

\bibitem[{\citenamefont{M{\"u}hlbauer et~al.}(2009)\citenamefont{M{\"u}hlbauer,
  Binz, Jonietz, Pfleiderer, Rosch, Neubauer, Georgii, and
  B{\"o}ni}}]{Muhlbauer2009}
\bibinfo{author}{\bibfnamefont{S.}~\bibnamefont{M{\"u}hlbauer}},
  \bibinfo{author}{\bibfnamefont{B.}~\bibnamefont{Binz}},
  \bibinfo{author}{\bibfnamefont{F.}~\bibnamefont{Jonietz}},
  \bibinfo{author}{\bibfnamefont{C.}~\bibnamefont{Pfleiderer}},
  \bibinfo{author}{\bibfnamefont{A.}~\bibnamefont{Rosch}},
  \bibinfo{author}{\bibfnamefont{A.}~\bibnamefont{Neubauer}},
  \bibinfo{author}{\bibfnamefont{R.}~\bibnamefont{Georgii}}, \bibnamefont{and}
  \bibinfo{author}{\bibfnamefont{P.}~\bibnamefont{B{\"o}ni}},
  \bibinfo{journal}{Science} \textbf{\bibinfo{volume}{323}},
  \bibinfo{pages}{915} (\bibinfo{year}{2009}).

\bibitem[{\citenamefont{M{\"u}ller}(1965)}]{Muller1965}
\bibinfo{author}{\bibfnamefont{E.~W.} \bibnamefont{M{\"u}ller}},
  \bibinfo{journal}{Science} \textbf{\bibinfo{volume}{149}},
  \bibinfo{pages}{591} (\bibinfo{year}{1965}).

\bibitem[{\citenamefont{Binnig and Rohrer}(1987)}]{Binnig1987}
\bibinfo{author}{\bibfnamefont{G.}~\bibnamefont{Binnig}} \bibnamefont{and}
  \bibinfo{author}{\bibfnamefont{H.}~\bibnamefont{Rohrer}},
  \bibinfo{journal}{Rev Mod Phys} \textbf{\bibinfo{volume}{59}},
  \bibinfo{pages}{615} (\bibinfo{year}{1987}).

\bibitem[{\citenamefont{Stroscio and Eigler}(1991)}]{Stroscio1991}
\bibinfo{author}{\bibfnamefont{J.~A.} \bibnamefont{Stroscio}} \bibnamefont{and}
  \bibinfo{author}{\bibfnamefont{D.~M.} \bibnamefont{Eigler}},
  \bibinfo{journal}{Science} \textbf{\bibinfo{volume}{254}},
  \bibinfo{pages}{1319} (\bibinfo{year}{1991}).

\bibitem[{\citenamefont{Damascelli et~al.}(2003)\citenamefont{Damascelli,
  Hussain, and Shen}}]{Damascelli2003}
\bibinfo{author}{\bibfnamefont{A.}~\bibnamefont{Damascelli}},
  \bibinfo{author}{\bibfnamefont{Z.}~\bibnamefont{Hussain}}, \bibnamefont{and}
  \bibinfo{author}{\bibfnamefont{Z.-X.} \bibnamefont{Shen}},
  \bibinfo{journal}{Rev Mod Phys} \textbf{\bibinfo{volume}{75}},
  \bibinfo{pages}{473} (\bibinfo{year}{2003}).

\bibitem[{\citenamefont{Lobo-Checa et~al.}(2009)\citenamefont{Lobo-Checa,
  Matena, M{\"u}ller, Dil, Meier, Gade, Jung, and St{\"o}hr}}]{Lobo2009}
\bibinfo{author}{\bibfnamefont{J.}~\bibnamefont{Lobo-Checa}},
  \bibinfo{author}{\bibfnamefont{M.}~\bibnamefont{Matena}},
  \bibinfo{author}{\bibfnamefont{K.}~\bibnamefont{M{\"u}ller}},
  \bibinfo{author}{\bibfnamefont{J.~H.} \bibnamefont{Dil}},
  \bibinfo{author}{\bibfnamefont{F.}~\bibnamefont{Meier}},
  \bibinfo{author}{\bibfnamefont{L.~H.} \bibnamefont{Gade}},
  \bibinfo{author}{\bibfnamefont{T.~A.} \bibnamefont{Jung}}, \bibnamefont{and}
  \bibinfo{author}{\bibfnamefont{M.}~\bibnamefont{St{\"o}hr}},
  \bibinfo{journal}{Science} \textbf{\bibinfo{volume}{325}},
  \bibinfo{pages}{300} (\bibinfo{year}{2009}).

\bibitem[{\citenamefont{Gambardella et~al.}(2003)\citenamefont{Gambardella,
  Rusponi, Veronese, Dhesi, Grazioli, Dallmeyer, Cabria, Zeller, Dederichs,
  Kern et~al.}}]{Gambardella2003}
\bibinfo{author}{\bibfnamefont{P.}~\bibnamefont{Gambardella}},
  \bibinfo{author}{\bibfnamefont{S.}~\bibnamefont{Rusponi}},
  \bibinfo{author}{\bibfnamefont{M.}~\bibnamefont{Veronese}},
  \bibinfo{author}{\bibfnamefont{S.~S.} \bibnamefont{Dhesi}},
  \bibinfo{author}{\bibfnamefont{C.}~\bibnamefont{Grazioli}},
  \bibinfo{author}{\bibfnamefont{A.}~\bibnamefont{Dallmeyer}},
  \bibinfo{author}{\bibfnamefont{I.}~\bibnamefont{Cabria}},
  \bibinfo{author}{\bibfnamefont{R.}~\bibnamefont{Zeller}},
  \bibinfo{author}{\bibfnamefont{P.~H.} \bibnamefont{Dederichs}},
  \bibinfo{author}{\bibfnamefont{K.}~\bibnamefont{Kern}}, \bibnamefont{et~al.},
  \bibinfo{journal}{Science} \textbf{\bibinfo{volume}{300}},
  \bibinfo{pages}{1130} (\bibinfo{year}{2003}).

\bibitem[{\citenamefont{Klappenberger}(2014)}]{Klappenberger2014}
\bibinfo{author}{\bibfnamefont{F.}~\bibnamefont{Klappenberger}},
  \bibinfo{journal}{Prog Surf Sci} \textbf{\bibinfo{volume}{89}},
  \bibinfo{pages}{1} (\bibinfo{year}{2014}).

\bibitem[{\citenamefont{Galeotti et~al.}(2020)\citenamefont{Galeotti,
  De~Marchi, Hamzehpoor, MacLean, Rajeswara~Rao, Chen, Besteiro, Dettmann,
  Ferrari, Frezza et~al.}}]{Galeotti2020}
\bibinfo{author}{\bibfnamefont{G.}~\bibnamefont{Galeotti}},
  \bibinfo{author}{\bibfnamefont{F.}~\bibnamefont{De~Marchi}},
  \bibinfo{author}{\bibfnamefont{E.}~\bibnamefont{Hamzehpoor}},
  \bibinfo{author}{\bibfnamefont{O.}~\bibnamefont{MacLean}},
  \bibinfo{author}{\bibfnamefont{M.}~\bibnamefont{Rajeswara~Rao}},
  \bibinfo{author}{\bibfnamefont{Y.}~\bibnamefont{Chen}},
  \bibinfo{author}{\bibfnamefont{L.~V.} \bibnamefont{Besteiro}},
  \bibinfo{author}{\bibfnamefont{D.}~\bibnamefont{Dettmann}},
  \bibinfo{author}{\bibfnamefont{L.}~\bibnamefont{Ferrari}},
  \bibinfo{author}{\bibfnamefont{F.}~\bibnamefont{Frezza}},
  \bibnamefont{et~al.}, \bibinfo{journal}{Nat Mater}
  \textbf{\bibinfo{volume}{19}}, \bibinfo{pages}{874} (\bibinfo{year}{2020}).

\bibitem[{\citenamefont{Yin et~al.}(2021)\citenamefont{Yin, Pan, and
  Zahid~Hasan}}]{Yin2021}
\bibinfo{author}{\bibfnamefont{J.-X.} \bibnamefont{Yin}},
  \bibinfo{author}{\bibfnamefont{S.~H.} \bibnamefont{Pan}}, \bibnamefont{and}
  \bibinfo{author}{\bibfnamefont{M.}~\bibnamefont{Zahid~Hasan}},
  \bibinfo{journal}{Nature Reviews Physics} \textbf{\bibinfo{volume}{3}},
  \bibinfo{pages}{249} (\bibinfo{year}{2021}).

\bibitem[{\citenamefont{Barth et~al.}(2000)\citenamefont{Barth, Weckesser, Cai,
  Günter, Bürgi, Jeandupeux, and Kern}}]{Barth2000}
\bibinfo{author}{\bibfnamefont{J.~V.} \bibnamefont{Barth}},
  \bibinfo{author}{\bibfnamefont{J.}~\bibnamefont{Weckesser}},
  \bibinfo{author}{\bibfnamefont{C.}~\bibnamefont{Cai}},
  \bibinfo{author}{\bibfnamefont{P.}~\bibnamefont{Günter}},
  \bibinfo{author}{\bibfnamefont{L.}~\bibnamefont{Bürgi}},
  \bibinfo{author}{\bibfnamefont{O.}~\bibnamefont{Jeandupeux}},
  \bibnamefont{and} \bibinfo{author}{\bibfnamefont{K.}~\bibnamefont{Kern}},
  \bibinfo{journal}{Angew Chem Int Ed} \textbf{\bibinfo{volume}{39}},
  \bibinfo{pages}{1230} (\bibinfo{year}{2000}).

\bibitem[{\citenamefont{Theobald et~al.}(2003)\citenamefont{Theobald, Oxtoby,
  Phillips, Champness, and Beton}}]{Beton2003}
\bibinfo{author}{\bibfnamefont{J.~A.} \bibnamefont{Theobald}},
  \bibinfo{author}{\bibfnamefont{N.~S.} \bibnamefont{Oxtoby}},
  \bibinfo{author}{\bibfnamefont{M.~A.} \bibnamefont{Phillips}},
  \bibinfo{author}{\bibfnamefont{N.~R.} \bibnamefont{Champness}},
  \bibnamefont{and} \bibinfo{author}{\bibfnamefont{P.~H.} \bibnamefont{Beton}},
  \bibinfo{journal}{Nature} \textbf{\bibinfo{volume}{424}},
  \bibinfo{pages}{1029} (\bibinfo{year}{2003}).

\bibitem[{\citenamefont{Yokoyama et~al.}(2001)\citenamefont{Yokoyama, Yokoyama,
  Kamikado, Okuno, and Mashiko}}]{Yokoyama2001}
\bibinfo{author}{\bibfnamefont{T.}~\bibnamefont{Yokoyama}},
  \bibinfo{author}{\bibfnamefont{S.}~\bibnamefont{Yokoyama}},
  \bibinfo{author}{\bibfnamefont{T.}~\bibnamefont{Kamikado}},
  \bibinfo{author}{\bibfnamefont{Y.}~\bibnamefont{Okuno}}, \bibnamefont{and}
  \bibinfo{author}{\bibfnamefont{S.}~\bibnamefont{Mashiko}},
  \bibinfo{journal}{Nature} \textbf{\bibinfo{volume}{413}},
  \bibinfo{pages}{619} (\bibinfo{year}{2001}).

\bibitem[{\citenamefont{Stepanow et~al.}(2004)\citenamefont{Stepanow,
  Lingenfelder, Dmitriev, Spillmann, Delvigne, Lin, Deng, Cai, Barth, and
  Kern}}]{Stepanow2004}
\bibinfo{author}{\bibfnamefont{S.}~\bibnamefont{Stepanow}},
  \bibinfo{author}{\bibfnamefont{M.}~\bibnamefont{Lingenfelder}},
  \bibinfo{author}{\bibfnamefont{A.}~\bibnamefont{Dmitriev}},
  \bibinfo{author}{\bibfnamefont{H.}~\bibnamefont{Spillmann}},
  \bibinfo{author}{\bibfnamefont{E.}~\bibnamefont{Delvigne}},
  \bibinfo{author}{\bibfnamefont{N.}~\bibnamefont{Lin}},
  \bibinfo{author}{\bibfnamefont{X.}~\bibnamefont{Deng}},
  \bibinfo{author}{\bibfnamefont{C.}~\bibnamefont{Cai}},
  \bibinfo{author}{\bibfnamefont{J.~V.} \bibnamefont{Barth}}, \bibnamefont{and}
  \bibinfo{author}{\bibfnamefont{K.}~\bibnamefont{Kern}}, \bibinfo{journal}{Nat
  Mater} \textbf{\bibinfo{volume}{3}}, \bibinfo{pages}{229}
  (\bibinfo{year}{2004}).

\bibitem[{\citenamefont{Barth et~al.}(2005)\citenamefont{Barth, Costantini, and
  Kern}}]{Barth2005}
\bibinfo{author}{\bibfnamefont{J.~V.} \bibnamefont{Barth}},
  \bibinfo{author}{\bibfnamefont{G.}~\bibnamefont{Costantini}},
  \bibnamefont{and} \bibinfo{author}{\bibfnamefont{K.}~\bibnamefont{Kern}},
  \bibinfo{journal}{Nature} \textbf{\bibinfo{volume}{437}},
  \bibinfo{pages}{671} (\bibinfo{year}{2005}).

\bibitem[{\citenamefont{K{\"u}hnle}(2009)}]{Kuhnle2009}
\bibinfo{author}{\bibfnamefont{A.}~\bibnamefont{K{\"u}hnle}},
  \bibinfo{journal}{Current Opinion in Colloid \& Interface Science}
  \textbf{\bibinfo{volume}{14}}, \bibinfo{pages}{157} (\bibinfo{year}{2009}).

\bibitem[{\citenamefont{Kudernac et~al.}(2009)\citenamefont{Kudernac, Lei,
  Elemans, and De~Feyter}}]{Kudernac2009}
\bibinfo{author}{\bibfnamefont{T.}~\bibnamefont{Kudernac}},
  \bibinfo{author}{\bibfnamefont{S.}~\bibnamefont{Lei}},
  \bibinfo{author}{\bibfnamefont{J.~A. A.~W.} \bibnamefont{Elemans}},
  \bibnamefont{and}
  \bibinfo{author}{\bibfnamefont{S.}~\bibnamefont{De~Feyter}},
  \bibinfo{journal}{Chem Soc Rev} \textbf{\bibinfo{volume}{38}},
  \bibinfo{pages}{402} (\bibinfo{year}{2009}).

\bibitem[{\citenamefont{Dong et~al.}(2016{\natexlab{a}})\citenamefont{Dong,
  Gao, and Lin}}]{NianLin2016}
\bibinfo{author}{\bibfnamefont{L.}~\bibnamefont{Dong}},
  \bibinfo{author}{\bibfnamefont{Z.}~\bibnamefont{Gao}}, \bibnamefont{and}
  \bibinfo{author}{\bibfnamefont{N.}~\bibnamefont{Lin}}, \bibinfo{journal}{Prog
  Surf Sci} \textbf{\bibinfo{volume}{91}}, \bibinfo{pages}{101}
  (\bibinfo{year}{2016}{\natexlab{a}}).

\bibitem[{\citenamefont{Goronzy et~al.}(2018)\citenamefont{Goronzy, Ebrahimi,
  Rosei, {Arramel}, Fang, De~Feyter, Tait, Wang, Beton, Wee
  et~al.}}]{Goronzy2018}
\bibinfo{author}{\bibfnamefont{D.~P.} \bibnamefont{Goronzy}},
  \bibinfo{author}{\bibfnamefont{M.}~\bibnamefont{Ebrahimi}},
  \bibinfo{author}{\bibfnamefont{F.}~\bibnamefont{Rosei}},
  \bibinfo{author}{\bibnamefont{{Arramel}}},
  \bibinfo{author}{\bibfnamefont{Y.}~\bibnamefont{Fang}},
  \bibinfo{author}{\bibfnamefont{S.}~\bibnamefont{De~Feyter}},
  \bibinfo{author}{\bibfnamefont{S.~L.} \bibnamefont{Tait}},
  \bibinfo{author}{\bibfnamefont{C.}~\bibnamefont{Wang}},
  \bibinfo{author}{\bibfnamefont{P.~H.} \bibnamefont{Beton}},
  \bibinfo{author}{\bibfnamefont{A.~T.~S.} \bibnamefont{Wee}},
  \bibnamefont{et~al.}, \bibinfo{journal}{ACS Nano}
  \textbf{\bibinfo{volume}{12}}, \bibinfo{pages}{7445} (\bibinfo{year}{2018}).

\bibitem[{\citenamefont{Xing et~al.}(2019)\citenamefont{Xing, Peng, Li, and
  Wu}}]{Xing2019}
\bibinfo{author}{\bibfnamefont{L.}~\bibnamefont{Xing}},
  \bibinfo{author}{\bibfnamefont{Z.}~\bibnamefont{Peng}},
  \bibinfo{author}{\bibfnamefont{W.}~\bibnamefont{Li}}, \bibnamefont{and}
  \bibinfo{author}{\bibfnamefont{K.}~\bibnamefont{Wu}}, \bibinfo{journal}{Acc
  Chem Res} \textbf{\bibinfo{volume}{52}}, \bibinfo{pages}{1048}
  (\bibinfo{year}{2019}).

\bibitem[{\citenamefont{Stöhr et~al.}(2007)\citenamefont{Stöhr, Wahl,
  Spillmann, Gade, and Jung}}]{Stohr2007}
\bibinfo{author}{\bibfnamefont{M.}~\bibnamefont{Stöhr}},
  \bibinfo{author}{\bibfnamefont{M.}~\bibnamefont{Wahl}},
  \bibinfo{author}{\bibfnamefont{H.}~\bibnamefont{Spillmann}},
  \bibinfo{author}{\bibfnamefont{L.}~\bibnamefont{Gade}}, \bibnamefont{and}
  \bibinfo{author}{\bibfnamefont{T.}~\bibnamefont{Jung}},
  \bibinfo{journal}{Small} \textbf{\bibinfo{volume}{3}}, \bibinfo{pages}{1336}
  (\bibinfo{year}{2007}).

\bibitem[{\citenamefont{K{\"u}hne et~al.}(2010)\citenamefont{K{\"u}hne,
  Klappenberger, Krenner, Klyatskaya, Ruben, and Barth}}]{Barth2010}
\bibinfo{author}{\bibfnamefont{D.}~\bibnamefont{K{\"u}hne}},
  \bibinfo{author}{\bibfnamefont{F.}~\bibnamefont{Klappenberger}},
  \bibinfo{author}{\bibfnamefont{W.}~\bibnamefont{Krenner}},
  \bibinfo{author}{\bibfnamefont{S.}~\bibnamefont{Klyatskaya}},
  \bibinfo{author}{\bibfnamefont{M.}~\bibnamefont{Ruben}}, \bibnamefont{and}
  \bibinfo{author}{\bibfnamefont{J.~V.} \bibnamefont{Barth}},
  \bibinfo{journal}{Proceedings of the National Academy of Sciences}
  \textbf{\bibinfo{volume}{107}}, \bibinfo{pages}{21332}
  (\bibinfo{year}{2010}).

\bibitem[{\citenamefont{Pivetta et~al.}(2013)\citenamefont{Pivetta, Pacchioni,
  Schlickum, Barth, and Brune}}]{Pivetta2013}
\bibinfo{author}{\bibfnamefont{M.}~\bibnamefont{Pivetta}},
  \bibinfo{author}{\bibfnamefont{G.~E.} \bibnamefont{Pacchioni}},
  \bibinfo{author}{\bibfnamefont{U.}~\bibnamefont{Schlickum}},
  \bibinfo{author}{\bibfnamefont{J.~V.} \bibnamefont{Barth}}, \bibnamefont{and}
  \bibinfo{author}{\bibfnamefont{H.}~\bibnamefont{Brune}},
  \bibinfo{journal}{Phys Rev Lett} \textbf{\bibinfo{volume}{110}},
  \bibinfo{pages}{086102} (\bibinfo{year}{2013}).

\bibitem[{\citenamefont{Nowakowska et~al.}(2016)\citenamefont{Nowakowska,
  W{\"a}ckerlin, Piquero-Zulaica, Nowakowski, Kawai, W{\"a}ckerlin, Matena,
  Nijs, Fatayer, Popova et~al.}}]{Nowakowska2016}
\bibinfo{author}{\bibfnamefont{S.}~\bibnamefont{Nowakowska}},
  \bibinfo{author}{\bibfnamefont{A.}~\bibnamefont{W{\"a}ckerlin}},
  \bibinfo{author}{\bibfnamefont{I.}~\bibnamefont{Piquero-Zulaica}},
  \bibinfo{author}{\bibfnamefont{J.}~\bibnamefont{Nowakowski}},
  \bibinfo{author}{\bibfnamefont{S.}~\bibnamefont{Kawai}},
  \bibinfo{author}{\bibfnamefont{C.}~\bibnamefont{W{\"a}ckerlin}},
  \bibinfo{author}{\bibfnamefont{M.}~\bibnamefont{Matena}},
  \bibinfo{author}{\bibfnamefont{T.}~\bibnamefont{Nijs}},
  \bibinfo{author}{\bibfnamefont{S.}~\bibnamefont{Fatayer}},
  \bibinfo{author}{\bibfnamefont{O.}~\bibnamefont{Popova}},
  \bibnamefont{et~al.}, \bibinfo{journal}{Small} \textbf{\bibinfo{volume}{12}},
  \bibinfo{pages}{3759} (\bibinfo{year}{2016}).

\bibitem[{\citenamefont{Zhang et~al.}(2015)\citenamefont{Zhang, Lyu, Chen, Lin,
  Liu, Liu, and Lin}}]{NianLin2015}
\bibinfo{author}{\bibfnamefont{R.}~\bibnamefont{Zhang}},
  \bibinfo{author}{\bibfnamefont{G.}~\bibnamefont{Lyu}},
  \bibinfo{author}{\bibfnamefont{C.}~\bibnamefont{Chen}},
  \bibinfo{author}{\bibfnamefont{T.}~\bibnamefont{Lin}},
  \bibinfo{author}{\bibfnamefont{J.}~\bibnamefont{Liu}},
  \bibinfo{author}{\bibfnamefont{P.~N.} \bibnamefont{Liu}}, \bibnamefont{and}
  \bibinfo{author}{\bibfnamefont{N.}~\bibnamefont{Lin}}, \bibinfo{journal}{ACS
  Nano} \textbf{\bibinfo{volume}{9}}, \bibinfo{pages}{8547}
  (\bibinfo{year}{2015}).

\bibitem[{\citenamefont{Nowakowska et~al.}(2015)\citenamefont{Nowakowska,
  W{\"a}ckerlin, Kawai, Ivas, Nowakowski, Fatayer, W{\"a}ckerlin, Nijs, Meyer,
  Bj{\"o}rk et~al.}}]{Nowakowska2015}
\bibinfo{author}{\bibfnamefont{S.}~\bibnamefont{Nowakowska}},
  \bibinfo{author}{\bibfnamefont{A.}~\bibnamefont{W{\"a}ckerlin}},
  \bibinfo{author}{\bibfnamefont{S.}~\bibnamefont{Kawai}},
  \bibinfo{author}{\bibfnamefont{T.}~\bibnamefont{Ivas}},
  \bibinfo{author}{\bibfnamefont{J.}~\bibnamefont{Nowakowski}},
  \bibinfo{author}{\bibfnamefont{S.}~\bibnamefont{Fatayer}},
  \bibinfo{author}{\bibfnamefont{C.}~\bibnamefont{W{\"a}ckerlin}},
  \bibinfo{author}{\bibfnamefont{T.}~\bibnamefont{Nijs}},
  \bibinfo{author}{\bibfnamefont{E.}~\bibnamefont{Meyer}},
  \bibinfo{author}{\bibfnamefont{J.}~\bibnamefont{Bj{\"o}rk}},
  \bibnamefont{et~al.}, \bibinfo{journal}{Nat Commun}
  \textbf{\bibinfo{volume}{6}}, \bibinfo{pages}{6071} (\bibinfo{year}{2015}).

\bibitem[{\citenamefont{Teyssandier et~al.}(2016)\citenamefont{Teyssandier,
  Feyter, and Mali}}]{DeFeyter2016}
\bibinfo{author}{\bibfnamefont{J.}~\bibnamefont{Teyssandier}},
  \bibinfo{author}{\bibfnamefont{S.~D.} \bibnamefont{Feyter}},
  \bibnamefont{and} \bibinfo{author}{\bibfnamefont{K.~S.} \bibnamefont{Mali}},
  \bibinfo{journal}{Chem Commun} \textbf{\bibinfo{volume}{52}},
  \bibinfo{pages}{11465} (\bibinfo{year}{2016}).

\bibitem[{\citenamefont{Kocić et~al.}(2019)\citenamefont{Kocić, Blank,
  Abufager, Lorente, Decurtins, Liu, and Repp}}]{Repp2019}
\bibinfo{author}{\bibfnamefont{N.}~\bibnamefont{Kocić}},
  \bibinfo{author}{\bibfnamefont{D.}~\bibnamefont{Blank}},
  \bibinfo{author}{\bibfnamefont{P.}~\bibnamefont{Abufager}},
  \bibinfo{author}{\bibfnamefont{N.}~\bibnamefont{Lorente}},
  \bibinfo{author}{\bibfnamefont{S.}~\bibnamefont{Decurtins}},
  \bibinfo{author}{\bibfnamefont{S.-X.} \bibnamefont{Liu}}, \bibnamefont{and}
  \bibinfo{author}{\bibfnamefont{J.}~\bibnamefont{Repp}},
  \bibinfo{journal}{Nano Lett} \textbf{\bibinfo{volume}{19}},
  \bibinfo{pages}{2750} (\bibinfo{year}{2019}).

\bibitem[{\citenamefont{Gutzler et~al.}(2015)\citenamefont{Gutzler, Stepanow,
  Grumelli, Lingenfelder, and Kern}}]{Kern2015}
\bibinfo{author}{\bibfnamefont{R.}~\bibnamefont{Gutzler}},
  \bibinfo{author}{\bibfnamefont{S.}~\bibnamefont{Stepanow}},
  \bibinfo{author}{\bibfnamefont{D.}~\bibnamefont{Grumelli}},
  \bibinfo{author}{\bibfnamefont{M.}~\bibnamefont{Lingenfelder}},
  \bibnamefont{and} \bibinfo{author}{\bibfnamefont{K.}~\bibnamefont{Kern}},
  \bibinfo{journal}{Acc Chem Res} \textbf{\bibinfo{volume}{48}},
  \bibinfo{pages}{2132} (\bibinfo{year}{2015}).

\bibitem[{\citenamefont{$\check{C}$echal
  et~al.}(2016)\citenamefont{$\check{C}$echal, Kley, Pétuya, Schramm, Ruben,
  Stepanow, Arnau, and Kern}}]{Stepanow2016}
\bibinfo{author}{\bibfnamefont{J.}~\bibnamefont{$\check{C}$echal}},
  \bibinfo{author}{\bibfnamefont{C.~S.} \bibnamefont{Kley}},
  \bibinfo{author}{\bibfnamefont{R.}~\bibnamefont{Pétuya}},
  \bibinfo{author}{\bibfnamefont{F.}~\bibnamefont{Schramm}},
  \bibinfo{author}{\bibfnamefont{M.}~\bibnamefont{Ruben}},
  \bibinfo{author}{\bibfnamefont{S.}~\bibnamefont{Stepanow}},
  \bibinfo{author}{\bibfnamefont{A.}~\bibnamefont{Arnau}}, \bibnamefont{and}
  \bibinfo{author}{\bibfnamefont{K.}~\bibnamefont{Kern}}, \bibinfo{journal}{The
  Journal of Physical Chemistry C} \textbf{\bibinfo{volume}{120}},
  \bibinfo{pages}{18622} (\bibinfo{year}{2016}).

\bibitem[{\citenamefont{\'Ecija et~al.}(2018)\citenamefont{\'Ecija, Urgel,
  Seitsonen, Auwärter, and Barth}}]{Ecija2018}
\bibinfo{author}{\bibfnamefont{D.}~\bibnamefont{\'Ecija}},
  \bibinfo{author}{\bibfnamefont{J.~I.} \bibnamefont{Urgel}},
  \bibinfo{author}{\bibfnamefont{A.~P.} \bibnamefont{Seitsonen}},
  \bibinfo{author}{\bibfnamefont{W.}~\bibnamefont{Auwärter}},
  \bibnamefont{and} \bibinfo{author}{\bibfnamefont{J.~V.} \bibnamefont{Barth}},
  \bibinfo{journal}{Acc Chem Res} \textbf{\bibinfo{volume}{51}},
  \bibinfo{pages}{365} (\bibinfo{year}{2018}).

\bibitem[{\citenamefont{Umbach et~al.}(2012)\citenamefont{Umbach, Bernien,
  Hermanns, Krüger, Sessi, Fernandez-Torrente, Stoll, Pascual, Franke, and
  Kuch}}]{Umbach2012}
\bibinfo{author}{\bibfnamefont{T.~R.} \bibnamefont{Umbach}},
  \bibinfo{author}{\bibfnamefont{M.}~\bibnamefont{Bernien}},
  \bibinfo{author}{\bibfnamefont{C.~F.} \bibnamefont{Hermanns}},
  \bibinfo{author}{\bibfnamefont{A.}~\bibnamefont{Krüger}},
  \bibinfo{author}{\bibfnamefont{V.}~\bibnamefont{Sessi}},
  \bibinfo{author}{\bibfnamefont{I.}~\bibnamefont{Fernandez-Torrente}},
  \bibinfo{author}{\bibfnamefont{P.}~\bibnamefont{Stoll}},
  \bibinfo{author}{\bibfnamefont{J.~I.} \bibnamefont{Pascual}},
  \bibinfo{author}{\bibfnamefont{K.~J.} \bibnamefont{Franke}},
  \bibnamefont{and} \bibinfo{author}{\bibfnamefont{W.}~\bibnamefont{Kuch}},
  \bibinfo{journal}{Phys Rev Lett} \textbf{\bibinfo{volume}{109}},
  \bibinfo{pages}{267207} (\bibinfo{year}{2012}).

\bibitem[{\citenamefont{Abdurakhmanova
  et~al.}(2013)\citenamefont{Abdurakhmanova, Tseng, Langner, Kley, Sessi,
  Stepanow, and Kern}}]{Stepanow2013}
\bibinfo{author}{\bibfnamefont{N.}~\bibnamefont{Abdurakhmanova}},
  \bibinfo{author}{\bibfnamefont{T.-C.} \bibnamefont{Tseng}},
  \bibinfo{author}{\bibfnamefont{A.}~\bibnamefont{Langner}},
  \bibinfo{author}{\bibfnamefont{C.~S.} \bibnamefont{Kley}},
  \bibinfo{author}{\bibfnamefont{V.}~\bibnamefont{Sessi}},
  \bibinfo{author}{\bibfnamefont{S.}~\bibnamefont{Stepanow}}, \bibnamefont{and}
  \bibinfo{author}{\bibfnamefont{K.}~\bibnamefont{Kern}},
  \bibinfo{journal}{Phys Rev Lett} \textbf{\bibinfo{volume}{110}},
  \bibinfo{pages}{027202} (\bibinfo{year}{2013}).

\bibitem[{\citenamefont{Gao et~al.}(2020)\citenamefont{Gao, Gao, Hua, Liu,
  Huang, and Lin}}]{Gao2020}
\bibinfo{author}{\bibfnamefont{Z.}~\bibnamefont{Gao}},
  \bibinfo{author}{\bibfnamefont{Y.}~\bibnamefont{Gao}},
  \bibinfo{author}{\bibfnamefont{M.}~\bibnamefont{Hua}},
  \bibinfo{author}{\bibfnamefont{J.}~\bibnamefont{Liu}},
  \bibinfo{author}{\bibfnamefont{L.}~\bibnamefont{Huang}}, \bibnamefont{and}
  \bibinfo{author}{\bibfnamefont{N.}~\bibnamefont{Lin}}, \bibinfo{journal}{The
  Journal of Physical Chemistry C} \textbf{\bibinfo{volume}{124}},
  \bibinfo{pages}{27017} (\bibinfo{year}{2020}).

\bibitem[{\citenamefont{H{\"o}tger et~al.}(2019)\citenamefont{H{\"o}tger,
  Etzkorn, Morchutt, Wurster, Dreiser, Stepanow, Grumelli, Gutzler, and
  Kern}}]{Hotger2019}
\bibinfo{author}{\bibfnamefont{D.}~\bibnamefont{H{\"o}tger}},
  \bibinfo{author}{\bibfnamefont{M.}~\bibnamefont{Etzkorn}},
  \bibinfo{author}{\bibfnamefont{C.}~\bibnamefont{Morchutt}},
  \bibinfo{author}{\bibfnamefont{B.}~\bibnamefont{Wurster}},
  \bibinfo{author}{\bibfnamefont{J.}~\bibnamefont{Dreiser}},
  \bibinfo{author}{\bibfnamefont{S.}~\bibnamefont{Stepanow}},
  \bibinfo{author}{\bibfnamefont{D.}~\bibnamefont{Grumelli}},
  \bibinfo{author}{\bibfnamefont{R.}~\bibnamefont{Gutzler}}, \bibnamefont{and}
  \bibinfo{author}{\bibfnamefont{K.}~\bibnamefont{Kern}},
  \bibinfo{journal}{Phys Chem Chem Phys} \textbf{\bibinfo{volume}{21}},
  \bibinfo{pages}{2587} (\bibinfo{year}{2019}).

\bibitem[{\citenamefont{Li et~al.}(2012)\citenamefont{Li, Xiao, Shubina, Chen,
  Shi, Schmid, Steinrück, Gottfried, and Lin}}]{Gottfried2012}
\bibinfo{author}{\bibfnamefont{Y.}~\bibnamefont{Li}},
  \bibinfo{author}{\bibfnamefont{J.}~\bibnamefont{Xiao}},
  \bibinfo{author}{\bibfnamefont{T.~E.} \bibnamefont{Shubina}},
  \bibinfo{author}{\bibfnamefont{M.}~\bibnamefont{Chen}},
  \bibinfo{author}{\bibfnamefont{Z.}~\bibnamefont{Shi}},
  \bibinfo{author}{\bibfnamefont{M.}~\bibnamefont{Schmid}},
  \bibinfo{author}{\bibfnamefont{H.-P.} \bibnamefont{Steinrück}},
  \bibinfo{author}{\bibfnamefont{J.~M.} \bibnamefont{Gottfried}},
  \bibnamefont{and} \bibinfo{author}{\bibfnamefont{N.}~\bibnamefont{Lin}},
  \bibinfo{journal}{J Am Chem Soc} \textbf{\bibinfo{volume}{134}},
  \bibinfo{pages}{6401} (\bibinfo{year}{2012}).

\bibitem[{\citenamefont{Urgel et~al.}(2016{\natexlab{a}})\citenamefont{Urgel,
  Écija, Lyu, Zhang, Palma, Auwärter, Lin, and Barth}}]{Urgel2016}
\bibinfo{author}{\bibfnamefont{J.~I.} \bibnamefont{Urgel}},
  \bibinfo{author}{\bibfnamefont{D.}~\bibnamefont{Écija}},
  \bibinfo{author}{\bibfnamefont{G.}~\bibnamefont{Lyu}},
  \bibinfo{author}{\bibfnamefont{R.}~\bibnamefont{Zhang}},
  \bibinfo{author}{\bibfnamefont{C.-A.} \bibnamefont{Palma}},
  \bibinfo{author}{\bibfnamefont{W.}~\bibnamefont{Auwärter}},
  \bibinfo{author}{\bibfnamefont{N.}~\bibnamefont{Lin}}, \bibnamefont{and}
  \bibinfo{author}{\bibfnamefont{J.~V.} \bibnamefont{Barth}},
  \bibinfo{journal}{Nat Chem} \textbf{\bibinfo{volume}{8}},
  \bibinfo{pages}{657} (\bibinfo{year}{2016}{\natexlab{a}}).

\bibitem[{\citenamefont{Zhang et~al.}(2018)\citenamefont{Zhang, Paszkiewicz,
  Du, Zhang, Lin, Chen, Klyatskaya, Ruben, Seitsonen, Barth
  et~al.}}]{Zhang2018}
\bibinfo{author}{\bibfnamefont{Y.-Q.} \bibnamefont{Zhang}},
  \bibinfo{author}{\bibfnamefont{M.}~\bibnamefont{Paszkiewicz}},
  \bibinfo{author}{\bibfnamefont{P.}~\bibnamefont{Du}},
  \bibinfo{author}{\bibfnamefont{L.}~\bibnamefont{Zhang}},
  \bibinfo{author}{\bibfnamefont{T.}~\bibnamefont{Lin}},
  \bibinfo{author}{\bibfnamefont{Z.}~\bibnamefont{Chen}},
  \bibinfo{author}{\bibfnamefont{S.}~\bibnamefont{Klyatskaya}},
  \bibinfo{author}{\bibfnamefont{M.}~\bibnamefont{Ruben}},
  \bibinfo{author}{\bibfnamefont{A.~P.} \bibnamefont{Seitsonen}},
  \bibinfo{author}{\bibfnamefont{J.~V.} \bibnamefont{Barth}},
  \bibnamefont{et~al.}, \bibinfo{journal}{Nat Chem}
  \textbf{\bibinfo{volume}{10}}, \bibinfo{pages}{296} (\bibinfo{year}{2018}).

\bibitem[{\citenamefont{Yan et~al.}(2017)\citenamefont{Yan, Kuang, Zhang,
  Shang, Liu, and Lin}}]{NianLin2017}
\bibinfo{author}{\bibfnamefont{L.}~\bibnamefont{Yan}},
  \bibinfo{author}{\bibfnamefont{G.}~\bibnamefont{Kuang}},
  \bibinfo{author}{\bibfnamefont{Q.}~\bibnamefont{Zhang}},
  \bibinfo{author}{\bibfnamefont{X.}~\bibnamefont{Shang}},
  \bibinfo{author}{\bibfnamefont{P.~N.} \bibnamefont{Liu}}, \bibnamefont{and}
  \bibinfo{author}{\bibfnamefont{N.}~\bibnamefont{Lin}},
  \bibinfo{journal}{Faraday Discuss} \textbf{\bibinfo{volume}{204}},
  \bibinfo{pages}{111} (\bibinfo{year}{2017}).

\bibitem[{\citenamefont{Zhang et~al.}(2016{\natexlab{a}})\citenamefont{Zhang,
  Wang, Huang, Cui, Wang, Du, Gao, and Liu}}]{Zhang2016}
\bibinfo{author}{\bibfnamefont{L.~Z.} \bibnamefont{Zhang}},
  \bibinfo{author}{\bibfnamefont{Z.~F.} \bibnamefont{Wang}},
  \bibinfo{author}{\bibfnamefont{B.}~\bibnamefont{Huang}},
  \bibinfo{author}{\bibfnamefont{B.}~\bibnamefont{Cui}},
  \bibinfo{author}{\bibfnamefont{Z.}~\bibnamefont{Wang}},
  \bibinfo{author}{\bibfnamefont{S.~X.} \bibnamefont{Du}},
  \bibinfo{author}{\bibfnamefont{H.-J.} \bibnamefont{Gao}}, \bibnamefont{and}
  \bibinfo{author}{\bibfnamefont{F.}~\bibnamefont{Liu}}, \bibinfo{journal}{Nano
  Lett} \textbf{\bibinfo{volume}{16}}, \bibinfo{pages}{2072}
  (\bibinfo{year}{2016}{\natexlab{a}}).

\bibitem[{\citenamefont{Kumar et~al.}(2018)\citenamefont{Kumar, Banerjee,
  Foster, and Liljeroth}}]{Kumar2018}
\bibinfo{author}{\bibfnamefont{A.}~\bibnamefont{Kumar}},
  \bibinfo{author}{\bibfnamefont{K.}~\bibnamefont{Banerjee}},
  \bibinfo{author}{\bibfnamefont{A.~S.} \bibnamefont{Foster}},
  \bibnamefont{and}
  \bibinfo{author}{\bibfnamefont{P.}~\bibnamefont{Liljeroth}},
  \bibinfo{journal}{Nano Lett} \textbf{\bibinfo{volume}{18}},
  \bibinfo{pages}{5596} (\bibinfo{year}{2018}).

\bibitem[{\citenamefont{Gao et~al.}(2019)\citenamefont{Gao, Hsu, Liu, Chuang,
  Zhang, Xia, Xu, Huang, Jin, Liu et~al.}}]{Gao2019}
\bibinfo{author}{\bibfnamefont{Z.}~\bibnamefont{Gao}},
  \bibinfo{author}{\bibfnamefont{C.-H.} \bibnamefont{Hsu}},
  \bibinfo{author}{\bibfnamefont{J.}~\bibnamefont{Liu}},
  \bibinfo{author}{\bibfnamefont{F.-C.} \bibnamefont{Chuang}},
  \bibinfo{author}{\bibfnamefont{R.}~\bibnamefont{Zhang}},
  \bibinfo{author}{\bibfnamefont{B.}~\bibnamefont{Xia}},
  \bibinfo{author}{\bibfnamefont{H.}~\bibnamefont{Xu}},
  \bibinfo{author}{\bibfnamefont{L.}~\bibnamefont{Huang}},
  \bibinfo{author}{\bibfnamefont{Q.}~\bibnamefont{Jin}},
  \bibinfo{author}{\bibfnamefont{P.~N.} \bibnamefont{Liu}},
  \bibnamefont{et~al.}, \bibinfo{journal}{Nanoscale}
  \textbf{\bibinfo{volume}{11}}, \bibinfo{pages}{878} (\bibinfo{year}{2019}).

\bibitem[{\citenamefont{Sun et~al.}(2018)\citenamefont{Sun, Tan, Feng, Zhao,
  and Petek}}]{Sun2018}
\bibinfo{author}{\bibfnamefont{H.}~\bibnamefont{Sun}},
  \bibinfo{author}{\bibfnamefont{S.}~\bibnamefont{Tan}},
  \bibinfo{author}{\bibfnamefont{M.}~\bibnamefont{Feng}},
  \bibinfo{author}{\bibfnamefont{J.}~\bibnamefont{Zhao}}, \bibnamefont{and}
  \bibinfo{author}{\bibfnamefont{H.}~\bibnamefont{Petek}},
  \bibinfo{journal}{The Journal of Physical Chemistry C}
  \textbf{\bibinfo{volume}{122}}, \bibinfo{pages}{18659}
  (\bibinfo{year}{2018}).

\bibitem[{\citenamefont{Dong et~al.}(2016{\natexlab{b}})\citenamefont{Dong,
  Kim, Er, Rappe, and Shenoy}}]{Dong2016}
\bibinfo{author}{\bibfnamefont{L.}~\bibnamefont{Dong}},
  \bibinfo{author}{\bibfnamefont{Y.}~\bibnamefont{Kim}},
  \bibinfo{author}{\bibfnamefont{D.}~\bibnamefont{Er}},
  \bibinfo{author}{\bibfnamefont{A.~M.} \bibnamefont{Rappe}}, \bibnamefont{and}
  \bibinfo{author}{\bibfnamefont{V.~B.} \bibnamefont{Shenoy}},
  \bibinfo{journal}{Phys Rev Lett} \textbf{\bibinfo{volume}{116}},
  \bibinfo{pages}{096601} (\bibinfo{year}{2016}{\natexlab{b}}).

\bibitem[{\citenamefont{Wang et~al.}(2013{\natexlab{a}})\citenamefont{Wang,
  Liu, and Liu}}]{Wang2013}
\bibinfo{author}{\bibfnamefont{Z.~F.} \bibnamefont{Wang}},
  \bibinfo{author}{\bibfnamefont{Z.}~\bibnamefont{Liu}}, \bibnamefont{and}
  \bibinfo{author}{\bibfnamefont{F.}~\bibnamefont{Liu}}, \bibinfo{journal}{Nat
  Commun} \textbf{\bibinfo{volume}{4}}, \bibinfo{pages}{1471}
  (\bibinfo{year}{2013}{\natexlab{a}}).

\bibitem[{\citenamefont{Hernández-López
  et~al.}(2021)\citenamefont{Hernández-López, Piquero-Zulaica, Downing,
  Piantek, Fujii, Serrate, Ortega, Bartolomé, and Lobo-Checa}}]{Hernandez2021}
\bibinfo{author}{\bibfnamefont{L.}~\bibnamefont{Hernández-López}},
  \bibinfo{author}{\bibfnamefont{I.}~\bibnamefont{Piquero-Zulaica}},
  \bibinfo{author}{\bibfnamefont{C.~A.} \bibnamefont{Downing}},
  \bibinfo{author}{\bibfnamefont{M.}~\bibnamefont{Piantek}},
  \bibinfo{author}{\bibfnamefont{J.}~\bibnamefont{Fujii}},
  \bibinfo{author}{\bibfnamefont{D.}~\bibnamefont{Serrate}},
  \bibinfo{author}{\bibfnamefont{J.~E.} \bibnamefont{Ortega}},
  \bibinfo{author}{\bibfnamefont{F.}~\bibnamefont{Bartolomé}},
  \bibnamefont{and}
  \bibinfo{author}{\bibfnamefont{J.}~\bibnamefont{Lobo-Checa}},
  \bibinfo{journal}{Nanoscale} \textbf{\bibinfo{volume}{13}},
  \bibinfo{pages}{5216} (\bibinfo{year}{2021}).

\bibitem[{\citenamefont{Jiang et~al.}(2021)\citenamefont{Jiang, Ni, and
  Liu}}]{Jiang2021}
\bibinfo{author}{\bibfnamefont{W.}~\bibnamefont{Jiang}},
  \bibinfo{author}{\bibfnamefont{X.}~\bibnamefont{Ni}}, \bibnamefont{and}
  \bibinfo{author}{\bibfnamefont{F.}~\bibnamefont{Liu}}, \bibinfo{journal}{Acc
  Chem Res} \textbf{\bibinfo{volume}{54}}, \bibinfo{pages}{416}
  (\bibinfo{year}{2021}).

\bibitem[{\citenamefont{Gomes et~al.}(2012)\citenamefont{Gomes, Mar, Ko,
  Guinea, and Manoharan}}]{Gomes2012}
\bibinfo{author}{\bibfnamefont{K.~K.} \bibnamefont{Gomes}},
  \bibinfo{author}{\bibfnamefont{W.}~\bibnamefont{Mar}},
  \bibinfo{author}{\bibfnamefont{W.}~\bibnamefont{Ko}},
  \bibinfo{author}{\bibfnamefont{F.}~\bibnamefont{Guinea}}, \bibnamefont{and}
  \bibinfo{author}{\bibfnamefont{H.~C.} \bibnamefont{Manoharan}},
  \bibinfo{journal}{Nature} \textbf{\bibinfo{volume}{483}},
  \bibinfo{pages}{306} (\bibinfo{year}{2012}).

\bibitem[{\citenamefont{Slot et~al.}(2017)\citenamefont{Slot, Gardenier,
  Jacobse, van Miert, Kempkes, Zevenhuizen, Smith, Vanmaekelbergh, and
  Swart}}]{Slot2017}
\bibinfo{author}{\bibfnamefont{M.~R.} \bibnamefont{Slot}},
  \bibinfo{author}{\bibfnamefont{T.~S.} \bibnamefont{Gardenier}},
  \bibinfo{author}{\bibfnamefont{P.~H.} \bibnamefont{Jacobse}},
  \bibinfo{author}{\bibfnamefont{G.~C.~P.} \bibnamefont{van Miert}},
  \bibinfo{author}{\bibfnamefont{S.~N.} \bibnamefont{Kempkes}},
  \bibinfo{author}{\bibfnamefont{S.~J.~M.} \bibnamefont{Zevenhuizen}},
  \bibinfo{author}{\bibfnamefont{C.~M.} \bibnamefont{Smith}},
  \bibinfo{author}{\bibfnamefont{D.}~\bibnamefont{Vanmaekelbergh}},
  \bibnamefont{and} \bibinfo{author}{\bibfnamefont{I.}~\bibnamefont{Swart}},
  \bibinfo{journal}{Nat Phys} \textbf{\bibinfo{volume}{13}},
  \bibinfo{pages}{672} (\bibinfo{year}{2017}).

\bibitem[{\citenamefont{Kempkes
  et~al.}(2019{\natexlab{a}})\citenamefont{Kempkes, Slot, van~den Broeke,
  Capiod, Benalcazar, Vanmaekelbergh, Bercioux, Swart, and
  Morais~Smith}}]{Kempkes2019a}
\bibinfo{author}{\bibfnamefont{S.~N.} \bibnamefont{Kempkes}},
  \bibinfo{author}{\bibfnamefont{M.~R.} \bibnamefont{Slot}},
  \bibinfo{author}{\bibfnamefont{J.~J.} \bibnamefont{van~den Broeke}},
  \bibinfo{author}{\bibfnamefont{P.}~\bibnamefont{Capiod}},
  \bibinfo{author}{\bibfnamefont{W.~A.} \bibnamefont{Benalcazar}},
  \bibinfo{author}{\bibfnamefont{D.}~\bibnamefont{Vanmaekelbergh}},
  \bibinfo{author}{\bibfnamefont{D.}~\bibnamefont{Bercioux}},
  \bibinfo{author}{\bibfnamefont{I.}~\bibnamefont{Swart}}, \bibnamefont{and}
  \bibinfo{author}{\bibfnamefont{C.}~\bibnamefont{Morais~Smith}},
  \bibinfo{journal}{Nat Mater} \textbf{\bibinfo{volume}{18}},
  \bibinfo{pages}{1292} (\bibinfo{year}{2019}{\natexlab{a}}).

\bibitem[{\citenamefont{Freeney
  et~al.}(2020{\natexlab{a}})\citenamefont{Freeney, van~den Broeke, Harsveld
  van~der Veen, Swart, and Morais~Smith}}]{Swart2020}
\bibinfo{author}{\bibfnamefont{S.~E.} \bibnamefont{Freeney}},
  \bibinfo{author}{\bibfnamefont{J.~J.} \bibnamefont{van~den Broeke}},
  \bibinfo{author}{\bibfnamefont{A.~J.~J.} \bibnamefont{Harsveld van~der
  Veen}}, \bibinfo{author}{\bibfnamefont{I.}~\bibnamefont{Swart}},
  \bibnamefont{and}
  \bibinfo{author}{\bibfnamefont{C.}~\bibnamefont{Morais~Smith}},
  \bibinfo{journal}{Phys Rev Lett} \textbf{\bibinfo{volume}{124}},
  \bibinfo{pages}{236404} (\bibinfo{year}{2020}{\natexlab{a}}).

\bibitem[{\citenamefont{Kempkes
  et~al.}(2019{\natexlab{b}})\citenamefont{Kempkes, Slot, Freeney, Zevenhuizen,
  Vanmaekelbergh, Swart, and Smith}}]{Kempkes2019}
\bibinfo{author}{\bibfnamefont{S.~N.} \bibnamefont{Kempkes}},
  \bibinfo{author}{\bibfnamefont{M.~R.} \bibnamefont{Slot}},
  \bibinfo{author}{\bibfnamefont{S.~E.} \bibnamefont{Freeney}},
  \bibinfo{author}{\bibfnamefont{S.~J.~M.} \bibnamefont{Zevenhuizen}},
  \bibinfo{author}{\bibfnamefont{D.}~\bibnamefont{Vanmaekelbergh}},
  \bibinfo{author}{\bibfnamefont{I.}~\bibnamefont{Swart}}, \bibnamefont{and}
  \bibinfo{author}{\bibfnamefont{C.~M.} \bibnamefont{Smith}},
  \bibinfo{journal}{Nat Phys} \textbf{\bibinfo{volume}{15}},
  \bibinfo{pages}{127} (\bibinfo{year}{2019}{\natexlab{b}}).

\bibitem[{\citenamefont{Collins et~al.}(2017)\citenamefont{Collins, Witte,
  Silverman, Green, and Gomes}}]{Collins2017}
\bibinfo{author}{\bibfnamefont{L.~C.} \bibnamefont{Collins}},
  \bibinfo{author}{\bibfnamefont{T.~G.} \bibnamefont{Witte}},
  \bibinfo{author}{\bibfnamefont{R.}~\bibnamefont{Silverman}},
  \bibinfo{author}{\bibfnamefont{D.~B.} \bibnamefont{Green}}, \bibnamefont{and}
  \bibinfo{author}{\bibfnamefont{K.~K.} \bibnamefont{Gomes}},
  \bibinfo{journal}{Nat Commun} \textbf{\bibinfo{volume}{8}},
  \bibinfo{pages}{15961} (\bibinfo{year}{2017}).

\bibitem[{\citenamefont{Cao et~al.}(2018{\natexlab{a}})\citenamefont{Cao,
  Fatemi, Fang, Watanabe, Taniguchi, Kaxiras, and Jarillo-Herrero}}]{Cao2018a}
\bibinfo{author}{\bibfnamefont{Y.}~\bibnamefont{Cao}},
  \bibinfo{author}{\bibfnamefont{V.}~\bibnamefont{Fatemi}},
  \bibinfo{author}{\bibfnamefont{S.}~\bibnamefont{Fang}},
  \bibinfo{author}{\bibfnamefont{K.}~\bibnamefont{Watanabe}},
  \bibinfo{author}{\bibfnamefont{T.}~\bibnamefont{Taniguchi}},
  \bibinfo{author}{\bibfnamefont{E.}~\bibnamefont{Kaxiras}}, \bibnamefont{and}
  \bibinfo{author}{\bibfnamefont{P.}~\bibnamefont{Jarillo-Herrero}},
  \bibinfo{journal}{Nature} \textbf{\bibinfo{volume}{556}}, \bibinfo{pages}{43}
  (\bibinfo{year}{2018}{\natexlab{a}}).

\bibitem[{\citenamefont{Cao et~al.}(2018{\natexlab{b}})\citenamefont{Cao,
  Fatemi, Demir, Fang, Tomarken, Luo, Sanchez-Yamagishi, Watanabe, Taniguchi,
  Kaxiras et~al.}}]{Cao2018b}
\bibinfo{author}{\bibfnamefont{Y.}~\bibnamefont{Cao}},
  \bibinfo{author}{\bibfnamefont{V.}~\bibnamefont{Fatemi}},
  \bibinfo{author}{\bibfnamefont{A.}~\bibnamefont{Demir}},
  \bibinfo{author}{\bibfnamefont{S.}~\bibnamefont{Fang}},
  \bibinfo{author}{\bibfnamefont{S.~L.} \bibnamefont{Tomarken}},
  \bibinfo{author}{\bibfnamefont{J.~Y.} \bibnamefont{Luo}},
  \bibinfo{author}{\bibfnamefont{J.~D.} \bibnamefont{Sanchez-Yamagishi}},
  \bibinfo{author}{\bibfnamefont{K.}~\bibnamefont{Watanabe}},
  \bibinfo{author}{\bibfnamefont{T.}~\bibnamefont{Taniguchi}},
  \bibinfo{author}{\bibfnamefont{E.}~\bibnamefont{Kaxiras}},
  \bibnamefont{et~al.}, \bibinfo{journal}{Nature}
  \textbf{\bibinfo{volume}{556}}, \bibinfo{pages}{80}
  (\bibinfo{year}{2018}{\natexlab{b}}).

\bibitem[{\citenamefont{Ketterle}(2002)}]{Ketterle2002}
\bibinfo{author}{\bibfnamefont{W.}~\bibnamefont{Ketterle}},
  \bibinfo{journal}{Rev Mod Phys} \textbf{\bibinfo{volume}{74}},
  \bibinfo{pages}{1131} (\bibinfo{year}{2002}).

\bibitem[{\citenamefont{Zapf et~al.}(2014)\citenamefont{Zapf, Jaime, and
  Batista}}]{Zapf2016}
\bibinfo{author}{\bibfnamefont{V.}~\bibnamefont{Zapf}},
  \bibinfo{author}{\bibfnamefont{M.}~\bibnamefont{Jaime}}, \bibnamefont{and}
  \bibinfo{author}{\bibfnamefont{C.~D.} \bibnamefont{Batista}},
  \bibinfo{journal}{Rev Mod Phys} \textbf{\bibinfo{volume}{86}},
  \bibinfo{pages}{563} (\bibinfo{year}{2014}).

\bibitem[{\citenamefont{Park and Louie}(2009{\natexlab{a}})}]{Louie2009}
\bibinfo{author}{\bibfnamefont{C.-H.} \bibnamefont{Park}} \bibnamefont{and}
  \bibinfo{author}{\bibfnamefont{S.~G.} \bibnamefont{Louie}},
  \bibinfo{journal}{Nano Lett} \textbf{\bibinfo{volume}{9}},
  \bibinfo{pages}{1793} (\bibinfo{year}{2009}{\natexlab{a}}).

\bibitem[{\citenamefont{Polini et~al.}(2013)\citenamefont{Polini, Guinea,
  Lewenstein, Manoharan, and Pellegrini}}]{Polini2013}
\bibinfo{author}{\bibfnamefont{M.}~\bibnamefont{Polini}},
  \bibinfo{author}{\bibfnamefont{F.}~\bibnamefont{Guinea}},
  \bibinfo{author}{\bibfnamefont{M.}~\bibnamefont{Lewenstein}},
  \bibinfo{author}{\bibfnamefont{H.~C.} \bibnamefont{Manoharan}},
  \bibnamefont{and}
  \bibinfo{author}{\bibfnamefont{V.}~\bibnamefont{Pellegrini}},
  \bibinfo{journal}{Nat Nanotechnol} \textbf{\bibinfo{volume}{8}},
  \bibinfo{pages}{625} (\bibinfo{year}{2013}).

\bibitem[{\citenamefont{Leykam et~al.}(2018)\citenamefont{Leykam, Andreanov,
  and Flach}}]{Leykam2018}
\bibinfo{author}{\bibfnamefont{D.}~\bibnamefont{Leykam}},
  \bibinfo{author}{\bibfnamefont{A.}~\bibnamefont{Andreanov}},
  \bibnamefont{and} \bibinfo{author}{\bibfnamefont{S.}~\bibnamefont{Flach}},
  \bibinfo{journal}{Advances in Physics: X} \textbf{\bibinfo{volume}{3}},
  \bibinfo{pages}{1473052} (\bibinfo{year}{2018}).

\bibitem[{\citenamefont{Greiner et~al.}(2002)\citenamefont{Greiner, Mandel,
  Esslinger, Hänsch, and Bloch}}]{Greiner2002}
\bibinfo{author}{\bibfnamefont{M.}~\bibnamefont{Greiner}},
  \bibinfo{author}{\bibfnamefont{O.}~\bibnamefont{Mandel}},
  \bibinfo{author}{\bibfnamefont{T.}~\bibnamefont{Esslinger}},
  \bibinfo{author}{\bibfnamefont{T.~W.} \bibnamefont{Hänsch}},
  \bibnamefont{and} \bibinfo{author}{\bibfnamefont{I.}~\bibnamefont{Bloch}},
  \bibinfo{journal}{Nature} \textbf{\bibinfo{volume}{415}}, \bibinfo{pages}{39}
  (\bibinfo{year}{2002}).

\bibitem[{\citenamefont{Duke}(2003)}]{Duke2003}
\bibinfo{author}{\bibfnamefont{C.~B.} \bibnamefont{Duke}},
  \bibinfo{journal}{Proceedings of the National Academy of Sciences}
  \textbf{\bibinfo{volume}{100}}, \bibinfo{pages}{3858} (\bibinfo{year}{2003}).

\bibitem[{\citenamefont{Gross et~al.}(2004)\citenamefont{Gross, Moresco, Savio,
  Gourdon, Joachim, and Rieder}}]{Gross2004}
\bibinfo{author}{\bibfnamefont{L.}~\bibnamefont{Gross}},
  \bibinfo{author}{\bibfnamefont{F.}~\bibnamefont{Moresco}},
  \bibinfo{author}{\bibfnamefont{L.}~\bibnamefont{Savio}},
  \bibinfo{author}{\bibfnamefont{A.}~\bibnamefont{Gourdon}},
  \bibinfo{author}{\bibfnamefont{C.}~\bibnamefont{Joachim}}, \bibnamefont{and}
  \bibinfo{author}{\bibfnamefont{K.-H.} \bibnamefont{Rieder}},
  \bibinfo{journal}{Phys Rev Lett} \textbf{\bibinfo{volume}{93}},
  \bibinfo{pages}{056103} (\bibinfo{year}{2004}).

\bibitem[{\citenamefont{Pennec et~al.}(2007)\citenamefont{Pennec, Auwärter,
  Schiffrin, Weber-Bargioni, Riemann, and Barth}}]{Pennec2007}
\bibinfo{author}{\bibfnamefont{Y.}~\bibnamefont{Pennec}},
  \bibinfo{author}{\bibfnamefont{W.}~\bibnamefont{Auwärter}},
  \bibinfo{author}{\bibfnamefont{A.}~\bibnamefont{Schiffrin}},
  \bibinfo{author}{\bibfnamefont{A.}~\bibnamefont{Weber-Bargioni}},
  \bibinfo{author}{\bibfnamefont{A.}~\bibnamefont{Riemann}}, \bibnamefont{and}
  \bibinfo{author}{\bibfnamefont{J.~V.} \bibnamefont{Barth}},
  \bibinfo{journal}{Nat Nanotechnol} \textbf{\bibinfo{volume}{2}},
  \bibinfo{pages}{99} (\bibinfo{year}{2007}).

\bibitem[{\citenamefont{Klappenberger et~al.}(2009)\citenamefont{Klappenberger,
  K{\"u}hne, Krenner, Silanes, Arnau, Garc\'{\i}a~de Abajo, Klyatskaya, Ruben,
  and Barth}}]{Barth2009}
\bibinfo{author}{\bibfnamefont{F.}~\bibnamefont{Klappenberger}},
  \bibinfo{author}{\bibfnamefont{D.}~\bibnamefont{K{\"u}hne}},
  \bibinfo{author}{\bibfnamefont{W.}~\bibnamefont{Krenner}},
  \bibinfo{author}{\bibfnamefont{I.}~\bibnamefont{Silanes}},
  \bibinfo{author}{\bibfnamefont{A.}~\bibnamefont{Arnau}},
  \bibinfo{author}{\bibfnamefont{F.~J.} \bibnamefont{Garc\'{\i}a~de Abajo}},
  \bibinfo{author}{\bibfnamefont{S.}~\bibnamefont{Klyatskaya}},
  \bibinfo{author}{\bibfnamefont{M.}~\bibnamefont{Ruben}}, \bibnamefont{and}
  \bibinfo{author}{\bibfnamefont{J.~V.} \bibnamefont{Barth}},
  \bibinfo{journal}{Nano Lett} \textbf{\bibinfo{volume}{9}},
  \bibinfo{pages}{3509} (\bibinfo{year}{2009}).

\bibitem[{\citenamefont{Cheng et~al.}(2010)\citenamefont{Cheng, Wyrick, Luo,
  Sun, Kim, Zhu, Lu, Kim, Einstein, and Bartels}}]{Bartels2010}
\bibinfo{author}{\bibfnamefont{Z.}~\bibnamefont{Cheng}},
  \bibinfo{author}{\bibfnamefont{J.}~\bibnamefont{Wyrick}},
  \bibinfo{author}{\bibfnamefont{M.}~\bibnamefont{Luo}},
  \bibinfo{author}{\bibfnamefont{D.}~\bibnamefont{Sun}},
  \bibinfo{author}{\bibfnamefont{D.}~\bibnamefont{Kim}},
  \bibinfo{author}{\bibfnamefont{Y.}~\bibnamefont{Zhu}},
  \bibinfo{author}{\bibfnamefont{W.}~\bibnamefont{Lu}},
  \bibinfo{author}{\bibfnamefont{K.}~\bibnamefont{Kim}},
  \bibinfo{author}{\bibfnamefont{T.~L.} \bibnamefont{Einstein}},
  \bibnamefont{and} \bibinfo{author}{\bibfnamefont{L.}~\bibnamefont{Bartels}},
  \bibinfo{journal}{Phys Rev Lett} \textbf{\bibinfo{volume}{105}},
  \bibinfo{pages}{066104} (\bibinfo{year}{2010}).

\bibitem[{\citenamefont{Wang et~al.}(2013{\natexlab{b}})\citenamefont{Wang,
  Wang, Tan, Li, Shi, Kuang, Liu, Louie, and Lin}}]{NianLin2013b}
\bibinfo{author}{\bibfnamefont{S.}~\bibnamefont{Wang}},
  \bibinfo{author}{\bibfnamefont{W.}~\bibnamefont{Wang}},
  \bibinfo{author}{\bibfnamefont{L.~Z.} \bibnamefont{Tan}},
  \bibinfo{author}{\bibfnamefont{X.~G.} \bibnamefont{Li}},
  \bibinfo{author}{\bibfnamefont{Z.}~\bibnamefont{Shi}},
  \bibinfo{author}{\bibfnamefont{G.}~\bibnamefont{Kuang}},
  \bibinfo{author}{\bibfnamefont{P.~N.} \bibnamefont{Liu}},
  \bibinfo{author}{\bibfnamefont{S.~G.} \bibnamefont{Louie}}, \bibnamefont{and}
  \bibinfo{author}{\bibfnamefont{N.}~\bibnamefont{Lin}}, \bibinfo{journal}{Phys
  Rev B: Condens Matter Mater Phys} \textbf{\bibinfo{volume}{88}},
  \bibinfo{pages}{245430} (\bibinfo{year}{2013}{\natexlab{b}}).

\bibitem[{\citenamefont{Müller et~al.}(2016)\citenamefont{Müller, Enache, and
  Stöhr}}]{Stohr2016}
\bibinfo{author}{\bibfnamefont{K.}~\bibnamefont{Müller}},
  \bibinfo{author}{\bibfnamefont{M.}~\bibnamefont{Enache}}, \bibnamefont{and}
  \bibinfo{author}{\bibfnamefont{M.}~\bibnamefont{Stöhr}}, \bibinfo{journal}{J
  Phys : Condens Matter} \textbf{\bibinfo{volume}{28}}, \bibinfo{pages}{153003}
  (\bibinfo{year}{2016}).

\bibitem[{\citenamefont{Mart\'{\i}n-Jim\'enez
  et~al.}(2019)\citenamefont{Mart\'{\i}n-Jim\'enez, Gallego, Miranda, and
  Otero}}]{Martinez2019}
\bibinfo{author}{\bibfnamefont{A.}~\bibnamefont{Mart\'{\i}n-Jim\'enez}},
  \bibinfo{author}{\bibfnamefont{J.~M.} \bibnamefont{Gallego}},
  \bibinfo{author}{\bibfnamefont{R.}~\bibnamefont{Miranda}}, \bibnamefont{and}
  \bibinfo{author}{\bibfnamefont{R.}~\bibnamefont{Otero}},
  \bibinfo{journal}{Phys Rev Lett} \textbf{\bibinfo{volume}{122}},
  \bibinfo{pages}{176801} (\bibinfo{year}{2019}).

\bibitem[{\citenamefont{Ashcroft and Mermin}(1976)}]{Ashcroft1976}
\bibinfo{author}{\bibfnamefont{N.~W.} \bibnamefont{Ashcroft}} \bibnamefont{and}
  \bibinfo{author}{\bibfnamefont{N.~D.} \bibnamefont{Mermin}},
  \emph{\bibinfo{title}{Solid state physics}} (\bibinfo{publisher}{Holt,
  Rinehart and Winston}, \bibinfo{address}{New York}, \bibinfo{year}{1976}).

\bibitem[{\citenamefont{Brennan}(1999)}]{Brennan1999}
\bibinfo{author}{\bibfnamefont{K.~F.} \bibnamefont{Brennan}},
  \emph{\bibinfo{title}{The {Physics} of {Semiconductors}: {With}
  {Applications} to {Optoelectronic} {Devices}}} (\bibinfo{publisher}{Cambridge
  University Press}, \bibinfo{year}{1999}), \bibinfo{edition}{1st} ed.

\bibitem[{\citenamefont{Yoffe}(2001)}]{Yoffe2001}
\bibinfo{author}{\bibfnamefont{A.~D.} \bibnamefont{Yoffe}},
  \bibinfo{journal}{Adv Phys} \textbf{\bibinfo{volume}{50}}, \bibinfo{pages}{1}
  (\bibinfo{year}{2001}).

\bibitem[{\citenamefont{Zwanenburg et~al.}(2013)\citenamefont{Zwanenburg,
  Dzurak, Morello, Simmons, Hollenberg, Klimeck, Rogge, Coppersmith, and
  Eriksson}}]{Floris2013}
\bibinfo{author}{\bibfnamefont{F.~A.} \bibnamefont{Zwanenburg}},
  \bibinfo{author}{\bibfnamefont{A.~S.} \bibnamefont{Dzurak}},
  \bibinfo{author}{\bibfnamefont{A.}~\bibnamefont{Morello}},
  \bibinfo{author}{\bibfnamefont{M.~Y.} \bibnamefont{Simmons}},
  \bibinfo{author}{\bibfnamefont{L.~C.~L.} \bibnamefont{Hollenberg}},
  \bibinfo{author}{\bibfnamefont{G.}~\bibnamefont{Klimeck}},
  \bibinfo{author}{\bibfnamefont{S.}~\bibnamefont{Rogge}},
  \bibinfo{author}{\bibfnamefont{S.~N.} \bibnamefont{Coppersmith}},
  \bibnamefont{and} \bibinfo{author}{\bibfnamefont{M.~A.}
  \bibnamefont{Eriksson}}, \bibinfo{journal}{Rev Mod Phys}
  \textbf{\bibinfo{volume}{85}}, \bibinfo{pages}{961} (\bibinfo{year}{2013}).

\bibitem[{\citenamefont{van Wees et~al.}(1988)\citenamefont{van Wees, van
  Houten, Beenakker, Williamson, Kouwenhoven, van~der Marel, and
  Foxon}}]{Vanwees1988}
\bibinfo{author}{\bibfnamefont{B.~J.} \bibnamefont{van Wees}},
  \bibinfo{author}{\bibfnamefont{H.}~\bibnamefont{van Houten}},
  \bibinfo{author}{\bibfnamefont{C.~W.~J.} \bibnamefont{Beenakker}},
  \bibinfo{author}{\bibfnamefont{J.~G.} \bibnamefont{Williamson}},
  \bibinfo{author}{\bibfnamefont{L.~P.} \bibnamefont{Kouwenhoven}},
  \bibinfo{author}{\bibfnamefont{D.}~\bibnamefont{van~der Marel}},
  \bibnamefont{and} \bibinfo{author}{\bibfnamefont{C.~T.} \bibnamefont{Foxon}},
  \bibinfo{journal}{Phys Rev Lett} \textbf{\bibinfo{volume}{60}},
  \bibinfo{pages}{848} (\bibinfo{year}{1988}).

\bibitem[{\citenamefont{Kanisawa et~al.}(2001)\citenamefont{Kanisawa, Butcher,
  Yamaguchi, and Hirayama}}]{Kanisawa2001}
\bibinfo{author}{\bibfnamefont{K.}~\bibnamefont{Kanisawa}},
  \bibinfo{author}{\bibfnamefont{M.~J.} \bibnamefont{Butcher}},
  \bibinfo{author}{\bibfnamefont{H.}~\bibnamefont{Yamaguchi}},
  \bibnamefont{and} \bibinfo{author}{\bibfnamefont{Y.}~\bibnamefont{Hirayama}},
  \bibinfo{journal}{Phys Rev Lett} \textbf{\bibinfo{volume}{86}},
  \bibinfo{pages}{3384} (\bibinfo{year}{2001}).

\bibitem[{\citenamefont{Sato and Ando}(2017)}]{Sato2017}
\bibinfo{author}{\bibfnamefont{M.}~\bibnamefont{Sato}} \bibnamefont{and}
  \bibinfo{author}{\bibfnamefont{Y.}~\bibnamefont{Ando}}, \bibinfo{journal}{Rep
  Prog Phys} \textbf{\bibinfo{volume}{80}}, \bibinfo{pages}{076501}
  (\bibinfo{year}{2017}).

\bibitem[{\citenamefont{Nayak et~al.}(2008)\citenamefont{Nayak, Simon, Stern,
  Freedman, and Das~Sarma}}]{Nayak2008}
\bibinfo{author}{\bibfnamefont{C.}~\bibnamefont{Nayak}},
  \bibinfo{author}{\bibfnamefont{S.~H.} \bibnamefont{Simon}},
  \bibinfo{author}{\bibfnamefont{A.}~\bibnamefont{Stern}},
  \bibinfo{author}{\bibfnamefont{M.}~\bibnamefont{Freedman}}, \bibnamefont{and}
  \bibinfo{author}{\bibfnamefont{S.}~\bibnamefont{Das~Sarma}},
  \bibinfo{journal}{Rev Mod Phys} \textbf{\bibinfo{volume}{80}},
  \bibinfo{pages}{1083} (\bibinfo{year}{2008}).

\bibitem[{\citenamefont{Choi et~al.}(2019)\citenamefont{Choi, Lorente, Wiebe,
  von Bergmann, Otte, and Heinrich}}]{Choi2019}
\bibinfo{author}{\bibfnamefont{D.-J.} \bibnamefont{Choi}},
  \bibinfo{author}{\bibfnamefont{N.}~\bibnamefont{Lorente}},
  \bibinfo{author}{\bibfnamefont{J.}~\bibnamefont{Wiebe}},
  \bibinfo{author}{\bibfnamefont{K.}~\bibnamefont{von Bergmann}},
  \bibinfo{author}{\bibfnamefont{A.~F.} \bibnamefont{Otte}}, \bibnamefont{and}
  \bibinfo{author}{\bibfnamefont{A.~J.} \bibnamefont{Heinrich}},
  \bibinfo{journal}{Rev Mod Phys} \textbf{\bibinfo{volume}{91}},
  \bibinfo{pages}{041001} (\bibinfo{year}{2019}).

\bibitem[{\citenamefont{Harrison}(2005)}]{Harrison2005}
\bibinfo{author}{\bibfnamefont{P.}~\bibnamefont{Harrison}},
  \emph{\bibinfo{title}{Quantum {Wells}, {Wires} and {Dots}: {Theoretical} and
  {Computational} {Physics} of {Semiconductor} {Nanostructures}}}
  (\bibinfo{publisher}{John Wiley \& Sons, Ltd}, \bibinfo{address}{Chichester,
  UK}, \bibinfo{year}{2005}).

\bibitem[{\citenamefont{van~der Wiel et~al.}(2002)\citenamefont{van~der Wiel,
  De~Franceschi, Elzerman, Fujisawa, Tarucha, and
  Kouwenhoven}}]{Vanderwiel2002}
\bibinfo{author}{\bibfnamefont{W.~G.} \bibnamefont{van~der Wiel}},
  \bibinfo{author}{\bibfnamefont{S.}~\bibnamefont{De~Franceschi}},
  \bibinfo{author}{\bibfnamefont{J.~M.} \bibnamefont{Elzerman}},
  \bibinfo{author}{\bibfnamefont{T.}~\bibnamefont{Fujisawa}},
  \bibinfo{author}{\bibfnamefont{S.}~\bibnamefont{Tarucha}}, \bibnamefont{and}
  \bibinfo{author}{\bibfnamefont{L.~P.} \bibnamefont{Kouwenhoven}},
  \bibinfo{journal}{Rev Mod Phys} \textbf{\bibinfo{volume}{75}},
  \bibinfo{pages}{1} (\bibinfo{year}{2002}).

\bibitem[{\citenamefont{Hanson et~al.}(2007)\citenamefont{Hanson, Kouwenhoven,
  Petta, Tarucha, and Vandersypen}}]{Hanson2007}
\bibinfo{author}{\bibfnamefont{R.}~\bibnamefont{Hanson}},
  \bibinfo{author}{\bibfnamefont{L.~P.} \bibnamefont{Kouwenhoven}},
  \bibinfo{author}{\bibfnamefont{J.~R.} \bibnamefont{Petta}},
  \bibinfo{author}{\bibfnamefont{S.}~\bibnamefont{Tarucha}}, \bibnamefont{and}
  \bibinfo{author}{\bibfnamefont{L.~M.~K.} \bibnamefont{Vandersypen}},
  \bibinfo{journal}{Rev Mod Phys} \textbf{\bibinfo{volume}{79}},
  \bibinfo{pages}{1217} (\bibinfo{year}{2007}).

\bibitem[{\citenamefont{Pekola et~al.}(2013)\citenamefont{Pekola, Saira, Maisi,
  Kemppinen, M\"ott\"onen, Pashkin, and Averin}}]{Pekola2013}
\bibinfo{author}{\bibfnamefont{J.~P.} \bibnamefont{Pekola}},
  \bibinfo{author}{\bibfnamefont{O.-P.} \bibnamefont{Saira}},
  \bibinfo{author}{\bibfnamefont{V.~F.} \bibnamefont{Maisi}},
  \bibinfo{author}{\bibfnamefont{A.}~\bibnamefont{Kemppinen}},
  \bibinfo{author}{\bibfnamefont{M.}~\bibnamefont{M\"ott\"onen}},
  \bibinfo{author}{\bibfnamefont{Y.~A.} \bibnamefont{Pashkin}},
  \bibnamefont{and} \bibinfo{author}{\bibfnamefont{D.~V.}
  \bibnamefont{Averin}}, \bibinfo{journal}{Rev Mod Phys}
  \textbf{\bibinfo{volume}{85}}, \bibinfo{pages}{1421} (\bibinfo{year}{2013}).

\bibitem[{\citenamefont{Wharam et~al.}(1988)\citenamefont{Wharam, Thornton,
  Newbury, Pepper, Ahmed, Frost, Hasko, Peacock, Ritchie, and
  Jones}}]{Wharam1988}
\bibinfo{author}{\bibfnamefont{D.~A.} \bibnamefont{Wharam}},
  \bibinfo{author}{\bibfnamefont{T.~J.} \bibnamefont{Thornton}},
  \bibinfo{author}{\bibfnamefont{R.}~\bibnamefont{Newbury}},
  \bibinfo{author}{\bibfnamefont{M.}~\bibnamefont{Pepper}},
  \bibinfo{author}{\bibfnamefont{H.}~\bibnamefont{Ahmed}},
  \bibinfo{author}{\bibfnamefont{J.~E.~F.} \bibnamefont{Frost}},
  \bibinfo{author}{\bibfnamefont{D.~G.} \bibnamefont{Hasko}},
  \bibinfo{author}{\bibfnamefont{D.~C.} \bibnamefont{Peacock}},
  \bibinfo{author}{\bibfnamefont{D.~A.} \bibnamefont{Ritchie}},
  \bibnamefont{and} \bibinfo{author}{\bibfnamefont{G.~A.~C.}
  \bibnamefont{Jones}}, \bibinfo{journal}{J Phys C: Solid State Phys}
  \textbf{\bibinfo{volume}{21}}, \bibinfo{pages}{L209} (\bibinfo{year}{1988}).

\bibitem[{\citenamefont{Ohnishi et~al.}(1998)\citenamefont{Ohnishi, Kondo, and
  Takayanagi}}]{Ohnishi1998}
\bibinfo{author}{\bibfnamefont{H.}~\bibnamefont{Ohnishi}},
  \bibinfo{author}{\bibfnamefont{Y.}~\bibnamefont{Kondo}}, \bibnamefont{and}
  \bibinfo{author}{\bibfnamefont{K.}~\bibnamefont{Takayanagi}},
  \bibinfo{journal}{Nature} \textbf{\bibinfo{volume}{395}},
  \bibinfo{pages}{780} (\bibinfo{year}{1998}).

\bibitem[{\citenamefont{Tarucha et~al.}(1996)\citenamefont{Tarucha, Austing,
  Honda, van~der Hage, and Kouwenhoven}}]{Tarucha1996}
\bibinfo{author}{\bibfnamefont{S.}~\bibnamefont{Tarucha}},
  \bibinfo{author}{\bibfnamefont{D.~G.} \bibnamefont{Austing}},
  \bibinfo{author}{\bibfnamefont{T.}~\bibnamefont{Honda}},
  \bibinfo{author}{\bibfnamefont{R.~J.} \bibnamefont{van~der Hage}},
  \bibnamefont{and} \bibinfo{author}{\bibfnamefont{L.~P.}
  \bibnamefont{Kouwenhoven}}, \bibinfo{journal}{Phys Rev Lett}
  \textbf{\bibinfo{volume}{77}}, \bibinfo{pages}{3613} (\bibinfo{year}{1996}).

\bibitem[{\citenamefont{Kagan and Murray}(2015)}]{Kagan2015}
\bibinfo{author}{\bibfnamefont{C.~R.} \bibnamefont{Kagan}} \bibnamefont{and}
  \bibinfo{author}{\bibfnamefont{C.~B.} \bibnamefont{Murray}},
  \bibinfo{journal}{Nat Nanotechnol} \textbf{\bibinfo{volume}{10}},
  \bibinfo{pages}{1013} (\bibinfo{year}{2015}).

\bibitem[{\citenamefont{Kagan et~al.}(2016)\citenamefont{Kagan, Lifshitz,
  Sargent, and Talapin}}]{Kagan2016}
\bibinfo{author}{\bibfnamefont{C.~R.} \bibnamefont{Kagan}},
  \bibinfo{author}{\bibfnamefont{E.}~\bibnamefont{Lifshitz}},
  \bibinfo{author}{\bibfnamefont{E.~H.} \bibnamefont{Sargent}},
  \bibnamefont{and} \bibinfo{author}{\bibfnamefont{D.~V.}
  \bibnamefont{Talapin}}, \bibinfo{journal}{Science}
  \textbf{\bibinfo{volume}{353}}, \bibinfo{pages}{aac5523}
  (\bibinfo{year}{2016}).

\bibitem[{\citenamefont{Broome et~al.}(2018)\citenamefont{Broome, Gorman,
  House, Hile, Keizer, Keith, Hill, Watson, Baker, Hollenberg
  et~al.}}]{Broome2018}
\bibinfo{author}{\bibfnamefont{M.~A.} \bibnamefont{Broome}},
  \bibinfo{author}{\bibfnamefont{S.~K.} \bibnamefont{Gorman}},
  \bibinfo{author}{\bibfnamefont{M.~G.} \bibnamefont{House}},
  \bibinfo{author}{\bibfnamefont{S.~J.} \bibnamefont{Hile}},
  \bibinfo{author}{\bibfnamefont{J.~G.} \bibnamefont{Keizer}},
  \bibinfo{author}{\bibfnamefont{D.}~\bibnamefont{Keith}},
  \bibinfo{author}{\bibfnamefont{C.~D.} \bibnamefont{Hill}},
  \bibinfo{author}{\bibfnamefont{T.~F.} \bibnamefont{Watson}},
  \bibinfo{author}{\bibfnamefont{W.~J.} \bibnamefont{Baker}},
  \bibinfo{author}{\bibfnamefont{L.~C.~L.} \bibnamefont{Hollenberg}},
  \bibnamefont{et~al.}, \bibinfo{journal}{Nat Commun}
  \textbf{\bibinfo{volume}{9}}, \bibinfo{pages}{980} (\bibinfo{year}{2018}).

\bibitem[{\citenamefont{Leon et~al.}(2020)\citenamefont{Leon, Yang, Hwang,
  Lemyre, Tanttu, Huang, Chan, Tan, Hudson, Itoh et~al.}}]{Leon2020}
\bibinfo{author}{\bibfnamefont{R.~C.~C.} \bibnamefont{Leon}},
  \bibinfo{author}{\bibfnamefont{C.~H.} \bibnamefont{Yang}},
  \bibinfo{author}{\bibfnamefont{J.~C.~C.} \bibnamefont{Hwang}},
  \bibinfo{author}{\bibfnamefont{J.~C.} \bibnamefont{Lemyre}},
  \bibinfo{author}{\bibfnamefont{T.}~\bibnamefont{Tanttu}},
  \bibinfo{author}{\bibfnamefont{W.}~\bibnamefont{Huang}},
  \bibinfo{author}{\bibfnamefont{K.~W.} \bibnamefont{Chan}},
  \bibinfo{author}{\bibfnamefont{K.~Y.} \bibnamefont{Tan}},
  \bibinfo{author}{\bibfnamefont{F.~E.} \bibnamefont{Hudson}},
  \bibinfo{author}{\bibfnamefont{K.~M.} \bibnamefont{Itoh}},
  \bibnamefont{et~al.}, \bibinfo{journal}{Nat Commun}
  \textbf{\bibinfo{volume}{11}}, \bibinfo{pages}{797} (\bibinfo{year}{2020}).

\bibitem[{\citenamefont{Walkup et~al.}(2020)\citenamefont{Walkup, Ghahari,
  Guti\'errez, Watanabe, Taniguchi, Zhitenev, and Stroscio}}]{Walkup2020}
\bibinfo{author}{\bibfnamefont{D.}~\bibnamefont{Walkup}},
  \bibinfo{author}{\bibfnamefont{F.}~\bibnamefont{Ghahari}},
  \bibinfo{author}{\bibfnamefont{C.}~\bibnamefont{Guti\'errez}},
  \bibinfo{author}{\bibfnamefont{K.}~\bibnamefont{Watanabe}},
  \bibinfo{author}{\bibfnamefont{T.}~\bibnamefont{Taniguchi}},
  \bibinfo{author}{\bibfnamefont{N.~B.} \bibnamefont{Zhitenev}},
  \bibnamefont{and} \bibinfo{author}{\bibfnamefont{J.~A.}
  \bibnamefont{Stroscio}}, \bibinfo{journal}{Phys Rev B: Condens Matter Mater
  Phys} \textbf{\bibinfo{volume}{101}}, \bibinfo{pages}{035428}
  (\bibinfo{year}{2020}).

\bibitem[{\citenamefont{Tokura et~al.}(2017)\citenamefont{Tokura, Kawasaki, and
  Nagaosa}}]{Tokura2017}
\bibinfo{author}{\bibfnamefont{Y.}~\bibnamefont{Tokura}},
  \bibinfo{author}{\bibfnamefont{M.}~\bibnamefont{Kawasaki}}, \bibnamefont{and}
  \bibinfo{author}{\bibfnamefont{N.}~\bibnamefont{Nagaosa}},
  \bibinfo{journal}{Nat Phys} \textbf{\bibinfo{volume}{13}},
  \bibinfo{pages}{1056} (\bibinfo{year}{2017}).

\bibitem[{\citenamefont{Keimer and Moore}(2017)}]{Keimer2017}
\bibinfo{author}{\bibfnamefont{B.}~\bibnamefont{Keimer}} \bibnamefont{and}
  \bibinfo{author}{\bibfnamefont{J.~E.} \bibnamefont{Moore}},
  \bibinfo{journal}{Nat Phys} \textbf{\bibinfo{volume}{13}},
  \bibinfo{pages}{1045} (\bibinfo{year}{2017}).

\bibitem[{\citenamefont{Huang et~al.}(2017)\citenamefont{Huang, Clark,
  Navarro-Moratalla, Klein, Cheng, Seyler, Zhong, Schmidgall, McGuire, Cobden
  et~al.}}]{Huang2017}
\bibinfo{author}{\bibfnamefont{B.}~\bibnamefont{Huang}},
  \bibinfo{author}{\bibfnamefont{G.}~\bibnamefont{Clark}},
  \bibinfo{author}{\bibfnamefont{E.}~\bibnamefont{Navarro-Moratalla}},
  \bibinfo{author}{\bibfnamefont{D.~R.} \bibnamefont{Klein}},
  \bibinfo{author}{\bibfnamefont{R.}~\bibnamefont{Cheng}},
  \bibinfo{author}{\bibfnamefont{K.~L.} \bibnamefont{Seyler}},
  \bibinfo{author}{\bibfnamefont{D.}~\bibnamefont{Zhong}},
  \bibinfo{author}{\bibfnamefont{E.}~\bibnamefont{Schmidgall}},
  \bibinfo{author}{\bibfnamefont{M.~A.} \bibnamefont{McGuire}},
  \bibinfo{author}{\bibfnamefont{D.~H.} \bibnamefont{Cobden}},
  \bibnamefont{et~al.}, \bibinfo{journal}{Nature}
  \textbf{\bibinfo{volume}{546}}, \bibinfo{pages}{270} (\bibinfo{year}{2017}).

\bibitem[{\citenamefont{Gong et~al.}(2017)\citenamefont{Gong, Li, Li, Ji,
  Stern, Xia, Cao, Bao, Wang, Wang et~al.}}]{Gong2017}
\bibinfo{author}{\bibfnamefont{C.}~\bibnamefont{Gong}},
  \bibinfo{author}{\bibfnamefont{L.}~\bibnamefont{Li}},
  \bibinfo{author}{\bibfnamefont{Z.}~\bibnamefont{Li}},
  \bibinfo{author}{\bibfnamefont{H.}~\bibnamefont{Ji}},
  \bibinfo{author}{\bibfnamefont{A.}~\bibnamefont{Stern}},
  \bibinfo{author}{\bibfnamefont{Y.}~\bibnamefont{Xia}},
  \bibinfo{author}{\bibfnamefont{T.}~\bibnamefont{Cao}},
  \bibinfo{author}{\bibfnamefont{W.}~\bibnamefont{Bao}},
  \bibinfo{author}{\bibfnamefont{C.}~\bibnamefont{Wang}},
  \bibinfo{author}{\bibfnamefont{Y.}~\bibnamefont{Wang}}, \bibnamefont{et~al.},
  \bibinfo{journal}{Nature} \textbf{\bibinfo{volume}{546}},
  \bibinfo{pages}{265} (\bibinfo{year}{2017}).

\bibitem[{\citenamefont{Andrei and MacDonald}(2020)}]{Andrei2020}
\bibinfo{author}{\bibfnamefont{E.~Y.} \bibnamefont{Andrei}} \bibnamefont{and}
  \bibinfo{author}{\bibfnamefont{A.~H.} \bibnamefont{MacDonald}},
  \bibinfo{journal}{Nat Mater} \textbf{\bibinfo{volume}{19}},
  \bibinfo{pages}{1265} (\bibinfo{year}{2020}).

\bibitem[{\citenamefont{Kezilebieke et~al.}(2020)\citenamefont{Kezilebieke,
  Huda, Vaňo, Aapro, Ganguli, Silveira, Głodzik, Foster, Ojanen, and
  Liljeroth}}]{Liljeroth2020}
\bibinfo{author}{\bibfnamefont{S.}~\bibnamefont{Kezilebieke}},
  \bibinfo{author}{\bibfnamefont{M.~N.} \bibnamefont{Huda}},
  \bibinfo{author}{\bibfnamefont{V.}~\bibnamefont{Vaňo}},
  \bibinfo{author}{\bibfnamefont{M.}~\bibnamefont{Aapro}},
  \bibinfo{author}{\bibfnamefont{S.~C.} \bibnamefont{Ganguli}},
  \bibinfo{author}{\bibfnamefont{O.~J.} \bibnamefont{Silveira}},
  \bibinfo{author}{\bibfnamefont{S.}~\bibnamefont{Głodzik}},
  \bibinfo{author}{\bibfnamefont{A.~S.} \bibnamefont{Foster}},
  \bibinfo{author}{\bibfnamefont{T.}~\bibnamefont{Ojanen}}, \bibnamefont{and}
  \bibinfo{author}{\bibfnamefont{P.}~\bibnamefont{Liljeroth}},
  \bibinfo{journal}{Nature} \textbf{\bibinfo{volume}{588}},
  \bibinfo{pages}{424} (\bibinfo{year}{2020}).

\bibitem[{\citenamefont{Li et~al.}(2021)\citenamefont{Li, Li, Naik, Xie, Li,
  Wang, Regan, Wang, Zhao, Zhao et~al.}}]{Crommie2021}
\bibinfo{author}{\bibfnamefont{H.}~\bibnamefont{Li}},
  \bibinfo{author}{\bibfnamefont{S.}~\bibnamefont{Li}},
  \bibinfo{author}{\bibfnamefont{M.~H.} \bibnamefont{Naik}},
  \bibinfo{author}{\bibfnamefont{J.}~\bibnamefont{Xie}},
  \bibinfo{author}{\bibfnamefont{X.}~\bibnamefont{Li}},
  \bibinfo{author}{\bibfnamefont{J.}~\bibnamefont{Wang}},
  \bibinfo{author}{\bibfnamefont{E.}~\bibnamefont{Regan}},
  \bibinfo{author}{\bibfnamefont{D.}~\bibnamefont{Wang}},
  \bibinfo{author}{\bibfnamefont{W.}~\bibnamefont{Zhao}},
  \bibinfo{author}{\bibfnamefont{S.}~\bibnamefont{Zhao}}, \bibnamefont{et~al.},
  \bibinfo{journal}{Nat Mater}  (\bibinfo{year}{2021}).

\bibitem[{\citenamefont{Reinert et~al.}(2001)\citenamefont{Reinert, Nicolay,
  Schmidt, Ehm, and H{\"u}fner}}]{Reinert2001}
\bibinfo{author}{\bibfnamefont{F.}~\bibnamefont{Reinert}},
  \bibinfo{author}{\bibfnamefont{G.}~\bibnamefont{Nicolay}},
  \bibinfo{author}{\bibfnamefont{S.}~\bibnamefont{Schmidt}},
  \bibinfo{author}{\bibfnamefont{D.}~\bibnamefont{Ehm}}, \bibnamefont{and}
  \bibinfo{author}{\bibfnamefont{S.}~\bibnamefont{H{\"u}fner}},
  \bibinfo{journal}{Phys Rev B: Condens Matter Mater Phys}
  \textbf{\bibinfo{volume}{63}}, \bibinfo{pages}{115415}
  (\bibinfo{year}{2001}).

\bibitem[{\citenamefont{Shockley}(1939)}]{Shockley1939}
\bibinfo{author}{\bibfnamefont{W.}~\bibnamefont{Shockley}},
  \bibinfo{journal}{Phys Rev} \textbf{\bibinfo{volume}{56}},
  \bibinfo{pages}{317} (\bibinfo{year}{1939}).

\bibitem[{\citenamefont{Kevan and Gaylord}(1987)}]{Kevan1987}
\bibinfo{author}{\bibfnamefont{S.~D.} \bibnamefont{Kevan}} \bibnamefont{and}
  \bibinfo{author}{\bibfnamefont{R.~H.} \bibnamefont{Gaylord}},
  \bibinfo{journal}{Phys Rev B} \textbf{\bibinfo{volume}{36}},
  \bibinfo{pages}{5809} (\bibinfo{year}{1987}).

\bibitem[{\citenamefont{Paniago
  et~al.}(1995{\natexlab{a}})\citenamefont{Paniago, Matzdorf, Meister, and
  Goldmann}}]{Paniago1995b}
\bibinfo{author}{\bibfnamefont{R.}~\bibnamefont{Paniago}},
  \bibinfo{author}{\bibfnamefont{R.}~\bibnamefont{Matzdorf}},
  \bibinfo{author}{\bibfnamefont{G.}~\bibnamefont{Meister}}, \bibnamefont{and}
  \bibinfo{author}{\bibfnamefont{A.}~\bibnamefont{Goldmann}},
  \bibinfo{journal}{Surf Sci} \textbf{\bibinfo{volume}{331-333}},
  \bibinfo{pages}{1233} (\bibinfo{year}{1995}{\natexlab{a}}).

\bibitem[{\citenamefont{Reinert and Hüfner}(2005)}]{Reinert2005}
\bibinfo{author}{\bibfnamefont{F.}~\bibnamefont{Reinert}} \bibnamefont{and}
  \bibinfo{author}{\bibfnamefont{S.}~\bibnamefont{Hüfner}},
  \bibinfo{journal}{New J Phys} \textbf{\bibinfo{volume}{7}},
  \bibinfo{pages}{97} (\bibinfo{year}{2005}).

\bibitem[{\citenamefont{Malterre et~al.}(2007)\citenamefont{Malterre, Kierren,
  Fagot-Revurat, Pons, Tejeda, Didiot, Cercellier, and
  Bendounan}}]{Malterre2007}
\bibinfo{author}{\bibfnamefont{D.}~\bibnamefont{Malterre}},
  \bibinfo{author}{\bibfnamefont{B.}~\bibnamefont{Kierren}},
  \bibinfo{author}{\bibfnamefont{Y.}~\bibnamefont{Fagot-Revurat}},
  \bibinfo{author}{\bibfnamefont{S.}~\bibnamefont{Pons}},
  \bibinfo{author}{\bibfnamefont{A.}~\bibnamefont{Tejeda}},
  \bibinfo{author}{\bibfnamefont{C.}~\bibnamefont{Didiot}},
  \bibinfo{author}{\bibfnamefont{H.}~\bibnamefont{Cercellier}},
  \bibnamefont{and}
  \bibinfo{author}{\bibfnamefont{A.}~\bibnamefont{Bendounan}},
  \bibinfo{journal}{New J Phys} \textbf{\bibinfo{volume}{9}},
  \bibinfo{pages}{391} (\bibinfo{year}{2007}).

\bibitem[{\citenamefont{Oka et~al.}(2014)\citenamefont{Oka, Brovko, Corbetta,
  Stepanyuk, Sander, and Kirschner}}]{Hirofumi2014}
\bibinfo{author}{\bibfnamefont{H.}~\bibnamefont{Oka}},
  \bibinfo{author}{\bibfnamefont{O.~O.} \bibnamefont{Brovko}},
  \bibinfo{author}{\bibfnamefont{M.}~\bibnamefont{Corbetta}},
  \bibinfo{author}{\bibfnamefont{V.~S.} \bibnamefont{Stepanyuk}},
  \bibinfo{author}{\bibfnamefont{D.}~\bibnamefont{Sander}}, \bibnamefont{and}
  \bibinfo{author}{\bibfnamefont{J.}~\bibnamefont{Kirschner}},
  \bibinfo{journal}{Rev Mod Phys} \textbf{\bibinfo{volume}{86}},
  \bibinfo{pages}{1127} (\bibinfo{year}{2014}).

\bibitem[{\citenamefont{Tamai et~al.}(2013)\citenamefont{Tamai, Meevasana,
  King, Nicholson, de~la Torre, Rozbicki, and Baumberger}}]{Tamai2013}
\bibinfo{author}{\bibfnamefont{A.}~\bibnamefont{Tamai}},
  \bibinfo{author}{\bibfnamefont{W.}~\bibnamefont{Meevasana}},
  \bibinfo{author}{\bibfnamefont{P.~D.~C.} \bibnamefont{King}},
  \bibinfo{author}{\bibfnamefont{C.~W.} \bibnamefont{Nicholson}},
  \bibinfo{author}{\bibfnamefont{A.}~\bibnamefont{de~la Torre}},
  \bibinfo{author}{\bibfnamefont{E.}~\bibnamefont{Rozbicki}}, \bibnamefont{and}
  \bibinfo{author}{\bibfnamefont{F.}~\bibnamefont{Baumberger}},
  \bibinfo{journal}{Phys Rev B: Condens Matter Mater Phys}
  \textbf{\bibinfo{volume}{87}}, \bibinfo{pages}{075113}
  (\bibinfo{year}{2013}).

\bibitem[{\citenamefont{Paniago
  et~al.}(1995{\natexlab{b}})\citenamefont{Paniago, Matzdorf, Meister, and
  Goldmann}}]{Paniago1995}
\bibinfo{author}{\bibfnamefont{R.}~\bibnamefont{Paniago}},
  \bibinfo{author}{\bibfnamefont{R.}~\bibnamefont{Matzdorf}},
  \bibinfo{author}{\bibfnamefont{G.}~\bibnamefont{Meister}}, \bibnamefont{and}
  \bibinfo{author}{\bibfnamefont{A.}~\bibnamefont{Goldmann}},
  \bibinfo{journal}{Surf Sci} \textbf{\bibinfo{volume}{336}},
  \bibinfo{pages}{113} (\bibinfo{year}{1995}{\natexlab{b}}).

\bibitem[{\citenamefont{LaShell et~al.}(1996)\citenamefont{LaShell, McDougall,
  and Jensen}}]{LaShell1996}
\bibinfo{author}{\bibfnamefont{S.}~\bibnamefont{LaShell}},
  \bibinfo{author}{\bibfnamefont{B.~A.} \bibnamefont{McDougall}},
  \bibnamefont{and} \bibinfo{author}{\bibfnamefont{E.}~\bibnamefont{Jensen}},
  \bibinfo{journal}{Phys Rev Lett} \textbf{\bibinfo{volume}{77}},
  \bibinfo{pages}{3419} (\bibinfo{year}{1996}).

\bibitem[{\citenamefont{Yan et~al.}(2015)\citenamefont{Yan, Stadtmüller, Haag,
  Jakobs, Seidel, Jungkenn, Mathias, Cinchetti, Aeschlimann, and
  Felser}}]{Yan2015}
\bibinfo{author}{\bibfnamefont{B.}~\bibnamefont{Yan}},
  \bibinfo{author}{\bibfnamefont{B.}~\bibnamefont{Stadtmüller}},
  \bibinfo{author}{\bibfnamefont{N.}~\bibnamefont{Haag}},
  \bibinfo{author}{\bibfnamefont{S.}~\bibnamefont{Jakobs}},
  \bibinfo{author}{\bibfnamefont{J.}~\bibnamefont{Seidel}},
  \bibinfo{author}{\bibfnamefont{D.}~\bibnamefont{Jungkenn}},
  \bibinfo{author}{\bibfnamefont{S.}~\bibnamefont{Mathias}},
  \bibinfo{author}{\bibfnamefont{M.}~\bibnamefont{Cinchetti}},
  \bibinfo{author}{\bibfnamefont{M.}~\bibnamefont{Aeschlimann}},
  \bibnamefont{and} \bibinfo{author}{\bibfnamefont{C.}~\bibnamefont{Felser}},
  \bibinfo{journal}{Nat Commun} \textbf{\bibinfo{volume}{6}},
  \bibinfo{pages}{10167} (\bibinfo{year}{2015}).

\bibitem[{\citenamefont{Tersoff and Hamann}(1985)}]{Tersoff1985}
\bibinfo{author}{\bibfnamefont{J.}~\bibnamefont{Tersoff}} \bibnamefont{and}
  \bibinfo{author}{\bibfnamefont{D.~R.} \bibnamefont{Hamann}},
  \bibinfo{journal}{Phys Rev B} \textbf{\bibinfo{volume}{31}},
  \bibinfo{pages}{805} (\bibinfo{year}{1985}).

\bibitem[{\citenamefont{Crommie
  et~al.}(1993{\natexlab{a}})\citenamefont{Crommie, Lutz, and
  Eigler}}]{Crommie1993}
\bibinfo{author}{\bibfnamefont{M.~F.} \bibnamefont{Crommie}},
  \bibinfo{author}{\bibfnamefont{C.~P.} \bibnamefont{Lutz}}, \bibnamefont{and}
  \bibinfo{author}{\bibfnamefont{D.~M.} \bibnamefont{Eigler}},
  \bibinfo{journal}{Nature} \textbf{\bibinfo{volume}{363}},
  \bibinfo{pages}{524} (\bibinfo{year}{1993}{\natexlab{a}}).

\bibitem[{\citenamefont{Crommie
  et~al.}(1993{\natexlab{b}})\citenamefont{Crommie, Lutz, and
  Eigler}}]{Crommie1993a}
\bibinfo{author}{\bibfnamefont{M.~F.} \bibnamefont{Crommie}},
  \bibinfo{author}{\bibfnamefont{C.~P.} \bibnamefont{Lutz}}, \bibnamefont{and}
  \bibinfo{author}{\bibfnamefont{D.~M.} \bibnamefont{Eigler}},
  \bibinfo{journal}{Science} \textbf{\bibinfo{volume}{262}},
  \bibinfo{pages}{218} (\bibinfo{year}{1993}{\natexlab{b}}).

\bibitem[{\citenamefont{Bertel and Memmel}(1996)}]{Memmel1996}
\bibinfo{author}{\bibfnamefont{E.}~\bibnamefont{Bertel}} \bibnamefont{and}
  \bibinfo{author}{\bibfnamefont{N.}~\bibnamefont{Memmel}},
  \bibinfo{journal}{Applied Physics A Materials Science and Processing}
  \textbf{\bibinfo{volume}{63}}, \bibinfo{pages}{523} (\bibinfo{year}{1996}).

\bibitem[{\citenamefont{Kulawik et~al.}(2005)\citenamefont{Kulawik, Rust,
  Heyde, Nilius, Mantooth, Weiss, and Freund}}]{Kulawik2005}
\bibinfo{author}{\bibfnamefont{M.}~\bibnamefont{Kulawik}},
  \bibinfo{author}{\bibfnamefont{H.-P.} \bibnamefont{Rust}},
  \bibinfo{author}{\bibfnamefont{M.}~\bibnamefont{Heyde}},
  \bibinfo{author}{\bibfnamefont{N.}~\bibnamefont{Nilius}},
  \bibinfo{author}{\bibfnamefont{B.~A.} \bibnamefont{Mantooth}},
  \bibinfo{author}{\bibfnamefont{P.~S.} \bibnamefont{Weiss}}, \bibnamefont{and}
  \bibinfo{author}{\bibfnamefont{H.-J.} \bibnamefont{Freund}},
  \bibinfo{journal}{Surf Sci} \textbf{\bibinfo{volume}{590}},
  \bibinfo{pages}{L253} (\bibinfo{year}{2005}).

\bibitem[{\citenamefont{Hasegawa and Avouris}(1993)}]{Hasegawa1993}
\bibinfo{author}{\bibfnamefont{Y.}~\bibnamefont{Hasegawa}} \bibnamefont{and}
  \bibinfo{author}{\bibfnamefont{P.}~\bibnamefont{Avouris}},
  \bibinfo{journal}{Phys Rev Lett} \textbf{\bibinfo{volume}{71}},
  \bibinfo{pages}{1071} (\bibinfo{year}{1993}).

\bibitem[{\citenamefont{Li et~al.}(1999)\citenamefont{Li, Schneider, Crampin,
  and Berndt}}]{Berndt1999}
\bibinfo{author}{\bibfnamefont{J.}~\bibnamefont{Li}},
  \bibinfo{author}{\bibfnamefont{W.-D.} \bibnamefont{Schneider}},
  \bibinfo{author}{\bibfnamefont{S.}~\bibnamefont{Crampin}}, \bibnamefont{and}
  \bibinfo{author}{\bibfnamefont{R.}~\bibnamefont{Berndt}},
  \bibinfo{journal}{Surf Sci} \textbf{\bibinfo{volume}{422}},
  \bibinfo{pages}{95} (\bibinfo{year}{1999}).

\bibitem[{\citenamefont{Rodary et~al.}(2007)\citenamefont{Rodary, Sander, Liu,
  Zhao, Niebergall, Stepanyuk, Bruno, and Kirschner}}]{Rodary2007}
\bibinfo{author}{\bibfnamefont{G.}~\bibnamefont{Rodary}},
  \bibinfo{author}{\bibfnamefont{D.}~\bibnamefont{Sander}},
  \bibinfo{author}{\bibfnamefont{H.}~\bibnamefont{Liu}},
  \bibinfo{author}{\bibfnamefont{H.}~\bibnamefont{Zhao}},
  \bibinfo{author}{\bibfnamefont{L.}~\bibnamefont{Niebergall}},
  \bibinfo{author}{\bibfnamefont{V.~S.} \bibnamefont{Stepanyuk}},
  \bibinfo{author}{\bibfnamefont{P.}~\bibnamefont{Bruno}}, \bibnamefont{and}
  \bibinfo{author}{\bibfnamefont{J.}~\bibnamefont{Kirschner}},
  \bibinfo{journal}{Phys Rev B: Condens Matter Mater Phys}
  \textbf{\bibinfo{volume}{75}}, \bibinfo{pages}{233412}
  (\bibinfo{year}{2007}).

\bibitem[{\citenamefont{Avouris and Lyo}(1994)}]{Avouris1994}
\bibinfo{author}{\bibfnamefont{P.}~\bibnamefont{Avouris}} \bibnamefont{and}
  \bibinfo{author}{\bibfnamefont{I.-W.} \bibnamefont{Lyo}},
  \bibinfo{journal}{Science} \textbf{\bibinfo{volume}{264}},
  \bibinfo{pages}{942} (\bibinfo{year}{1994}).

\bibitem[{\citenamefont{B{\"u}rgi et~al.}(1998)\citenamefont{B{\"u}rgi,
  Jeandupeux, Hirstein, Brune, and Kern}}]{Burgi1998}
\bibinfo{author}{\bibfnamefont{L.}~\bibnamefont{B{\"u}rgi}},
  \bibinfo{author}{\bibfnamefont{O.}~\bibnamefont{Jeandupeux}},
  \bibinfo{author}{\bibfnamefont{A.}~\bibnamefont{Hirstein}},
  \bibinfo{author}{\bibfnamefont{H.}~\bibnamefont{Brune}}, \bibnamefont{and}
  \bibinfo{author}{\bibfnamefont{K.}~\bibnamefont{Kern}},
  \bibinfo{journal}{Phys Rev Lett} \textbf{\bibinfo{volume}{81}},
  \bibinfo{pages}{5370} (\bibinfo{year}{1998}).

\bibitem[{\citenamefont{Forster et~al.}(2004)\citenamefont{Forster, H{\"u}fner,
  and Reinert}}]{Forster2004}
\bibinfo{author}{\bibfnamefont{F.}~\bibnamefont{Forster}},
  \bibinfo{author}{\bibfnamefont{S.}~\bibnamefont{H{\"u}fner}},
  \bibnamefont{and} \bibinfo{author}{\bibfnamefont{F.}~\bibnamefont{Reinert}},
  \bibinfo{journal}{The Journal of Physical Chemistry B}
  \textbf{\bibinfo{volume}{108}}, \bibinfo{pages}{14692}
  (\bibinfo{year}{2004}).

\bibitem[{\citenamefont{Friedel}(1958)}]{Friedel1958}
\bibinfo{author}{\bibfnamefont{J.}~\bibnamefont{Friedel}}, \bibinfo{journal}{Il
  Nuovo Cimento} \textbf{\bibinfo{volume}{7}}, \bibinfo{pages}{287}
  (\bibinfo{year}{1958}).

\bibitem[{\citenamefont{Sprunger et~al.}(1997)\citenamefont{Sprunger, Petersen,
  Plummer, Lægsgaard, and Besenbacher}}]{Sprunger1997}
\bibinfo{author}{\bibfnamefont{P.~T.} \bibnamefont{Sprunger}},
  \bibinfo{author}{\bibfnamefont{L.}~\bibnamefont{Petersen}},
  \bibinfo{author}{\bibfnamefont{E.~W.} \bibnamefont{Plummer}},
  \bibinfo{author}{\bibfnamefont{E.}~\bibnamefont{Lægsgaard}},
  \bibnamefont{and}
  \bibinfo{author}{\bibfnamefont{F.}~\bibnamefont{Besenbacher}},
  \bibinfo{journal}{Science} \textbf{\bibinfo{volume}{275}},
  \bibinfo{pages}{1764} (\bibinfo{year}{1997}).

\bibitem[{\citenamefont{Repp et~al.}(2000)\citenamefont{Repp, Moresco, Meyer,
  Rieder, Hyldgaard, and Persson}}]{Repp2000}
\bibinfo{author}{\bibfnamefont{J.}~\bibnamefont{Repp}},
  \bibinfo{author}{\bibfnamefont{F.}~\bibnamefont{Moresco}},
  \bibinfo{author}{\bibfnamefont{G.}~\bibnamefont{Meyer}},
  \bibinfo{author}{\bibfnamefont{K.-H.} \bibnamefont{Rieder}},
  \bibinfo{author}{\bibfnamefont{P.}~\bibnamefont{Hyldgaard}},
  \bibnamefont{and} \bibinfo{author}{\bibfnamefont{M.}~\bibnamefont{Persson}},
  \bibinfo{journal}{Phys Rev Lett} \textbf{\bibinfo{volume}{85}},
  \bibinfo{pages}{2981} (\bibinfo{year}{2000}).

\bibitem[{\citenamefont{Knorr et~al.}(2002)\citenamefont{Knorr, Brune, Epple,
  Hirstein, Schneider, and Kern}}]{Knorr2002}
\bibinfo{author}{\bibfnamefont{N.}~\bibnamefont{Knorr}},
  \bibinfo{author}{\bibfnamefont{H.}~\bibnamefont{Brune}},
  \bibinfo{author}{\bibfnamefont{M.}~\bibnamefont{Epple}},
  \bibinfo{author}{\bibfnamefont{A.}~\bibnamefont{Hirstein}},
  \bibinfo{author}{\bibfnamefont{M.~A.} \bibnamefont{Schneider}},
  \bibnamefont{and} \bibinfo{author}{\bibfnamefont{K.}~\bibnamefont{Kern}},
  \bibinfo{journal}{Phys Rev B} \textbf{\bibinfo{volume}{65}},
  \bibinfo{pages}{115420} (\bibinfo{year}{2002}).

\bibitem[{\citenamefont{Ternes et~al.}(2010)\citenamefont{Ternes, Pivetta,
  Patthey, and Schneider}}]{Ternes2010}
\bibinfo{author}{\bibfnamefont{M.}~\bibnamefont{Ternes}},
  \bibinfo{author}{\bibfnamefont{M.}~\bibnamefont{Pivetta}},
  \bibinfo{author}{\bibfnamefont{F.}~\bibnamefont{Patthey}}, \bibnamefont{and}
  \bibinfo{author}{\bibfnamefont{W.-D.} \bibnamefont{Schneider}},
  \bibinfo{journal}{Prog Surf Sci} \textbf{\bibinfo{volume}{85}},
  \bibinfo{pages}{1} (\bibinfo{year}{2010}).

\bibitem[{\citenamefont{Han and Weiss}(2012)}]{Weiss2012}
\bibinfo{author}{\bibfnamefont{P.}~\bibnamefont{Han}} \bibnamefont{and}
  \bibinfo{author}{\bibfnamefont{P.~S.} \bibnamefont{Weiss}},
  \bibinfo{journal}{Surf Sci Rep} \textbf{\bibinfo{volume}{67}},
  \bibinfo{pages}{19} (\bibinfo{year}{2012}).

\bibitem[{\citenamefont{Ding et~al.}(2007)\citenamefont{Ding, Stepanyuk,
  Ignatiev, Negulyaev, Niebergall, Wasniowska, Gao, Bruno, and
  Kirschner}}]{Ding2007}
\bibinfo{author}{\bibfnamefont{H.~F.} \bibnamefont{Ding}},
  \bibinfo{author}{\bibfnamefont{V.~S.} \bibnamefont{Stepanyuk}},
  \bibinfo{author}{\bibfnamefont{P.~A.} \bibnamefont{Ignatiev}},
  \bibinfo{author}{\bibfnamefont{N.~N.} \bibnamefont{Negulyaev}},
  \bibinfo{author}{\bibfnamefont{L.}~\bibnamefont{Niebergall}},
  \bibinfo{author}{\bibfnamefont{M.}~\bibnamefont{Wasniowska}},
  \bibinfo{author}{\bibfnamefont{C.~L.} \bibnamefont{Gao}},
  \bibinfo{author}{\bibfnamefont{P.}~\bibnamefont{Bruno}}, \bibnamefont{and}
  \bibinfo{author}{\bibfnamefont{J.}~\bibnamefont{Kirschner}},
  \bibinfo{journal}{Phys Rev B} \textbf{\bibinfo{volume}{76}},
  \bibinfo{pages}{033409} (\bibinfo{year}{2007}).

\bibitem[{\citenamefont{Silly et~al.}(2004)\citenamefont{Silly, Pivetta,
  Ternes, Patthey, Pelz, and Schneider}}]{Silly2004}
\bibinfo{author}{\bibfnamefont{F.}~\bibnamefont{Silly}},
  \bibinfo{author}{\bibfnamefont{M.}~\bibnamefont{Pivetta}},
  \bibinfo{author}{\bibfnamefont{M.}~\bibnamefont{Ternes}},
  \bibinfo{author}{\bibfnamefont{F.}~\bibnamefont{Patthey}},
  \bibinfo{author}{\bibfnamefont{J.~P.} \bibnamefont{Pelz}}, \bibnamefont{and}
  \bibinfo{author}{\bibfnamefont{W.-D.} \bibnamefont{Schneider}},
  \bibinfo{journal}{Phys Rev Lett} \textbf{\bibinfo{volume}{92}},
  \bibinfo{pages}{016101} (\bibinfo{year}{2004}).

\bibitem[{\citenamefont{Crommie
  et~al.}(1995{\natexlab{a}})\citenamefont{Crommie, Lutz, Eigler, and
  Heller}}]{Crommie1995}
\bibinfo{author}{\bibfnamefont{M.~F.} \bibnamefont{Crommie}},
  \bibinfo{author}{\bibfnamefont{C.~P.} \bibnamefont{Lutz}},
  \bibinfo{author}{\bibfnamefont{D.~M.} \bibnamefont{Eigler}},
  \bibnamefont{and} \bibinfo{author}{\bibfnamefont{E.~J.}
  \bibnamefont{Heller}}, \bibinfo{journal}{Physica D}
  \textbf{\bibinfo{volume}{83}}, \bibinfo{pages}{98}
  (\bibinfo{year}{1995}{\natexlab{a}}).

\bibitem[{\citenamefont{Kliewer et~al.}(2001)\citenamefont{Kliewer, Berndt, and
  Crampin}}]{Kliewer2001}
\bibinfo{author}{\bibfnamefont{J.}~\bibnamefont{Kliewer}},
  \bibinfo{author}{\bibfnamefont{R.}~\bibnamefont{Berndt}}, \bibnamefont{and}
  \bibinfo{author}{\bibfnamefont{S.}~\bibnamefont{Crampin}},
  \bibinfo{journal}{New J Phys} \textbf{\bibinfo{volume}{3}},
  \bibinfo{pages}{22} (\bibinfo{year}{2001}).

\bibitem[{\citenamefont{Li et~al.}(1998)\citenamefont{Li, Schneider, Berndt,
  and Crampin}}]{Berndt1998}
\bibinfo{author}{\bibfnamefont{J.}~\bibnamefont{Li}},
  \bibinfo{author}{\bibfnamefont{W.-D.} \bibnamefont{Schneider}},
  \bibinfo{author}{\bibfnamefont{R.}~\bibnamefont{Berndt}}, \bibnamefont{and}
  \bibinfo{author}{\bibfnamefont{S.}~\bibnamefont{Crampin}},
  \bibinfo{journal}{Phys Rev Lett} \textbf{\bibinfo{volume}{80}},
  \bibinfo{pages}{3332} (\bibinfo{year}{1998}).

\bibitem[{\citenamefont{Jensen et~al.}(2005)\citenamefont{Jensen, Kröger,
  Berndt, and Crampin}}]{Jensen2005}
\bibinfo{author}{\bibfnamefont{H.}~\bibnamefont{Jensen}},
  \bibinfo{author}{\bibfnamefont{J.}~\bibnamefont{Kröger}},
  \bibinfo{author}{\bibfnamefont{R.}~\bibnamefont{Berndt}}, \bibnamefont{and}
  \bibinfo{author}{\bibfnamefont{S.}~\bibnamefont{Crampin}},
  \bibinfo{journal}{Phys Rev B} \textbf{\bibinfo{volume}{71}},
  \bibinfo{pages}{155417} (\bibinfo{year}{2005}).

\bibitem[{\citenamefont{Shchyrba
  et~al.}(2014{\natexlab{a}})\citenamefont{Shchyrba, Martens, W{\"a}ckerlin,
  Matena, Ivas, Wadepohl, St{\"o}hr, Jung, and Gade}}]{Wackerlin2014}
\bibinfo{author}{\bibfnamefont{A.}~\bibnamefont{Shchyrba}},
  \bibinfo{author}{\bibfnamefont{S.~C.} \bibnamefont{Martens}},
  \bibinfo{author}{\bibfnamefont{C.}~\bibnamefont{W{\"a}ckerlin}},
  \bibinfo{author}{\bibfnamefont{M.}~\bibnamefont{Matena}},
  \bibinfo{author}{\bibfnamefont{T.}~\bibnamefont{Ivas}},
  \bibinfo{author}{\bibfnamefont{H.}~\bibnamefont{Wadepohl}},
  \bibinfo{author}{\bibfnamefont{M.}~\bibnamefont{St{\"o}hr}},
  \bibinfo{author}{\bibfnamefont{T.~A.} \bibnamefont{Jung}}, \bibnamefont{and}
  \bibinfo{author}{\bibfnamefont{L.~H.} \bibnamefont{Gade}},
  \bibinfo{journal}{Chem Commun} \textbf{\bibinfo{volume}{50}},
  \bibinfo{pages}{7628} (\bibinfo{year}{2014}{\natexlab{a}}).

\bibitem[{\citenamefont{Ortega et~al.}(2018)\citenamefont{Ortega, Vasseur,
  Piquero-Zulaica, Matencio, Valbuena, Rault, Schiller, Corso, Mugarza, and
  Lobo-Checa}}]{Ortega2018}
\bibinfo{author}{\bibfnamefont{J.~E.} \bibnamefont{Ortega}},
  \bibinfo{author}{\bibfnamefont{G.}~\bibnamefont{Vasseur}},
  \bibinfo{author}{\bibfnamefont{I.}~\bibnamefont{Piquero-Zulaica}},
  \bibinfo{author}{\bibfnamefont{S.}~\bibnamefont{Matencio}},
  \bibinfo{author}{\bibfnamefont{M.~A.} \bibnamefont{Valbuena}},
  \bibinfo{author}{\bibfnamefont{J.~E.} \bibnamefont{Rault}},
  \bibinfo{author}{\bibfnamefont{F.}~\bibnamefont{Schiller}},
  \bibinfo{author}{\bibfnamefont{M.}~\bibnamefont{Corso}},
  \bibinfo{author}{\bibfnamefont{A.}~\bibnamefont{Mugarza}}, \bibnamefont{and}
  \bibinfo{author}{\bibfnamefont{J.}~\bibnamefont{Lobo-Checa}},
  \bibinfo{journal}{New J Phys} \textbf{\bibinfo{volume}{20}},
  \bibinfo{pages}{073010} (\bibinfo{year}{2018}).

\bibitem[{\citenamefont{Mugarza and Ortega}(2003)}]{Mugarza2003}
\bibinfo{author}{\bibfnamefont{A.}~\bibnamefont{Mugarza}} \bibnamefont{and}
  \bibinfo{author}{\bibfnamefont{J.~E.} \bibnamefont{Ortega}},
  \bibinfo{journal}{J Phys : Condens Matter} \textbf{\bibinfo{volume}{15}},
  \bibinfo{pages}{S3281} (\bibinfo{year}{2003}).

\bibitem[{\citenamefont{Fern\'andez et~al.}(2016)\citenamefont{Fern\'andez,
  Moro-Lagares, Serrate, and Aligia}}]{Fernandez2016}
\bibinfo{author}{\bibfnamefont{J.}~\bibnamefont{Fern\'andez}},
  \bibinfo{author}{\bibfnamefont{M.}~\bibnamefont{Moro-Lagares}},
  \bibinfo{author}{\bibfnamefont{D.}~\bibnamefont{Serrate}}, \bibnamefont{and}
  \bibinfo{author}{\bibfnamefont{A.~A.} \bibnamefont{Aligia}},
  \bibinfo{journal}{Phys Rev B} \textbf{\bibinfo{volume}{94}},
  \bibinfo{pages}{075408} (\bibinfo{year}{2016}).

\bibitem[{\citenamefont{Ortega et~al.}(2013)\citenamefont{Ortega, Lobo-Checa,
  Peschel, Schirone, Abd El-Fattah, Matena, Schiller, Borghetti, Gambardella,
  and Mugarza}}]{Ortega2013}
\bibinfo{author}{\bibfnamefont{J.~E.} \bibnamefont{Ortega}},
  \bibinfo{author}{\bibfnamefont{J.}~\bibnamefont{Lobo-Checa}},
  \bibinfo{author}{\bibfnamefont{G.}~\bibnamefont{Peschel}},
  \bibinfo{author}{\bibfnamefont{S.}~\bibnamefont{Schirone}},
  \bibinfo{author}{\bibfnamefont{Z.~M.} \bibnamefont{Abd El-Fattah}},
  \bibinfo{author}{\bibfnamefont{M.}~\bibnamefont{Matena}},
  \bibinfo{author}{\bibfnamefont{F.}~\bibnamefont{Schiller}},
  \bibinfo{author}{\bibfnamefont{P.}~\bibnamefont{Borghetti}},
  \bibinfo{author}{\bibfnamefont{P.}~\bibnamefont{Gambardella}},
  \bibnamefont{and} \bibinfo{author}{\bibfnamefont{A.}~\bibnamefont{Mugarza}},
  \bibinfo{journal}{Phys Rev B} \textbf{\bibinfo{volume}{87}},
  \bibinfo{pages}{115425} (\bibinfo{year}{2013}).

\bibitem[{\citenamefont{Negulyaev et~al.}(2008)\citenamefont{Negulyaev,
  Stepanyuk, Niebergall, Bruno, Hergert, Repp, Rieder, and
  Meyer}}]{Negulyaev2008}
\bibinfo{author}{\bibfnamefont{N.~N.} \bibnamefont{Negulyaev}},
  \bibinfo{author}{\bibfnamefont{V.~S.} \bibnamefont{Stepanyuk}},
  \bibinfo{author}{\bibfnamefont{L.}~\bibnamefont{Niebergall}},
  \bibinfo{author}{\bibfnamefont{P.}~\bibnamefont{Bruno}},
  \bibinfo{author}{\bibfnamefont{W.}~\bibnamefont{Hergert}},
  \bibinfo{author}{\bibfnamefont{J.}~\bibnamefont{Repp}},
  \bibinfo{author}{\bibfnamefont{K.-H.} \bibnamefont{Rieder}},
  \bibnamefont{and} \bibinfo{author}{\bibfnamefont{G.}~\bibnamefont{Meyer}},
  \bibinfo{journal}{Phys Rev Lett} \textbf{\bibinfo{volume}{101}},
  \bibinfo{pages}{226601} (\bibinfo{year}{2008}).

\bibitem[{\citenamefont{Mugarza et~al.}(2006)\citenamefont{Mugarza, Schiller,
  Kuntze, Cord\'on, Ruiz-Os\'es, and Ortega}}]{Mugarza2006}
\bibinfo{author}{\bibfnamefont{A.}~\bibnamefont{Mugarza}},
  \bibinfo{author}{\bibfnamefont{F.}~\bibnamefont{Schiller}},
  \bibinfo{author}{\bibfnamefont{J.}~\bibnamefont{Kuntze}},
  \bibinfo{author}{\bibfnamefont{J.}~\bibnamefont{Cord\'on}},
  \bibinfo{author}{\bibfnamefont{M.}~\bibnamefont{Ruiz-Os\'es}},
  \bibnamefont{and} \bibinfo{author}{\bibfnamefont{J.~E.}
  \bibnamefont{Ortega}}, \bibinfo{journal}{J Phys : Condens Matter}
  \textbf{\bibinfo{volume}{18}}, \bibinfo{pages}{S27} (\bibinfo{year}{2006}).

\bibitem[{\citenamefont{Baumberger et~al.}(2002)\citenamefont{Baumberger,
  Greber, Delley, and Osterwalder}}]{Baumberger2002}
\bibinfo{author}{\bibfnamefont{F.}~\bibnamefont{Baumberger}},
  \bibinfo{author}{\bibfnamefont{T.}~\bibnamefont{Greber}},
  \bibinfo{author}{\bibfnamefont{B.}~\bibnamefont{Delley}}, \bibnamefont{and}
  \bibinfo{author}{\bibfnamefont{J.}~\bibnamefont{Osterwalder}},
  \bibinfo{journal}{Phys Rev Lett} \textbf{\bibinfo{volume}{88}},
  \bibinfo{pages}{237601} (\bibinfo{year}{2002}).

\bibitem[{\citenamefont{Baumberger et~al.}(2004)\citenamefont{Baumberger,
  Hengsberger, Muntwiler, Shi, Krempasky, Patthey, Osterwalder, and
  Greber}}]{Baumberger2004}
\bibinfo{author}{\bibfnamefont{F.}~\bibnamefont{Baumberger}},
  \bibinfo{author}{\bibfnamefont{M.}~\bibnamefont{Hengsberger}},
  \bibinfo{author}{\bibfnamefont{M.}~\bibnamefont{Muntwiler}},
  \bibinfo{author}{\bibfnamefont{M.}~\bibnamefont{Shi}},
  \bibinfo{author}{\bibfnamefont{J.}~\bibnamefont{Krempasky}},
  \bibinfo{author}{\bibfnamefont{L.}~\bibnamefont{Patthey}},
  \bibinfo{author}{\bibfnamefont{J.}~\bibnamefont{Osterwalder}},
  \bibnamefont{and} \bibinfo{author}{\bibfnamefont{T.}~\bibnamefont{Greber}},
  \bibinfo{journal}{Phys Rev Lett} \textbf{\bibinfo{volume}{92}},
  \bibinfo{pages}{196805} (\bibinfo{year}{2004}).

\bibitem[{\citenamefont{Shiraki et~al.}(2004)\citenamefont{Shiraki, Fujisawa,
  Nantoh, and Kawai}}]{Kawai2004}
\bibinfo{author}{\bibfnamefont{S.}~\bibnamefont{Shiraki}},
  \bibinfo{author}{\bibfnamefont{H.}~\bibnamefont{Fujisawa}},
  \bibinfo{author}{\bibfnamefont{M.}~\bibnamefont{Nantoh}}, \bibnamefont{and}
  \bibinfo{author}{\bibfnamefont{M.}~\bibnamefont{Kawai}},
  \bibinfo{journal}{Phys Rev Lett} \textbf{\bibinfo{volume}{92}},
  \bibinfo{pages}{096102} (\bibinfo{year}{2004}).

\bibitem[{\citenamefont{Mitsuoka and Tamura}(2011)}]{Mitsuoka2011}
\bibinfo{author}{\bibfnamefont{S.}~\bibnamefont{Mitsuoka}} \bibnamefont{and}
  \bibinfo{author}{\bibfnamefont{A.}~\bibnamefont{Tamura}}, \bibinfo{journal}{J
  Phys : Condens Matter} \textbf{\bibinfo{volume}{23}}, \bibinfo{pages}{045008}
  (\bibinfo{year}{2011}).

\bibitem[{\citenamefont{Ortega et~al.}(2011)\citenamefont{Ortega, Corso, Abd-el
  Fattah, Goiri, and Schiller}}]{Ortega2011}
\bibinfo{author}{\bibfnamefont{J.~E.} \bibnamefont{Ortega}},
  \bibinfo{author}{\bibfnamefont{M.}~\bibnamefont{Corso}},
  \bibinfo{author}{\bibfnamefont{Z.~M.} \bibnamefont{Abd-el Fattah}},
  \bibinfo{author}{\bibfnamefont{E.~A.} \bibnamefont{Goiri}}, \bibnamefont{and}
  \bibinfo{author}{\bibfnamefont{F.}~\bibnamefont{Schiller}},
  \bibinfo{journal}{Phys Rev B: Condens Matter Mater Phys}
  \textbf{\bibinfo{volume}{83}}, \bibinfo{pages}{085411}
  (\bibinfo{year}{2011}).

\bibitem[{\citenamefont{Crommie
  et~al.}(1995{\natexlab{b}})\citenamefont{Crommie, Lutz, Eigler, and
  Heller}}]{Crommie1995a}
\bibinfo{author}{\bibfnamefont{M.~F.} \bibnamefont{Crommie}},
  \bibinfo{author}{\bibfnamefont{C.~P.} \bibnamefont{Lutz}},
  \bibinfo{author}{\bibfnamefont{D.~M.} \bibnamefont{Eigler}},
  \bibnamefont{and} \bibinfo{author}{\bibfnamefont{E.~J.}
  \bibnamefont{Heller}}, \bibinfo{journal}{Surf Rev Lett}
  \textbf{\bibinfo{volume}{02}}, \bibinfo{pages}{127}
  (\bibinfo{year}{1995}{\natexlab{b}}).

\bibitem[{\citenamefont{Heller et~al.}(1994)\citenamefont{Heller, Crommie,
  Lutz, and Eigler}}]{Heller1994}
\bibinfo{author}{\bibfnamefont{E.~J.} \bibnamefont{Heller}},
  \bibinfo{author}{\bibfnamefont{M.~F.} \bibnamefont{Crommie}},
  \bibinfo{author}{\bibfnamefont{C.~P.} \bibnamefont{Lutz}}, \bibnamefont{and}
  \bibinfo{author}{\bibfnamefont{D.~M.} \bibnamefont{Eigler}},
  \bibinfo{journal}{Nature} \textbf{\bibinfo{volume}{369}},
  \bibinfo{pages}{464} (\bibinfo{year}{1994}).

\bibitem[{\citenamefont{Braun and Rieder}(2002)}]{Braun2002}
\bibinfo{author}{\bibfnamefont{K.-F.} \bibnamefont{Braun}} \bibnamefont{and}
  \bibinfo{author}{\bibfnamefont{K.-H.} \bibnamefont{Rieder}},
  \bibinfo{journal}{Phys Rev Lett} \textbf{\bibinfo{volume}{88}},
  \bibinfo{pages}{096801} (\bibinfo{year}{2002}).

\bibitem[{\citenamefont{Lagoute et~al.}(2005)\citenamefont{Lagoute, Liu, and
  F\"olsch}}]{Lagoute2005}
\bibinfo{author}{\bibfnamefont{J.}~\bibnamefont{Lagoute}},
  \bibinfo{author}{\bibfnamefont{X.}~\bibnamefont{Liu}}, \bibnamefont{and}
  \bibinfo{author}{\bibfnamefont{S.}~\bibnamefont{F\"olsch}},
  \bibinfo{journal}{Phys Rev Lett} \textbf{\bibinfo{volume}{95}},
  \bibinfo{pages}{136801} (\bibinfo{year}{2005}).

\bibitem[{\citenamefont{Kumagai and Tamura}(2008)}]{Kumagai2008}
\bibinfo{author}{\bibfnamefont{T.}~\bibnamefont{Kumagai}} \bibnamefont{and}
  \bibinfo{author}{\bibfnamefont{A.}~\bibnamefont{Tamura}}, \bibinfo{journal}{J
  Phys Soc Jpn} \textbf{\bibinfo{volume}{77}}, \bibinfo{pages}{014601}
  (\bibinfo{year}{2008}).

\bibitem[{\citenamefont{Crampin and Bryant}(1996)}]{Crampin1996}
\bibinfo{author}{\bibfnamefont{S.}~\bibnamefont{Crampin}} \bibnamefont{and}
  \bibinfo{author}{\bibfnamefont{O.~R.} \bibnamefont{Bryant}},
  \bibinfo{journal}{Phys Rev B} \textbf{\bibinfo{volume}{54}},
  \bibinfo{pages}{R17367} (\bibinfo{year}{1996}).

\bibitem[{\citenamefont{Kumagai and Tamura}(2009)}]{Kumagai2009}
\bibinfo{author}{\bibfnamefont{T.}~\bibnamefont{Kumagai}} \bibnamefont{and}
  \bibinfo{author}{\bibfnamefont{A.}~\bibnamefont{Tamura}}, \bibinfo{journal}{J
  Phys : Condens Matter} \textbf{\bibinfo{volume}{21}}, \bibinfo{pages}{225004}
  (\bibinfo{year}{2009}).

\bibitem[{\citenamefont{Tatsumi et~al.}(2018)\citenamefont{Tatsumi, Mitsuoka,
  and Tamura}}]{Tatsumi2018}
\bibinfo{author}{\bibfnamefont{Y.}~\bibnamefont{Tatsumi}},
  \bibinfo{author}{\bibfnamefont{S.}~\bibnamefont{Mitsuoka}}, \bibnamefont{and}
  \bibinfo{author}{\bibfnamefont{A.}~\bibnamefont{Tamura}},
  \bibinfo{journal}{Surf Rev Lett} \textbf{\bibinfo{volume}{25}},
  \bibinfo{pages}{1850091} (\bibinfo{year}{2018}).

\bibitem[{\citenamefont{Harbury and Porod}(1996)}]{Harbury1996}
\bibinfo{author}{\bibfnamefont{H.~K.} \bibnamefont{Harbury}} \bibnamefont{and}
  \bibinfo{author}{\bibfnamefont{W.}~\bibnamefont{Porod}},
  \bibinfo{journal}{Phys Rev B} \textbf{\bibinfo{volume}{53}},
  \bibinfo{pages}{15455} (\bibinfo{year}{1996}).

\bibitem[{\citenamefont{Rahachou and Zozoulenko}(2004)}]{Rahachou2004}
\bibinfo{author}{\bibfnamefont{A.~I.} \bibnamefont{Rahachou}} \bibnamefont{and}
  \bibinfo{author}{\bibfnamefont{I.~V.} \bibnamefont{Zozoulenko}},
  \bibinfo{journal}{Phys Rev B} \textbf{\bibinfo{volume}{70}},
  \bibinfo{pages}{233409} (\bibinfo{year}{2004}).

\bibitem[{\citenamefont{Fiete and Heller}(2003)}]{Fiete2003}
\bibinfo{author}{\bibfnamefont{G.~A.} \bibnamefont{Fiete}} \bibnamefont{and}
  \bibinfo{author}{\bibfnamefont{E.~J.} \bibnamefont{Heller}},
  \bibinfo{journal}{Rev Mod Phys} \textbf{\bibinfo{volume}{75}},
  \bibinfo{pages}{933} (\bibinfo{year}{2003}).

\bibitem[{\citenamefont{Niebergall et~al.}(2006)\citenamefont{Niebergall,
  Rodary, Ding, Sander, Stepanyuk, Bruno, and Kirschner}}]{Niebergall2006}
\bibinfo{author}{\bibfnamefont{L.}~\bibnamefont{Niebergall}},
  \bibinfo{author}{\bibfnamefont{G.}~\bibnamefont{Rodary}},
  \bibinfo{author}{\bibfnamefont{H.~F.} \bibnamefont{Ding}},
  \bibinfo{author}{\bibfnamefont{D.}~\bibnamefont{Sander}},
  \bibinfo{author}{\bibfnamefont{V.~S.} \bibnamefont{Stepanyuk}},
  \bibinfo{author}{\bibfnamefont{P.}~\bibnamefont{Bruno}}, \bibnamefont{and}
  \bibinfo{author}{\bibfnamefont{J.}~\bibnamefont{Kirschner}},
  \bibinfo{journal}{Phys Rev B} \textbf{\bibinfo{volume}{74}},
  \bibinfo{pages}{195436} (\bibinfo{year}{2006}).

\bibitem[{\citenamefont{Crampin et~al.}(2005)\citenamefont{Crampin, Jensen,
  Kröger, Limot, and Berndt}}]{Crampin2005}
\bibinfo{author}{\bibfnamefont{S.}~\bibnamefont{Crampin}},
  \bibinfo{author}{\bibfnamefont{H.}~\bibnamefont{Jensen}},
  \bibinfo{author}{\bibfnamefont{J.}~\bibnamefont{Kröger}},
  \bibinfo{author}{\bibfnamefont{L.}~\bibnamefont{Limot}}, \bibnamefont{and}
  \bibinfo{author}{\bibfnamefont{R.}~\bibnamefont{Berndt}},
  \bibinfo{journal}{Phys Rev B} \textbf{\bibinfo{volume}{72}},
  \bibinfo{pages}{035443} (\bibinfo{year}{2005}).

\bibitem[{\citenamefont{Ternes et~al.}(2004)\citenamefont{Ternes, Weber,
  Pivetta, Patthey, Pelz, Giamarchi, Mila, and Schneider}}]{Ternes2004}
\bibinfo{author}{\bibfnamefont{M.}~\bibnamefont{Ternes}},
  \bibinfo{author}{\bibfnamefont{C.}~\bibnamefont{Weber}},
  \bibinfo{author}{\bibfnamefont{M.}~\bibnamefont{Pivetta}},
  \bibinfo{author}{\bibfnamefont{F.}~\bibnamefont{Patthey}},
  \bibinfo{author}{\bibfnamefont{J.~P.} \bibnamefont{Pelz}},
  \bibinfo{author}{\bibfnamefont{T.}~\bibnamefont{Giamarchi}},
  \bibinfo{author}{\bibfnamefont{F.}~\bibnamefont{Mila}}, \bibnamefont{and}
  \bibinfo{author}{\bibfnamefont{W.-D.} \bibnamefont{Schneider}},
  \bibinfo{journal}{Phys Rev Lett} \textbf{\bibinfo{volume}{93}},
  \bibinfo{pages}{146805} (\bibinfo{year}{2004}).

\bibitem[{\citenamefont{J\"ack et~al.}(2021)\citenamefont{J\"ack, Zinser,
  K\"onig, Wissing, Schmidt, Donath, Kern, and Ast}}]{Jack2020}
\bibinfo{author}{\bibfnamefont{B.}~\bibnamefont{J\"ack}},
  \bibinfo{author}{\bibfnamefont{F.}~\bibnamefont{Zinser}},
  \bibinfo{author}{\bibfnamefont{E.~J.} \bibnamefont{K\"onig}},
  \bibinfo{author}{\bibfnamefont{S.~N.~P.} \bibnamefont{Wissing}},
  \bibinfo{author}{\bibfnamefont{A.~B.} \bibnamefont{Schmidt}},
  \bibinfo{author}{\bibfnamefont{M.}~\bibnamefont{Donath}},
  \bibinfo{author}{\bibfnamefont{K.}~\bibnamefont{Kern}}, \bibnamefont{and}
  \bibinfo{author}{\bibfnamefont{C.~R.} \bibnamefont{Ast}},
  \bibinfo{journal}{Phys. Rev. Research} \textbf{\bibinfo{volume}{3}},
  \bibinfo{pages}{013022} (\bibinfo{year}{2021}).

\bibitem[{\citenamefont{Stilp et~al.}(2021)\citenamefont{Stilp, Bereczuk,
  Berwanger, Mundigl, Richter, and Giessibl}}]{Stilp2021}
\bibinfo{author}{\bibfnamefont{F.}~\bibnamefont{Stilp}},
  \bibinfo{author}{\bibfnamefont{A.}~\bibnamefont{Bereczuk}},
  \bibinfo{author}{\bibfnamefont{J.}~\bibnamefont{Berwanger}},
  \bibinfo{author}{\bibfnamefont{N.}~\bibnamefont{Mundigl}},
  \bibinfo{author}{\bibfnamefont{K.}~\bibnamefont{Richter}}, \bibnamefont{and}
  \bibinfo{author}{\bibfnamefont{F.~J.} \bibnamefont{Giessibl}},
  \bibinfo{journal}{Science} \textbf{\bibinfo{volume}{372}},
  \bibinfo{pages}{1196} (\bibinfo{year}{2021}).

\bibitem[{\citenamefont{Manoharan et~al.}(2000)\citenamefont{Manoharan, Lutz,
  and Eigler}}]{Manoharan2000}
\bibinfo{author}{\bibfnamefont{H.~C.} \bibnamefont{Manoharan}},
  \bibinfo{author}{\bibfnamefont{C.~P.} \bibnamefont{Lutz}}, \bibnamefont{and}
  \bibinfo{author}{\bibfnamefont{D.~M.} \bibnamefont{Eigler}},
  \bibinfo{journal}{Nature} \textbf{\bibinfo{volume}{403}},
  \bibinfo{pages}{512} (\bibinfo{year}{2000}).

\bibitem[{\citenamefont{Oka et~al.}(2010)\citenamefont{Oka, Ignatiev, Wedekind,
  Rodary, Niebergall, Stepanyuk, Sander, and Kirschner}}]{Oka2010}
\bibinfo{author}{\bibfnamefont{H.}~\bibnamefont{Oka}},
  \bibinfo{author}{\bibfnamefont{P.~A.} \bibnamefont{Ignatiev}},
  \bibinfo{author}{\bibfnamefont{S.}~\bibnamefont{Wedekind}},
  \bibinfo{author}{\bibfnamefont{G.}~\bibnamefont{Rodary}},
  \bibinfo{author}{\bibfnamefont{L.}~\bibnamefont{Niebergall}},
  \bibinfo{author}{\bibfnamefont{V.~S.} \bibnamefont{Stepanyuk}},
  \bibinfo{author}{\bibfnamefont{D.}~\bibnamefont{Sander}}, \bibnamefont{and}
  \bibinfo{author}{\bibfnamefont{J.}~\bibnamefont{Kirschner}},
  \bibinfo{journal}{Science} \textbf{\bibinfo{volume}{327}},
  \bibinfo{pages}{843} (\bibinfo{year}{2010}).

\bibitem[{\citenamefont{Stepanyuk et~al.}(2005)\citenamefont{Stepanyuk,
  Niebergall, Hergert, and Bruno}}]{Stepanyuk2005}
\bibinfo{author}{\bibfnamefont{V.~S.} \bibnamefont{Stepanyuk}},
  \bibinfo{author}{\bibfnamefont{L.}~\bibnamefont{Niebergall}},
  \bibinfo{author}{\bibfnamefont{W.}~\bibnamefont{Hergert}}, \bibnamefont{and}
  \bibinfo{author}{\bibfnamefont{P.}~\bibnamefont{Bruno}},
  \bibinfo{journal}{Phys Rev Lett} \textbf{\bibinfo{volume}{94}},
  \bibinfo{pages}{187201} (\bibinfo{year}{2005}).

\bibitem[{\citenamefont{Rossi and Morr}(2006)}]{Rossi2006}
\bibinfo{author}{\bibfnamefont{E.}~\bibnamefont{Rossi}} \bibnamefont{and}
  \bibinfo{author}{\bibfnamefont{D.~K.} \bibnamefont{Morr}},
  \bibinfo{journal}{Phys Rev Lett} \textbf{\bibinfo{volume}{97}},
  \bibinfo{pages}{236602} (\bibinfo{year}{2006}).

\bibitem[{\citenamefont{Figgins et~al.}(2019)\citenamefont{Figgins, Mattos,
  Mar, Chen, Manoharan, and Morr}}]{Morr2019}
\bibinfo{author}{\bibfnamefont{J.}~\bibnamefont{Figgins}},
  \bibinfo{author}{\bibfnamefont{L.~S.} \bibnamefont{Mattos}},
  \bibinfo{author}{\bibfnamefont{W.}~\bibnamefont{Mar}},
  \bibinfo{author}{\bibfnamefont{Y.-T.} \bibnamefont{Chen}},
  \bibinfo{author}{\bibfnamefont{H.~C.} \bibnamefont{Manoharan}},
  \bibnamefont{and} \bibinfo{author}{\bibfnamefont{D.~K.} \bibnamefont{Morr}},
  \bibinfo{journal}{Nat Commun} \textbf{\bibinfo{volume}{10}},
  \bibinfo{pages}{5588} (\bibinfo{year}{2019}).

\bibitem[{\citenamefont{Li et~al.}(2020{\natexlab{a}})\citenamefont{Li, Li,
  Miao, Sun, Chen, Han, and Ding}}]{Li2020}
\bibinfo{author}{\bibfnamefont{Q.}~\bibnamefont{Li}},
  \bibinfo{author}{\bibfnamefont{X.}~\bibnamefont{Li}},
  \bibinfo{author}{\bibfnamefont{B.}~\bibnamefont{Miao}},
  \bibinfo{author}{\bibfnamefont{L.}~\bibnamefont{Sun}},
  \bibinfo{author}{\bibfnamefont{G.}~\bibnamefont{Chen}},
  \bibinfo{author}{\bibfnamefont{P.}~\bibnamefont{Han}}, \bibnamefont{and}
  \bibinfo{author}{\bibfnamefont{H.}~\bibnamefont{Ding}}, \bibinfo{journal}{Nat
  Commun} \textbf{\bibinfo{volume}{11}}, \bibinfo{pages}{1400}
  (\bibinfo{year}{2020}{\natexlab{a}}).

\bibitem[{\citenamefont{Li et~al.}(2020{\natexlab{b}})\citenamefont{Li, Cao,
  and Ding}}]{Li2020b}
\bibinfo{author}{\bibfnamefont{Q.}~\bibnamefont{Li}},
  \bibinfo{author}{\bibfnamefont{R.}~\bibnamefont{Cao}}, \bibnamefont{and}
  \bibinfo{author}{\bibfnamefont{H.}~\bibnamefont{Ding}},
  \bibinfo{journal}{Appl Phys Lett} \textbf{\bibinfo{volume}{117}},
  \bibinfo{pages}{060501} (\bibinfo{year}{2020}{\natexlab{b}}).

\bibitem[{\citenamefont{Pietzsch et~al.}(2006)\citenamefont{Pietzsch, Okatov,
  Kubetzka, Bode, Heinze, Lichtenstein, and Wiesendanger}}]{Pietzsch2006}
\bibinfo{author}{\bibfnamefont{O.}~\bibnamefont{Pietzsch}},
  \bibinfo{author}{\bibfnamefont{S.}~\bibnamefont{Okatov}},
  \bibinfo{author}{\bibfnamefont{A.}~\bibnamefont{Kubetzka}},
  \bibinfo{author}{\bibfnamefont{M.}~\bibnamefont{Bode}},
  \bibinfo{author}{\bibfnamefont{S.}~\bibnamefont{Heinze}},
  \bibinfo{author}{\bibfnamefont{A.}~\bibnamefont{Lichtenstein}},
  \bibnamefont{and}
  \bibinfo{author}{\bibfnamefont{R.}~\bibnamefont{Wiesendanger}},
  \bibinfo{journal}{Phys Rev Lett} \textbf{\bibinfo{volume}{96}},
  \bibinfo{pages}{237203} (\bibinfo{year}{2006}).

\bibitem[{\citenamefont{Pascual et~al.}(2004)\citenamefont{Pascual, Bihlmayer,
  Koroteev, Rust, Ceballos, Hansmann, Horn, Chulkov, Bl\"ugel, Echenique
  et~al.}}]{Pascual2004}
\bibinfo{author}{\bibfnamefont{J.~I.} \bibnamefont{Pascual}},
  \bibinfo{author}{\bibfnamefont{G.}~\bibnamefont{Bihlmayer}},
  \bibinfo{author}{\bibfnamefont{Y.~M.} \bibnamefont{Koroteev}},
  \bibinfo{author}{\bibfnamefont{H.-P.} \bibnamefont{Rust}},
  \bibinfo{author}{\bibfnamefont{G.}~\bibnamefont{Ceballos}},
  \bibinfo{author}{\bibfnamefont{M.}~\bibnamefont{Hansmann}},
  \bibinfo{author}{\bibfnamefont{K.}~\bibnamefont{Horn}},
  \bibinfo{author}{\bibfnamefont{E.~V.} \bibnamefont{Chulkov}},
  \bibinfo{author}{\bibfnamefont{S.}~\bibnamefont{Bl\"ugel}},
  \bibinfo{author}{\bibfnamefont{P.~M.} \bibnamefont{Echenique}},
  \bibnamefont{et~al.}, \bibinfo{journal}{Phys Rev Lett}
  \textbf{\bibinfo{volume}{93}}, \bibinfo{pages}{196802}
  (\bibinfo{year}{2004}).

\bibitem[{\citenamefont{Zhang et~al.}(2020{\natexlab{a}})\citenamefont{Zhang,
  Li, Pei, Liu, and Chen}}]{Zhang2020b}
\bibinfo{author}{\bibfnamefont{C.}~\bibnamefont{Zhang}},
  \bibinfo{author}{\bibfnamefont{Y.}~\bibnamefont{Li}},
  \bibinfo{author}{\bibfnamefont{D.}~\bibnamefont{Pei}},
  \bibinfo{author}{\bibfnamefont{Z.}~\bibnamefont{Liu}}, \bibnamefont{and}
  \bibinfo{author}{\bibfnamefont{Y.}~\bibnamefont{Chen}},
  \bibinfo{journal}{Annu Rev Mater Res} \textbf{\bibinfo{volume}{50}},
  \bibinfo{pages}{131} (\bibinfo{year}{2020}{\natexlab{a}}).

\bibitem[{\citenamefont{Sobota et~al.}(2021)\citenamefont{Sobota, He, and
  Shen}}]{Sobota2020}
\bibinfo{author}{\bibfnamefont{J.~A.} \bibnamefont{Sobota}},
  \bibinfo{author}{\bibfnamefont{Y.}~\bibnamefont{He}}, \bibnamefont{and}
  \bibinfo{author}{\bibfnamefont{Z.-X.} \bibnamefont{Shen}},
  \bibinfo{journal}{Rev Mod Phys} \textbf{\bibinfo{volume}{93}},
  \bibinfo{pages}{025006} (\bibinfo{year}{2021}).

\bibitem[{\citenamefont{Seo et~al.}(2010)\citenamefont{Seo, Roushan,
  Beidenkopf, Hor, Cava, and Yazdani}}]{Yazdani2010}
\bibinfo{author}{\bibfnamefont{J.}~\bibnamefont{Seo}},
  \bibinfo{author}{\bibfnamefont{P.}~\bibnamefont{Roushan}},
  \bibinfo{author}{\bibfnamefont{H.}~\bibnamefont{Beidenkopf}},
  \bibinfo{author}{\bibfnamefont{Y.~S.} \bibnamefont{Hor}},
  \bibinfo{author}{\bibfnamefont{R.~J.} \bibnamefont{Cava}}, \bibnamefont{and}
  \bibinfo{author}{\bibfnamefont{A.}~\bibnamefont{Yazdani}},
  \bibinfo{journal}{Nature} \textbf{\bibinfo{volume}{466}},
  \bibinfo{pages}{343} (\bibinfo{year}{2010}).

\bibitem[{\citenamefont{Zhang et~al.}(2009)\citenamefont{Zhang, Cheng, Chen,
  Jia, Ma, He, Wang, Zhang, Dai, Fang et~al.}}]{Zhang2009}
\bibinfo{author}{\bibfnamefont{T.}~\bibnamefont{Zhang}},
  \bibinfo{author}{\bibfnamefont{P.}~\bibnamefont{Cheng}},
  \bibinfo{author}{\bibfnamefont{X.}~\bibnamefont{Chen}},
  \bibinfo{author}{\bibfnamefont{J.-F.} \bibnamefont{Jia}},
  \bibinfo{author}{\bibfnamefont{X.}~\bibnamefont{Ma}},
  \bibinfo{author}{\bibfnamefont{K.}~\bibnamefont{He}},
  \bibinfo{author}{\bibfnamefont{L.}~\bibnamefont{Wang}},
  \bibinfo{author}{\bibfnamefont{H.}~\bibnamefont{Zhang}},
  \bibinfo{author}{\bibfnamefont{X.}~\bibnamefont{Dai}},
  \bibinfo{author}{\bibfnamefont{Z.}~\bibnamefont{Fang}}, \bibnamefont{et~al.},
  \bibinfo{journal}{Phys Rev Lett} \textbf{\bibinfo{volume}{103}},
  \bibinfo{pages}{266803} (\bibinfo{year}{2009}).

\bibitem[{\citenamefont{Chen et~al.}(2019)\citenamefont{Chen, Jiang, Peng,
  Zhang, Chang, Feng, Fu, Zheng, Zhang, Wang et~al.}}]{Chen2019}
\bibinfo{author}{\bibfnamefont{M.}~\bibnamefont{Chen}},
  \bibinfo{author}{\bibfnamefont{Y.-P.} \bibnamefont{Jiang}},
  \bibinfo{author}{\bibfnamefont{J.}~\bibnamefont{Peng}},
  \bibinfo{author}{\bibfnamefont{H.}~\bibnamefont{Zhang}},
  \bibinfo{author}{\bibfnamefont{C.-Z.} \bibnamefont{Chang}},
  \bibinfo{author}{\bibfnamefont{X.}~\bibnamefont{Feng}},
  \bibinfo{author}{\bibfnamefont{Z.}~\bibnamefont{Fu}},
  \bibinfo{author}{\bibfnamefont{F.}~\bibnamefont{Zheng}},
  \bibinfo{author}{\bibfnamefont{P.}~\bibnamefont{Zhang}},
  \bibinfo{author}{\bibfnamefont{L.}~\bibnamefont{Wang}}, \bibnamefont{et~al.},
  \bibinfo{journal}{Sci Adv} \textbf{\bibinfo{volume}{5}},
  \bibinfo{pages}{eaaw3988} (\bibinfo{year}{2019}).

\bibitem[{\citenamefont{Avraham et~al.}(2020)\citenamefont{Avraham,
  Kumar~Nayak, Steinbok, Norris, Fu, Sun, Qi, Pan, Isaeva, Zeugner
  et~al.}}]{Avraham2020}
\bibinfo{author}{\bibfnamefont{N.}~\bibnamefont{Avraham}},
  \bibinfo{author}{\bibfnamefont{A.}~\bibnamefont{Kumar~Nayak}},
  \bibinfo{author}{\bibfnamefont{A.}~\bibnamefont{Steinbok}},
  \bibinfo{author}{\bibfnamefont{A.}~\bibnamefont{Norris}},
  \bibinfo{author}{\bibfnamefont{H.}~\bibnamefont{Fu}},
  \bibinfo{author}{\bibfnamefont{Y.}~\bibnamefont{Sun}},
  \bibinfo{author}{\bibfnamefont{Y.}~\bibnamefont{Qi}},
  \bibinfo{author}{\bibfnamefont{L.}~\bibnamefont{Pan}},
  \bibinfo{author}{\bibfnamefont{A.}~\bibnamefont{Isaeva}},
  \bibinfo{author}{\bibfnamefont{A.}~\bibnamefont{Zeugner}},
  \bibnamefont{et~al.}, \bibinfo{journal}{Nat Mater}
  \textbf{\bibinfo{volume}{19}}, \bibinfo{pages}{610} (\bibinfo{year}{2020}).

\bibitem[{\citenamefont{Sessi et~al.}(2016)\citenamefont{Sessi, Di~Sante,
  Szczerbakow, Glott, Wilfert, Schmidt, Bathon, Dziawa, Greiter, Neupert
  et~al.}}]{Sessi2016}
\bibinfo{author}{\bibfnamefont{P.}~\bibnamefont{Sessi}},
  \bibinfo{author}{\bibfnamefont{D.}~\bibnamefont{Di~Sante}},
  \bibinfo{author}{\bibfnamefont{A.}~\bibnamefont{Szczerbakow}},
  \bibinfo{author}{\bibfnamefont{F.}~\bibnamefont{Glott}},
  \bibinfo{author}{\bibfnamefont{S.}~\bibnamefont{Wilfert}},
  \bibinfo{author}{\bibfnamefont{H.}~\bibnamefont{Schmidt}},
  \bibinfo{author}{\bibfnamefont{T.}~\bibnamefont{Bathon}},
  \bibinfo{author}{\bibfnamefont{P.}~\bibnamefont{Dziawa}},
  \bibinfo{author}{\bibfnamefont{M.}~\bibnamefont{Greiter}},
  \bibinfo{author}{\bibfnamefont{T.}~\bibnamefont{Neupert}},
  \bibnamefont{et~al.}, \bibinfo{journal}{Science}
  \textbf{\bibinfo{volume}{354}}, \bibinfo{pages}{1269} (\bibinfo{year}{2016}).

\bibitem[{\citenamefont{Potter and Lee}(2012)}]{Lee2012}
\bibinfo{author}{\bibfnamefont{A.~C.} \bibnamefont{Potter}} \bibnamefont{and}
  \bibinfo{author}{\bibfnamefont{P.~A.} \bibnamefont{Lee}},
  \bibinfo{journal}{Phys Rev B} \textbf{\bibinfo{volume}{85}},
  \bibinfo{pages}{094516} (\bibinfo{year}{2012}).

\bibitem[{\citenamefont{Wei et~al.}(2019)\citenamefont{Wei, Manna, Eich, Lee,
  and Moodera}}]{Wei2019}
\bibinfo{author}{\bibfnamefont{P.}~\bibnamefont{Wei}},
  \bibinfo{author}{\bibfnamefont{S.}~\bibnamefont{Manna}},
  \bibinfo{author}{\bibfnamefont{M.}~\bibnamefont{Eich}},
  \bibinfo{author}{\bibfnamefont{P.}~\bibnamefont{Lee}}, \bibnamefont{and}
  \bibinfo{author}{\bibfnamefont{J.}~\bibnamefont{Moodera}},
  \bibinfo{journal}{Phys Rev Lett} \textbf{\bibinfo{volume}{122}},
  \bibinfo{pages}{247002} (\bibinfo{year}{2019}).

\bibitem[{\citenamefont{Manna et~al.}(2020)\citenamefont{Manna, Wei, Xie, Law,
  Lee, and Moodera}}]{Manna2020}
\bibinfo{author}{\bibfnamefont{S.}~\bibnamefont{Manna}},
  \bibinfo{author}{\bibfnamefont{P.}~\bibnamefont{Wei}},
  \bibinfo{author}{\bibfnamefont{Y.}~\bibnamefont{Xie}},
  \bibinfo{author}{\bibfnamefont{K.~T.} \bibnamefont{Law}},
  \bibinfo{author}{\bibfnamefont{P.~A.} \bibnamefont{Lee}}, \bibnamefont{and}
  \bibinfo{author}{\bibfnamefont{J.~S.} \bibnamefont{Moodera}},
  \bibinfo{journal}{Proceedings of the National Academy of Sciences}
  \textbf{\bibinfo{volume}{117}}, \bibinfo{pages}{8775} (\bibinfo{year}{2020}).

\bibitem[{\citenamefont{Moon et~al.}(2008)\citenamefont{Moon, Mattos, Foster,
  Zeltzer, Ko, and Manoharan}}]{Moon2008}
\bibinfo{author}{\bibfnamefont{C.~R.} \bibnamefont{Moon}},
  \bibinfo{author}{\bibfnamefont{L.~S.} \bibnamefont{Mattos}},
  \bibinfo{author}{\bibfnamefont{B.~K.} \bibnamefont{Foster}},
  \bibinfo{author}{\bibfnamefont{G.}~\bibnamefont{Zeltzer}},
  \bibinfo{author}{\bibfnamefont{W.}~\bibnamefont{Ko}}, \bibnamefont{and}
  \bibinfo{author}{\bibfnamefont{H.~C.} \bibnamefont{Manoharan}},
  \bibinfo{journal}{Science} \textbf{\bibinfo{volume}{319}},
  \bibinfo{pages}{782} (\bibinfo{year}{2008}).

\bibitem[{\citenamefont{Moon et~al.}(2009)\citenamefont{Moon, Mattos, Foster,
  Zeltzer, and Manoharan}}]{Manoharan2009}
\bibinfo{author}{\bibfnamefont{C.~R.} \bibnamefont{Moon}},
  \bibinfo{author}{\bibfnamefont{L.~S.} \bibnamefont{Mattos}},
  \bibinfo{author}{\bibfnamefont{B.~K.} \bibnamefont{Foster}},
  \bibinfo{author}{\bibfnamefont{G.}~\bibnamefont{Zeltzer}}, \bibnamefont{and}
  \bibinfo{author}{\bibfnamefont{H.~C.} \bibnamefont{Manoharan}},
  \bibinfo{journal}{Nat Nanotechnol} \textbf{\bibinfo{volume}{4}},
  \bibinfo{pages}{167} (\bibinfo{year}{2009}).

\bibitem[{\citenamefont{Park and Louie}(2009{\natexlab{b}})}]{Park2009}
\bibinfo{author}{\bibfnamefont{C.-H.} \bibnamefont{Park}} \bibnamefont{and}
  \bibinfo{author}{\bibfnamefont{S.~G.} \bibnamefont{Louie}},
  \bibinfo{journal}{Nano Lett} \textbf{\bibinfo{volume}{9}},
  \bibinfo{pages}{1793} (\bibinfo{year}{2009}{\natexlab{b}}).

\bibitem[{\citenamefont{Singha et~al.}(2011)\citenamefont{Singha, Gibertini,
  Karmakar, Yuan, Polini, Vignale, Katsnelson, Pinczuk, Pfeiffer, West
  et~al.}}]{Singha2011}
\bibinfo{author}{\bibfnamefont{A.}~\bibnamefont{Singha}},
  \bibinfo{author}{\bibfnamefont{M.}~\bibnamefont{Gibertini}},
  \bibinfo{author}{\bibfnamefont{B.}~\bibnamefont{Karmakar}},
  \bibinfo{author}{\bibfnamefont{S.}~\bibnamefont{Yuan}},
  \bibinfo{author}{\bibfnamefont{M.}~\bibnamefont{Polini}},
  \bibinfo{author}{\bibfnamefont{G.}~\bibnamefont{Vignale}},
  \bibinfo{author}{\bibfnamefont{M.~I.} \bibnamefont{Katsnelson}},
  \bibinfo{author}{\bibfnamefont{A.}~\bibnamefont{Pinczuk}},
  \bibinfo{author}{\bibfnamefont{L.~N.} \bibnamefont{Pfeiffer}},
  \bibinfo{author}{\bibfnamefont{K.~W.} \bibnamefont{West}},
  \bibnamefont{et~al.}, \bibinfo{journal}{Science}
  \textbf{\bibinfo{volume}{332}}, \bibinfo{pages}{1176} (\bibinfo{year}{2011}).

\bibitem[{\citenamefont{Yan and Liljeroth}(2019)}]{Yan2019}
\bibinfo{author}{\bibfnamefont{L.}~\bibnamefont{Yan}} \bibnamefont{and}
  \bibinfo{author}{\bibfnamefont{P.}~\bibnamefont{Liljeroth}},
  \bibinfo{journal}{Advances in Physics: X} \textbf{\bibinfo{volume}{4}},
  \bibinfo{pages}{1651672} (\bibinfo{year}{2019}).

\bibitem[{\citenamefont{Khajetoorians et~al.}(2019)\citenamefont{Khajetoorians,
  Wegner, Otte, and Swart}}]{Khajetoorians2019}
\bibinfo{author}{\bibfnamefont{A.~A.} \bibnamefont{Khajetoorians}},
  \bibinfo{author}{\bibfnamefont{D.}~\bibnamefont{Wegner}},
  \bibinfo{author}{\bibfnamefont{A.~F.} \bibnamefont{Otte}}, \bibnamefont{and}
  \bibinfo{author}{\bibfnamefont{I.}~\bibnamefont{Swart}},
  \bibinfo{journal}{Nature Reviews Physics} \textbf{\bibinfo{volume}{1}},
  \bibinfo{pages}{703} (\bibinfo{year}{2019}).

\bibitem[{\citenamefont{Yan et~al.}(2019)\citenamefont{Yan, Hua, Zhang, Ngai,
  Guo, Wu, Wang, and Lin}}]{Yan2019a}
\bibinfo{author}{\bibfnamefont{L.}~\bibnamefont{Yan}},
  \bibinfo{author}{\bibfnamefont{M.}~\bibnamefont{Hua}},
  \bibinfo{author}{\bibfnamefont{Q.}~\bibnamefont{Zhang}},
  \bibinfo{author}{\bibfnamefont{T.~U.} \bibnamefont{Ngai}},
  \bibinfo{author}{\bibfnamefont{Z.}~\bibnamefont{Guo}},
  \bibinfo{author}{\bibfnamefont{T.~C.} \bibnamefont{Wu}},
  \bibinfo{author}{\bibfnamefont{T.}~\bibnamefont{Wang}}, \bibnamefont{and}
  \bibinfo{author}{\bibfnamefont{N.}~\bibnamefont{Lin}}, \bibinfo{journal}{New
  J Phys} \textbf{\bibinfo{volume}{21}}, \bibinfo{pages}{083005}
  (\bibinfo{year}{2019}).

\bibitem[{\citenamefont{Yue et~al.}(2020)\citenamefont{Yue, Zhou, Geng, Sun,
  Arita, Shimada, Cheng, Chen, Meng, Wu et~al.}}]{Yue2020}
\bibinfo{author}{\bibfnamefont{S.}~\bibnamefont{Yue}},
  \bibinfo{author}{\bibfnamefont{H.}~\bibnamefont{Zhou}},
  \bibinfo{author}{\bibfnamefont{D.}~\bibnamefont{Geng}},
  \bibinfo{author}{\bibfnamefont{Z.}~\bibnamefont{Sun}},
  \bibinfo{author}{\bibfnamefont{M.}~\bibnamefont{Arita}},
  \bibinfo{author}{\bibfnamefont{K.}~\bibnamefont{Shimada}},
  \bibinfo{author}{\bibfnamefont{P.}~\bibnamefont{Cheng}},
  \bibinfo{author}{\bibfnamefont{L.}~\bibnamefont{Chen}},
  \bibinfo{author}{\bibfnamefont{S.}~\bibnamefont{Meng}},
  \bibinfo{author}{\bibfnamefont{K.}~\bibnamefont{Wu}}, \bibnamefont{et~al.},
  \bibinfo{journal}{Phys Rev B} \textbf{\bibinfo{volume}{102}},
  \bibinfo{pages}{201401} (\bibinfo{year}{2020}).

\bibitem[{\citenamefont{Telychko et~al.}(2021)\citenamefont{Telychko, Li,
  Mutombo, Soler-Polo, Peng, Su, Song, Koh, Edmonds, Jelínek
  et~al.}}]{Pavel2021}
\bibinfo{author}{\bibfnamefont{M.}~\bibnamefont{Telychko}},
  \bibinfo{author}{\bibfnamefont{G.}~\bibnamefont{Li}},
  \bibinfo{author}{\bibfnamefont{P.}~\bibnamefont{Mutombo}},
  \bibinfo{author}{\bibfnamefont{D.}~\bibnamefont{Soler-Polo}},
  \bibinfo{author}{\bibfnamefont{X.}~\bibnamefont{Peng}},
  \bibinfo{author}{\bibfnamefont{J.}~\bibnamefont{Su}},
  \bibinfo{author}{\bibfnamefont{S.}~\bibnamefont{Song}},
  \bibinfo{author}{\bibfnamefont{M.~J.} \bibnamefont{Koh}},
  \bibinfo{author}{\bibfnamefont{M.}~\bibnamefont{Edmonds}},
  \bibinfo{author}{\bibfnamefont{P.}~\bibnamefont{Jelínek}},
  \bibnamefont{et~al.}, \bibinfo{journal}{Sci Adv}
  \textbf{\bibinfo{volume}{7}}, \bibinfo{pages}{eabf0269}
  (\bibinfo{year}{2021}).

\bibitem[{\citenamefont{Wang et~al.}(2014)\citenamefont{Wang, Tan, Wang, Louie,
  and Lin}}]{Wang2014}
\bibinfo{author}{\bibfnamefont{S.}~\bibnamefont{Wang}},
  \bibinfo{author}{\bibfnamefont{L.~Z.} \bibnamefont{Tan}},
  \bibinfo{author}{\bibfnamefont{W.}~\bibnamefont{Wang}},
  \bibinfo{author}{\bibfnamefont{S.~G.} \bibnamefont{Louie}}, \bibnamefont{and}
  \bibinfo{author}{\bibfnamefont{N.}~\bibnamefont{Lin}}, \bibinfo{journal}{Phys
  Rev Lett} \textbf{\bibinfo{volume}{113}}, \bibinfo{pages}{196803}
  (\bibinfo{year}{2014}).

\bibitem[{\citenamefont{{Nathan Guisinger} et~al.}(2021)\citenamefont{{Nathan
  Guisinger}, {Daniel Trainer}, {Srilok Srinivasan}, {Brandon Fisher}, {Yuan
  Zhang}, {Constance Pfeiffer}, {Saw Hla}, and {Pierre Darancet}}}]{Hla2021}
\bibinfo{author}{\bibnamefont{{Nathan Guisinger}}},
  \bibinfo{author}{\bibnamefont{{Daniel Trainer}}},
  \bibinfo{author}{\bibnamefont{{Srilok Srinivasan}}},
  \bibinfo{author}{\bibnamefont{{Brandon Fisher}}},
  \bibinfo{author}{\bibnamefont{{Yuan Zhang}}},
  \bibinfo{author}{\bibnamefont{{Constance Pfeiffer}}},
  \bibinfo{author}{\bibnamefont{{Saw Hla}}}, \bibnamefont{and}
  \bibinfo{author}{\bibnamefont{{Pierre Darancet}}}, \bibinfo{journal}{Research
  Square},  \bibinfo{pages}{DOI: 10.21203/rs.3.rs-409243/v1} (\bibinfo{year}{2021}).

\bibitem[{\citenamefont{Jiang et~al.}(2020)\citenamefont{Jiang, Zhang, Wang,
  Liu, and Low}}]{Jiang2020}
\bibinfo{author}{\bibfnamefont{W.}~\bibnamefont{Jiang}},
  \bibinfo{author}{\bibfnamefont{S.}~\bibnamefont{Zhang}},
  \bibinfo{author}{\bibfnamefont{Z.}~\bibnamefont{Wang}},
  \bibinfo{author}{\bibfnamefont{F.}~\bibnamefont{Liu}}, \bibnamefont{and}
  \bibinfo{author}{\bibfnamefont{T.}~\bibnamefont{Low}}, \bibinfo{journal}{Nano
  Lett} \textbf{\bibinfo{volume}{20}}, \bibinfo{pages}{1959}
  (\bibinfo{year}{2020}).

\bibitem[{\citenamefont{Slot et~al.}(2019)\citenamefont{Slot, Kempkes, Knol,
  van Weerdenburg, van~den Broeke, Wegner, Vanmaekelbergh, Khajetoorians,
  Morais~Smith, and Swart}}]{Slot2019}
\bibinfo{author}{\bibfnamefont{M.~R.} \bibnamefont{Slot}},
  \bibinfo{author}{\bibfnamefont{S.~N.} \bibnamefont{Kempkes}},
  \bibinfo{author}{\bibfnamefont{E.~J.} \bibnamefont{Knol}},
  \bibinfo{author}{\bibfnamefont{W.~M.~J.} \bibnamefont{van Weerdenburg}},
  \bibinfo{author}{\bibfnamefont{J.~J.} \bibnamefont{van~den Broeke}},
  \bibinfo{author}{\bibfnamefont{D.}~\bibnamefont{Wegner}},
  \bibinfo{author}{\bibfnamefont{D.}~\bibnamefont{Vanmaekelbergh}},
  \bibinfo{author}{\bibfnamefont{A.~A.} \bibnamefont{Khajetoorians}},
  \bibinfo{author}{\bibfnamefont{C.}~\bibnamefont{Morais~Smith}},
  \bibnamefont{and} \bibinfo{author}{\bibfnamefont{I.}~\bibnamefont{Swart}},
  \bibinfo{journal}{Phys Rev X} \textbf{\bibinfo{volume}{9}},
  \bibinfo{pages}{011009} (\bibinfo{year}{2019}).

\bibitem[{\citenamefont{Gardenier et~al.}(2020)\citenamefont{Gardenier, van~den
  Broeke, Moes, Swart, Delerue, Slot, Smith, and
  Vanmaekelbergh}}]{Gardenier2020}
\bibinfo{author}{\bibfnamefont{T.~S.} \bibnamefont{Gardenier}},
  \bibinfo{author}{\bibfnamefont{J.~J.} \bibnamefont{van~den Broeke}},
  \bibinfo{author}{\bibfnamefont{J.~R.} \bibnamefont{Moes}},
  \bibinfo{author}{\bibfnamefont{I.}~\bibnamefont{Swart}},
  \bibinfo{author}{\bibfnamefont{C.}~\bibnamefont{Delerue}},
  \bibinfo{author}{\bibfnamefont{M.~R.} \bibnamefont{Slot}},
  \bibinfo{author}{\bibfnamefont{C.~M.} \bibnamefont{Smith}}, \bibnamefont{and}
  \bibinfo{author}{\bibfnamefont{D.}~\bibnamefont{Vanmaekelbergh}},
  \bibinfo{journal}{ACS Nano} \textbf{\bibinfo{volume}{14}},
  \bibinfo{pages}{13638} (\bibinfo{year}{2020}).

\bibitem[{\citenamefont{Freeney
  et~al.}(2020{\natexlab{b}})\citenamefont{Freeney, Borman, Harteveld, and
  Swart}}]{Freeney2020}
\bibinfo{author}{\bibfnamefont{S.}~\bibnamefont{Freeney}},
  \bibinfo{author}{\bibfnamefont{S.}~\bibnamefont{Borman}},
  \bibinfo{author}{\bibfnamefont{J.}~\bibnamefont{Harteveld}},
  \bibnamefont{and} \bibinfo{author}{\bibfnamefont{I.}~\bibnamefont{Swart}},
  \bibinfo{journal}{SciPost Phys.} \textbf{\bibinfo{volume}{9}},
  \bibinfo{pages}{85} (\bibinfo{year}{2020}{\natexlab{b}}).

\bibitem[{\citenamefont{Brovko and Stepanyuk}(2012)}]{Stepanyuk2012}
\bibinfo{author}{\bibfnamefont{O.~O.} \bibnamefont{Brovko}} \bibnamefont{and}
  \bibinfo{author}{\bibfnamefont{V.~S.} \bibnamefont{Stepanyuk}},
  \bibinfo{journal}{Appl Phys Lett} \textbf{\bibinfo{volume}{100}},
  \bibinfo{pages}{163112} (\bibinfo{year}{2012}).

\bibitem[{\citenamefont{Seufert et~al.}(2013)\citenamefont{Seufert,
  Auw{\"a}rter, Garc\'{\i}a~de Abajo, Ecija, Vijayaraghavan, Joshi, and
  Barth}}]{Seufert2013}
\bibinfo{author}{\bibfnamefont{K.}~\bibnamefont{Seufert}},
  \bibinfo{author}{\bibfnamefont{W.}~\bibnamefont{Auw{\"a}rter}},
  \bibinfo{author}{\bibfnamefont{F.~J.} \bibnamefont{Garc\'{\i}a~de Abajo}},
  \bibinfo{author}{\bibfnamefont{D.}~\bibnamefont{Ecija}},
  \bibinfo{author}{\bibfnamefont{S.}~\bibnamefont{Vijayaraghavan}},
  \bibinfo{author}{\bibfnamefont{S.}~\bibnamefont{Joshi}}, \bibnamefont{and}
  \bibinfo{author}{\bibfnamefont{J.~V.} \bibnamefont{Barth}},
  \bibinfo{journal}{Nano Lett} \textbf{\bibinfo{volume}{13}},
  \bibinfo{pages}{6130} (\bibinfo{year}{2013}).

\bibitem[{\citenamefont{Huang et~al.}(2011)\citenamefont{Huang, Wong, Chen, and
  Wee}}]{Huang2011}
\bibinfo{author}{\bibfnamefont{H.}~\bibnamefont{Huang}},
  \bibinfo{author}{\bibfnamefont{S.~L.} \bibnamefont{Wong}},
  \bibinfo{author}{\bibfnamefont{W.}~\bibnamefont{Chen}}, \bibnamefont{and}
  \bibinfo{author}{\bibfnamefont{A.~T.~S.} \bibnamefont{Wee}},
  \bibinfo{journal}{J Phys D: Appl Phys} \textbf{\bibinfo{volume}{44}},
  \bibinfo{pages}{464005} (\bibinfo{year}{2011}).

\bibitem[{\citenamefont{Auw\"arter et~al.}(2010)\citenamefont{Auw\"arter,
  Seufert, Klappenberger, Reichert, Weber-Bargioni, Verdini, Cvetko,
  Dell'Angela, Floreano, Cossaro et~al.}}]{Auwarter2010}
\bibinfo{author}{\bibfnamefont{W.}~\bibnamefont{Auw\"arter}},
  \bibinfo{author}{\bibfnamefont{K.}~\bibnamefont{Seufert}},
  \bibinfo{author}{\bibfnamefont{F.}~\bibnamefont{Klappenberger}},
  \bibinfo{author}{\bibfnamefont{J.}~\bibnamefont{Reichert}},
  \bibinfo{author}{\bibfnamefont{A.}~\bibnamefont{Weber-Bargioni}},
  \bibinfo{author}{\bibfnamefont{A.}~\bibnamefont{Verdini}},
  \bibinfo{author}{\bibfnamefont{D.}~\bibnamefont{Cvetko}},
  \bibinfo{author}{\bibfnamefont{M.}~\bibnamefont{Dell'Angela}},
  \bibinfo{author}{\bibfnamefont{L.}~\bibnamefont{Floreano}},
  \bibinfo{author}{\bibfnamefont{A.}~\bibnamefont{Cossaro}},
  \bibnamefont{et~al.}, \bibinfo{journal}{Phys Rev B}
  \textbf{\bibinfo{volume}{81}}, \bibinfo{pages}{245403}
  (\bibinfo{year}{2010}).

\bibitem[{\citenamefont{Galbraith et~al.}(2014)\citenamefont{Galbraith, Marks,
  Tonner, and Höfer}}]{Hofer2013}
\bibinfo{author}{\bibfnamefont{M.~C.~E.} \bibnamefont{Galbraith}},
  \bibinfo{author}{\bibfnamefont{M.}~\bibnamefont{Marks}},
  \bibinfo{author}{\bibfnamefont{R.}~\bibnamefont{Tonner}}, \bibnamefont{and}
  \bibinfo{author}{\bibfnamefont{U.}~\bibnamefont{Höfer}},
  \bibinfo{journal}{The Journal of Physical Chemistry Letters}
  \textbf{\bibinfo{volume}{5}}, \bibinfo{pages}{50} (\bibinfo{year}{2014}).

\bibitem[{\citenamefont{Caplins et~al.}(2014)\citenamefont{Caplins, Suich,
  Shearer, and Harris}}]{Caplins2014}
\bibinfo{author}{\bibfnamefont{B.~W.} \bibnamefont{Caplins}},
  \bibinfo{author}{\bibfnamefont{D.~E.} \bibnamefont{Suich}},
  \bibinfo{author}{\bibfnamefont{A.~J.} \bibnamefont{Shearer}},
  \bibnamefont{and} \bibinfo{author}{\bibfnamefont{C.~B.}
  \bibnamefont{Harris}}, \bibinfo{journal}{The Journal of Physical Chemistry
  Letters} \textbf{\bibinfo{volume}{5}}, \bibinfo{pages}{1679}
  (\bibinfo{year}{2014}).

\bibitem[{\citenamefont{Morgenstern et~al.}(2002)\citenamefont{Morgenstern,
  Braun, and Rieder}}]{Karina2002}
\bibinfo{author}{\bibfnamefont{K.}~\bibnamefont{Morgenstern}},
  \bibinfo{author}{\bibfnamefont{K.-F.} \bibnamefont{Braun}}, \bibnamefont{and}
  \bibinfo{author}{\bibfnamefont{K.-H.} \bibnamefont{Rieder}},
  \bibinfo{journal}{Phys Rev Lett} \textbf{\bibinfo{volume}{89}},
  \bibinfo{pages}{226801} (\bibinfo{year}{2002}).

\bibitem[{\citenamefont{Neuhold and Horn}(1997)}]{Horn1997}
\bibinfo{author}{\bibfnamefont{G.}~\bibnamefont{Neuhold}} \bibnamefont{and}
  \bibinfo{author}{\bibfnamefont{K.}~\bibnamefont{Horn}},
  \bibinfo{journal}{Phys Rev Lett} \textbf{\bibinfo{volume}{78}},
  \bibinfo{pages}{1327} (\bibinfo{year}{1997}).

\bibitem[{\citenamefont{Piquero-Zulaica
  et~al.}(2017{\natexlab{a}})\citenamefont{Piquero-Zulaica, Lobo-Checa,
  Sadeghi, El-Fattah, Mitsui, Okamoto, Pawlak, Meier, Arnau, Ortega
  et~al.}}]{Piquero2017}
\bibinfo{author}{\bibfnamefont{I.}~\bibnamefont{Piquero-Zulaica}},
  \bibinfo{author}{\bibfnamefont{J.}~\bibnamefont{Lobo-Checa}},
  \bibinfo{author}{\bibfnamefont{A.}~\bibnamefont{Sadeghi}},
  \bibinfo{author}{\bibfnamefont{Z.~M.~A.} \bibnamefont{El-Fattah}},
  \bibinfo{author}{\bibfnamefont{C.}~\bibnamefont{Mitsui}},
  \bibinfo{author}{\bibfnamefont{T.}~\bibnamefont{Okamoto}},
  \bibinfo{author}{\bibfnamefont{R.}~\bibnamefont{Pawlak}},
  \bibinfo{author}{\bibfnamefont{T.}~\bibnamefont{Meier}},
  \bibinfo{author}{\bibfnamefont{A.}~\bibnamefont{Arnau}},
  \bibinfo{author}{\bibfnamefont{J.~E.} \bibnamefont{Ortega}},
  \bibnamefont{et~al.}, \bibinfo{journal}{Nat Commun}
  \textbf{\bibinfo{volume}{8}}, \bibinfo{pages}{787}
  (\bibinfo{year}{2017}{\natexlab{a}}).

\bibitem[{\citenamefont{Udhardt et~al.}(2017)\citenamefont{Udhardt, Otto, Kern,
  Lüftner, Huempfner, Kirchhuebel, Sojka, Meissner, Schröter, Forker
  et~al.}}]{Udhardt2017}
\bibinfo{author}{\bibfnamefont{C.}~\bibnamefont{Udhardt}},
  \bibinfo{author}{\bibfnamefont{F.}~\bibnamefont{Otto}},
  \bibinfo{author}{\bibfnamefont{C.}~\bibnamefont{Kern}},
  \bibinfo{author}{\bibfnamefont{D.}~\bibnamefont{Lüftner}},
  \bibinfo{author}{\bibfnamefont{T.}~\bibnamefont{Huempfner}},
  \bibinfo{author}{\bibfnamefont{T.}~\bibnamefont{Kirchhuebel}},
  \bibinfo{author}{\bibfnamefont{F.}~\bibnamefont{Sojka}},
  \bibinfo{author}{\bibfnamefont{M.}~\bibnamefont{Meissner}},
  \bibinfo{author}{\bibfnamefont{B.}~\bibnamefont{Schröter}},
  \bibinfo{author}{\bibfnamefont{R.}~\bibnamefont{Forker}},
  \bibnamefont{et~al.}, \bibinfo{journal}{The Journal of Physical Chemistry C}
  \textbf{\bibinfo{volume}{121}}, \bibinfo{pages}{12285}
  (\bibinfo{year}{2017}).

\bibitem[{\citenamefont{Wintjes et~al.}(2010)\citenamefont{Wintjes, Lobo-Checa,
  Hornung, Samuely, Diederich, and Jung}}]{Wintjes2010}
\bibinfo{author}{\bibfnamefont{N.}~\bibnamefont{Wintjes}},
  \bibinfo{author}{\bibfnamefont{J.}~\bibnamefont{Lobo-Checa}},
  \bibinfo{author}{\bibfnamefont{J.}~\bibnamefont{Hornung}},
  \bibinfo{author}{\bibfnamefont{T.}~\bibnamefont{Samuely}},
  \bibinfo{author}{\bibfnamefont{F.}~\bibnamefont{Diederich}},
  \bibnamefont{and} \bibinfo{author}{\bibfnamefont{T.~A.} \bibnamefont{Jung}},
  \bibinfo{journal}{J Am Chem Soc} \textbf{\bibinfo{volume}{132}},
  \bibinfo{pages}{7306} (\bibinfo{year}{2010}).

\bibitem[{\citenamefont{Stadtmüller et~al.}(2014)\citenamefont{Stadtmüller,
  Lüftner, Willenbockel, Reinisch, Sueyoshi, Koller, Soubatch, Ramsey,
  Puschnig, Tautz et~al.}}]{Stadtmuller2014}
\bibinfo{author}{\bibfnamefont{B.}~\bibnamefont{Stadtmüller}},
  \bibinfo{author}{\bibfnamefont{D.}~\bibnamefont{Lüftner}},
  \bibinfo{author}{\bibfnamefont{M.}~\bibnamefont{Willenbockel}},
  \bibinfo{author}{\bibfnamefont{E.~M.} \bibnamefont{Reinisch}},
  \bibinfo{author}{\bibfnamefont{T.}~\bibnamefont{Sueyoshi}},
  \bibinfo{author}{\bibfnamefont{G.}~\bibnamefont{Koller}},
  \bibinfo{author}{\bibfnamefont{S.}~\bibnamefont{Soubatch}},
  \bibinfo{author}{\bibfnamefont{M.~G.} \bibnamefont{Ramsey}},
  \bibinfo{author}{\bibfnamefont{P.}~\bibnamefont{Puschnig}},
  \bibinfo{author}{\bibfnamefont{F.~S.} \bibnamefont{Tautz}},
  \bibnamefont{et~al.}, \bibinfo{journal}{Nat Commun}
  \textbf{\bibinfo{volume}{5}}, \bibinfo{pages}{3685} (\bibinfo{year}{2014}).

\bibitem[{\citenamefont{Goiri et~al.}(2016)\citenamefont{Goiri, Borghetti,
  El-Sayed, Ortega, and de~Oteyza}}]{Goiri2015}
\bibinfo{author}{\bibfnamefont{E.}~\bibnamefont{Goiri}},
  \bibinfo{author}{\bibfnamefont{P.}~\bibnamefont{Borghetti}},
  \bibinfo{author}{\bibfnamefont{A.}~\bibnamefont{El-Sayed}},
  \bibinfo{author}{\bibfnamefont{J.~E.} \bibnamefont{Ortega}},
  \bibnamefont{and} \bibinfo{author}{\bibfnamefont{D.~G.}
  \bibnamefont{de~Oteyza}}, \bibinfo{journal}{Adv Mater}
  \textbf{\bibinfo{volume}{28}}, \bibinfo{pages}{1340} (\bibinfo{year}{2016}).

\bibitem[{\citenamefont{Girovsky
  et~al.}(2017{\natexlab{a}})\citenamefont{Girovsky, Nowakowski, Ali,
  Baljozovic, Rossmann, Nijs, Aeby, Nowakowska, Siewert, Srivastava
  et~al.}}]{Girovsky2017}
\bibinfo{author}{\bibfnamefont{J.}~\bibnamefont{Girovsky}},
  \bibinfo{author}{\bibfnamefont{J.}~\bibnamefont{Nowakowski}},
  \bibinfo{author}{\bibfnamefont{M.~E.} \bibnamefont{Ali}},
  \bibinfo{author}{\bibfnamefont{M.}~\bibnamefont{Baljozovic}},
  \bibinfo{author}{\bibfnamefont{H.~R.} \bibnamefont{Rossmann}},
  \bibinfo{author}{\bibfnamefont{T.}~\bibnamefont{Nijs}},
  \bibinfo{author}{\bibfnamefont{E.~A.} \bibnamefont{Aeby}},
  \bibinfo{author}{\bibfnamefont{S.}~\bibnamefont{Nowakowska}},
  \bibinfo{author}{\bibfnamefont{D.}~\bibnamefont{Siewert}},
  \bibinfo{author}{\bibfnamefont{G.}~\bibnamefont{Srivastava}},
  \bibnamefont{et~al.}, \bibinfo{journal}{Nat Commun}
  \textbf{\bibinfo{volume}{8}}, \bibinfo{pages}{15388}
  (\bibinfo{year}{2017}{\natexlab{a}}).

\bibitem[{\citenamefont{Kalff et~al.}(2016)\citenamefont{Kalff, Rebergen,
  Fahrenfort, Girovsky, Toskovic, Lado, Fernández-Rossier, and
  Otte}}]{Kalff2016}
\bibinfo{author}{\bibfnamefont{F.~E.} \bibnamefont{Kalff}},
  \bibinfo{author}{\bibfnamefont{M.~P.} \bibnamefont{Rebergen}},
  \bibinfo{author}{\bibfnamefont{E.}~\bibnamefont{Fahrenfort}},
  \bibinfo{author}{\bibfnamefont{J.}~\bibnamefont{Girovsky}},
  \bibinfo{author}{\bibfnamefont{R.}~\bibnamefont{Toskovic}},
  \bibinfo{author}{\bibfnamefont{J.~L.} \bibnamefont{Lado}},
  \bibinfo{author}{\bibfnamefont{J.}~\bibnamefont{Fernández-Rossier}},
  \bibnamefont{and} \bibinfo{author}{\bibfnamefont{A.~F.} \bibnamefont{Otte}},
  \bibinfo{journal}{Nat Nanotechnol} \textbf{\bibinfo{volume}{11}},
  \bibinfo{pages}{926} (\bibinfo{year}{2016}).

\bibitem[{\citenamefont{Drost et~al.}(2017)\citenamefont{Drost, Ojanen, Harju,
  and Liljeroth}}]{Drost2017}
\bibinfo{author}{\bibfnamefont{R.}~\bibnamefont{Drost}},
  \bibinfo{author}{\bibfnamefont{T.}~\bibnamefont{Ojanen}},
  \bibinfo{author}{\bibfnamefont{A.}~\bibnamefont{Harju}}, \bibnamefont{and}
  \bibinfo{author}{\bibfnamefont{P.}~\bibnamefont{Liljeroth}},
  \bibinfo{journal}{Nat Phys} \textbf{\bibinfo{volume}{13}},
  \bibinfo{pages}{668} (\bibinfo{year}{2017}).

\bibitem[{\citenamefont{Huda et~al.}(2020)\citenamefont{Huda, Kezilebieke,
  Ojanen, Drost, and Liljeroth}}]{Huda2020}
\bibinfo{author}{\bibfnamefont{M.~N.} \bibnamefont{Huda}},
  \bibinfo{author}{\bibfnamefont{S.}~\bibnamefont{Kezilebieke}},
  \bibinfo{author}{\bibfnamefont{T.}~\bibnamefont{Ojanen}},
  \bibinfo{author}{\bibfnamefont{R.}~\bibnamefont{Drost}}, \bibnamefont{and}
  \bibinfo{author}{\bibfnamefont{P.}~\bibnamefont{Liljeroth}},
  \bibinfo{journal}{npj Quantum Materials} \textbf{\bibinfo{volume}{5}},
  \bibinfo{pages}{17} (\bibinfo{year}{2020}).

\bibitem[{\citenamefont{Girovsky
  et~al.}(2017{\natexlab{b}})\citenamefont{Girovsky, Lado, Kalff, Fahrenfort,
  Peters, Fernández-Rossier, and Otte}}]{Otte2017}
\bibinfo{author}{\bibfnamefont{J.}~\bibnamefont{Girovsky}},
  \bibinfo{author}{\bibfnamefont{J.~L.} \bibnamefont{Lado}},
  \bibinfo{author}{\bibfnamefont{F.~E.} \bibnamefont{Kalff}},
  \bibinfo{author}{\bibfnamefont{E.}~\bibnamefont{Fahrenfort}},
  \bibinfo{author}{\bibfnamefont{L.~J. J.~M.} \bibnamefont{Peters}},
  \bibinfo{author}{\bibfnamefont{J.}~\bibnamefont{Fernández-Rossier}},
  \bibnamefont{and} \bibinfo{author}{\bibfnamefont{A.~F.} \bibnamefont{Otte}},
  \bibinfo{journal}{SciPost Phys.} \textbf{\bibinfo{volume}{2}},
  \bibinfo{pages}{020} (\bibinfo{year}{2017}{\natexlab{b}}).

\bibitem[{\citenamefont{Barth}(2007)}]{Barth2007}
\bibinfo{author}{\bibfnamefont{J.~V.} \bibnamefont{Barth}},
  \bibinfo{journal}{Annu Rev Phys Chem} \textbf{\bibinfo{volume}{58}},
  \bibinfo{pages}{375} (\bibinfo{year}{2007}).

\bibitem[{\citenamefont{Weckesser et~al.}(2001)\citenamefont{Weckesser,
  De~Vita, Barth, Cai, and Kern}}]{Kern2001}
\bibinfo{author}{\bibfnamefont{J.}~\bibnamefont{Weckesser}},
  \bibinfo{author}{\bibfnamefont{A.}~\bibnamefont{De~Vita}},
  \bibinfo{author}{\bibfnamefont{J.~V.} \bibnamefont{Barth}},
  \bibinfo{author}{\bibfnamefont{C.}~\bibnamefont{Cai}}, \bibnamefont{and}
  \bibinfo{author}{\bibfnamefont{K.}~\bibnamefont{Kern}},
  \bibinfo{journal}{Phys Rev Lett} \textbf{\bibinfo{volume}{87}},
  \bibinfo{pages}{096101} (\bibinfo{year}{2001}).

\bibitem[{\citenamefont{Schiffrin et~al.}(2007)\citenamefont{Schiffrin,
  Riemann, Auw\"arter, Pennec, Weber-Bargioni, Cvetko, Cossaro, Morgante, and
  Barth}}]{Schiffrin2007}
\bibinfo{author}{\bibfnamefont{A.}~\bibnamefont{Schiffrin}},
  \bibinfo{author}{\bibfnamefont{A.}~\bibnamefont{Riemann}},
  \bibinfo{author}{\bibfnamefont{W.}~\bibnamefont{Auw\"arter}},
  \bibinfo{author}{\bibfnamefont{Y.}~\bibnamefont{Pennec}},
  \bibinfo{author}{\bibfnamefont{A.}~\bibnamefont{Weber-Bargioni}},
  \bibinfo{author}{\bibfnamefont{D.}~\bibnamefont{Cvetko}},
  \bibinfo{author}{\bibfnamefont{A.}~\bibnamefont{Cossaro}},
  \bibinfo{author}{\bibfnamefont{A.}~\bibnamefont{Morgante}}, \bibnamefont{and}
  \bibinfo{author}{\bibfnamefont{J.~V.} \bibnamefont{Barth}},
  \bibinfo{journal}{Proceedings of the National Academy of Sciences}
  \textbf{\bibinfo{volume}{104}}, \bibinfo{pages}{5279} (\bibinfo{year}{2007}).

\bibitem[{\citenamefont{Urgel et~al.}(2016{\natexlab{b}})\citenamefont{Urgel,
  Vijayaraghavan, Ecija, Auwärter, and Barth}}]{Urgel2016a}
\bibinfo{author}{\bibfnamefont{J.~I.} \bibnamefont{Urgel}},
  \bibinfo{author}{\bibfnamefont{S.}~\bibnamefont{Vijayaraghavan}},
  \bibinfo{author}{\bibfnamefont{D.}~\bibnamefont{Ecija}},
  \bibinfo{author}{\bibfnamefont{W.}~\bibnamefont{Auwärter}},
  \bibnamefont{and} \bibinfo{author}{\bibfnamefont{J.}~\bibnamefont{Barth}},
  \bibinfo{journal}{Surf Sci} \textbf{\bibinfo{volume}{643}},
  \bibinfo{pages}{87} (\bibinfo{year}{2016}{\natexlab{b}}).

\bibitem[{\citenamefont{Clair et~al.}(2005)\citenamefont{Clair, Pons, Brune,
  Kern, and Barth}}]{Clair2005}
\bibinfo{author}{\bibfnamefont{S.}~\bibnamefont{Clair}},
  \bibinfo{author}{\bibfnamefont{S.}~\bibnamefont{Pons}},
  \bibinfo{author}{\bibfnamefont{H.}~\bibnamefont{Brune}},
  \bibinfo{author}{\bibfnamefont{K.}~\bibnamefont{Kern}}, \bibnamefont{and}
  \bibinfo{author}{\bibfnamefont{J.~V.} \bibnamefont{Barth}},
  \bibinfo{journal}{Angew Chem Int Ed} \textbf{\bibinfo{volume}{44}},
  \bibinfo{pages}{7294} (\bibinfo{year}{2005}).

\bibitem[{\citenamefont{Krenner et~al.}(2013)\citenamefont{Krenner, K{\"u}hne,
  Klappenberger, and Barth}}]{Barth2013}
\bibinfo{author}{\bibfnamefont{W.}~\bibnamefont{Krenner}},
  \bibinfo{author}{\bibfnamefont{D.}~\bibnamefont{K{\"u}hne}},
  \bibinfo{author}{\bibfnamefont{F.}~\bibnamefont{Klappenberger}},
  \bibnamefont{and} \bibinfo{author}{\bibfnamefont{J.~V.} \bibnamefont{Barth}},
  \bibinfo{journal}{Sci Rep} \textbf{\bibinfo{volume}{3}},
  \bibinfo{pages}{1454} (\bibinfo{year}{2013}).

\bibitem[{\citenamefont{Chung et~al.}(2011)\citenamefont{Chung, Park, Kim,
  Yoon, Kim, Han, and Kahng}}]{Chung2011}
\bibinfo{author}{\bibfnamefont{K.-H.} \bibnamefont{Chung}},
  \bibinfo{author}{\bibfnamefont{J.}~\bibnamefont{Park}},
  \bibinfo{author}{\bibfnamefont{K.~Y.} \bibnamefont{Kim}},
  \bibinfo{author}{\bibfnamefont{J.~K.} \bibnamefont{Yoon}},
  \bibinfo{author}{\bibfnamefont{H.}~\bibnamefont{Kim}},
  \bibinfo{author}{\bibfnamefont{S.}~\bibnamefont{Han}}, \bibnamefont{and}
  \bibinfo{author}{\bibfnamefont{S.-J.} \bibnamefont{Kahng}},
  \bibinfo{journal}{Chem Commun} \textbf{\bibinfo{volume}{47}},
  \bibinfo{pages}{11492} (\bibinfo{year}{2011}).

\bibitem[{\citenamefont{Schouteden and Van~Haesendonck}(2012)}]{Schouteden2012}
\bibinfo{author}{\bibfnamefont{K.}~\bibnamefont{Schouteden}} \bibnamefont{and}
  \bibinfo{author}{\bibfnamefont{C.}~\bibnamefont{Van~Haesendonck}},
  \bibinfo{journal}{Phys Rev Lett} \textbf{\bibinfo{volume}{108}},
  \bibinfo{pages}{076806} (\bibinfo{year}{2012}).

\bibitem[{\citenamefont{Myroshnychenko
  et~al.}(2008)\citenamefont{Myroshnychenko, Carbó-Argibay, Pastoriza-Santos,
  Pérez-Juste, Liz-Marzán, and García~de Abajo}}]{Abajo2008}
\bibinfo{author}{\bibfnamefont{V.}~\bibnamefont{Myroshnychenko}},
  \bibinfo{author}{\bibfnamefont{E.}~\bibnamefont{Carbó-Argibay}},
  \bibinfo{author}{\bibfnamefont{I.}~\bibnamefont{Pastoriza-Santos}},
  \bibinfo{author}{\bibfnamefont{J.}~\bibnamefont{Pérez-Juste}},
  \bibinfo{author}{\bibfnamefont{L.~M.} \bibnamefont{Liz-Marzán}},
  \bibnamefont{and} \bibinfo{author}{\bibfnamefont{F.~J.}
  \bibnamefont{García~de Abajo}}, \bibinfo{journal}{Adv Mater}
  \textbf{\bibinfo{volume}{20}}, \bibinfo{pages}{4288} (\bibinfo{year}{2008}).

\bibitem[{\citenamefont{Klappenberger et~al.}(2011)\citenamefont{Klappenberger,
  K{\"u}hne, Krenner, Silanes, Arnau, Garc\'{\i}a~de Abajo, Klyatskaya, Ruben,
  and Barth}}]{Florian2011}
\bibinfo{author}{\bibfnamefont{F.}~\bibnamefont{Klappenberger}},
  \bibinfo{author}{\bibfnamefont{D.}~\bibnamefont{K{\"u}hne}},
  \bibinfo{author}{\bibfnamefont{W.}~\bibnamefont{Krenner}},
  \bibinfo{author}{\bibfnamefont{I.}~\bibnamefont{Silanes}},
  \bibinfo{author}{\bibfnamefont{A.}~\bibnamefont{Arnau}},
  \bibinfo{author}{\bibfnamefont{F.~J.} \bibnamefont{Garc\'{\i}a~de Abajo}},
  \bibinfo{author}{\bibfnamefont{S.}~\bibnamefont{Klyatskaya}},
  \bibinfo{author}{\bibfnamefont{M.}~\bibnamefont{Ruben}}, \bibnamefont{and}
  \bibinfo{author}{\bibfnamefont{J.~V.} \bibnamefont{Barth}},
  \bibinfo{journal}{Phys Rev Lett} \textbf{\bibinfo{volume}{106}},
  \bibinfo{pages}{026802} (\bibinfo{year}{2011}).

\bibitem[{\citenamefont{Wang et~al.}(2018)\citenamefont{Wang, Zhang, Jiang,
  Wang, and Hou}}]{Wang2018}
\bibinfo{author}{\bibfnamefont{H.}~\bibnamefont{Wang}},
  \bibinfo{author}{\bibfnamefont{X.}~\bibnamefont{Zhang}},
  \bibinfo{author}{\bibfnamefont{Z.}~\bibnamefont{Jiang}},
  \bibinfo{author}{\bibfnamefont{Y.}~\bibnamefont{Wang}}, \bibnamefont{and}
  \bibinfo{author}{\bibfnamefont{S.}~\bibnamefont{Hou}}, \bibinfo{journal}{Phys
  Rev B} \textbf{\bibinfo{volume}{97}}, \bibinfo{pages}{115451}
  (\bibinfo{year}{2018}).

\bibitem[{\citenamefont{Schlickum et~al.}(2008)\citenamefont{Schlickum, Decker,
  Klappenberger, Zoppellaro, Klyatskaya, Auwärter, Neppl, Kern, Brune, Ruben
  et~al.}}]{Schlickum2008}
\bibinfo{author}{\bibfnamefont{U.}~\bibnamefont{Schlickum}},
  \bibinfo{author}{\bibfnamefont{R.}~\bibnamefont{Decker}},
  \bibinfo{author}{\bibfnamefont{F.}~\bibnamefont{Klappenberger}},
  \bibinfo{author}{\bibfnamefont{G.}~\bibnamefont{Zoppellaro}},
  \bibinfo{author}{\bibfnamefont{S.}~\bibnamefont{Klyatskaya}},
  \bibinfo{author}{\bibfnamefont{W.}~\bibnamefont{Auwärter}},
  \bibinfo{author}{\bibfnamefont{S.}~\bibnamefont{Neppl}},
  \bibinfo{author}{\bibfnamefont{K.}~\bibnamefont{Kern}},
  \bibinfo{author}{\bibfnamefont{H.}~\bibnamefont{Brune}},
  \bibinfo{author}{\bibfnamefont{M.}~\bibnamefont{Ruben}},
  \bibnamefont{et~al.}, \bibinfo{journal}{J Am Chem Soc}
  \textbf{\bibinfo{volume}{130}}, \bibinfo{pages}{11778}
  (\bibinfo{year}{2008}).

\bibitem[{\citenamefont{Klyatskaya et~al.}(2011)\citenamefont{Klyatskaya,
  Klappenberger, Schlickum, Kühne, Marschall, Reichert, Decker, Krenner,
  Zoppellaro, Brune et~al.}}]{Klyatskaya2011}
\bibinfo{author}{\bibfnamefont{S.}~\bibnamefont{Klyatskaya}},
  \bibinfo{author}{\bibfnamefont{F.}~\bibnamefont{Klappenberger}},
  \bibinfo{author}{\bibfnamefont{U.}~\bibnamefont{Schlickum}},
  \bibinfo{author}{\bibfnamefont{D.}~\bibnamefont{Kühne}},
  \bibinfo{author}{\bibfnamefont{M.}~\bibnamefont{Marschall}},
  \bibinfo{author}{\bibfnamefont{J.}~\bibnamefont{Reichert}},
  \bibinfo{author}{\bibfnamefont{R.}~\bibnamefont{Decker}},
  \bibinfo{author}{\bibfnamefont{W.}~\bibnamefont{Krenner}},
  \bibinfo{author}{\bibfnamefont{G.}~\bibnamefont{Zoppellaro}},
  \bibinfo{author}{\bibfnamefont{H.}~\bibnamefont{Brune}},
  \bibnamefont{et~al.}, \bibinfo{journal}{Adv Funct Mater}
  \textbf{\bibinfo{volume}{21}}, \bibinfo{pages}{1230} (\bibinfo{year}{2011}).

\bibitem[{\citenamefont{Taber et~al.}(2016)\citenamefont{Taber, Gervasi, Mills,
  Kislitsyn, Darzi, Crowley, Jasti, and Nazin}}]{Nazin2016}
\bibinfo{author}{\bibfnamefont{B.~N.} \bibnamefont{Taber}},
  \bibinfo{author}{\bibfnamefont{C.~F.} \bibnamefont{Gervasi}},
  \bibinfo{author}{\bibfnamefont{J.~M.} \bibnamefont{Mills}},
  \bibinfo{author}{\bibfnamefont{D.~A.} \bibnamefont{Kislitsyn}},
  \bibinfo{author}{\bibfnamefont{E.~R.} \bibnamefont{Darzi}},
  \bibinfo{author}{\bibfnamefont{W.~G.} \bibnamefont{Crowley}},
  \bibinfo{author}{\bibfnamefont{R.}~\bibnamefont{Jasti}}, \bibnamefont{and}
  \bibinfo{author}{\bibfnamefont{G.~V.} \bibnamefont{Nazin}},
  \bibinfo{journal}{The Journal of Physical Chemistry Letters}
  \textbf{\bibinfo{volume}{7}}, \bibinfo{pages}{3073} (\bibinfo{year}{2016}).

\bibitem[{\citenamefont{Chen et~al.}(2017)\citenamefont{Chen, Shang, Wang, Wu,
  Kuttner, Hilt, Hieringer, and Gottfried}}]{Gottfried2016}
\bibinfo{author}{\bibfnamefont{M.}~\bibnamefont{Chen}},
  \bibinfo{author}{\bibfnamefont{J.}~\bibnamefont{Shang}},
  \bibinfo{author}{\bibfnamefont{Y.}~\bibnamefont{Wang}},
  \bibinfo{author}{\bibfnamefont{K.}~\bibnamefont{Wu}},
  \bibinfo{author}{\bibfnamefont{J.}~\bibnamefont{Kuttner}},
  \bibinfo{author}{\bibfnamefont{G.}~\bibnamefont{Hilt}},
  \bibinfo{author}{\bibfnamefont{W.}~\bibnamefont{Hieringer}},
  \bibnamefont{and} \bibinfo{author}{\bibfnamefont{J.~M.}
  \bibnamefont{Gottfried}}, \bibinfo{journal}{ACS Nano}
  \textbf{\bibinfo{volume}{11}}, \bibinfo{pages}{134} (\bibinfo{year}{2017}).

\bibitem[{\citenamefont{Hao et~al.}(2019)\citenamefont{Hao, Song, Yan, Zhang,
  Ruan, Sun, Lu, and Cai}}]{Hao2019}
\bibinfo{author}{\bibfnamefont{Z.}~\bibnamefont{Hao}},
  \bibinfo{author}{\bibfnamefont{L.}~\bibnamefont{Song}},
  \bibinfo{author}{\bibfnamefont{C.}~\bibnamefont{Yan}},
  \bibinfo{author}{\bibfnamefont{H.}~\bibnamefont{Zhang}},
  \bibinfo{author}{\bibfnamefont{Z.}~\bibnamefont{Ruan}},
  \bibinfo{author}{\bibfnamefont{S.}~\bibnamefont{Sun}},
  \bibinfo{author}{\bibfnamefont{J.}~\bibnamefont{Lu}}, \bibnamefont{and}
  \bibinfo{author}{\bibfnamefont{J.}~\bibnamefont{Cai}}, \bibinfo{journal}{Chem
  Commun} \textbf{\bibinfo{volume}{55}}, \bibinfo{pages}{10800}
  (\bibinfo{year}{2019}).

\bibitem[{\citenamefont{Reecht et~al.}(2013)\citenamefont{Reecht, Bulou,
  Scheurer, Speisser, Carri\`ere, Mathevet, and Schull}}]{Reecht2013}
\bibinfo{author}{\bibfnamefont{G.}~\bibnamefont{Reecht}},
  \bibinfo{author}{\bibfnamefont{H.}~\bibnamefont{Bulou}},
  \bibinfo{author}{\bibfnamefont{F.}~\bibnamefont{Scheurer}},
  \bibinfo{author}{\bibfnamefont{V.}~\bibnamefont{Speisser}},
  \bibinfo{author}{\bibfnamefont{B.}~\bibnamefont{Carri\`ere}},
  \bibinfo{author}{\bibfnamefont{F.}~\bibnamefont{Mathevet}}, \bibnamefont{and}
  \bibinfo{author}{\bibfnamefont{G.}~\bibnamefont{Schull}},
  \bibinfo{journal}{Phys Rev Lett} \textbf{\bibinfo{volume}{110}},
  \bibinfo{pages}{056802} (\bibinfo{year}{2013}).

\bibitem[{\citenamefont{Schlickum et~al.}(2007)\citenamefont{Schlickum, Decker,
  Klappenberger, Zoppellaro, Klyatskaya, Ruben, Silanes, Arnau, Kern, Brune
  et~al.}}]{Schlickum2007}
\bibinfo{author}{\bibfnamefont{U.}~\bibnamefont{Schlickum}},
  \bibinfo{author}{\bibfnamefont{R.}~\bibnamefont{Decker}},
  \bibinfo{author}{\bibfnamefont{F.}~\bibnamefont{Klappenberger}},
  \bibinfo{author}{\bibfnamefont{G.}~\bibnamefont{Zoppellaro}},
  \bibinfo{author}{\bibfnamefont{S.}~\bibnamefont{Klyatskaya}},
  \bibinfo{author}{\bibfnamefont{M.}~\bibnamefont{Ruben}},
  \bibinfo{author}{\bibfnamefont{I.}~\bibnamefont{Silanes}},
  \bibinfo{author}{\bibfnamefont{A.}~\bibnamefont{Arnau}},
  \bibinfo{author}{\bibfnamefont{K.}~\bibnamefont{Kern}},
  \bibinfo{author}{\bibfnamefont{H.}~\bibnamefont{Brune}},
  \bibnamefont{et~al.}, \bibinfo{journal}{Nano Lett}
  \textbf{\bibinfo{volume}{7}}, \bibinfo{pages}{3813} (\bibinfo{year}{2007}).

\bibitem[{\citenamefont{K{\"u}hne et~al.}(2009)\citenamefont{K{\"u}hne,
  Klappenberger, Decker, Schlickum, Brune, Klyatskaya, Ruben, and
  Barth}}]{Kuhne2009}
\bibinfo{author}{\bibfnamefont{D.}~\bibnamefont{K{\"u}hne}},
  \bibinfo{author}{\bibfnamefont{F.}~\bibnamefont{Klappenberger}},
  \bibinfo{author}{\bibfnamefont{R.}~\bibnamefont{Decker}},
  \bibinfo{author}{\bibfnamefont{U.}~\bibnamefont{Schlickum}},
  \bibinfo{author}{\bibfnamefont{H.}~\bibnamefont{Brune}},
  \bibinfo{author}{\bibfnamefont{S.}~\bibnamefont{Klyatskaya}},
  \bibinfo{author}{\bibfnamefont{M.}~\bibnamefont{Ruben}}, \bibnamefont{and}
  \bibinfo{author}{\bibfnamefont{J.~V.} \bibnamefont{Barth}},
  \bibinfo{journal}{J Am Chem Soc} \textbf{\bibinfo{volume}{131}},
  \bibinfo{pages}{3881} (\bibinfo{year}{2009}).

\bibitem[{\citenamefont{Pacchioni et~al.}(2015)\citenamefont{Pacchioni,
  Pivetta, and Brune}}]{Pacchioni2015}
\bibinfo{author}{\bibfnamefont{G.~E.} \bibnamefont{Pacchioni}},
  \bibinfo{author}{\bibfnamefont{M.}~\bibnamefont{Pivetta}}, \bibnamefont{and}
  \bibinfo{author}{\bibfnamefont{H.}~\bibnamefont{Brune}},
  \bibinfo{journal}{The Journal of Physical Chemistry C}
  \textbf{\bibinfo{volume}{119}}, \bibinfo{pages}{25442}
  (\bibinfo{year}{2015}).

\bibitem[{\citenamefont{Limot et~al.}(2005)\citenamefont{Limot, Pehlke,
  Kröger, and Berndt}}]{Limot2005}
\bibinfo{author}{\bibfnamefont{L.}~\bibnamefont{Limot}},
  \bibinfo{author}{\bibfnamefont{E.}~\bibnamefont{Pehlke}},
  \bibinfo{author}{\bibfnamefont{J.}~\bibnamefont{Kröger}}, \bibnamefont{and}
  \bibinfo{author}{\bibfnamefont{R.}~\bibnamefont{Berndt}},
  \bibinfo{journal}{Phys Rev Lett} \textbf{\bibinfo{volume}{94}},
  \bibinfo{pages}{036805} (\bibinfo{year}{2005}).

\bibitem[{\citenamefont{Madhavan et~al.}(2001)\citenamefont{Madhavan, Chen,
  Jamneala, Crommie, and Wingreen}}]{Crommie2001}
\bibinfo{author}{\bibfnamefont{V.}~\bibnamefont{Madhavan}},
  \bibinfo{author}{\bibfnamefont{W.}~\bibnamefont{Chen}},
  \bibinfo{author}{\bibfnamefont{T.}~\bibnamefont{Jamneala}},
  \bibinfo{author}{\bibfnamefont{M.~F.} \bibnamefont{Crommie}},
  \bibnamefont{and} \bibinfo{author}{\bibfnamefont{N.~S.}
  \bibnamefont{Wingreen}}, \bibinfo{journal}{Phys Rev B}
  \textbf{\bibinfo{volume}{64}}, \bibinfo{pages}{165412}
  (\bibinfo{year}{2001}).

\bibitem[{\citenamefont{Liu et~al.}(2006)\citenamefont{Liu, Matsuda, Hobara,
  and Hasegawa}}]{Liu2006}
\bibinfo{author}{\bibfnamefont{C.}~\bibnamefont{Liu}},
  \bibinfo{author}{\bibfnamefont{I.}~\bibnamefont{Matsuda}},
  \bibinfo{author}{\bibfnamefont{R.}~\bibnamefont{Hobara}}, \bibnamefont{and}
  \bibinfo{author}{\bibfnamefont{S.}~\bibnamefont{Hasegawa}},
  \bibinfo{journal}{Phys Rev Lett} \textbf{\bibinfo{volume}{96}},
  \bibinfo{pages}{036803} (\bibinfo{year}{2006}).

\bibitem[{\citenamefont{Olsson et~al.}(2004)\citenamefont{Olsson, Persson,
  Borisov, Gauyacq, Lagoute, and Fölsch}}]{Olsson2004}
\bibinfo{author}{\bibfnamefont{F.~E.} \bibnamefont{Olsson}},
  \bibinfo{author}{\bibfnamefont{M.}~\bibnamefont{Persson}},
  \bibinfo{author}{\bibfnamefont{A.~G.} \bibnamefont{Borisov}},
  \bibinfo{author}{\bibfnamefont{J.-P.} \bibnamefont{Gauyacq}},
  \bibinfo{author}{\bibfnamefont{J.}~\bibnamefont{Lagoute}}, \bibnamefont{and}
  \bibinfo{author}{\bibfnamefont{S.}~\bibnamefont{Fölsch}},
  \bibinfo{journal}{Phys Rev Lett} \textbf{\bibinfo{volume}{93}},
  \bibinfo{pages}{206803} (\bibinfo{year}{2004}).

\bibitem[{\citenamefont{Piquero-Zulaica
  et~al.}(2019{\natexlab{a}})\citenamefont{Piquero-Zulaica, Li, Abd El-Fattah,
  Solianyk, Gallardo, Monjas, Hirsch, Arnau, Ortega, Stöhr
  et~al.}}]{Piquero2019b}
\bibinfo{author}{\bibfnamefont{I.}~\bibnamefont{Piquero-Zulaica}},
  \bibinfo{author}{\bibfnamefont{J.}~\bibnamefont{Li}},
  \bibinfo{author}{\bibfnamefont{Z.~M.} \bibnamefont{Abd El-Fattah}},
  \bibinfo{author}{\bibfnamefont{L.}~\bibnamefont{Solianyk}},
  \bibinfo{author}{\bibfnamefont{I.}~\bibnamefont{Gallardo}},
  \bibinfo{author}{\bibfnamefont{L.}~\bibnamefont{Monjas}},
  \bibinfo{author}{\bibfnamefont{A.~K.~H.} \bibnamefont{Hirsch}},
  \bibinfo{author}{\bibfnamefont{A.}~\bibnamefont{Arnau}},
  \bibinfo{author}{\bibfnamefont{J.~E.} \bibnamefont{Ortega}},
  \bibinfo{author}{\bibfnamefont{M.}~\bibnamefont{Stöhr}},
  \bibnamefont{et~al.}, \bibinfo{journal}{Nanoscale}
  \textbf{\bibinfo{volume}{11}}, \bibinfo{pages}{23132}
  (\bibinfo{year}{2019}{\natexlab{a}}).

\bibitem[{\citenamefont{Faraggi et~al.}(2012)\citenamefont{Faraggi, Jiang,
  Gonzalez-Lakunza, Langner, Stepanow, Kern, and Arnau}}]{Faraggi2012}
\bibinfo{author}{\bibfnamefont{M.~N.} \bibnamefont{Faraggi}},
  \bibinfo{author}{\bibfnamefont{N.}~\bibnamefont{Jiang}},
  \bibinfo{author}{\bibfnamefont{N.}~\bibnamefont{Gonzalez-Lakunza}},
  \bibinfo{author}{\bibfnamefont{A.}~\bibnamefont{Langner}},
  \bibinfo{author}{\bibfnamefont{S.}~\bibnamefont{Stepanow}},
  \bibinfo{author}{\bibfnamefont{K.}~\bibnamefont{Kern}}, \bibnamefont{and}
  \bibinfo{author}{\bibfnamefont{A.}~\bibnamefont{Arnau}},
  \bibinfo{journal}{The Journal of Physical Chemistry C}
  \textbf{\bibinfo{volume}{116}}, \bibinfo{pages}{24558}
  (\bibinfo{year}{2012}).

\bibitem[{\citenamefont{Colazzo et~al.}(2019)\citenamefont{Colazzo, Mohammed,
  Gallardo, Abd El-Fattah, Pomposo, Jelínek, and de~Oteyza}}]{Colazzo2019}
\bibinfo{author}{\bibfnamefont{L.}~\bibnamefont{Colazzo}},
  \bibinfo{author}{\bibfnamefont{M.~S.~G.} \bibnamefont{Mohammed}},
  \bibinfo{author}{\bibfnamefont{A.}~\bibnamefont{Gallardo}},
  \bibinfo{author}{\bibfnamefont{Z.~M.} \bibnamefont{Abd El-Fattah}},
  \bibinfo{author}{\bibfnamefont{J.~A.} \bibnamefont{Pomposo}},
  \bibinfo{author}{\bibfnamefont{P.}~\bibnamefont{Jelínek}}, \bibnamefont{and}
  \bibinfo{author}{\bibfnamefont{D.~G.} \bibnamefont{de~Oteyza}},
  \bibinfo{journal}{Nanoscale} \textbf{\bibinfo{volume}{11}},
  \bibinfo{pages}{15567} (\bibinfo{year}{2019}).

\bibitem[{\citenamefont{Wang et~al.}(2019)\citenamefont{Wang, Xue, Li, Wu, Li,
  Hou, and Wang}}]{Wang2019}
\bibinfo{author}{\bibfnamefont{Y.}~\bibnamefont{Wang}},
  \bibinfo{author}{\bibfnamefont{N.}~\bibnamefont{Xue}},
  \bibinfo{author}{\bibfnamefont{R.}~\bibnamefont{Li}},
  \bibinfo{author}{\bibfnamefont{T.}~\bibnamefont{Wu}},
  \bibinfo{author}{\bibfnamefont{N.}~\bibnamefont{Li}},
  \bibinfo{author}{\bibfnamefont{S.}~\bibnamefont{Hou}}, \bibnamefont{and}
  \bibinfo{author}{\bibfnamefont{Y.}~\bibnamefont{Wang}},
  \bibinfo{journal}{ChemPhysChem} \textbf{\bibinfo{volume}{20}},
  \bibinfo{pages}{2262} (\bibinfo{year}{2019}).

\bibitem[{\citenamefont{Piquero-Zulaica
  et~al.}(2019{\natexlab{b}})\citenamefont{Piquero-Zulaica, Sadeghi, Kherelden,
  Hua, Liu, Kuang, Yan, Ortega, El-Fattah, Azizi et~al.}}]{Piquero2019c}
\bibinfo{author}{\bibfnamefont{I.}~\bibnamefont{Piquero-Zulaica}},
  \bibinfo{author}{\bibfnamefont{A.}~\bibnamefont{Sadeghi}},
  \bibinfo{author}{\bibfnamefont{M.}~\bibnamefont{Kherelden}},
  \bibinfo{author}{\bibfnamefont{M.}~\bibnamefont{Hua}},
  \bibinfo{author}{\bibfnamefont{J.}~\bibnamefont{Liu}},
  \bibinfo{author}{\bibfnamefont{G.}~\bibnamefont{Kuang}},
  \bibinfo{author}{\bibfnamefont{L.}~\bibnamefont{Yan}},
  \bibinfo{author}{\bibfnamefont{J.~E.} \bibnamefont{Ortega}},
  \bibinfo{author}{\bibfnamefont{Z.~M.~A.} \bibnamefont{El-Fattah}},
  \bibinfo{author}{\bibfnamefont{B.}~\bibnamefont{Azizi}},
  \bibnamefont{et~al.}, \bibinfo{journal}{Phys Rev Lett}
  \textbf{\bibinfo{volume}{123}}, \bibinfo{pages}{266805}
  (\bibinfo{year}{2019}{\natexlab{b}}).

\bibitem[{\citenamefont{Ecija et~al.}(2013)\citenamefont{Ecija, Urgel,
  Papageorgiou, Joshi, Auwarter, Seitsonen, Klyatskaya, Ruben, Fischer,
  Vijayaraghavan et~al.}}]{Ecija2013}
\bibinfo{author}{\bibfnamefont{D.}~\bibnamefont{Ecija}},
  \bibinfo{author}{\bibfnamefont{J.~I.} \bibnamefont{Urgel}},
  \bibinfo{author}{\bibfnamefont{A.~C.} \bibnamefont{Papageorgiou}},
  \bibinfo{author}{\bibfnamefont{S.}~\bibnamefont{Joshi}},
  \bibinfo{author}{\bibfnamefont{W.}~\bibnamefont{Auwarter}},
  \bibinfo{author}{\bibfnamefont{A.~P.} \bibnamefont{Seitsonen}},
  \bibinfo{author}{\bibfnamefont{S.}~\bibnamefont{Klyatskaya}},
  \bibinfo{author}{\bibfnamefont{M.}~\bibnamefont{Ruben}},
  \bibinfo{author}{\bibfnamefont{S.}~\bibnamefont{Fischer}},
  \bibinfo{author}{\bibfnamefont{S.}~\bibnamefont{Vijayaraghavan}},
  \bibnamefont{et~al.}, \bibinfo{journal}{Proceedings of the National Academy
  of Sciences} \textbf{\bibinfo{volume}{110}}, \bibinfo{pages}{6678}
  (\bibinfo{year}{2013}).

\bibitem[{\citenamefont{Matena et~al.}(2014)\citenamefont{Matena, Bj{\"o}rk,
  Wahl, Lee, Zegenhagen, Gade, Jung, Persson, and Stöhr}}]{Matena2014}
\bibinfo{author}{\bibfnamefont{M.}~\bibnamefont{Matena}},
  \bibinfo{author}{\bibfnamefont{J.}~\bibnamefont{Bj{\"o}rk}},
  \bibinfo{author}{\bibfnamefont{M.}~\bibnamefont{Wahl}},
  \bibinfo{author}{\bibfnamefont{T.-L.} \bibnamefont{Lee}},
  \bibinfo{author}{\bibfnamefont{J.}~\bibnamefont{Zegenhagen}},
  \bibinfo{author}{\bibfnamefont{L.~H.} \bibnamefont{Gade}},
  \bibinfo{author}{\bibfnamefont{T.~A.} \bibnamefont{Jung}},
  \bibinfo{author}{\bibfnamefont{M.}~\bibnamefont{Persson}}, \bibnamefont{and}
  \bibinfo{author}{\bibfnamefont{M.}~\bibnamefont{Stöhr}},
  \bibinfo{journal}{Phys Rev B: Condens Matter Mater Phys}
  \textbf{\bibinfo{volume}{90}}, \bibinfo{pages}{125408}
  (\bibinfo{year}{2014}).

\bibitem[{\citenamefont{Shchyrba
  et~al.}(2014{\natexlab{b}})\citenamefont{Shchyrba, W{\"a}ckerlin, Nowakowski,
  Nowakowska, Bj{\"o}rk, Fatayer, Girovsky, Nijs, Martens, Kleibert
  et~al.}}]{Gade2014}
\bibinfo{author}{\bibfnamefont{A.}~\bibnamefont{Shchyrba}},
  \bibinfo{author}{\bibfnamefont{C.}~\bibnamefont{W{\"a}ckerlin}},
  \bibinfo{author}{\bibfnamefont{J.}~\bibnamefont{Nowakowski}},
  \bibinfo{author}{\bibfnamefont{S.}~\bibnamefont{Nowakowska}},
  \bibinfo{author}{\bibfnamefont{J.}~\bibnamefont{Bj{\"o}rk}},
  \bibinfo{author}{\bibfnamefont{S.}~\bibnamefont{Fatayer}},
  \bibinfo{author}{\bibfnamefont{J.}~\bibnamefont{Girovsky}},
  \bibinfo{author}{\bibfnamefont{T.}~\bibnamefont{Nijs}},
  \bibinfo{author}{\bibfnamefont{S.~C.} \bibnamefont{Martens}},
  \bibinfo{author}{\bibfnamefont{A.}~\bibnamefont{Kleibert}},
  \bibnamefont{et~al.}, \bibinfo{journal}{J Am Chem Soc}
  \textbf{\bibinfo{volume}{136}}, \bibinfo{pages}{9355}
  (\bibinfo{year}{2014}{\natexlab{b}}).

\bibitem[{\citenamefont{Piquero-Zulaica
  et~al.}(2017{\natexlab{b}})\citenamefont{Piquero-Zulaica, Nowakowska, Ortega,
  St{\"o}hr, Gade, Jung, and Lobo-Checa}}]{Piquero2016}
\bibinfo{author}{\bibfnamefont{I.}~\bibnamefont{Piquero-Zulaica}},
  \bibinfo{author}{\bibfnamefont{S.}~\bibnamefont{Nowakowska}},
  \bibinfo{author}{\bibfnamefont{J.~E.} \bibnamefont{Ortega}},
  \bibinfo{author}{\bibfnamefont{M.}~\bibnamefont{St{\"o}hr}},
  \bibinfo{author}{\bibfnamefont{L.~H.} \bibnamefont{Gade}},
  \bibinfo{author}{\bibfnamefont{T.~A.} \bibnamefont{Jung}}, \bibnamefont{and}
  \bibinfo{author}{\bibfnamefont{J.}~\bibnamefont{Lobo-Checa}},
  \bibinfo{journal}{Appl Surf Sci} \textbf{\bibinfo{volume}{391}},
  \bibinfo{pages}{39} (\bibinfo{year}{2017}{\natexlab{b}}).

\bibitem[{\citenamefont{Piquero-Zulaica
  et~al.}(2019{\natexlab{c}})\citenamefont{Piquero-Zulaica, Abd El-Fattah,
  Popova, Kawai, Nowakowska, Matena, Enache, Stöhr, Tejeda, Taleb
  et~al.}}]{Piquero2019a}
\bibinfo{author}{\bibfnamefont{I.}~\bibnamefont{Piquero-Zulaica}},
  \bibinfo{author}{\bibfnamefont{Z.~M.} \bibnamefont{Abd El-Fattah}},
  \bibinfo{author}{\bibfnamefont{O.}~\bibnamefont{Popova}},
  \bibinfo{author}{\bibfnamefont{S.}~\bibnamefont{Kawai}},
  \bibinfo{author}{\bibfnamefont{S.}~\bibnamefont{Nowakowska}},
  \bibinfo{author}{\bibfnamefont{M.}~\bibnamefont{Matena}},
  \bibinfo{author}{\bibfnamefont{M.}~\bibnamefont{Enache}},
  \bibinfo{author}{\bibfnamefont{M.}~\bibnamefont{Stöhr}},
  \bibinfo{author}{\bibfnamefont{A.}~\bibnamefont{Tejeda}},
  \bibinfo{author}{\bibfnamefont{A.}~\bibnamefont{Taleb}},
  \bibnamefont{et~al.}, \bibinfo{journal}{New J Phys}
  \textbf{\bibinfo{volume}{21}}, \bibinfo{pages}{053004}
  (\bibinfo{year}{2019}{\natexlab{c}}).

\bibitem[{\citenamefont{Kep$\breve{c}$ija
  et~al.}(2015)\citenamefont{Kep$\breve{c}$ija, Huang, Klappenberger, and
  Barth}}]{Barth2015}
\bibinfo{author}{\bibfnamefont{N.}~\bibnamefont{Kep$\breve{c}$ija}},
  \bibinfo{author}{\bibfnamefont{T.-J.} \bibnamefont{Huang}},
  \bibinfo{author}{\bibfnamefont{F.}~\bibnamefont{Klappenberger}},
  \bibnamefont{and} \bibinfo{author}{\bibfnamefont{J.~V.} \bibnamefont{Barth}},
  \bibinfo{journal}{The Journal of Chemical Physics}
  \textbf{\bibinfo{volume}{142}}, \bibinfo{pages}{101931}
  (\bibinfo{year}{2015}).

\bibitem[{\citenamefont{Kawai et~al.}(2016)\citenamefont{Kawai, Foster,
  Bj{\"o}rkman, Nowakowska, Bj{\"o}rk, Canova, Gade, Jung, and
  Meyer}}]{Kawai2016}
\bibinfo{author}{\bibfnamefont{S.}~\bibnamefont{Kawai}},
  \bibinfo{author}{\bibfnamefont{A.~S.} \bibnamefont{Foster}},
  \bibinfo{author}{\bibfnamefont{T.}~\bibnamefont{Bj{\"o}rkman}},
  \bibinfo{author}{\bibfnamefont{S.}~\bibnamefont{Nowakowska}},
  \bibinfo{author}{\bibfnamefont{J.}~\bibnamefont{Bj{\"o}rk}},
  \bibinfo{author}{\bibfnamefont{F.~F.} \bibnamefont{Canova}},
  \bibinfo{author}{\bibfnamefont{L.~H.} \bibnamefont{Gade}},
  \bibinfo{author}{\bibfnamefont{T.~A.} \bibnamefont{Jung}}, \bibnamefont{and}
  \bibinfo{author}{\bibfnamefont{E.}~\bibnamefont{Meyer}},
  \bibinfo{journal}{Nat Commun} \textbf{\bibinfo{volume}{7}},
  \bibinfo{pages}{11559} (\bibinfo{year}{2016}).

\bibitem[{\citenamefont{Malterre et~al.}(2011)\citenamefont{Malterre, Kierren,
  Fagot-Revurat, Didiot, Garc\'{\i}a~de Abajo, Schiller, Cord\'on, and
  Ortega}}]{Malterre2011}
\bibinfo{author}{\bibfnamefont{D.}~\bibnamefont{Malterre}},
  \bibinfo{author}{\bibfnamefont{B.}~\bibnamefont{Kierren}},
  \bibinfo{author}{\bibfnamefont{Y.}~\bibnamefont{Fagot-Revurat}},
  \bibinfo{author}{\bibfnamefont{C.}~\bibnamefont{Didiot}},
  \bibinfo{author}{\bibfnamefont{F.~J.} \bibnamefont{Garc\'{\i}a~de Abajo}},
  \bibinfo{author}{\bibfnamefont{F.}~\bibnamefont{Schiller}},
  \bibinfo{author}{\bibfnamefont{J.}~\bibnamefont{Cord\'on}}, \bibnamefont{and}
  \bibinfo{author}{\bibfnamefont{J.~E.} \bibnamefont{Ortega}},
  \bibinfo{journal}{New J Phys} \textbf{\bibinfo{volume}{13}},
  \bibinfo{pages}{013026} (\bibinfo{year}{2011}).

\bibitem[{\citenamefont{Kawai et~al.}(2015)\citenamefont{Kawai, Sadeghi, Xu,
  Peng, Orita, Otera, Goedecker, and Meyer}}]{Kawai2015}
\bibinfo{author}{\bibfnamefont{S.}~\bibnamefont{Kawai}},
  \bibinfo{author}{\bibfnamefont{A.}~\bibnamefont{Sadeghi}},
  \bibinfo{author}{\bibfnamefont{F.}~\bibnamefont{Xu}},
  \bibinfo{author}{\bibfnamefont{L.}~\bibnamefont{Peng}},
  \bibinfo{author}{\bibfnamefont{A.}~\bibnamefont{Orita}},
  \bibinfo{author}{\bibfnamefont{J.}~\bibnamefont{Otera}},
  \bibinfo{author}{\bibfnamefont{S.}~\bibnamefont{Goedecker}},
  \bibnamefont{and} \bibinfo{author}{\bibfnamefont{E.}~\bibnamefont{Meyer}},
  \bibinfo{journal}{ACS Nano} \textbf{\bibinfo{volume}{9}},
  \bibinfo{pages}{2574} (\bibinfo{year}{2015}).

\bibitem[{\citenamefont{Han et~al.}(2017)\citenamefont{Han, Czap, Chiang, Xu,
  Wagner, Wei, Zhang, Wu, and Ho}}]{Ho2017}
\bibinfo{author}{\bibfnamefont{Z.}~\bibnamefont{Han}},
  \bibinfo{author}{\bibfnamefont{G.}~\bibnamefont{Czap}},
  \bibinfo{author}{\bibfnamefont{C.-l.} \bibnamefont{Chiang}},
  \bibinfo{author}{\bibfnamefont{C.}~\bibnamefont{Xu}},
  \bibinfo{author}{\bibfnamefont{P.~J.} \bibnamefont{Wagner}},
  \bibinfo{author}{\bibfnamefont{X.}~\bibnamefont{Wei}},
  \bibinfo{author}{\bibfnamefont{Y.}~\bibnamefont{Zhang}},
  \bibinfo{author}{\bibfnamefont{R.}~\bibnamefont{Wu}}, \bibnamefont{and}
  \bibinfo{author}{\bibfnamefont{W.}~\bibnamefont{Ho}},
  \bibinfo{journal}{Science} \textbf{\bibinfo{volume}{358}},
  \bibinfo{pages}{206} (\bibinfo{year}{2017}).

\bibitem[{\citenamefont{Mukherjee et~al.}(2019)\citenamefont{Mukherjee,
  Sanz-Matias, Velpula, Waghray, Ivasenko, Bilbao, Harvey, Mali, and
  De~Feyter}}]{Feyter2019}
\bibinfo{author}{\bibfnamefont{A.}~\bibnamefont{Mukherjee}},
  \bibinfo{author}{\bibfnamefont{A.}~\bibnamefont{Sanz-Matias}},
  \bibinfo{author}{\bibfnamefont{G.}~\bibnamefont{Velpula}},
  \bibinfo{author}{\bibfnamefont{D.}~\bibnamefont{Waghray}},
  \bibinfo{author}{\bibfnamefont{O.}~\bibnamefont{Ivasenko}},
  \bibinfo{author}{\bibfnamefont{N.}~\bibnamefont{Bilbao}},
  \bibinfo{author}{\bibfnamefont{J.~N.} \bibnamefont{Harvey}},
  \bibinfo{author}{\bibfnamefont{K.~S.} \bibnamefont{Mali}}, \bibnamefont{and}
  \bibinfo{author}{\bibfnamefont{S.}~\bibnamefont{De~Feyter}},
  \bibinfo{journal}{Chem Sci} \textbf{\bibinfo{volume}{10}},
  \bibinfo{pages}{3881} (\bibinfo{year}{2019}).

\bibitem[{\citenamefont{Shang et~al.}(2015)\citenamefont{Shang, Wang, Chen,
  Dai, Zhou, Kuttner, Hilt, Shao, Gottfried, and Wu}}]{Shang2015}
\bibinfo{author}{\bibfnamefont{J.}~\bibnamefont{Shang}},
  \bibinfo{author}{\bibfnamefont{Y.}~\bibnamefont{Wang}},
  \bibinfo{author}{\bibfnamefont{M.}~\bibnamefont{Chen}},
  \bibinfo{author}{\bibfnamefont{J.}~\bibnamefont{Dai}},
  \bibinfo{author}{\bibfnamefont{X.}~\bibnamefont{Zhou}},
  \bibinfo{author}{\bibfnamefont{J.}~\bibnamefont{Kuttner}},
  \bibinfo{author}{\bibfnamefont{G.}~\bibnamefont{Hilt}},
  \bibinfo{author}{\bibfnamefont{X.}~\bibnamefont{Shao}},
  \bibinfo{author}{\bibfnamefont{J.~M.} \bibnamefont{Gottfried}},
  \bibnamefont{and} \bibinfo{author}{\bibfnamefont{K.}~\bibnamefont{Wu}},
  \bibinfo{journal}{Nat Chem} \textbf{\bibinfo{volume}{7}},
  \bibinfo{pages}{389} (\bibinfo{year}{2015}).

\bibitem[{\citenamefont{Zhou et~al.}(2020)\citenamefont{Zhou, Liu, Feng, Shao,
  Zeng, Wang, Feng, and Liu}}]{Zhou2020}
\bibinfo{author}{\bibfnamefont{C.-S.} \bibnamefont{Zhou}},
  \bibinfo{author}{\bibfnamefont{X.-R.} \bibnamefont{Liu}},
  \bibinfo{author}{\bibfnamefont{Y.}~\bibnamefont{Feng}},
  \bibinfo{author}{\bibfnamefont{X.}~\bibnamefont{Shao}},
  \bibinfo{author}{\bibfnamefont{M.}~\bibnamefont{Zeng}},
  \bibinfo{author}{\bibfnamefont{K.}~\bibnamefont{Wang}},
  \bibinfo{author}{\bibfnamefont{M.}~\bibnamefont{Feng}}, \bibnamefont{and}
  \bibinfo{author}{\bibfnamefont{C.}~\bibnamefont{Liu}}, \bibinfo{journal}{Appl
  Phys Lett} \textbf{\bibinfo{volume}{117}}, \bibinfo{pages}{191601}
  (\bibinfo{year}{2020}).

\bibitem[{\citenamefont{Zhang and Zhao}(2015)}]{Zhao2015}
\bibinfo{author}{\bibfnamefont{X.}~\bibnamefont{Zhang}} \bibnamefont{and}
  \bibinfo{author}{\bibfnamefont{M.}~\bibnamefont{Zhao}}, \bibinfo{journal}{Sci
  Rep} \textbf{\bibinfo{volume}{5}}, \bibinfo{pages}{14098}
  (\bibinfo{year}{2015}).

\bibitem[{\citenamefont{Echenique et~al.}(2004)\citenamefont{Echenique, Berndt,
  Chulkov, Fauster, Goldmann, and H{\"o}fer}}]{Echenique2004}
\bibinfo{author}{\bibfnamefont{P.}~\bibnamefont{Echenique}},
  \bibinfo{author}{\bibfnamefont{R.}~\bibnamefont{Berndt}},
  \bibinfo{author}{\bibfnamefont{E.}~\bibnamefont{Chulkov}},
  \bibinfo{author}{\bibfnamefont{T.}~\bibnamefont{Fauster}},
  \bibinfo{author}{\bibfnamefont{A.}~\bibnamefont{Goldmann}}, \bibnamefont{and}
  \bibinfo{author}{\bibfnamefont{U.}~\bibnamefont{H{\"o}fer}},
  \bibinfo{journal}{Surf Sci Rep} \textbf{\bibinfo{volume}{52}},
  \bibinfo{pages}{219} (\bibinfo{year}{2004}).

\bibitem[{\citenamefont{Garc\'{\i}a~de Abajo
  et~al.}(2010)\citenamefont{Garc\'{\i}a~de Abajo, Cord\'on, Corso, Schiller,
  and Ortega}}]{GarciaAbajo2010}
\bibinfo{author}{\bibfnamefont{F.~J.} \bibnamefont{Garc\'{\i}a~de Abajo}},
  \bibinfo{author}{\bibfnamefont{J.}~\bibnamefont{Cord\'on}},
  \bibinfo{author}{\bibfnamefont{M.}~\bibnamefont{Corso}},
  \bibinfo{author}{\bibfnamefont{F.}~\bibnamefont{Schiller}}, \bibnamefont{and}
  \bibinfo{author}{\bibfnamefont{J.~E.} \bibnamefont{Ortega}},
  \bibinfo{journal}{Nanoscale} \textbf{\bibinfo{volume}{2}},
  \bibinfo{pages}{717} (\bibinfo{year}{2010}).

\bibitem[{\citenamefont{Abd El-Fattah et~al.}(2019)\citenamefont{Abd El-Fattah,
  Kher-Elden, Piquero-Zulaica, de~Abajo, and Ortega}}]{Zakaria2019}
\bibinfo{author}{\bibfnamefont{Z.~M.} \bibnamefont{Abd El-Fattah}},
  \bibinfo{author}{\bibfnamefont{M.~A.} \bibnamefont{Kher-Elden}},
  \bibinfo{author}{\bibfnamefont{I.}~\bibnamefont{Piquero-Zulaica}},
  \bibinfo{author}{\bibfnamefont{F.~J.~G.} \bibnamefont{de~Abajo}},
  \bibnamefont{and} \bibinfo{author}{\bibfnamefont{J.~E.}
  \bibnamefont{Ortega}}, \bibinfo{journal}{Phys Rev B}
  \textbf{\bibinfo{volume}{99}}, \bibinfo{pages}{115443}
  (\bibinfo{year}{2019}).

\bibitem[{\citenamefont{Piquero-Zulaica
  et~al.}(2018)\citenamefont{Piquero-Zulaica, Garcia-Lekue, Colazzo, Krug,
  Mohammed, Abd El-Fattah, Gottfried, de~Oteyza, Ortega, and
  Lobo-Checa}}]{Piquero2018}
\bibinfo{author}{\bibfnamefont{I.}~\bibnamefont{Piquero-Zulaica}},
  \bibinfo{author}{\bibfnamefont{A.}~\bibnamefont{Garcia-Lekue}},
  \bibinfo{author}{\bibfnamefont{L.}~\bibnamefont{Colazzo}},
  \bibinfo{author}{\bibfnamefont{C.~K.} \bibnamefont{Krug}},
  \bibinfo{author}{\bibfnamefont{M.~S.~G.} \bibnamefont{Mohammed}},
  \bibinfo{author}{\bibfnamefont{Z.~M.} \bibnamefont{Abd El-Fattah}},
  \bibinfo{author}{\bibfnamefont{J.~M.} \bibnamefont{Gottfried}},
  \bibinfo{author}{\bibfnamefont{D.~G.} \bibnamefont{de~Oteyza}},
  \bibinfo{author}{\bibfnamefont{J.~E.} \bibnamefont{Ortega}},
  \bibnamefont{and}
  \bibinfo{author}{\bibfnamefont{J.}~\bibnamefont{Lobo-Checa}},
  \bibinfo{journal}{ACS Nano} \textbf{\bibinfo{volume}{12}},
  \bibinfo{pages}{10537} (\bibinfo{year}{2018}).

\bibitem[{\citenamefont{Kher-Elden et~al.}(2020)\citenamefont{Kher-Elden,
  Piquero-Zulaica, Abd El-Aziz, Ortega, and Abd El-Fattah}}]{Zakaria2020}
\bibinfo{author}{\bibfnamefont{M.~A.} \bibnamefont{Kher-Elden}},
  \bibinfo{author}{\bibfnamefont{I.}~\bibnamefont{Piquero-Zulaica}},
  \bibinfo{author}{\bibfnamefont{K.~M.} \bibnamefont{Abd El-Aziz}},
  \bibinfo{author}{\bibfnamefont{J.~E.} \bibnamefont{Ortega}},
  \bibnamefont{and} \bibinfo{author}{\bibfnamefont{Z.~M.} \bibnamefont{Abd
  El-Fattah}}, \bibinfo{journal}{RSC Adv} \textbf{\bibinfo{volume}{10}},
  \bibinfo{pages}{33844} (\bibinfo{year}{2020}).

\bibitem[{\citenamefont{Franceschetti and Zunger}(1999)}]{Franceschetti1999}
\bibinfo{author}{\bibfnamefont{A.}~\bibnamefont{Franceschetti}}
  \bibnamefont{and} \bibinfo{author}{\bibfnamefont{A.}~\bibnamefont{Zunger}},
  \bibinfo{journal}{Nature} \textbf{\bibinfo{volume}{402}}, \bibinfo{pages}{60}
  (\bibinfo{year}{1999}).

\bibitem[{\citenamefont{Ourmazd}(2020)}]{Ourmazd2020}
\bibinfo{author}{\bibfnamefont{A.}~\bibnamefont{Ourmazd}},
  \bibinfo{journal}{Nature Reviews Physics} \textbf{\bibinfo{volume}{2}},
  \bibinfo{pages}{342} (\bibinfo{year}{2020}).

\bibitem[{\citenamefont{Hörmann et~al.}(2019)\citenamefont{Hörmann, Jeindl,
  Egger, Scherbela, and Hofmann}}]{Hormann2019}
\bibinfo{author}{\bibfnamefont{L.}~\bibnamefont{Hörmann}},
  \bibinfo{author}{\bibfnamefont{A.}~\bibnamefont{Jeindl}},
  \bibinfo{author}{\bibfnamefont{A.~T.} \bibnamefont{Egger}},
  \bibinfo{author}{\bibfnamefont{M.}~\bibnamefont{Scherbela}},
  \bibnamefont{and} \bibinfo{author}{\bibfnamefont{O.~T.}
  \bibnamefont{Hofmann}}, \bibinfo{journal}{Comput Phys Commun}
  \textbf{\bibinfo{volume}{244}}, \bibinfo{pages}{143} (\bibinfo{year}{2019}).

\bibitem[{\citenamefont{Alldritt et~al.}(2020)\citenamefont{Alldritt, Hapala,
  Oinonen, Urtev, Krejci, Federici~Canova, Kannala, Schulz, Liljeroth, and
  Foster}}]{Alldritt2020}
\bibinfo{author}{\bibfnamefont{B.}~\bibnamefont{Alldritt}},
  \bibinfo{author}{\bibfnamefont{P.}~\bibnamefont{Hapala}},
  \bibinfo{author}{\bibfnamefont{N.}~\bibnamefont{Oinonen}},
  \bibinfo{author}{\bibfnamefont{F.}~\bibnamefont{Urtev}},
  \bibinfo{author}{\bibfnamefont{O.}~\bibnamefont{Krejci}},
  \bibinfo{author}{\bibfnamefont{F.}~\bibnamefont{Federici~Canova}},
  \bibinfo{author}{\bibfnamefont{J.}~\bibnamefont{Kannala}},
  \bibinfo{author}{\bibfnamefont{F.}~\bibnamefont{Schulz}},
  \bibinfo{author}{\bibfnamefont{P.}~\bibnamefont{Liljeroth}},
  \bibnamefont{and} \bibinfo{author}{\bibfnamefont{A.~S.}
  \bibnamefont{Foster}}, \bibinfo{journal}{Sci Adv}
  \textbf{\bibinfo{volume}{6}}, \bibinfo{pages}{eaay6913}
  (\bibinfo{year}{2020}).

\bibitem[{\citenamefont{Krull et~al.}(2020)\citenamefont{Krull, Hirsch, Rother,
  Schiffrin, and Krull}}]{Krull2020}
\bibinfo{author}{\bibfnamefont{A.}~\bibnamefont{Krull}},
  \bibinfo{author}{\bibfnamefont{P.}~\bibnamefont{Hirsch}},
  \bibinfo{author}{\bibfnamefont{C.}~\bibnamefont{Rother}},
  \bibinfo{author}{\bibfnamefont{A.}~\bibnamefont{Schiffrin}},
  \bibnamefont{and} \bibinfo{author}{\bibfnamefont{C.}~\bibnamefont{Krull}},
  \bibinfo{journal}{Communications Physics} \textbf{\bibinfo{volume}{3}},
  \bibinfo{pages}{54} (\bibinfo{year}{2020}).

\bibitem[{\citenamefont{Blunt et~al.}(2010)\citenamefont{Blunt, Russell,
  Champness, and Beton}}]{Beton2010}
\bibinfo{author}{\bibfnamefont{M.~O.} \bibnamefont{Blunt}},
  \bibinfo{author}{\bibfnamefont{J.~C.} \bibnamefont{Russell}},
  \bibinfo{author}{\bibfnamefont{N.~R.} \bibnamefont{Champness}},
  \bibnamefont{and} \bibinfo{author}{\bibfnamefont{P.~H.} \bibnamefont{Beton}},
  \bibinfo{journal}{Chem Commun} \textbf{\bibinfo{volume}{46}},
  \bibinfo{pages}{7157} (\bibinfo{year}{2010}).

\bibitem[{\citenamefont{Kliewer et~al.}(2000)\citenamefont{Kliewer, Berndt, and
  Crampin}}]{Kliewer2000}
\bibinfo{author}{\bibfnamefont{J.}~\bibnamefont{Kliewer}},
  \bibinfo{author}{\bibfnamefont{R.}~\bibnamefont{Berndt}}, \bibnamefont{and}
  \bibinfo{author}{\bibfnamefont{S.}~\bibnamefont{Crampin}},
  \bibinfo{journal}{Phys Rev Lett} \textbf{\bibinfo{volume}{85}},
  \bibinfo{pages}{4936} (\bibinfo{year}{2000}).

\bibitem[{\citenamefont{Schiffrin et~al.}(2008)\citenamefont{Schiffrin,
  Reichert, Auwärter, Jahnz, Pennec, Weber-Bargioni, Stepanyuk, Niebergall,
  Bruno, and Barth}}]{Schiffrin2008}
\bibinfo{author}{\bibfnamefont{A.}~\bibnamefont{Schiffrin}},
  \bibinfo{author}{\bibfnamefont{J.}~\bibnamefont{Reichert}},
  \bibinfo{author}{\bibfnamefont{W.}~\bibnamefont{Auwärter}},
  \bibinfo{author}{\bibfnamefont{G.}~\bibnamefont{Jahnz}},
  \bibinfo{author}{\bibfnamefont{Y.}~\bibnamefont{Pennec}},
  \bibinfo{author}{\bibfnamefont{A.}~\bibnamefont{Weber-Bargioni}},
  \bibinfo{author}{\bibfnamefont{V.~S.} \bibnamefont{Stepanyuk}},
  \bibinfo{author}{\bibfnamefont{L.}~\bibnamefont{Niebergall}},
  \bibinfo{author}{\bibfnamefont{P.}~\bibnamefont{Bruno}}, \bibnamefont{and}
  \bibinfo{author}{\bibfnamefont{J.~V.} \bibnamefont{Barth}},
  \bibinfo{journal}{Phys Rev B} \textbf{\bibinfo{volume}{78}},
  \bibinfo{pages}{035424} (\bibinfo{year}{2008}).

\bibitem[{\citenamefont{Negulyaev et~al.}(2009)\citenamefont{Negulyaev,
  Stepanyuk, Niebergall, Bruno, Auw\"arter, Pennec, Jahnz, and
  Barth}}]{Negulyaev2009}
\bibinfo{author}{\bibfnamefont{N.~N.} \bibnamefont{Negulyaev}},
  \bibinfo{author}{\bibfnamefont{V.~S.} \bibnamefont{Stepanyuk}},
  \bibinfo{author}{\bibfnamefont{L.}~\bibnamefont{Niebergall}},
  \bibinfo{author}{\bibfnamefont{P.}~\bibnamefont{Bruno}},
  \bibinfo{author}{\bibfnamefont{W.}~\bibnamefont{Auw\"arter}},
  \bibinfo{author}{\bibfnamefont{Y.}~\bibnamefont{Pennec}},
  \bibinfo{author}{\bibfnamefont{G.}~\bibnamefont{Jahnz}}, \bibnamefont{and}
  \bibinfo{author}{\bibfnamefont{J.~V.} \bibnamefont{Barth}},
  \bibinfo{journal}{Phys Rev B} \textbf{\bibinfo{volume}{79}},
  \bibinfo{pages}{195411} (\bibinfo{year}{2009}).

\bibitem[{\citenamefont{Pawin et~al.}(2006)\citenamefont{Pawin, Wong, Kwon, and
  Bartels}}]{Pawin2006}
\bibinfo{author}{\bibfnamefont{G.}~\bibnamefont{Pawin}},
  \bibinfo{author}{\bibfnamefont{K.~L.} \bibnamefont{Wong}},
  \bibinfo{author}{\bibfnamefont{K.-Y.} \bibnamefont{Kwon}}, \bibnamefont{and}
  \bibinfo{author}{\bibfnamefont{L.}~\bibnamefont{Bartels}},
  \bibinfo{journal}{Science} \textbf{\bibinfo{volume}{313}},
  \bibinfo{pages}{961} (\bibinfo{year}{2006}).

\bibitem[{\citenamefont{Wang et~al.}(2009)\citenamefont{Wang, Ge, Manzano,
  Kröger, Berndt, Hofer, Tang, and Cerda}}]{Wang09}
\bibinfo{author}{\bibfnamefont{Y.}~\bibnamefont{Wang}},
  \bibinfo{author}{\bibfnamefont{X.}~\bibnamefont{Ge}},
  \bibinfo{author}{\bibfnamefont{C.}~\bibnamefont{Manzano}},
  \bibinfo{author}{\bibfnamefont{J.}~\bibnamefont{Kröger}},
  \bibinfo{author}{\bibfnamefont{R.}~\bibnamefont{Berndt}},
  \bibinfo{author}{\bibfnamefont{W.~A.} \bibnamefont{Hofer}},
  \bibinfo{author}{\bibfnamefont{H.}~\bibnamefont{Tang}}, \bibnamefont{and}
  \bibinfo{author}{\bibfnamefont{J.}~\bibnamefont{Cerda}}, \bibinfo{journal}{J
  Am Chem Soc} \textbf{\bibinfo{volume}{131}}, \bibinfo{pages}{10400}
  (\bibinfo{year}{2009}).

\bibitem[{\citenamefont{Wyrick et~al.}(2011)\citenamefont{Wyrick, Kim, Sun,
  Cheng, Lu, Zhu, Berland, Kim, Rotenberg, Luo et~al.}}]{Wyrick2011}
\bibinfo{author}{\bibfnamefont{J.}~\bibnamefont{Wyrick}},
  \bibinfo{author}{\bibfnamefont{D.-H.} \bibnamefont{Kim}},
  \bibinfo{author}{\bibfnamefont{D.}~\bibnamefont{Sun}},
  \bibinfo{author}{\bibfnamefont{Z.}~\bibnamefont{Cheng}},
  \bibinfo{author}{\bibfnamefont{W.}~\bibnamefont{Lu}},
  \bibinfo{author}{\bibfnamefont{Y.}~\bibnamefont{Zhu}},
  \bibinfo{author}{\bibfnamefont{K.}~\bibnamefont{Berland}},
  \bibinfo{author}{\bibfnamefont{Y.~S.} \bibnamefont{Kim}},
  \bibinfo{author}{\bibfnamefont{E.}~\bibnamefont{Rotenberg}},
  \bibinfo{author}{\bibfnamefont{M.}~\bibnamefont{Luo}}, \bibnamefont{et~al.},
  \bibinfo{journal}{Nano Lett} \textbf{\bibinfo{volume}{11}},
  \bibinfo{pages}{2944} (\bibinfo{year}{2011}).

\bibitem[{\citenamefont{Einstein et~al.}(2018)\citenamefont{Einstein, Bartels,
  and Morales-Cifuentes}}]{Einstein2018}
\bibinfo{author}{\bibfnamefont{T.~L.} \bibnamefont{Einstein}},
  \bibinfo{author}{\bibfnamefont{L.}~\bibnamefont{Bartels}}, \bibnamefont{and}
  \bibinfo{author}{\bibfnamefont{J.~R.} \bibnamefont{Morales-Cifuentes}},
  \bibinfo{journal}{e-J Surf Sci Nanotechnol} \textbf{\bibinfo{volume}{16}},
  \bibinfo{pages}{201} (\bibinfo{year}{2018}).

\bibitem[{\citenamefont{Ahsan et~al.}(2018)\citenamefont{Ahsan, Mousavi, Nijs,
  Nowakowska, Popova, Wäckerlin, Björk, Gade, and Jung}}]{Jung2018}
\bibinfo{author}{\bibfnamefont{A.}~\bibnamefont{Ahsan}},
  \bibinfo{author}{\bibfnamefont{S.~F.} \bibnamefont{Mousavi}},
  \bibinfo{author}{\bibfnamefont{T.}~\bibnamefont{Nijs}},
  \bibinfo{author}{\bibfnamefont{S.}~\bibnamefont{Nowakowska}},
  \bibinfo{author}{\bibfnamefont{O.}~\bibnamefont{Popova}},
  \bibinfo{author}{\bibfnamefont{A.}~\bibnamefont{Wäckerlin}},
  \bibinfo{author}{\bibfnamefont{J.}~\bibnamefont{Björk}},
  \bibinfo{author}{\bibfnamefont{L.~H.} \bibnamefont{Gade}}, \bibnamefont{and}
  \bibinfo{author}{\bibfnamefont{T.~A.} \bibnamefont{Jung}},
  \bibinfo{journal}{Small} p. \bibinfo{pages}{1803169} (\bibinfo{year}{2018}).

\bibitem[{\citenamefont{Ahsan et~al.}(2019)\citenamefont{Ahsan,
  Fatemeh~Mousavi, Nijs, Nowakowska, Popova, Wäckerlin, Björk, Gade, and
  Jung}}]{Jung2019}
\bibinfo{author}{\bibfnamefont{A.}~\bibnamefont{Ahsan}},
  \bibinfo{author}{\bibfnamefont{S.}~\bibnamefont{Fatemeh~Mousavi}},
  \bibinfo{author}{\bibfnamefont{T.}~\bibnamefont{Nijs}},
  \bibinfo{author}{\bibfnamefont{S.}~\bibnamefont{Nowakowska}},
  \bibinfo{author}{\bibfnamefont{O.}~\bibnamefont{Popova}},
  \bibinfo{author}{\bibfnamefont{A.}~\bibnamefont{Wäckerlin}},
  \bibinfo{author}{\bibfnamefont{J.}~\bibnamefont{Björk}},
  \bibinfo{author}{\bibfnamefont{L.~H.} \bibnamefont{Gade}}, \bibnamefont{and}
  \bibinfo{author}{\bibfnamefont{T.~A.} \bibnamefont{Jung}},
  \bibinfo{journal}{Nanoscale} \textbf{\bibinfo{volume}{11}},
  \bibinfo{pages}{4895} (\bibinfo{year}{2019}).

\bibitem[{\citenamefont{Seyller et~al.}(1998)\citenamefont{Seyller, Caragiu,
  Diehl, Kaukasoina, and Lindroos}}]{Seyller1998}
\bibinfo{author}{\bibfnamefont{T.}~\bibnamefont{Seyller}},
  \bibinfo{author}{\bibfnamefont{M.}~\bibnamefont{Caragiu}},
  \bibinfo{author}{\bibfnamefont{R.}~\bibnamefont{Diehl}},
  \bibinfo{author}{\bibfnamefont{P.}~\bibnamefont{Kaukasoina}},
  \bibnamefont{and} \bibinfo{author}{\bibfnamefont{M.}~\bibnamefont{Lindroos}},
  \bibinfo{journal}{Chem Phys Lett} \textbf{\bibinfo{volume}{291}},
  \bibinfo{pages}{567} (\bibinfo{year}{1998}).

\bibitem[{\citenamefont{Dil et~al.}(2008)\citenamefont{Dil, Lobo-Checa,
  Laskowski, Blaha, Berner, Osterwalder, and Greber}}]{Dil2008a}
\bibinfo{author}{\bibfnamefont{H.}~\bibnamefont{Dil}},
  \bibinfo{author}{\bibfnamefont{J.}~\bibnamefont{Lobo-Checa}},
  \bibinfo{author}{\bibfnamefont{R.}~\bibnamefont{Laskowski}},
  \bibinfo{author}{\bibfnamefont{P.}~\bibnamefont{Blaha}},
  \bibinfo{author}{\bibfnamefont{S.}~\bibnamefont{Berner}},
  \bibinfo{author}{\bibfnamefont{J.}~\bibnamefont{Osterwalder}},
  \bibnamefont{and} \bibinfo{author}{\bibfnamefont{T.}~\bibnamefont{Greber}},
  \bibinfo{journal}{Science} \textbf{\bibinfo{volume}{319}},
  \bibinfo{pages}{1824} (\bibinfo{year}{2008}).

\bibitem[{\citenamefont{Palma et~al.}(2014)\citenamefont{Palma, Björk, Rao,
  Kühne, Klappenberger, and Barth}}]{Palma2014}
\bibinfo{author}{\bibfnamefont{C.-A.} \bibnamefont{Palma}},
  \bibinfo{author}{\bibfnamefont{J.}~\bibnamefont{Björk}},
  \bibinfo{author}{\bibfnamefont{F.}~\bibnamefont{Rao}},
  \bibinfo{author}{\bibfnamefont{D.}~\bibnamefont{Kühne}},
  \bibinfo{author}{\bibfnamefont{F.}~\bibnamefont{Klappenberger}},
  \bibnamefont{and} \bibinfo{author}{\bibfnamefont{J.~V.} \bibnamefont{Barth}},
  \bibinfo{journal}{Nano Lett} \textbf{\bibinfo{volume}{14}},
  \bibinfo{pages}{4461} (\bibinfo{year}{2014}).

\bibitem[{\citenamefont{Palma et~al.}(2015)\citenamefont{Palma, Björk,
  Klappenberger, Arras, Kühne, Stafström, and Barth}}]{Palma2015}
\bibinfo{author}{\bibfnamefont{C.-A.} \bibnamefont{Palma}},
  \bibinfo{author}{\bibfnamefont{J.}~\bibnamefont{Björk}},
  \bibinfo{author}{\bibfnamefont{F.}~\bibnamefont{Klappenberger}},
  \bibinfo{author}{\bibfnamefont{E.}~\bibnamefont{Arras}},
  \bibinfo{author}{\bibfnamefont{D.}~\bibnamefont{Kühne}},
  \bibinfo{author}{\bibfnamefont{S.}~\bibnamefont{Stafström}},
  \bibnamefont{and} \bibinfo{author}{\bibfnamefont{J.~V.} \bibnamefont{Barth}},
  \bibinfo{journal}{Nat Commun} \textbf{\bibinfo{volume}{6}},
  \bibinfo{pages}{6210} (\bibinfo{year}{2015}).

\bibitem[{\citenamefont{Park et~al.}(2000)\citenamefont{Park, Ham, Kahng, Kuk,
  Miyake, Hata, and Shigekawa}}]{Park2000}
\bibinfo{author}{\bibfnamefont{J.-Y.} \bibnamefont{Park}},
  \bibinfo{author}{\bibfnamefont{U.~D.} \bibnamefont{Ham}},
  \bibinfo{author}{\bibfnamefont{S.-J.} \bibnamefont{Kahng}},
  \bibinfo{author}{\bibfnamefont{Y.}~\bibnamefont{Kuk}},
  \bibinfo{author}{\bibfnamefont{K.}~\bibnamefont{Miyake}},
  \bibinfo{author}{\bibfnamefont{K.}~\bibnamefont{Hata}}, \bibnamefont{and}
  \bibinfo{author}{\bibfnamefont{H.}~\bibnamefont{Shigekawa}},
  \bibinfo{journal}{Phys Rev B} \textbf{\bibinfo{volume}{62}},
  \bibinfo{pages}{R16341} (\bibinfo{year}{2000}).

\bibitem[{\citenamefont{H{\"o}vel et~al.}(2001)\citenamefont{H{\"o}vel, Grimm,
  and Reihl}}]{Hovel2001}
\bibinfo{author}{\bibfnamefont{H.}~\bibnamefont{H{\"o}vel}},
  \bibinfo{author}{\bibfnamefont{B.}~\bibnamefont{Grimm}}, \bibnamefont{and}
  \bibinfo{author}{\bibfnamefont{B.}~\bibnamefont{Reihl}},
  \bibinfo{journal}{Surf Sci} \textbf{\bibinfo{volume}{477}},
  \bibinfo{pages}{43} (\bibinfo{year}{2001}).

\bibitem[{\citenamefont{Abd El-Fattah et~al.}(2017)\citenamefont{Abd El-Fattah,
  Kher-Elden, Yassin, El-Okr, Ortega, and García~de Abajo}}]{Zaka2017}
\bibinfo{author}{\bibfnamefont{Z.~M.} \bibnamefont{Abd El-Fattah}},
  \bibinfo{author}{\bibfnamefont{M.~A.} \bibnamefont{Kher-Elden}},
  \bibinfo{author}{\bibfnamefont{O.}~\bibnamefont{Yassin}},
  \bibinfo{author}{\bibfnamefont{M.~M.} \bibnamefont{El-Okr}},
  \bibinfo{author}{\bibfnamefont{J.~E.} \bibnamefont{Ortega}},
  \bibnamefont{and} \bibinfo{author}{\bibfnamefont{F.~J.}
  \bibnamefont{García~de Abajo}}, \bibinfo{journal}{J Appl Phys}
  \textbf{\bibinfo{volume}{122}}, \bibinfo{pages}{195306}
  (\bibinfo{year}{2017}).

\bibitem[{\citenamefont{Repp et~al.}(2004)\citenamefont{Repp, Meyer, and
  Rieder}}]{Repp2004}
\bibinfo{author}{\bibfnamefont{J.}~\bibnamefont{Repp}},
  \bibinfo{author}{\bibfnamefont{G.}~\bibnamefont{Meyer}}, \bibnamefont{and}
  \bibinfo{author}{\bibfnamefont{K.-H.} \bibnamefont{Rieder}},
  \bibinfo{journal}{Phys Rev Lett} \textbf{\bibinfo{volume}{92}},
  \bibinfo{pages}{036803} (\bibinfo{year}{2004}).

\bibitem[{\citenamefont{Moreno et~al.}(2018)\citenamefont{Moreno, Vilas-Varela,
  Kretz, Garcia-Lekue, Costache, Paradinas, Panighel, Ceballos, Valenzuela,
  Pe{\~n}a et~al.}}]{Moreno2018}
\bibinfo{author}{\bibfnamefont{C.}~\bibnamefont{Moreno}},
  \bibinfo{author}{\bibfnamefont{M.}~\bibnamefont{Vilas-Varela}},
  \bibinfo{author}{\bibfnamefont{B.}~\bibnamefont{Kretz}},
  \bibinfo{author}{\bibfnamefont{A.}~\bibnamefont{Garcia-Lekue}},
  \bibinfo{author}{\bibfnamefont{M.~V.} \bibnamefont{Costache}},
  \bibinfo{author}{\bibfnamefont{M.}~\bibnamefont{Paradinas}},
  \bibinfo{author}{\bibfnamefont{M.}~\bibnamefont{Panighel}},
  \bibinfo{author}{\bibfnamefont{G.}~\bibnamefont{Ceballos}},
  \bibinfo{author}{\bibfnamefont{S.~O.} \bibnamefont{Valenzuela}},
  \bibinfo{author}{\bibfnamefont{D.}~\bibnamefont{Pe{\~n}a}},
  \bibnamefont{et~al.}, \bibinfo{journal}{Science}
  \textbf{\bibinfo{volume}{360}}, \bibinfo{pages}{199} (\bibinfo{year}{2018}).

\bibitem[{\citenamefont{Calogero et~al.}(2019)\citenamefont{Calogero, Papior,
  Kretz, Garcia-Lekue, Frederiksen, and Brandbyge}}]{Calogero2019}
\bibinfo{author}{\bibfnamefont{G.}~\bibnamefont{Calogero}},
  \bibinfo{author}{\bibfnamefont{N.~R.} \bibnamefont{Papior}},
  \bibinfo{author}{\bibfnamefont{B.}~\bibnamefont{Kretz}},
  \bibinfo{author}{\bibfnamefont{A.}~\bibnamefont{Garcia-Lekue}},
  \bibinfo{author}{\bibfnamefont{T.}~\bibnamefont{Frederiksen}},
  \bibnamefont{and}
  \bibinfo{author}{\bibfnamefont{M.}~\bibnamefont{Brandbyge}},
  \bibinfo{journal}{Nano Lett} \textbf{\bibinfo{volume}{19}},
  \bibinfo{pages}{576} (\bibinfo{year}{2019}).

\bibitem[{\citenamefont{Niesner and Fauster}(2014)}]{Fauster2014}
\bibinfo{author}{\bibfnamefont{D.}~\bibnamefont{Niesner}} \bibnamefont{and}
  \bibinfo{author}{\bibfnamefont{T.}~\bibnamefont{Fauster}},
  \bibinfo{journal}{J Phys : Condens Matter} \textbf{\bibinfo{volume}{26}},
  \bibinfo{pages}{393001} (\bibinfo{year}{2014}).

\bibitem[{\citenamefont{Kawai et~al.}(2021)\citenamefont{Kawai, Kher-Elden,
  Sadeghi, Abd El-Fattah, Sun, Izumi, Minakata, Takeda, and
  Lobo-Checa}}]{Kawai2021}
\bibinfo{author}{\bibfnamefont{S.}~\bibnamefont{Kawai}},
  \bibinfo{author}{\bibfnamefont{M.~A.} \bibnamefont{Kher-Elden}},
  \bibinfo{author}{\bibfnamefont{A.}~\bibnamefont{Sadeghi}},
  \bibinfo{author}{\bibfnamefont{Z.~M.} \bibnamefont{Abd El-Fattah}},
  \bibinfo{author}{\bibfnamefont{K.}~\bibnamefont{Sun}},
  \bibinfo{author}{\bibfnamefont{S.}~\bibnamefont{Izumi}},
  \bibinfo{author}{\bibfnamefont{S.}~\bibnamefont{Minakata}},
  \bibinfo{author}{\bibfnamefont{Y.}~\bibnamefont{Takeda}}, \bibnamefont{and}
  \bibinfo{author}{\bibfnamefont{J.}~\bibnamefont{Lobo-Checa}},
  \bibinfo{journal}{Nano Lett} \bibinfo{pages}{DOI: 10.1021/acs.nanolett.1c01200}
  (\bibinfo{year}{2021}).

\bibitem[{\citenamefont{Zhang et~al.}(2016{\natexlab{b}})\citenamefont{Zhang,
  Björk, Barth, and Klappenberger}}]{Zhang2016a}
\bibinfo{author}{\bibfnamefont{Y.-Q.} \bibnamefont{Zhang}},
  \bibinfo{author}{\bibfnamefont{J.}~\bibnamefont{Björk}},
  \bibinfo{author}{\bibfnamefont{J.~V.} \bibnamefont{Barth}}, \bibnamefont{and}
  \bibinfo{author}{\bibfnamefont{F.}~\bibnamefont{Klappenberger}},
  \bibinfo{journal}{Nano Lett} \textbf{\bibinfo{volume}{16}},
  \bibinfo{pages}{4274} (\bibinfo{year}{2016}{\natexlab{b}}).

\bibitem[{\citenamefont{Hieulle et~al.}(2018)\citenamefont{Hieulle,
  Carbonell-Sanrom{\`{a}}, Vilas-Varela, Garcia-Lekue, Guiti{\'{a}}n,
  Pe{\~{n}}a, and Pascual}}]{Hieulle2018}
\bibinfo{author}{\bibfnamefont{J.}~\bibnamefont{Hieulle}},
  \bibinfo{author}{\bibfnamefont{E.}~\bibnamefont{Carbonell-Sanrom{\`{a}}}},
  \bibinfo{author}{\bibfnamefont{M.}~\bibnamefont{Vilas-Varela}},
  \bibinfo{author}{\bibfnamefont{A.}~\bibnamefont{Garcia-Lekue}},
  \bibinfo{author}{\bibfnamefont{E.}~\bibnamefont{Guiti{\'{a}}n}},
  \bibinfo{author}{\bibfnamefont{D.}~\bibnamefont{Pe{\~{n}}a}},
  \bibnamefont{and} \bibinfo{author}{\bibfnamefont{J.~I.}
  \bibnamefont{Pascual}}, \bibinfo{journal}{Nano Lett}
  \textbf{\bibinfo{volume}{18}}, \bibinfo{pages}{418} (\bibinfo{year}{2018}).

\bibitem[{\citenamefont{Zhang et~al.}(2020{\natexlab{b}})\citenamefont{Zhang,
  Liu, Gao, Hua, Xia, Knecht, Papageorgiou, Reichert, Barth, Xu
  et~al.}}]{Zhang2020}
\bibinfo{author}{\bibfnamefont{R.}~\bibnamefont{Zhang}},
  \bibinfo{author}{\bibfnamefont{J.}~\bibnamefont{Liu}},
  \bibinfo{author}{\bibfnamefont{Y.}~\bibnamefont{Gao}},
  \bibinfo{author}{\bibfnamefont{M.}~\bibnamefont{Hua}},
  \bibinfo{author}{\bibfnamefont{B.}~\bibnamefont{Xia}},
  \bibinfo{author}{\bibfnamefont{P.}~\bibnamefont{Knecht}},
  \bibinfo{author}{\bibfnamefont{A.~C.} \bibnamefont{Papageorgiou}},
  \bibinfo{author}{\bibfnamefont{J.}~\bibnamefont{Reichert}},
  \bibinfo{author}{\bibfnamefont{J.~V.} \bibnamefont{Barth}},
  \bibinfo{author}{\bibfnamefont{H.}~\bibnamefont{Xu}}, \bibnamefont{et~al.},
  \bibinfo{journal}{Angew Chem Int Ed} \textbf{\bibinfo{volume}{59}},
  \bibinfo{pages}{2669} (\bibinfo{year}{2020}{\natexlab{b}}).

\bibitem[{\citenamefont{Zhang et~al.}(2017)\citenamefont{Zhang, Zhou, Cui,
  Zhao, and Liu}}]{Zhang2017}
\bibinfo{author}{\bibfnamefont{X.}~\bibnamefont{Zhang}},
  \bibinfo{author}{\bibfnamefont{Y.}~\bibnamefont{Zhou}},
  \bibinfo{author}{\bibfnamefont{B.}~\bibnamefont{Cui}},
  \bibinfo{author}{\bibfnamefont{M.}~\bibnamefont{Zhao}}, \bibnamefont{and}
  \bibinfo{author}{\bibfnamefont{F.}~\bibnamefont{Liu}}, \bibinfo{journal}{Nano
  Lett} \textbf{\bibinfo{volume}{17}}, \bibinfo{pages}{6166}
  (\bibinfo{year}{2017}).

\bibitem[{\citenamefont{Bieri et~al.}(2009)\citenamefont{Bieri, Treier, Cai,
  Aït-Mansour, Ruffieux, Gröning, Gröning, Kastler, Rieger, Feng
  et~al.}}]{Bieri2009}
\bibinfo{author}{\bibfnamefont{M.}~\bibnamefont{Bieri}},
  \bibinfo{author}{\bibfnamefont{M.}~\bibnamefont{Treier}},
  \bibinfo{author}{\bibfnamefont{J.}~\bibnamefont{Cai}},
  \bibinfo{author}{\bibfnamefont{K.}~\bibnamefont{Aït-Mansour}},
  \bibinfo{author}{\bibfnamefont{P.}~\bibnamefont{Ruffieux}},
  \bibinfo{author}{\bibfnamefont{O.}~\bibnamefont{Gröning}},
  \bibinfo{author}{\bibfnamefont{P.}~\bibnamefont{Gröning}},
  \bibinfo{author}{\bibfnamefont{M.}~\bibnamefont{Kastler}},
  \bibinfo{author}{\bibfnamefont{R.}~\bibnamefont{Rieger}},
  \bibinfo{author}{\bibfnamefont{X.}~\bibnamefont{Feng}}, \bibnamefont{et~al.},
  \bibinfo{journal}{Chem Commun} p. \bibinfo{pages}{6919}
  (\bibinfo{year}{2009}).

\bibitem[{\citenamefont{Jacobse et~al.}(2020)\citenamefont{Jacobse, McCurdy,
  Jiang, Rizzo, Veber, Butler, Zuzak, Louie, Fischer, and
  Crommie}}]{Jacobse2020}
\bibinfo{author}{\bibfnamefont{P.~H.} \bibnamefont{Jacobse}},
  \bibinfo{author}{\bibfnamefont{R.~D.} \bibnamefont{McCurdy}},
  \bibinfo{author}{\bibfnamefont{J.}~\bibnamefont{Jiang}},
  \bibinfo{author}{\bibfnamefont{D.~J.} \bibnamefont{Rizzo}},
  \bibinfo{author}{\bibfnamefont{G.}~\bibnamefont{Veber}},
  \bibinfo{author}{\bibfnamefont{P.}~\bibnamefont{Butler}},
  \bibinfo{author}{\bibfnamefont{R.}~\bibnamefont{Zuzak}},
  \bibinfo{author}{\bibfnamefont{S.~G.} \bibnamefont{Louie}},
  \bibinfo{author}{\bibfnamefont{F.~R.} \bibnamefont{Fischer}},
  \bibnamefont{and} \bibinfo{author}{\bibfnamefont{M.~F.}
  \bibnamefont{Crommie}}, \bibinfo{journal}{J Am Chem Soc}
  \textbf{\bibinfo{volume}{142}}, \bibinfo{pages}{13507}
  (\bibinfo{year}{2020}).

\bibitem[{\citenamefont{Gobbi et~al.}(2018)\citenamefont{Gobbi, Orgiu, and
  Samorì}}]{Gobbi2018}
\bibinfo{author}{\bibfnamefont{M.}~\bibnamefont{Gobbi}},
  \bibinfo{author}{\bibfnamefont{E.}~\bibnamefont{Orgiu}}, \bibnamefont{and}
  \bibinfo{author}{\bibfnamefont{P.}~\bibnamefont{Samorì}},
  \bibinfo{journal}{Adv Mater} \textbf{\bibinfo{volume}{30}},
  \bibinfo{pages}{1706103} (\bibinfo{year}{2018}).

\bibitem[{\citenamefont{Yu et~al.}(2020)\citenamefont{Yu, Chen, Liu, Mattioli,
  Sang, Shi, Huang, Shen, Li, Ding et~al.}}]{Yu2020}
\bibinfo{author}{\bibfnamefont{M.}~\bibnamefont{Yu}},
  \bibinfo{author}{\bibfnamefont{C.}~\bibnamefont{Chen}},
  \bibinfo{author}{\bibfnamefont{Q.}~\bibnamefont{Liu}},
  \bibinfo{author}{\bibfnamefont{C.}~\bibnamefont{Mattioli}},
  \bibinfo{author}{\bibfnamefont{H.}~\bibnamefont{Sang}},
  \bibinfo{author}{\bibfnamefont{G.}~\bibnamefont{Shi}},
  \bibinfo{author}{\bibfnamefont{W.}~\bibnamefont{Huang}},
  \bibinfo{author}{\bibfnamefont{K.}~\bibnamefont{Shen}},
  \bibinfo{author}{\bibfnamefont{Z.}~\bibnamefont{Li}},
  \bibinfo{author}{\bibfnamefont{P.}~\bibnamefont{Ding}}, \bibnamefont{et~al.},
  \bibinfo{journal}{Nat Chem} \textbf{\bibinfo{volume}{12}},
  \bibinfo{pages}{1035} (\bibinfo{year}{2020}).

\end{thebibliography}

\end{document}